\documentclass[12pt,a4paper]{article}
\pdfoutput=1
\usepackage{jheppub}
\usepackage{amsthm,amsbsy,amsfonts,mathrsfs,enumerate,float,wrapfig,amsmath,amssymb,cleveref,upgreek}
\usepackage[colorinlistoftodos,prependcaption,textsize=scriptsize]{todonotes}

\newcommand{\nn}{\nonumber}
\newcommand{\Emptyset}{\text{\o}}
\newcommand{\ft}{\mathfrak{t}}
\newcommand{\fq}{\mathfrak{q}}

\newcommand{\calN}{\mathcal{N}}
\newcommand{\calR}{\mathcal{R}}
\newcommand{\calM}{\mathcal{M}}

\newcommand{\calZ}{\mathcal{Z}}

\newcommand{\calO}{\mathcal{O}}

\newcommand{\oo}{\text{\o}}
\newcommand{\uupbeta}{\boldsymbol{\beta}}

\newcommand{\be}{\begin{equation}}
\newcommand{\ee}{\end{equation}}
\newcommand{\ba}{\begin{aligned}}
\newcommand{\ea}{\end{aligned}}
\newcommand{\f}{\frac}

\allowdisplaybreaks
\usepackage{tikz}
\usetikzlibrary{decorations.pathreplacing,decorations.markings}
\usetikzlibrary{arrows, arrows.meta}
\tikzset{
	on each segment/.style={
		decorate,
		decoration={
			show path construction,
			moveto code={},
			lineto code={
				\path [#1]
				(\tikzinputsegmentfirst) -- (\tikzinputsegmentlast);
			},
			curveto code={
				\path [#1] (\tikzinputsegmentfirst)
				.. controls
				(\tikzinputsegmentsupporta) and (\tikzinputsegmentsupportb)
				..
				(\tikzinputsegmentlast);
			},
			closepath code={
				\path [#1]
				(\tikzinputsegmentfirst) -- (\tikzinputsegmentlast);
			},
		},
	},
	mid arrow/.style={postaction={decorate,decoration={
				markings,
				mark=at position .65 with {\arrow[#1]{latex}}
	}}},
}
\title
{$DE$-type little strings from glued brane webs}
\author[a]{Sung-Soo Kim,}
\author[b]{Yuji Sugimoto,}
\author[a]{Xing-Yue Wei,}
\author[c]{Futoshi Yagi}
\affiliation[a]{School of Physics, University of Electronic Science and Technology of China,\\
No. 2006 Xiyuan Ave, West Hi-Tech Zone, Chengdu, Sichuan 611731, China}
\affiliation[b]
{Department of Physics, POSTECH, Pohang 37673, Korea}
\affiliation[c]{School of Mathematics, Southwest Jiaotong University,
West zone, High-tech district, Chengdu, Sichuan 611756, China
}
\emailAdd{sungsoo.kim@uestc.edu.cn}
\emailAdd{yujisugimoto@postech.ac.kr}
\emailAdd{xingyue\_wei@std.uestc.edu.cn}
\emailAdd{futoshi\_yagi@swjtu.edu.cn}

\abstract{
We propose brane web configurations for $D$-type and $E$-type $\mathcal{N}=(1,0)$ little string theories based on a trivalent or quadrivalent gluing of 5-brane web diagrams. Tri-/quadri-valent gluing is a powerful way of computing 5d/6d partition functions for supersymmetric gauge theories based on the topological vertex. We generalize the gluing techniques to little string theories by introducing a new compact direction and compute their supersymmetric partition functions on Omega-deformed $\mathbb{R}^4\times T^2$. As concrete examples, we consider little string theories arising from Type IIB NS5-branes probing $D_4$ or $D_5$ singularity. Their effective gauge theory descriptions as the affine $D_4$ or $D_5$ quiver gauge theory can be realized with quadrivalent or trivalent gluing, respectively. Based on these gluings of 5-brane webs, we compute their refined partition functions and compare them with the known results. We extend the computation of the partition function to little string theory engineered from IIB NS5-branes probing $E_6$ singularity based on a trivalent gluing. We also discuss the generalization to higher rank cases and the symmetries of the partition functions.
}

\begin{document}
\maketitle

\section{Introduction}\label{sec:intro}
The supersymmetric partition function provides a powerful way of understanding BPS spectra of higher dimensional field theories. Various computational tools for computing the supersymmetric partition functions on $\mathbb{R}^4\times S^1$ for 5d theories or $\mathbb{R}^4\times T^2$ for 6d theories have been developed, in particular, for theories of eight supercharges, such as 5d/6d superconformal theories (SCFTs) as well as  little string theories (LSTs). These partition functions are the Witten indices on the so-called $\Omega$-deformed $\mathbb{R}^4$ counting the BPS states which are BPS spectrum protected by supersymmetry and also weighted by electric charges and angular momenta. 

The computational tools include the ADHM construction of the instanton moduli space \cite{Atiyah:1978ri,Nekrasov:2002qd,Nekrasov:2003rj,Marino:2004cn,Nekrasov:2004vw,Fucito:2004gi,Hwang:2014uwa}. This leads to Nekrasov instanton partition functions for 5d $\mathcal{N}=1$ gauge theories in the Omega background. The partition functions count the BPS bound states of instantons on the Coulomb branch of the moduli space. A systematic way of computation has been developed for gauge theories with the classical gauge groups, but there are some limitations for those theories with hypermultiplets of higher dimensional representations or with higher Chern-Simons levels, 
or with exceptional gauge groups.  The ADHM construction is also applicable for 6d SCFTs, computing the elliptic genera of the self-dual strings in 6d SCFTs and LSTs in the tensor branch \cite{Haghighat:2013tka,Haghighat:2013gba,Kim:2014dza, Haghighat:2014vxa,Gadde:2015tra,Kim:2015gha, Kim:2016foj,  Kim:2017xan,Haghighat:2017vch,Kim:2018gak, Kim:2018gjo, Haghighat:2018dwe}.

The blowup method \cite{Nakajima:2003pg,Nakajima:2005fg, Gottsche:2006bm} is a systematic and very powerful tool for computing the partition function where the BPS partition function is obtained by solving the blowup equation which identifies the partition functions on $\Omega$-deformed $\mathbb{C}^2$ and that on the blown-up $\mathbb{P}^1$ in $\hat{\mathbb{C}}^2$ that can be smoothly blown-down to $\mathbb{C}^2$. The blowup method was generalized and extended to any 5d/6d gauge theory, quiver gauge theory, and 6d theory on a circle with/without a twist \cite{Keller:2012da, Kim:2019uqw, Kim:2020hhh, Kim:2021gyj} as well as extended \cite{Gu:2017ccq, Huang:2017mis, Kim:2020hhh} to any local Calabi-Yau 3-fold associated with geometric constructions of \cite{Jefferson:2018irk, Bhardwaj:2019fzv}.

Topological vertex method \cite{Aganagic:2003db,Iqbal:2007ii,Awata:2008ed} is yet another systematic tool for computing the partition function. It uses Type IIB 5-brane web diagrams \cite{Aharony:1997ju, Aharony:1997bh} and the partition functions are expressed as a sum of instanton contributions, that is a sum of Young diagrams associated with K\"ahler parameter for instanton fugacity. If a 5-brane web diagram for the supersymmetric field theory of interest is given, one can perform the topological vertex to obtain the corresponding partition function. Here, 5-brane webs used for topological vertex refer to not only toric web diagrams but also a particular class of non-toric ones which include generalized toric web diagrams and 5-brane webs with orientifolds (O5- or ON-planes) \cite{Kim:2017jqn, Hayashi:2020hhb, Nawata:2021dlk, Kim:2022dbr}.

It is worthy of noting that new 5-brane configurations have been obtained, including webs for 5d $G_2$ gauge theories \cite{Hayashi:2018bkd,Hayashi:2018lyv}, 5d gauge theories with rank-3 antisymmetric hypermultiplets \cite{Hayashi:2019yxj}, local $\mathbb{P}^2$ with an adjoint hyper or similar non-Lagrangian theories with an O7$^+$-plane  \cite{Kim:2020hhh}, and also periodic webs for 6d SCFTs of spiral shapes of E-string theory \cite{Kim:2015jba,Hayashi:2015fsa} and two O5-planes for those in the Higgsing and twisting sequences of 6d $D_N$ gauge theories \cite{Kim:2019dqn,Kim:2021cua}. There are, however, still many theories whose 5-brane webs are unknown, for instance, 5d SU(3)$_8$ theory. For such theories, conventional topological vertex method is not applicable. On the other hand, recently, new 5-brane web configurations for 5d and 6d SCFTs were proposed based on gluings of 5-brane webs in a trivalent or quadrivalent way, which is not realized on a two-dimensional $(p, q)$-plane but is extended to an ambient space such that theories are realized in a way similar to Dynkin diagrams for gauge groups of 5d gauge theories or quiver gauge theories of the shape of an affine Dynkin diagram for 6d conformal matters, and more with Higgsing of these two. We refer to them as the tri-/quadri-valent gluings \cite{Hayashi:2017jze, Hayashi:2021pcj}. These new 5-brane web configurations do not require orientifold planes but can realize much more 5d/6d SCFTs than conventional 5-brane webs on the $(p,q)$-plane can do, which includes 5d exceptional gauge theories or 6d conformal matter of exceptional symmetries. The computation of the partition functions based on these tri-/quadri-valent gluings also agrees with the known result. It is however an extension of such gluings to little string theories not yet developed.

Little string theories are non-gravitational string theories in six dimensions \cite{Seiberg:1997zk,Berkooz:1997cq,Losev:1997hx,Dijkgraaf:1996cv,Dijkgraaf:1996hk}, which arise in the limit of the string coupling $g_s\to 0$ \cite{Seiberg:1997zk,Berkooz:1997cq,Dijkgraaf:1996cv,Dijkgraaf:1996hk,Blum:1997mm,Intriligator:1997dh,Brunner:1997gf,Hanany:1997gh} and have a scale associated with $1/\sqrt{\alpha'}$ or the tensions of little strings. A large class of LSTs has been constructed from F-theory on an elliptically fibered non-compact Calabi-Yau threefolds \cite{Bershadsky:1996nu,Aspinwall:1996vc,Aspinwall:1997ye,Bhardwaj:2015oru}, and the classification of LSTs is discussed in \cite{Bhardwaj:2015oru,Bhardwaj:2019hhd}. LSTs can be regarded as an affine extension of 6d SCFTs. In other words, on tensor branch, one can realize a certain class of LSTs as  6d quiver gauge theories of the shape of affine Dynkin diagrams. For a necklace and a $D$-type affine quiver cases, one can depict them with a conventional $(p,q)$ 5-brane web. A necklace quiver is for $A_N$-type LSTs and the corresponding 5-brane webs have two periodic directions from which T-duality of LSTs can be realized an $SL(2,\mathbb{Z})$ symmetry of 5-brane web with two periodicities \cite{Kim:2015gha}. 5-brane webs for $A$-type LSTs are studied from the perspective of triality and for computing the partition functions \cite{Hohenegger:2015btj, Hohenegger:2016eqy, Hohenegger:2016yuv, Bastian:2017ing, Bastian:2017ary,Bastian:2018dfu}. For $D$-type LSTs, it is also possible to study them with 5-brane webs of 
two ON-branes \cite{Sen:1997kz, Sen:1998ii, Kapustin:1998fa, Hanany:1999sj}, which describes an affine $D$-type Dynkin quiver with a periodic direction with NS5-branes. T-duality is realized again as the S-dual relation between 5-brane webs with two ON-planes and with two O5-planes. $E$-type LSTs still remain as a challenge for conventional $(p,q)$ 5-brane webs.

In this paper, we explore a possible extension of tri-/quadri-valent gluing to 6d little string theories, which makes it possible to describe LSTs engineered from $k$ IIB NS5-branes probing $D_N, E_N$-type singularities as glued 5-brane webs. (For simplicity, we refer to them as rank-$k$ $D_N, E_N$-type LSTs.) Tri-/quadri-valent gluings of 6d ($G$, $G$) conformal matters realize the 6d SCFTs as a quiver theory of the shape of affine Dynkin diagram associated with symmetry group $G$, where each node corresponds to a usual 5-brane web. Notice that the glued branes, therefore, have external 5-branes which are extended to infinity. So, it is not good for describing LSTs. To make the tri-/quadri-valent gluing to describe LSTs, we propose the following prescription. We further glue the external 5-branes such that two external 5-branes that are a part of a single strip diagram are glued together as if they are compactified on a circle. All the external 5-branes are glued in this bivalent way but are required to be compactified on the {\it same} circle. In this way, we can introduce another periodicity for the resultant glued webs. In other words, the periodicity, which is necessary for a 5d affine quiver to be LSTs, comes from compactifying all the external 5-branes from a tri-/quadri-valent glued configuration. The resulting 5-brane configurations are hence an affine $D$-, and $E$-type quiver equipped with a periodic direction, and therefore they are 5-brane configurations that can describe a certain class of LSTs associated with affine $D,E$-extensions of 6d SCFTs. We then apply the refined topological vertex method to our proposed webs to compute the supersymmetric partition functions for LSTs. As illustrative examples, we explicitly compute the quadrivalent gluing for $D_4$-type little string theory and the trivalent gluing for $D_5$-type little string theory. We also compute the partition function for $E_6$-type LST based on the trivalent gluing.  We then discuss the symmetry of the theories on tensor branch in connection with Weyl reflection symmetry associated with external NS-charged 5-branes, and based on the Weyl symmetry we expand the partition functions of $D_4$-, $D_5$- and $E_6$-type LSTs. Our results for $D_4$-, $D_5$-type LSTs reproduce the known elliptic genus results \cite{Kim:2017xan} for these two theories obtained based on the ADHM construction where T-duality was checked, implying our proposal is also of the T-duality in it. It is worth noting that our computation based on tri-/quadri-valent glugings is performed in a different expansion compared to \cite{Kim:2017xan}, which manifests global symmetry of the theories.

The organization of the paper is as follows. In section \ref{sec:gluings}, we provide a short review of the topological vertex method and discuss our setup for conventions and notations. 
In section \ref{sec:LSTandselfGluing}, we discuss our proposal for tri-/quadri-valent gluings for rank-$1$ $D$-type and $E$-type LSTs. We then propose a general expression for the partition function of a $D_N$- or $E_N$-type LST with general rank which can be constructed from tri-/quadri- valent gluings in section \ref{sec:higherRanks}. Detailed computations and symmetries are discussed in section \ref{sec:Symmetry}.  We summarize our results and possible further direction to pursue in section \ref{sec:conclusion}. In appendices, we discuss some of the computational details, and we also include known results of $A$-type little string theories from the literature.

\bigskip

\section{Topological vertex and gluings 5-brane webs}\label{sec:gluings}
We start with a short review on topological vertex formalism \cite{Aganagic:2003db,Iqbal:2007ii} and then discuss tri-/quadri-valent gluings \cite{Hayashi:2017jze, Hayashi:2021pcj}. We then discuss an interesting relation between trivalent gluing and 5-brane webs with ON-planes. 

\subsection{Refined topological vertex}
The topological vertex is our main tool for computing the partition function of 5d $\mathcal{N}=1$ supersymmetric gauge theories on a circle or 6d $\mathcal{N}=(1,0)$ theories on a torus based on Type IIB 5-brane webs. The partition function captures the BPS spectrum of these theories on $\mathbb{R}^4\times S^1$ or $\mathbb{R}^4\times T^2$, which is the Witten index on the $\Omega$-deformed $\mathbb{R}^4$, counting the number of the BPS states weighted by their electric charges and angular momenta. Here, we denote the $\Omega$-background parameters $\epsilon_1, \epsilon_2$ and define their fugacities, $\fq= e^{-{\boldsymbol{\beta}}\epsilon_1}$ and $\ft= e^{\uupbeta\epsilon_2}$, where  ${\boldsymbol{\beta}}$  is the circumference of $S^1$.

\begin{figure}[t]
	\centering
\includegraphics[scale=0.7]{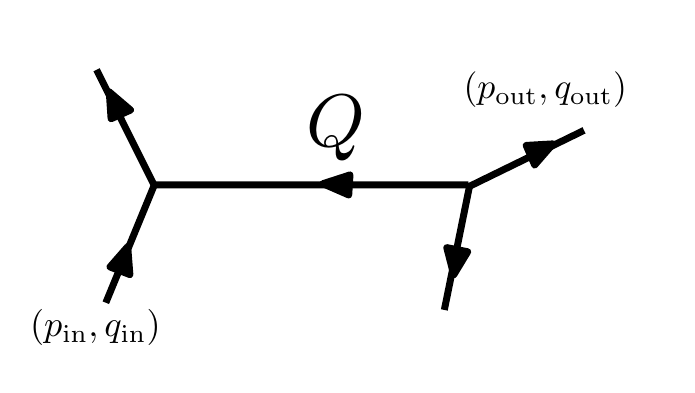}
	\caption{A convention of the charge of external legs. Here we define a K\"ahler parameter $Q$ along an internal line.}
\label{fig:gluing}
\end{figure}

To perform the computation of the partition function using the topological vertex, one needs to identify the edge factors and the vertex factors for a given 5-brane web.
First, an edge factor is given to each edge in the 5-brane web, and likewise a vertex factor is associated with each vertex in the 5-brane web. To give the edge factor, one needs to distinguish the framing factors along either with the preferred direction $f_{\mu}(\ft,\fq)$ or with the non-preferred directions $\tilde{f}_{\mu}(\ft,\fq)$. The preferred direction here is the direction associated with an expansion parameter such as the instanton fugacity, while the non-preferred directions are other directions that are not a preferred direction. With a Young diagram denoted by a Greek letter\footnote{In this paper we use Greek letters to define the Young diagrams.} $\mu$, these two framing factors take the following forms,
\begin{align}\label{eq:framing factor}
f_\mu(\ft,\fq)&=(-1)^{|\mu|}\ft^{\frac{||\mu^t||^2}{2}}\fq^{-\frac{||\mu||^2}{2}} = f_{\mu^t}(\fq,\ft)^{-1},
\cr
\tilde{f}_{\mu}(\ft,\fq)&=(-1)^{|\mu|}\ft^{\frac{||\mu^t||^2+|\mu|}{2}}\fq^{-\frac{||\mu||^2+|\mu|}{2}}= \tilde f_{\mu^t}(\fq,\ft)^{-1}\ ,	
\end{align}
where $\mu^t$ is the transpose of the Young diagram $\mu$,  $|\mu| =\sum^{\ell(\mu)}_{i=1} \mu_i$, and $||\mu||^2 =\sum^{\ell(\mu)}_{i=1} \mu_i^2$. The edge factor along the edge associated with the Young diagram $\mu$ is then defined as
\begin{equation}
\begin{aligned}\label{eq:edge factor}
    &(-Q)^{|\mu|} f^{\mathfrak{n}}_\mu (\ft,\fq)\quad
    \text{for a preferred direction},
\\
    &(-Q)^{|\mu|} \tilde f^{\mathfrak{n}}_\mu (\ft,\fq)\quad
    \text{for a non-preferred direction},
\end{aligned}
\end{equation}
where $Q$ is a K\"ahler parameter associated with the edge, the power $\mathfrak{n}$ is given as $\mathfrak{n}= p_\text{in}q_\text{out}-p_\text{out}q_\text{in}$ for a pair of incoming and outgoing charges of 5-branes, $(p_\text{in}, q_\text{in})$, and $(p_\text{out}, q_\text{out})$, which are connected to the edge of interest (see also figure \ref{fig:gluing}).
The vertex factor is defined as 
\begin{align}
C_{\lambda\mu\nu}(\ft,\!\fq)=\fq^{\frac{||\mu||^2+||\nu||^2}{2}}\ft^{-\frac{||\mu^t||^2}{2}}\tilde{Z}_\nu(\ft,\!\fq)\!\sum_{\eta}\!\Big(\frac{\fq}{\ft}\Big)^{\!\!\frac{|\eta|+|\lambda|-|\mu|}{2}}\!\!\! s_{\lambda^t/\eta}(\ft^{-\rho}\fq^{-\nu})s_{\mu/\eta}(\fq^{-\rho}\ft^{-\nu^t}),
\label{eq:vertex factor}
\end{align}
where the Young diagrams are chosen such that the associated arrows are all out-going.
The transpose of a Young diagram will be assigned when the arrow is opposite, and the last Young diagram $\nu$ in \eqref{eq:vertex factor} is the Young diagram along the preferred direction. The factor  $\tilde{Z}_{\nu}(\ft,\fq)$ is defined as 
\begin{align}
\tilde{Z}_{\nu}(\ft,\fq)= \prod_{i=1}^{\ell(\nu)}\prod_{j=1}^{\nu_i}\big(1-\ft^{\nu^t_j-i+1}\fq^{\nu_i-j}\big)^{-1},
\end{align}
and $s_{\lambda/\eta}(\bf{x})$ are the skew-Schur functions of a vector $\bf{x}$ with infinite components. For example, 
$\ft^{-\rho}\fq^{-\nu} = (\ft^\frac{1}{2} \fq^{-\nu_1},\ft^{\frac32}\fq^{-\nu_2}, \cdots)$. 

From them, the partition function can be obtained by summing over all the edge and the vertex factors of the 5-brane web of interest,
\begin{align}\label{eq:Z as Edge and Vertex}
Z = \sum_{\lambda,\mu,\nu,\cdots} \Big(\prod \text{Edge factor}\Big) \Big( \prod \text{Vertex factor}\Big)\ .
\end{align}
Note that in usual toric diagrams, we can evaluate the sum of Young diagrams along the non-preferred direction and get closed forms which may be interpreted as contributions coming from vector multiplets and hypermultiplets. Eventually, the partition function can be expressed as a sum of the Young diagrams along the preferred direction.

It is convenient to introduce the following functions:
\begin{align}
\calR_{\lambda\mu}(Q;\ft,\fq)&\equiv \! \prod_{i,j=1}^{\infty}\!\!\Big(\!1\!-\!Q \ft^{i-\frac{1}{2}-\lambda_j}\fq^{j-\frac{1}{2}-\mu_i}\!\Big)\!=\!\calM\big(Q\sqrt{\tfrac{\ft}{\fq}};\ft,\fq\big)^{-1}\calN_{\lambda^t\mu}\big(Q\sqrt{\tfrac{\ft}{\fq}};\ft,\fq\big), \label{eq:RMN}\\
\calM(Q;\ft,\fq)&\equiv \!\prod_{i,j=1}^{\infty}\big(1-Q \ft^{i-1}\fq^{j}\big)^{-1},\\
\calN_{\lambda\mu}(Q;\ft,\fq)&\equiv \! \prod_{i,j=1}^{\infty}\frac{1-Q\ft^{i-1-\lambda_j^t}\fq^{j-\mu_i}}{1-Q\ft^{i-1}\fq^j}\label{eq:curlyN}\\
&=\!\prod_{(i,j)\in\lambda}(1-Q\ft^{\mu_j^t-i}\fq^{\lambda_i-j+1})\prod_{(i,j)\in\mu}(1-Q\ft^{-\lambda_j^t+i-1}\fq^{-\mu_i+j})\ .\nonumber
\end{align}
With these functions, one can simplify or evaluate the Young diagram sums by applying the Cauchy identities, 
\begin{align}
&\sum_\lambda Q^{|\lambda|} s_{\lambda/\eta_1}(\ft^{-\rho}\fq^{-\nu_1}) s_{\lambda/\eta_2}(\fq^{-\rho}\ft^{-\nu_2}) \nonumber \\
&\quad = \mathcal{R}_{\nu_2 \nu_1}(Q;\ft,\fq)^{-1} \sum_\lambda Q^{|\eta_1|+|\eta_2|-|\lambda|} s_{\eta_2/\lambda}(\ft^{-\rho}\fq^{-\nu_1}) s_{\eta_1/\lambda}(\fq^{-\rho}\ft^{-\nu_2}) \label{eq:cauchy2}\ , \\
&\sum_\lambda Q^{|\lambda|} s_{\lambda/\eta_1^t}(\ft^{-\rho}\fq^{-\nu_1}) s_{\lambda^t/\eta_2}(\fq^{-\rho}\ft^{-\nu_2}) \nonumber \\
&\quad = \mathcal{R}_{\nu_2 \nu_1}(-Q;\ft, \fq) \sum_\lambda Q^{|\eta_1|+|\eta_2|-|\lambda|} s_{\eta_2^t/\lambda}(\ft^{-\rho}\fq^{-\nu_1}) s_{\eta_1/\lambda^t}(\fq^{-\rho}\ft^{-\nu_2}) \ \label{eq:cauchy2'}. 
\end{align}
It is also useful to implement an extended version of the Cauchy identities:
\begin{align}
&\sum_{\lambda}Q^{|\lambda|}s_{\lambda/\mu_1}(Q_1\ft^{-\rho}\fq^{-\nu_1})s_{\lambda^t/\mu_2}(Q_2\fq^{-\rho}\ft^{-\nu_2})\nonumber\\
&~=\calR_{\nu_2\nu_1}(-QQ_1Q_2)\sum_{\lambda}Q^{|\lambda|}s_{\mu_2^t/\lambda}(QQ_1\ft^{-\rho}\fq^{-\nu_1})s_{\mu_1^t/\lambda^t}(QQ_2\fq^{-\rho}\ft^{-\nu_2})\ , \label{eq:Cauchy-new1}\\
&\sum_{\lambda}Q^{|\lambda|}s_{\lambda/\mu_1}(Q_1\ft^{-\rho}\fq^{-\nu_1})s_{\lambda/\mu_2}(Q_2\fq^{-\rho}\ft^{-\nu_2})\nonumber\\
&~=\calR_{\nu_2\nu_1}(QQ_1Q_2)^{-1}\sum_{\lambda}Q^{|\lambda|}s_{\mu_2/\lambda}(QQ_1\ft^{-\rho}\fq^{-\nu_1})s_{\mu_1/\lambda}(QQ_2\fq^{-\rho}\ft^{-\nu_2})\ .\label{eq:Cauchy-new2}
\end{align}

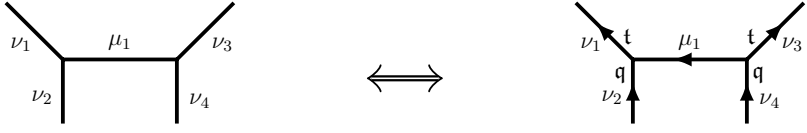
\begin{figure}[t]
	\centering
\begin{tikzpicture}[thick,scale=0.5, every node/.style={scale=0.5}]
\path [line width=0.5mm,draw=black,postaction={on each segment={mid arrow=black}}]

		(3+15,3) to (0+15,3)
		(3+15,1.3) to (3+15,3)
		(0+15,3) to (-1.5+15,4.5)
		(0+15,1.3) to (0+15,3)
		(3+15,3) to (4.5+15,4.5);
		
		\node[scale=1.5] at (1.5+15,3.45){$\mu_1$};
		\node[scale=1.5] at (-0.15+15,3.6){$\ft$};
		\node[scale=1.5] at (-0.33+15,2.6){$\fq$};
		\node[scale=1.5] at (3.1+15,3.6){$\ft$};
		\node[scale=1.5] at (3.3+15,2.6){$\fq$};
	
		\node[scale=1.5] at (-1.1+15,3.4){$\nu_1$};
		\node[scale=1.5] at (-0.55+15,2){$\nu_2$};
		\node[scale=1.5] at (3.6+15,1.9){$\nu_4$};
		\node[scale=1.5] at (4.2+15,3.4){$\nu_3$};
		
		\draw[line width=0.5mm,black]
		(3,3) to (0,3)
		(3,1.3) to (3,3)
		(0,3) to (-1.5,4.5)
		(0,1.3) to (0,3)
		(3,3) to (4.5,4.5);
		\node[scale=1.5] at (1.5,3.45){$\mu_1$};
		\node[scale=1.5] at (-1.1,3.4){$\nu_1$};
		\node[scale=1.5] at (-0.55,2){$\nu_2$};
		\node[scale=1.5] at (3.6,1.9){$\nu_4$};
		\node[scale=1.5] at (4.2,3.4){$\nu_3$};
		\draw[line width=0.3mm,>=Classical TikZ Rightarrow,double,<->] (8,2.5) to (10,2.5);
\end{tikzpicture}
\caption{Convention used in this paper. }
	\label{fig:arrowconvention}
\end{figure}

\paragraph{Convention.}
Following the convention introduced in \cite{Kim:2022dbr}, we choose the direction of the arrows on the edges to be chosen to be either pointing upward or to the left as shown in figure \ref{fig:arrowconvention}. The preferred direction is always taken to be horizontal. The $\ft,\fq$ factors are assigned such that the one above the preferred direction is $\ft$ and the one below is $\fq$. Throughout the paper, we take this convention, so that we will omit the arrows and the Omega deformation factors explicitly. We only denote the Young diagram associated with edges, as depicted in figure \ref{fig:arrowconvention}. %

\subsection{Tri-/quadri-valent gluing of 5-brane webs}
A trivalent or quadrivalent gluing is a 5-brane web-like construction realizing a geometric description of a given theory \cite{Hayashi:2017jze}. Its building block is 5-brane webs of parallel external edges, such as SU(2)$_{2\pi}$ or a 5-brane web for pure SU($N$)$_\kappa$ theory at the Chern-Simons level $\kappa=N$, denoted by SU($N$)$_N$, or those of fundamental hypermultiplets. Based on the topological vertex formalism, one can compute the partition function by gluing 5-brane webs. Here, the gluing means to take the Young diagram sums along the  parallel external edges, which makes a gauging of the external edges and hence increase the number of Coulomb branch. We will explain the tri-/quadri-valent gluing diagrammatically, and the detailed computation will be explained in Section \ref{sec:LSTandselfGluing} as applications of LSTs.

The simplest gluing would be  bivalent gluing which is to glue a pair of 5-brane webs with parallel external edges. It is, in fact, a commonly used technique in the topological vertex formalism to compute partition functions from 5-brane webs. For instance, consider a 5-brane web for the 5d ${\rm SU}(2)_\pi\times {\rm SU}(2)\times{\rm SU}(2)_\pi$ theory. The partition function can be computed from two 5-brane webs for SU(2) theory with two flavors, by gluing two sets of parallel edges which are two flavors of each SU(2) theory. We call such gluing involving two sets of parallel edges the bivalent gluing. 

As a generalization, three (and four) sets of parallel edges can be glued together. This gluing is called trivalent (and quadrivalent) gluing. Unlike two sets of parallel edges, gluing of more than three sets of parallel edges may not be depicted on a 5-brane plane. Many 5d/6d examples of the tri-/quadri-valent gluing technique have been discussed in~\cite{Hayashi:2017jze,Hayashi:2021pcj}, supporting that the gluing technique is another powerful way of computing the partition function especially for those theories which do not have a 5-brane realization yet, which includes 5d $F_4, E_{6,7,8}$ exceptional gauge theories and 6d $(E_N, E_N)$ conformal matters. We note that multivalent gluings of more than four sets parallel edges do not lead to a physically sensible theory.

\begin{figure}[t]
	\centering
	\includegraphics[scale=0.6]{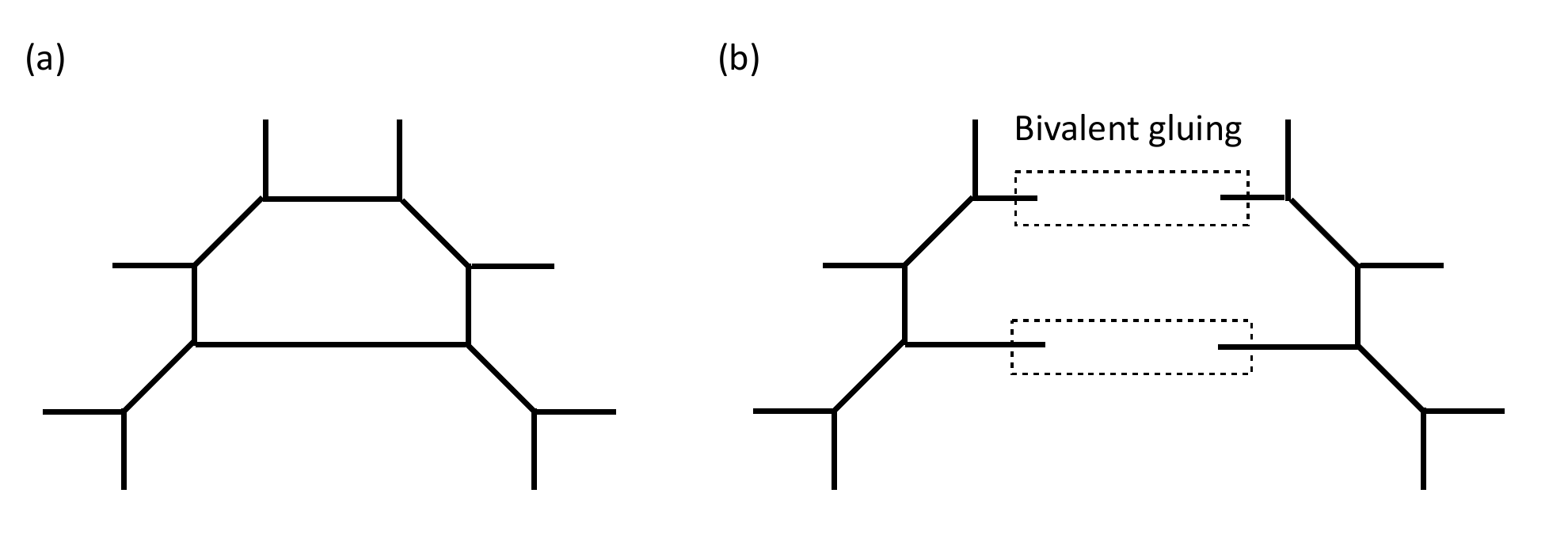}
	\caption{A bivalent gluing.}
	\label{fig:biguling}
\end{figure}
\begin{figure}[t]
	\centering	\includegraphics[scale=0.6]{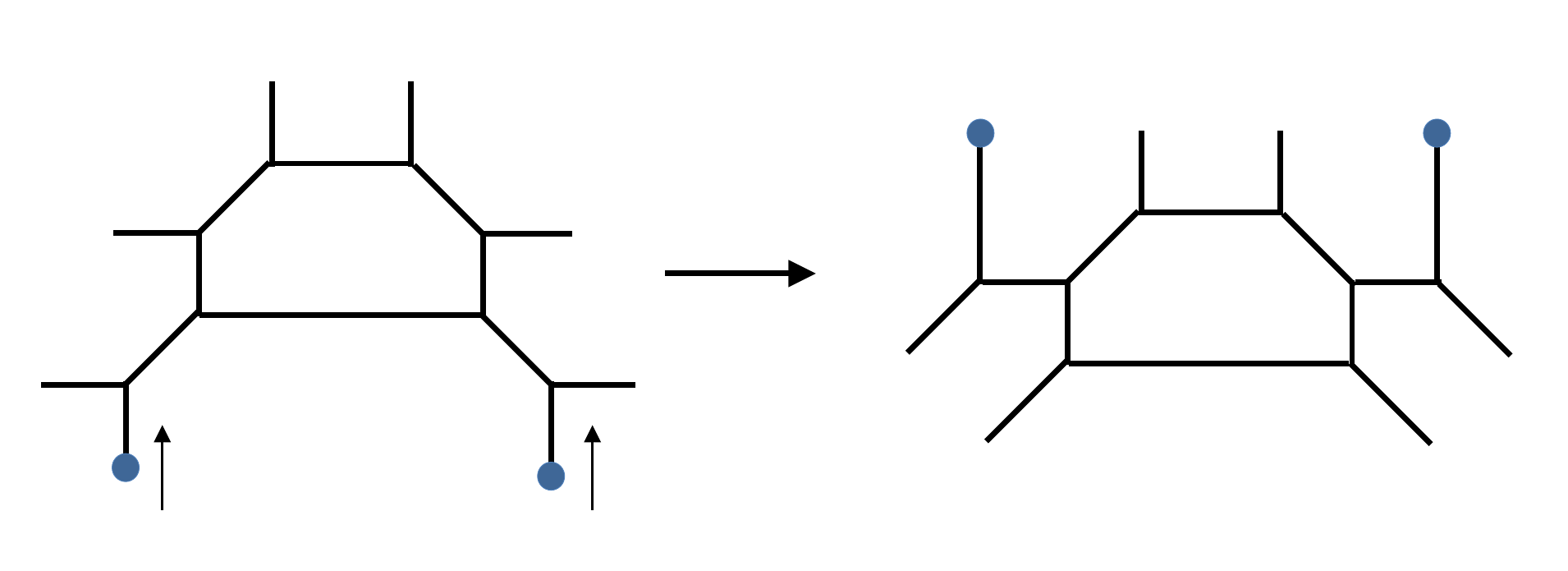}
 \caption{Equivalence between SU(2)+4$\mathbf{F}$ (left) and ${\rm SU}(1)-{\rm SU}(2)-{\rm SU}(1)$ (right) by vertical Hanany-Witten moves, where the blue dots are (0,1) 7-branes.}
	\label{fig:su2+4HW}
\end{figure}
\begin{figure}[t]
	\centering
	\includegraphics[scale=0.6]{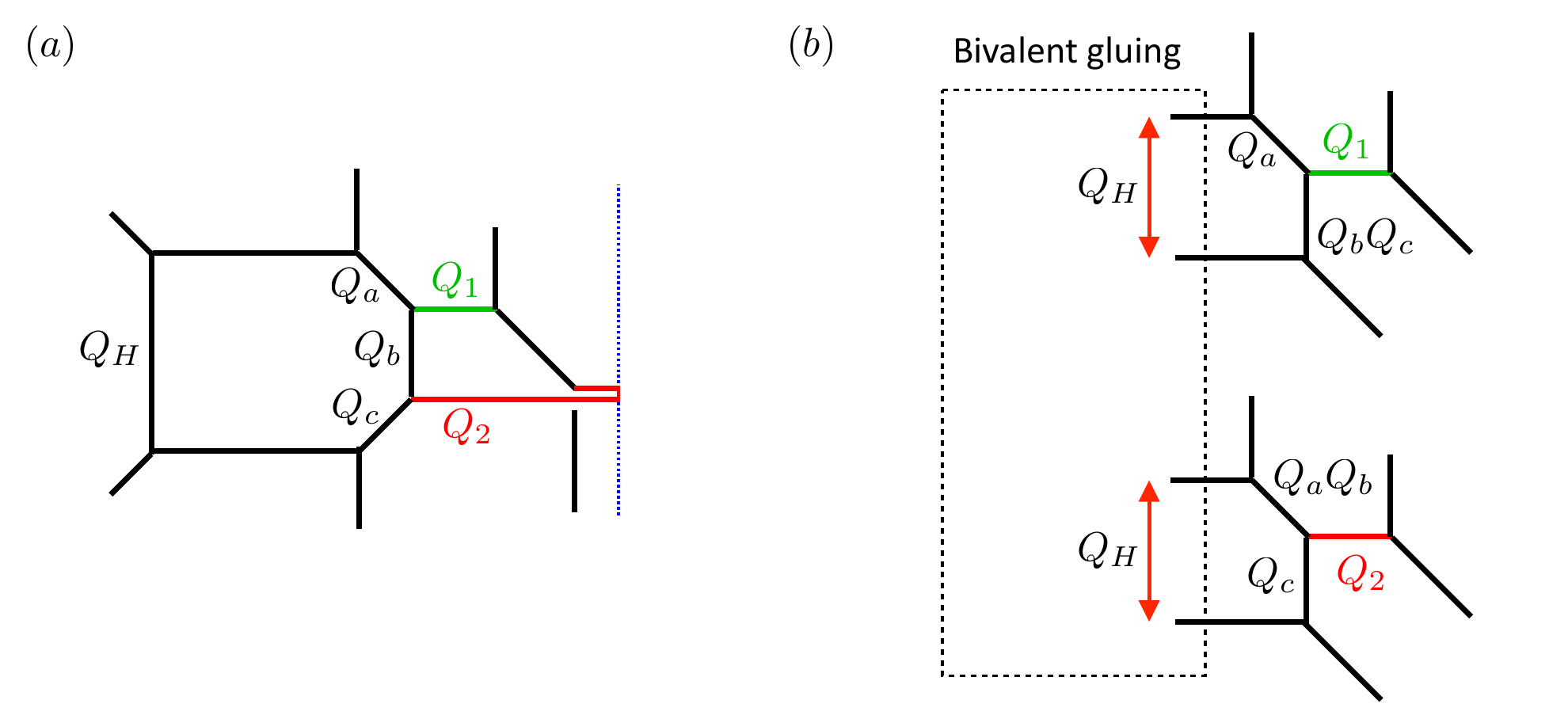}
	\caption{(a) An SU(1)$-$SU(2)$-$SU(1) 5-brane web with an ON-plane which is located on the right.  (b) A bivalent gluing configuration for SU(1)$-$SU(2)$-$SU(1) quiver theory.}
	\label{fig:ON-su1su2su1}
\end{figure}

We briefly demonstrate these gluing techniques with 5d SU($2$) theory with an even number of hypermultiplets in the fundamental representation (flavors).
First, consider a 5-brane web with 4 flavors given in figure \ref{fig:biguling}, in which we can vertically cut the 5-brane web by half, leading to two pieces of 5-brane web representing a free hypermultiplet. One can obviously glue together these free hypermultiplets to reconstruct a 5-brane web for 5d SU(2) gauge theory with 4 flavors, which is, in fact, what we usually do. We here note that one also represents two flavors as an ``SU(1)'' theory, as shown in figure \ref{fig:su2+4HW}. 
This means that we have a quiver gauge theory SU(1)$\,-\,$SU(2)$\,-\,$SU(1) which is a $D$-type quiver. 
It is also possible to construct the same theory on a 5-brane configuration with an ON-plane as shown in figure \ref{fig:ON-su1su2su1}. 
On a 5-brane web with an ON-plane, we can also apply a gluing of two SU(1) theories with two flavors by gluing the two flavors of one SU(1) theory with the two flavors of the other SU(1) theory, as shown in figure \ref{fig:ON-su1su2su1}. One needs to carefully assign the K\"ahler parameter map as given in figure \ref{fig:ON-su1su2su1} and also mode out the U(1) factor for each 5-brane piece.

\begin{figure}[t]
	\centering
\includegraphics[scale=0.5]{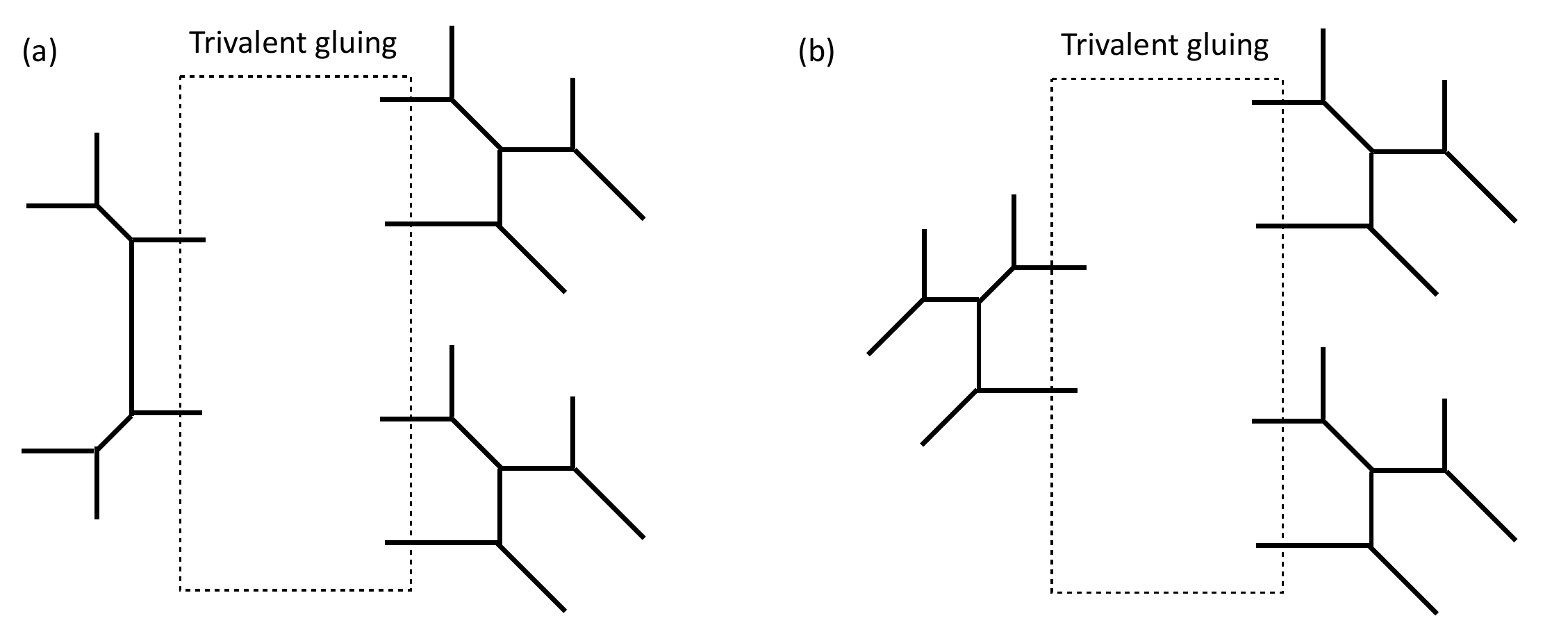}
\caption{SU(2)+6$\mathbf{F}$ as a trivalent gluing configuration discussed in \eqref{eq:quiver-su2+6F}.} 	\label{fig:tri-quiver-su2+6F}
\end{figure}

Now, consider a 5-brane web for SU(2) gauge theory with 6 flavors (SU(2)+6$\mathbf{F}$) by adding two flavors to the SU(1)$-$SU(2)$-$SU(1) quiver theory, the resulting web diagram is given in figure \ref{fig:tri-quiver-su2+6F}(a). On the other hand, since we know that we can add two flavors by putting an SU(1) theory, we find another realization of the web diagram of the SU(2)+6$\mathbf{F}$ theory given in figure \ref{fig:tri-quiver-su2+6F}(b). In any case, we have a $D$-type quiver theory and hence SU(2)+6$\mathbf{F}$ can be represented as follows:
\begin{align}
	\begin{tikzpicture}
		\draw[thick](-1.4,0)--(-0.6,0);
		\draw[thick](0.6,0.2)--(1.4,.8);
		\draw[thick](0.6,-0.2)--(1.4,-.8); 
		\node at (0,0){SU$(2)$};
		\node at (-1.8,0){[2]};
		\node at (2,1){SU$(1)$};
		\node at (2,-1){SU$(1)$};
		\node at (3,-1.15){,};
	\end{tikzpicture}
\qquad
	\begin{tikzpicture}
		\draw[thick](-1.4,0)--(-0.6,0);
		\draw[thick](0.6,0.2)--(1.4,.8);
		\draw[thick](0.6,-0.2)--(1.4,-.8); 
		\node at (0,0){SU$(2)$};
		\node at (-2,0){SU$(1)$};
		\node at (2,1){SU$(1)$};
		\node at (2,-1){SU$(1)$};
  		\node at (3,-1.15){.};
	\end{tikzpicture}
	\label{eq:quiver-su2+6F}
\end{align}
For a trivalent quiver with three SU(1)s, we can do a trivalent gluing of three SU(1)+2$\mathbf{F}$ where parallel external 5-branes associated with two flavors are glued. In other words, three 2$\mathbf{F}$ are gauged to become an SU(2) node in the middle of the trivalent quiver \eqref{eq:quiver-su2+6F}. One can show that the partition function based on this trivalent gluing correctly reproduces that of SU(2)$+6\mathbf{F}$.

\begin{figure}[t]
	\centering
	\includegraphics[scale=0.5]{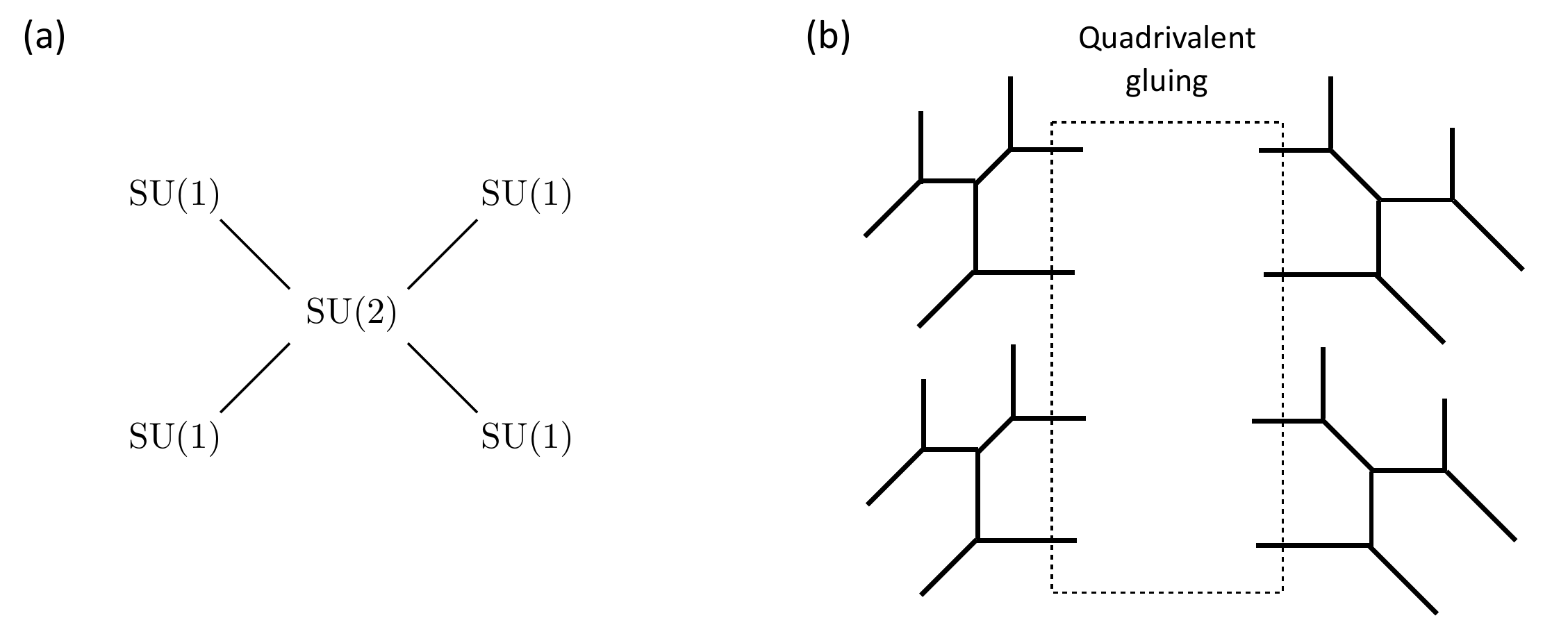}
	\caption{A quadrivalent gluing for SU(2)$+8\mathbf{F}$.} 	\label{fig:quadri-su2+8F}
\end{figure}
We can even consider a 5-brane web for SU(2) gauge theory with 8 flavors by introducing one more SU(1)$+2\mathbf{F}$,
which makes a quadrivalent quiver, as shown in figure \ref{fig:quadri-su2+8F}.   
As discussed, 
SU(1)-SU(2)-SU(1) of $D$-type quiver can be depicted as a 5-brane web with an ON-plane in figure \ref{fig:ON-su1su2su1}(a).
The quadrivalent quiver for SU(2)$+8\mathbf{F}$ having four SU(1)'s can also be realized in a similar way, but in the quadrivalent case we need to put two ON-planes. It is worth noting that topological vertex computation based on a 5-brane web with ON-plane(s) was recently developed in \cite{Kim:2022dbr}. We do not provide the computation detail, but we report that the partition function computed by using the refined topological vertex method based on this quadrivalent quiver for SU(2)$+8\mathbf{F}$ perfectly agrees with that based on a 5-brane web with two ON-planes obtained in \cite{Kim:2022dbr}.

We remark that the 5-brane web with two ON-planes has a periodic direction so that this corresponds to a 6d theory on a circle. In this case, it is the 6d E-string theory on a circle which is represented by an affine $D_4$ quiver theory in 5d description and hence  can be realized as a quadrivalent gluing.
In other words, a quadrivalent gluing gives rise to affine $D_4$ quiver in the 5d description, representing the 6d theory on a circle. 

We also note that the equivalence between tri-/quadri-valent gluing and 5-brane web with ON-plane(s) can be readily generalized to (affine) $D$-type quiver gauge theories with more than one Coulomb branch parameters. For instance, 6d $(D_N, D_N)$ conformal matter can be realized on a 5-brane with two ON-planes as an affine $D_{N\ge 5}$ quiver or as two trivalent gluings of the shape of affine $D_{N\ge 5}$ quiver. One can compute the corresponding partition function based on either a 5-brane web with two ON-planes or trivalent gluings. The parameter map between these two cases can be easily read off as a simple generalization of the map given in figure \ref{fig:ON-su1su2su1}.

\bigskip

\section{Little string theory from bivalent self-gluings}
\label{sec:LSTandselfGluing}
In this section, we propose a novel 5-brane configuration for a certain class of $\mathcal{N}=(1,0)$ LSTs. In conventional Type IIB 5-brane construction, some LSTs are realized on 5-brane webs that have two periodic structures where parallel external edges are identified or two same kinds of orientifolds are presented \cite{Brunner:1997gf, Hanany:1997gh}. As a result, 5-brane configurations describing LSTs have no external edges instead they make compact directions vertically and horizontally. T-duality of LSTs can be understood as an S-duality of such a 5-brane web. If one decompactifies one of these compact directions, one gets a 6d SCFT. Conversely, for a given 5-brane web for a 6d SCFT, one can identify parallel external edges to make a 5-brane configuration for 6d LSTs.

We extend this logic to tri-/quadri-valently glued 5-brane webs describing 6d SCFTs. A trivalent or quadrivalent gluing making a quiver of the shape of an affine Dynkin diagram (affine quiver) gives rise to a 5d realization of a 6d theory with a global symmetry associated with the Dynkin diagram of the affine quiver. For such 5-brane configuration to describe certain LSTs, we further generalize this construction to include bivalent ``self-gluings,'' such that all the external edges in the same quiver node are glued together. This makes an additional compact direction and leaves no external edges. In this section, based on the bivalent self-gluings, we discuss how to introduce these self-gluings to 5-brane configurations and compute the partition function for 6d $\mathcal{N}=(1,0)$ little string theories engineered from an NS5-brane probing $D$-type and $E$-type singularities, which we refer to as the $D$-type and $E$-type LSTs. For $D$-type LSTs, we also discuss equivalence between Type IIB 5-brane webs for LSTs and tri-/quadri-valent gluings for LSTs by presenting the K\"ahler parameter maps between these two brane webs.

\begin{figure}[t]
	\centering
	\includegraphics[scale=0.5]{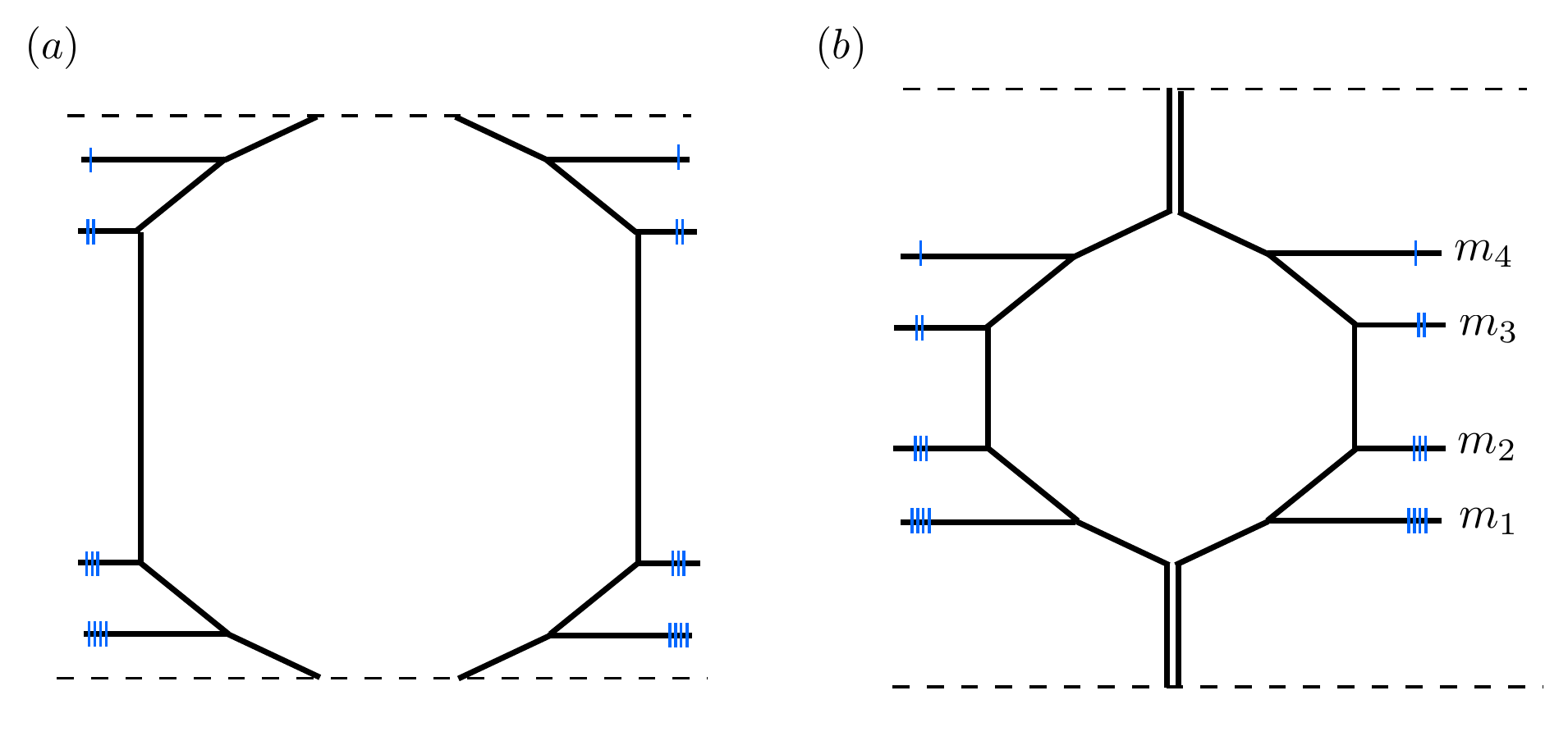}
	\caption{
 A Type IIB 5-brane configuration for the $D_4$-type LST which is of the shape of a circular quiver horizontally and has two O5-planes (dashed lines) vertically.  (a) The resulting configuration gives rise to a 6d SO(8)-Sp(0) circular quiver. (b) A deformed brane web by a generalized flop near two O5-planes.
}
\label{fig:D4_O5}
\end{figure}

\subsection{LST on \texorpdfstring{$D_4$}{D4} singularity}\label{sec:D4}

\begin{figure}[t]
	\centering
	\includegraphics[scale=0.45]{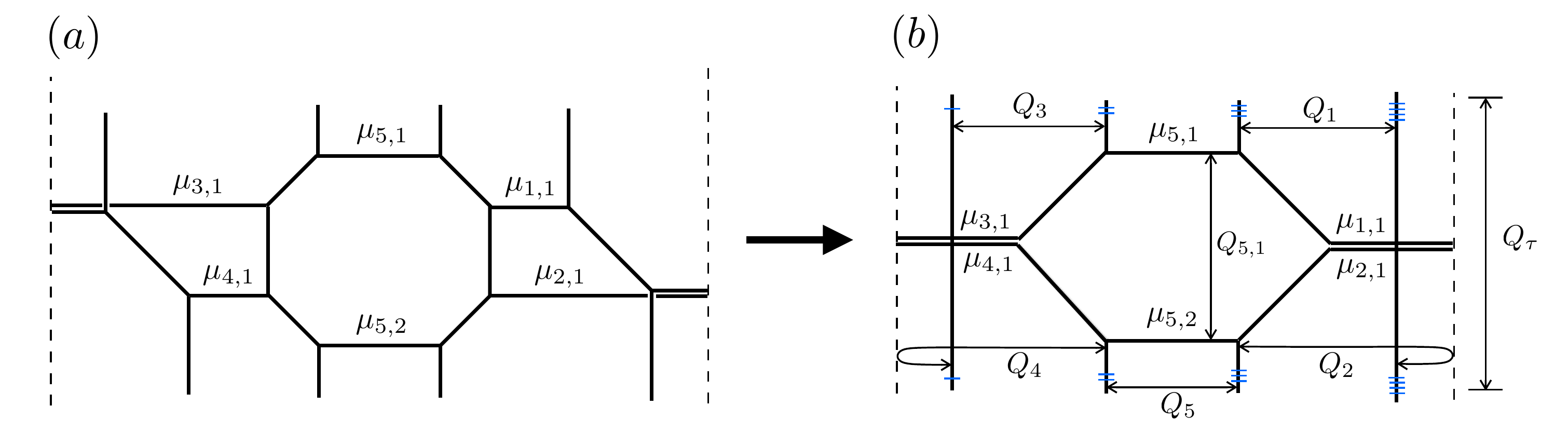}
	\caption{(a) A 5-brane web with two ON-planes giving rise to an affine $D_4$ quiver associated with ($D_4$, $D_4$) conformal matter or E-string theory. (b) A 5-brane web for 6d $D_4$-type LST which can be obtained from tuning of NS5-branes and identifying them so that, as a result, they provide an additional periodic (vertical) direction.}
	\label{fig:D4_ON}
\end{figure}

We start with the $D_4$-type little string theory which is the simplest $D$-type little string theory. Recall first that the $D_4$-type LST can be realized by a Type IIB 5-brane web diagram as depicted in figure \ref{fig:D4_O5}(a), which is an SO(8)-``Sp(0)" circular quiver due to the identification of the left four D5-branes with the right four D5-branes making the horizontal periodic direction, and two dashed lines in the web diagram represent two O5-planes making the vertical periodic direction. We note that the 5-brane configuration in figure \ref{fig:D4_O5}(a) can be deformed into the one in figure \ref{fig:D4_O5}(b), where we perform a ``generalized flop'' \cite{Hayashi:2017btw} near two O5-planes in transition from figure \ref{fig:D4_O5}(a) to figure \ref{fig:D4_O5}(b). 
We can further deform the 5-brane web in figure \ref{fig:D4_O5}(b) by shifting the locations of SO(8) color branes labeled with $m_1$ further down while the one with $m_4$ further up. We do this deformation to take an S-duality which gives rise to two ON-planes, which is known as a Type IIB 5-brane web with ON-planes describing an (affine) $D$-type quiver as shown in figure \ref{fig:D4_ON}(b) where two ON-planes are depicted by two vertical dashed lines. 
Notice that 
the web diagram in figure \ref{fig:D4_ON}(b) can also be obtained from that in figure \ref{fig:D4_ON}(a) which is a 5d affine $D_4$ quiver gauge theory by the following procedure: First, we shrink the distances between pairs of the edges associated with $\mu_{1,1}$ and $\mu_{2,1}$ that are near the right ON-plane and the edges associated with $\mu_{3,1}$ and $\mu_{4,1}$ that are near the left ON-plane in figure \ref{fig:D4_ON}(a). We then tune the horizontal coordinates of the two ends of each of the middle two vertical strips to be the same.
After that, 
we identify vertical branes by gluing the ends of each vertical strip to have an vertical periodic direction.
Therefore, 5-brane configuration in figure \ref{fig:D4_ON}(b) describes 6d affine $D_4$ quiver,
\begin{align*}
	\begin{tikzpicture}
		\draw[thick](-1.4,.8)--(-0.6,0.2);
		\draw[thick](-1.4,-.8)--(-0.6,-0.2);
		\draw[thick](0.6,0.2)--(1.4,.8);
		\draw[thick](0.6,-0.2)--(1.4,-.8); 
		\node at (0,0){SU$(2)$};
		\node at (-2,1){SU$(1)$};
		\node at (-2,-1){SU$(1)$};
		\node at (2,1){SU$(1)$};
		\node at (2,-1){SU$(1)$};
		\node at (3,-1.15){.};
	\end{tikzpicture}
\end{align*}
This affine structure gives rise to a periodic direction of the $D_4$-type LST and the other periodic direction is realized by the vertical compactification as shown in figure \ref{fig:D4_ON}(b).

We remark here that in figure \ref{fig:D4_ON}(a), the color branes associated with $\mu_{1,1}$ and $\mu_{2,1}$ connect to the rightmost vertical strip which leads to a bifundamental contribution that mixes $\mu_{1,1}$ and $\mu_{2,1}$ to the partition function, while the color branes $\mu_{1,1}$ and $\mu_{2,1}$ also connect to the second right vertical strip which leads to a vector contribution that also mixes $\mu_{1,1}$ and $\mu_{2,1}$ to the partition function, and these two contributions, in fact, cancel each other so that there are no factors that mix $\mu_{1,1}$ and $\mu_{2,1}$ \cite{Kim:2022dbr,Bourgine:2017rik}, as expected from a $D$-type quiver theory. 
The same analysis can be applied to the bifundamental and vector contributions from $\mu_{3,1}$ and $\mu_{4,1}$. As a result, one can factorize the partition function into five parts that correspond to each of the five gauge nodes, which turns out to be equivalent to the computation procedure from the quadrivalent gluing method.  

We expect that the cancellations of the bifundamental and vector contributions that mix $\mu_{1,1}$ with $\mu_{2,1}$ and $\mu_{3,1}$ with $\mu_{4,1}$ also happen for the LST in figure \ref{fig:D4_ON}(b) which can be understood as the following. Due to the identification of external edges the vertical strips are now periodic and we can use the method in \cite{Haghighat:2013gba} to compute such periodic strips which we summarized in Appendix \ref{app:formulas}. In particular, from \eqref{eq:Zofgeneralstrip} we know that there are infinitely many bifundamental and vectors contributions that mix $\mu_{1,1}$ with $\mu_{2,1}$ and $\mu_{3,1}$ with $\mu_{4,1}$. But due to the brane web structure in figure \ref{fig:D4_ON}(a), the color brane associated with $\mu_{1,1}$ always connects to the rightmost strip from the left side and the color brane associated with $\mu_{2,1}$ always connects to the rightmost strip from the right side in figure \ref{fig:D4_ON}(b). Then from \cite{Kim:2022dbr}, $\mu_{1,1}$ and $\mu_{2,1}$ belong to two SU gauge nodes respectively and we can see the cancellations of bifundamental and vector contributions that mix $\mu_{1,1}$ and $\mu_{2,1}$ by swapping infinitely many color branes(due to vertical periodicity) associated with $\mu_{1,1}$ and $\mu_{2,1}$ so that the two SU gauge nodes do not overlap. The same analysis goes for $\mu_{3,1}$ and $\mu_{4,1}$. After the cancellations, there are no factors in the partition function that mix $\mu_{1,1}$ with $\mu_{2,1}$ and $\mu_{3,1}$ with $\mu_{4,1}$ which also leads to the factorization of the partition function into five parts with each part correspond to a gauge node. 
This also means that the $D_4$-type LST, although ``vertically'' periodic\footnote{Here, by abuse of notation, we use the term, vertically periodic, to denote the periodicity arising from bivalent gluings of non-parallel edges and also to distinguish it from the other periodicity induced from affine quivers depicted horizontally.} compared to the E-string theory, can also be reconstructed by the quadrivalent gluing method. 
Therefore, we propose that one can realize the $D_4$-type LST of figure \ref{fig:D4_ON}(b) by the quadrivalent gluing in figure \ref{fig:D4_quad}(a). There are several things that need to be explained about this quadrivalent gluing diagram before we explain how to realize the vertical periodicity using the bivalent self-gluing: 
\begin{figure}[htbp]
	\centering
	\includegraphics[scale=0.45]{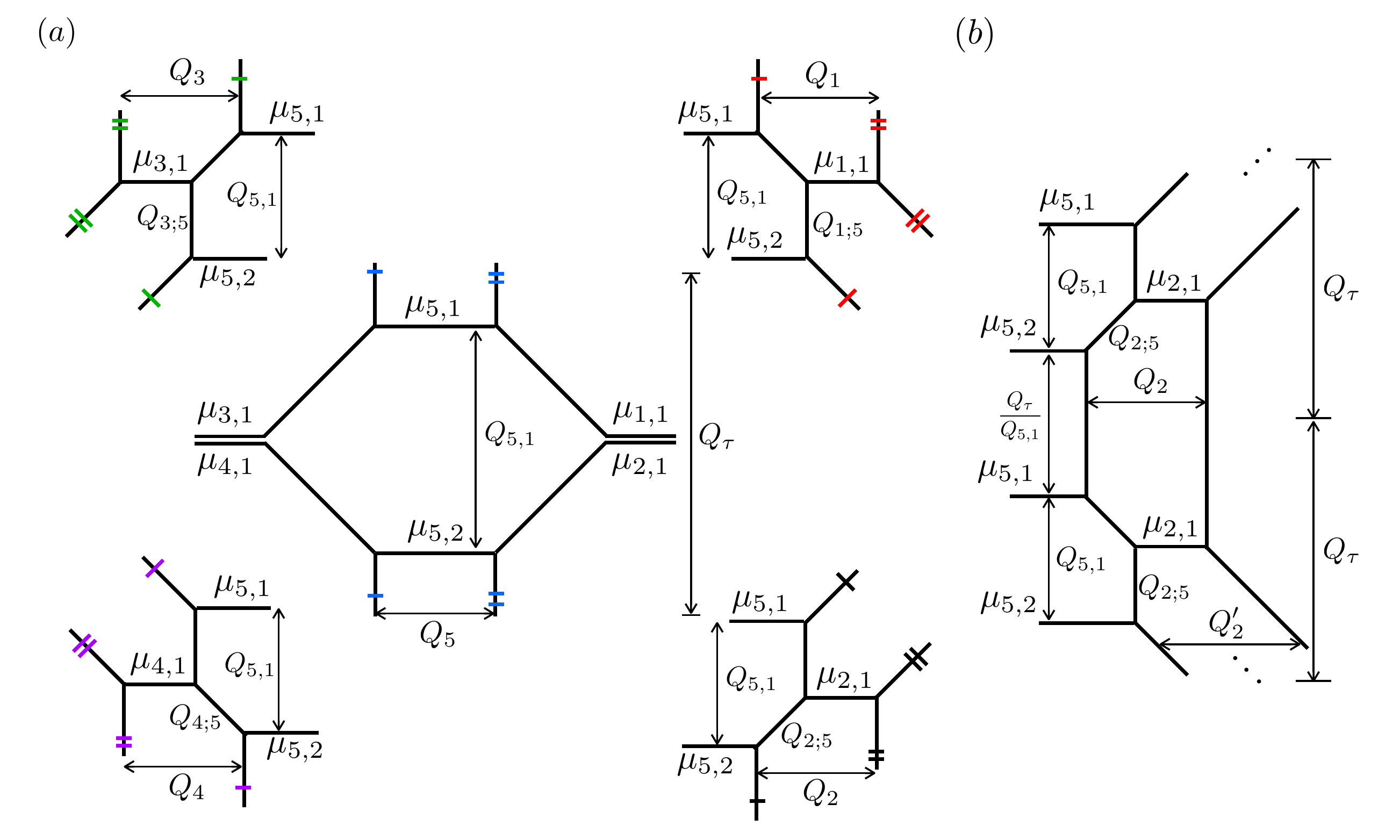}
	\caption{(a) A quadrivalent gluing realization of the $D_4$-type LST. The four subdiagrams in the corners correspond to four SU(1) gauge nodes, the middle subdiagram corresponds to SU(2) gauge node. (b) Gluing for non-parallel edges for the diagram on the bottom right of (a). The configuration shown corresponds to a gluing associated with a double period $Q_\tau$.}
	\label{fig:D4_quad}
\end{figure}
\begin{figure}[htbp]
    \centering
    \includegraphics[scale=0.4]{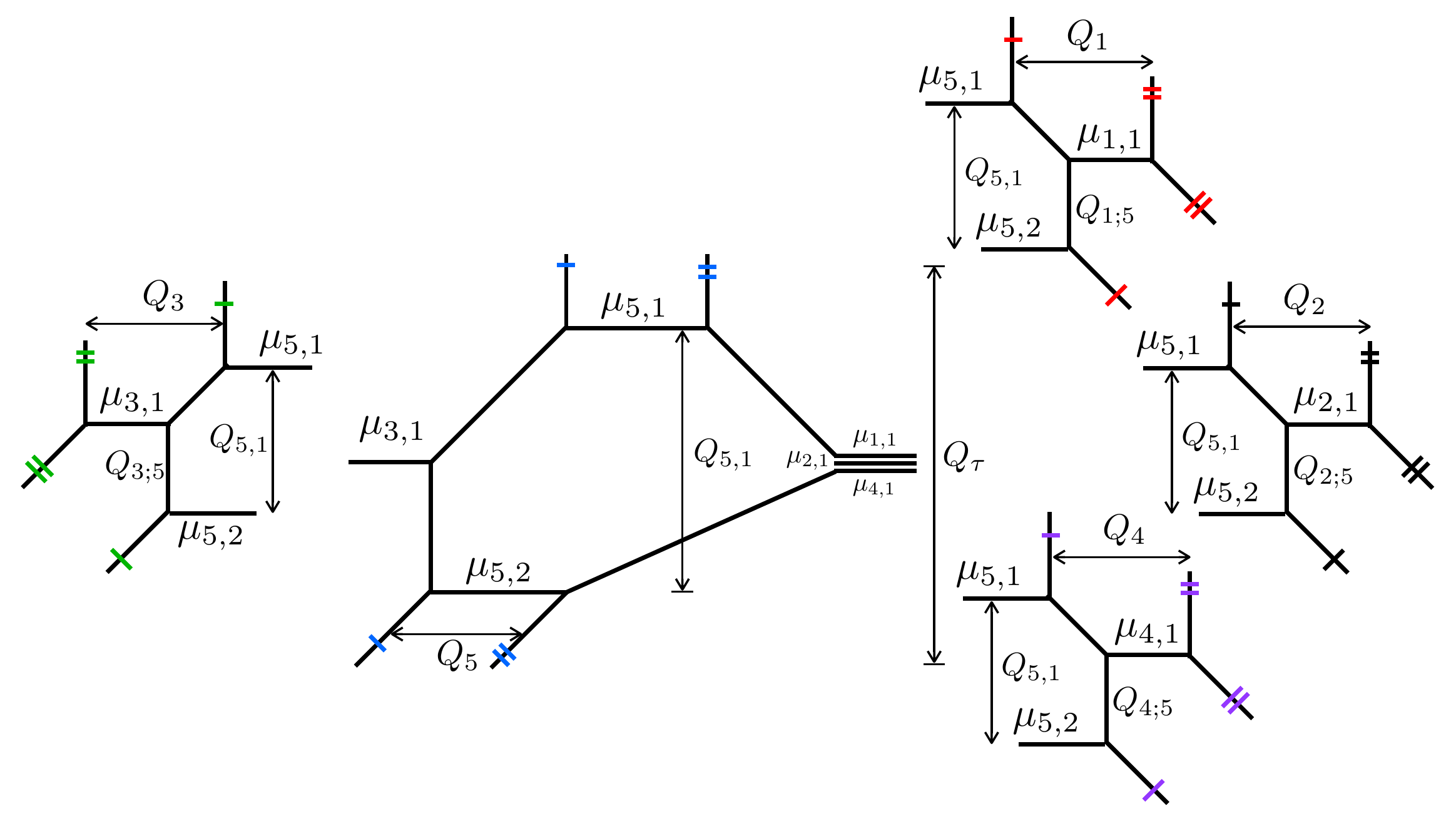}
    \caption{Another equivalent quadrivalent gluing realization of the $D_4$-type LST.}
    \label{fig:D4_quad2}
\end{figure}
\begin{enumerate}
\item [(1)]We did not explicitly show how to tri-/quadri-valently glue subdiagrams together in the previous section.
Here figure \ref{fig:D4_quad}(a) explicitly illustrates how to do this by also showing the subdiagram of the SU(2) gauge node in the middle of the figure, and this middle diagram looks exactly like the middle part of the ON-plane realization in figure \ref{fig:D4_ON}(b). 
Now we have five subdiagrams in figure \ref{fig:D4_quad}(a).
Each subdiagram corresponds to a gauge node of the affine $D_4$ quiver, and there are overlaps between two subdiagrams that correspond to two connected gauge nodes. 

\item [(2)]Each subdiagram in figure \ref{fig:D4_quad}(a) is equivalent to its $SL(2,\mathbb{Z})$ transformed diagram. For example, we can do an $SL(2,\mathbb{Z})$ transformation to make the shape of the bottom right subdiagram just like the shape of the top right subdiagram, but this should not affect the middle SU(2) subdiagram in which the top two NS-charged branes have to parallel to each other and also the bottom two NS-charged branes have to parallel to each other. 

\item [(3)] Although we draw two SU(1) 
subdiagrams on the left and the other two on the right, the tri-/quadri-valent gluing are not sensitive from where subdiagrams are glued.  
For example, we can also draw one SU(1) subdiagram on the left and three on the right as shown in figure \ref{fig:D4_quad2} in which the middle SU(2) subdiagram also accordingly changes. 
However, by using the topological vertex formalism for the quadrivalent gluing diagrams in figure \ref{fig:D4_quad}(a) and figure \ref{fig:D4_quad2}, it is easy to show that the two partition functions are the same. Here we omit the proof, but the reader can easily prove this by using the formula that we will obtain in section \ref{sec:generalformula}. 
\end{enumerate}

\paragraph{Bivalent self-gluing.}
We label the K\"ahler parameters and Young diagrams of figure \ref{fig:D4_quad}(a) in accord with figure \ref{fig:D4_ON}(b) and, in particular, the Young diagrams $\mu_{I,j}$ label the $j$-th color brane of the $I$-th gauge node. It is obvious that figure \ref{fig:D4_quad}(a) has affine $D_4$ quiver structure but it is less obvious to see the vertical periodicity from the figure. The way that one realizes the vertical periodicity is simply gluing the two end edges of each vertical strip for each subdiagram in figure \ref{fig:D4_quad}(a), which we already referred to as the bivalent self-gluing. 

Gluing non-parallel edges are not a conventional gluing, though it was discussed for a 5-brane configuration with two O5-planes \cite{Kim:2021cua}. So, we explain how to bivalently self-glue non-parallel lines. First, we pick the bottom right subdiagram in figure \ref{fig:D4_quad}(a), as an example. In figure \ref{fig:D4_quad}(b), we draw two adjacent periods of the bottom right subdiagram of figure \ref{fig:D4_quad}(a). The upper period is just like the original subdiagram in figure \ref{fig:D4_quad}(a), but because the two ends of each vertical strip in the bottom right subdiagram of figure \ref{fig:D4_quad}(a) are non-parallel, in order to draw the next lower period continuously, we have to tilt the original subdiagram by an $SL(2,\mathbb{Z})$ transformation which leads to the lower period shown in figure \ref{fig:D4_quad}(b). However, the $SL(2,\mathbb{Z})$ transformation does not change the K\"ahler parameters, so if the brane web in figure \ref{fig:D4_quad}(b) wants to be vertically periodic, the K\"ahler parameters should not change between different periods. In particular, $Q_2$ should be equal to $Q_2'$. In order to satisfy the periodicity constraint $Q_2=Q'_2$, $Q_{2;5}$ has to satisfy the condition
\begin{align}\label{eq:D4PDcond1}
    Q_{2;5}=Q_{5,1}^{\frac12}\ .
\end{align}
By the same reasoning, for the other three SU(1) subdiagrams, we also have
\begin{align}\label{eq:D4PDcond2}
    Q_{1;5}=Q_{3;5}=Q_{4;5}=Q_{5,1}^{\frac12}\ .
\end{align}
The middle SU(2) subdiagram in figure \ref{fig:D4_quad}(a) is automatically periodic once the periodicity conditions \eqref{eq:D4PDcond1} and \eqref{eq:D4PDcond2} are satisfied. We can see that the periodicity conditions of the quadrivalent gluing in figure \ref{fig:D4_quad}(a) are also consistent with the periodicity conditions coming from the brane web with two ON-planes in figure \ref{fig:D4_ON}(b). Then due to the vertical periodicity, in figure \ref{fig:D4_quad}(a) if we draw lines along each external NS-charged branes in each gauge node towards the center of Coulomb branch, the two intersection points will be just at the center of Coulomb branch, the distance between this two points is just the instanton factor of the gauge node, and it is just equal to the horizontal distance between the two parallel NS-charged branes of the gauge node.
So $Q_1,Q_2,Q_3,Q_4,Q_5$ are the instanton factors of the gauge node $1,2,3,4,5$ respectively in the 5d theory perspective, and they are also the tensor branch parameters of the 6d $D_4$-type LST. The horizontal period $u$ of the theory due to its affine $D_4$ quiver structure is then given by
\begin{align}\label{eq:horiofD4}
    u=Q_1Q_2Q_3Q_4Q_5^2\ .
\end{align}

\bigskip

Now we are ready to use the quadrivalent gluing diagram in figure \ref{fig:D4_quad}(a) to compute the partition function $Z^{D_4}$ of $D_4$-type LST which takes the following summation and product form,
\begin{multline}
Z^{D_4}=\sum_{\mu_{5,1}\mu_{5,2}}Z_{\mu_{5,1}\mu_{5,2}}^{\text{SU(2)}}\sum_{\mu_{1,1}}Z_{\mu_{5,1}\mu_{5,2}\mu_{1,1}}(Q_{5,1},Q_{\tau},Q_{1})\sum_{\mu_{2,1}}Z_{\mu_{5,1}\mu_{5,2}\mu_{2,1}}(Q_{5,1},Q_{\tau},Q_{2})\\
\times\sum_{\mu_{3,1}}Z_{\mu_{5,1}\mu_{5,2}\mu_{3,1}}(Q_{5,1},Q_{\tau},Q_{3})\sum_{\mu_{4,1}}Z_{\mu_{5,1}\mu_{5,2}\mu_{4,1}}(Q_{5,1},Q_{\tau},Q_{4}),
\label{eq:Z_D4}
\end{multline}
in which 
\begin{equation}
	Z_{\mu_{5,1}\mu_{5,2}}^{\text{SU(2)}}\equiv(-Q_{5})^{|\mu_{5,1}|+|\mu_{5,2}|}f_{\mu_{5,1}}(\ft,\fq)f_{\mu_{5,2}}(\ft,\fq)^{-1}\times
	\begin{tikzpicture}[scale=0.6,thick,baseline=-15]
		\draw (0,0)--++(-1.4,0) coordinate (a)--++(0,-1.4) coordinate (b)--++(1.4,0);
		\draw (a)--++(-0.7,0.7);
		\draw (b)--++(-0.7,-0.7);
		\node at (-0.6,0.33){$\mu_{5,1}$};
		\node at (-0.6,-1.06){$\mu_{5,2}$};
		\draw[<->] (0.2,0)--++(0,-1.4);
		\node at (0.9,-0.7){$Q_{5,1}$};
		\draw[<->] (-2.4,0.7)--++(0,-2.8);
		\node at (-2.85,-0.7){$Q_{\tau}$};
		\draw (-2.43+0.8,-2.27+0.5)++(-0.05,-0.05)--++(-0.14,0.14);
		\draw (-2.43+0.8,0.87-0.5)++(-0.05,0.05)--++(-0.14,-0.14);
	\end{tikzpicture}
	\times
	\begin{tikzpicture}[scale=.6,thick,baseline = -15]
		\draw (0,0)--++(1.4,0) coordinate (a)--++(0,-1.4) coordinate (b)--++(-1.4,0);
		\draw (a)--++(0.7,0.7);
		\draw (b)--++(0.7,-0.7);
		\node at (0.8,0.33){$\mu_{5,1}$};
		\node at (0.8,-1.06){$\mu_{5,2}$};
		\draw[<->] (-0.2,0)--++(0,-1.4);
		\node at (-0.85,-0.7){$Q_{5,1}$};
		\draw (2.43-0.8,-2.27+0.5)++(0.05,-0.05)--++(0.14,0.14);
		\draw (2.43-0.8,0.87-0.5)++(0.05,0.05)--++(0.14,-0.14);
		\draw[<->] (2.4,0.7)--++(0,-2.8);
		\node at (2.9,-0.7){$Q_{\tau}$};
	\end{tikzpicture},
	\label{eq:mid_SU2}
\end{equation}
where the edge factors of $\mu_{5,1},\mu_{5,2}$ come from the middle subdiagram of figure \ref{fig:D4_quad}(a), and
\begin{equation}
Z_{\mu_{5,1}\mu_{5,2}\mu_{1,1}}(Q_{5,1},Q_{\tau},Q_{1})\equiv
\begin{tikzpicture}[scale=.6,thick,baseline = -15]
	\draw (0,0)--++(1,0) coordinate (a)--++(0,1);
	\draw (a)--++(.7,-.7) coordinate (c)--++(0,-.7) coordinate (b)--++(-1.7,0);
	\draw (b)--++(1,-1);
	\draw (c)--++(1,0) coordinate (d)--++(0,1.3);
	\draw (d)--++(.7,-.7);
	\node at (0.4,0.33){$\mu_{5,1}$};
	\node at (0.8,-1.06){$\mu_{5,2}$};
	\node at (2.35,-1.1){$\mu_{1,1}$};
	\draw (0.9,0.8)--(1.1,0.8);
	\draw (2.43,-2.27)--++(0.14,0.14);
	\draw (2.6,0.4-0.2)--++(0.2,0);
	\draw (2.6,0.32-0.2)--++(0.2,0);
	\draw (2.7,-0.7)++(0.57-0.07,-0.57-0.07)--++(0.14,0.14);
	\draw (2.7,-0.7)++(0.51-0.07,-0.51-0.07)--++(0.14,0.14);
	\draw[<->] (-0.2,0)--++(0,-1.4);
	\node at (-0.9,-0.7){$Q_{5,1}$};
	\draw[<->] (1,0.47)--node[midway,anchor=south]{$Q_{1}$}++(1.7,0);
	\draw[<->] (3.55,0.7)--++(0,-2.8);
	\node at (4.1,-0.7){$Q_{\tau}$};
\end{tikzpicture}
\Bigg/
\begin{tikzpicture}[scale=.6,thick,baseline = -15]
	\draw (0,0)--++(1.4,0) coordinate (a)--++(0,-1.4) coordinate (b)--++(-1.4,0);
	\draw (a)--++(0.7,0.7);
	\draw (b)--++(0.7,-0.7);
	\node at (0.8,0.33){$\mu_{5,1}$};
	\node at (0.8,-1.06){$\mu_{5,2}$};
	\draw[<->] (-0.2,0)--++(0,-1.4);
	\node at (-0.9,-0.7){$Q_{5,1}$};
	\draw (2.43-0.8,-2.27+0.5)++(0.05,-0.05)--++(0.14,0.14);
	\draw (2.43-0.8,0.87-0.5)++(0.05,0.05)--++(0.14,-0.14);
	\draw[<->] (2.4,0.7)--++(0,-2.8);
	\node at (3,-0.7){$Q_{\tau}$};
\end{tikzpicture}\ .
\label{eq:sub_mu1}
\end{equation}
The other summands, $Z_{\mu_{5,1}\mu_{5,2}\mu_{2,1}},Z_{\mu_{5,1}\mu_{5,2}\mu_{3,1}},Z_{\mu_{5,1}\mu_{5,2}\mu_{4,1}}$, are defined in the same form as  $Z_{\mu_{5,1}\mu_{5,2}\mu_{1,1}}$ with $\mu_{1,1}, Q_1$ replaced by $\mu_{2,1}, Q_2, \mu_{3,1}, Q_3, \mu_{4,1}, Q_4$ respectively as they correspond to the four SU(1) subdiagrams of figure \ref{fig:D4_quad}(a) which are related to each other by $SL(2,\mathbb{Z})$ or a mirror reflection.

Let us first compute \eqref{eq:mid_SU2}. By the topological vertex formalism, we find
\begin{align}
&\begin{tikzpicture}[scale=.6,thick,baseline = -15]
	\draw (0,0)--++(1.4,0) coordinate (a)--++(0,-1.4) coordinate (b)--++(-1.4,0);
	\draw (a)--++(0.7,0.7);
	\draw (b)--++(0.7,-0.7);
	\node at (0.8,0.33){$\mu_{5,1}$};
	\node at (0.8,-1.06){$\mu_{5,2}$};
	\draw[<->] (-0.2,0)--++(0,-1.4);
	\node at (-0.9,-0.7){$Q_{5,1}$};
	\draw (2.43-0.8,-2.27+0.5)++(0.05,-0.05)--++(0.14,0.14);
	\draw (2.43-0.8,0.87-0.5)++(0.05,0.05)--++(0.14,-0.14);
	\draw[<->] (2.4,0.7)--++(0,-2.8);
	\node at (3,-0.7){$Q_{\tau}$};
\end{tikzpicture}=\sum_{\nu_1\nu_2}
\begin{tikzpicture}[scale=.6,thick,baseline = -15]
	\draw (0,0)--++(1.4,0) coordinate (a)--++(0,-1.4) coordinate (b)--++(-1.4,0);
	\draw (a)--++(0.7,0.7);
	\draw (b)--++(0.7,-0.7);
	\node at (0.8,0.33){$\mu_{5,1}$};
	\node at (0.8,-1.06){$\mu_{5,2}$};
	\draw[<->] (-0.2,0)--++(0,-1.4);
	\node at (-0.9,-0.7){$Q_{5,1}$};
	\node at (2.1,0){$\nu_1$};
	\node at (1.8,-0.7){$\nu_2$};
	\node at (2.1,-1.5){$\nu_1$};
	\draw[<->] (2.4,0.7)--++(0,-2.8);
	\node at (3,-0.7){$Q_{\tau}$};
	\draw (2.43-0.8,-2.27+0.5)++(0.05,-0.05)--++(0.14,0.14);
	\draw (2.43-0.8,0.87-0.5)++(0.05,0.05)--++(0.14,-0.14);
\end{tikzpicture}\nonumber\\
=&\sum_{\nu_1\nu_2}C_{\nu_1\nu_2^t\mu_{5,1}}(\ft,\fq)C_{\nu_2\nu_1^t\mu_{5,2}}(\ft,\fq)(-Q_{\tau}Q_{5,1}^{-1})^{|\nu_1|}\tilde{f}_{\nu_1}(\fq,\ft)^{-1}(-Q_{5,1})^{|\nu_2|}\tilde{f}_{\nu_2}(\fq,\ft)^{-1}\nonumber\\
=&\,\fq^{\frac{||\mu_{5,1}||^2+||\mu_{5,2}||^2}{2}}\tilde{Z}_{\mu_{5,1}}(\ft,\fq)\tilde{Z}_{\mu_{5,2}}(\ft,\fq)\sum_{\nu_i\eta_i}\left(Q_{\tau}Q_{5,1}^{-1}\sqrt{\frac{\ft}{\fq}}\right)^{|\nu_1|}\left(Q_{5,1}\sqrt{\frac{\ft}{\fq}}\right)^{|\nu_2|}\left(\frac{\fq}{\ft}\right)^{\frac{|\eta_1|+|\eta_2|}{2}}
\nonumber\\
&\times s_{\nu_1^t/\eta_1}(\ft^{-\rho}\fq^{-\mu_{5,1}})s_{\nu_2^t/\eta_1}(\fq^{-\rho}\ft^{-\mu_{5,1}^t})s_{\nu_2^t/\eta_2}(\ft^{-\rho}\fq^{-\mu_{5,2}})s_{\nu_1^t/\eta_2}(\fq^{-\rho}\ft^{-\mu_{5,2}^t}) 
\nonumber \\
=&~\fq^{\frac{||\mu_{5,1}||^2+||\mu_{5,2}||^2}{2}}\tilde{Z}_{\mu_{5,1}}(\ft,\fq)\tilde{Z}_{\mu_{5,2}}(\ft,\fq)\prod_{l=1}^{\infty}\frac{1}{1-Q_{\tau}^l}\nonumber\\
&\times\prod_{n=1}^{\infty}\frac{1}{\calR_{\mu_{5,2}^t\mu_{5,1}}(Q_{\tau}^nQ_{5,1}^{-1}\sqrt{\frac{\ft}{\fq}})\calR_{\mu_{5,1}^t\mu_{5,2}}(Q_{\tau}^{n-1}Q_{5,1}\sqrt{\frac{\ft}{\fq}})\calR_{\mu_{5,1}^t\mu_{5,1}}(Q_{\tau}^n\sqrt{\frac{\ft}{\fq}})\calR_{\mu_{5,2}^t\mu_{5,2}}(Q_{\tau}^n\sqrt{\frac{\ft}{\fq}})},
\label{eq:rt_half}
\end{align}
where in the last equality, we used the Cauchy identities repeatedly \cite{Haghighat:2013gba}.\footnote{See Appendix \ref{app:formulas} for more details on how to compute the vertical periodic strip.} 

We can immediately obtain the other half SU(2) subdiagram based on \eqref{eq:rt_half} since they are mirror images to each other,
\begin{align}
	&\begin{tikzpicture}[scale=.6,thick,baseline = -15]
		\draw (0,0)--++(-1.4,0) coordinate (a)--++(0,-1.4) coordinate (b)--++(1.4,0);
		\draw (a)--++(-0.7,0.7);
		\draw (b)--++(-0.7,-0.7);
		\node at (-0.6,0.33){$\mu_{5,1}$};
		\node at (-0.6,-1.06){$\mu_{5,2}$};
		\draw[<->] (0.2,0)--++(0,-1.4);
		\node at (1,-0.7){$Q_{5,1}$};
		\draw[<->] (-2.4,0.7)--++(0,-2.8);
		\node at (-2.95,-0.7){$Q_{\tau}$};
		\draw (-2.43+0.8,-2.27+0.5)++(-0.05,-0.05)--++(-0.14,0.14);
		\draw (-2.43+0.8,0.87-0.5)++(-0.05,0.05)--++(-0.14,-0.14);
	\end{tikzpicture}\nonumber\\
	&=\ft^{\frac{||\mu_{5,1}^t||^2+||\mu_{5,2}^t||^2}{2}}\tilde{Z}_{\mu_{5,1}^t}(\fq,\ft)\tilde{Z}_{\mu_{5,2}^t}(\fq,\ft)\prod_{l=1}^{\infty}\frac{1}{1-Q_{\tau}^l}\nonumber\\
	&\times\prod_{n=1}^{\infty}\frac{1}{\calR_{\mu_{5,2}^t\mu_{5,1}}(\frac{Q_{\tau}^n}{Q_{5,1}}\sqrt{\frac{\fq}{\ft}})\calR_{\mu_{5,1}^t\mu_{5,2}}(Q_{\tau}^{n-1}Q_{5,1}\sqrt{\frac{\fq}{\ft}})\calR_{\mu_{5,1}^t\mu_{5,1}}(Q_{\tau}^n\sqrt{\frac{\fq}{\ft}})\calR_{\mu_{5,2}^t\mu_{5,2}}(Q_{\tau}^n\sqrt{\frac{\fq}{\ft}})}.
	\label{eq:lf_half}
\end{align}
By substituting \eqref{eq:lf_half} and \eqref{eq:rt_half} into \eqref{eq:mid_SU2}, we get
\begin{align}\label{eq:D4ZSU21}
Z_{\mu_{5,1}\mu_{5,2}}^{\text{SU(2)}}=&\prod_{l=1}^{\infty}\frac{1}{(1-Q_{\tau}^l)^2}\prod_{n=1}^{\infty}\calM(Q_{\tau}^n)^2\calM(Q_{\tau}^n\tfrac{\ft}{\fq})^2\calM(Q_{\tau}^nQ_{5,1}^{-1})\calM(Q_{\tau}^nQ_{5,1}^{-1}\tfrac{\ft}{\fq})\nonumber\\
&\times
\calM(Q_{\tau}^{n-1}Q_{5,1})\calM(Q_{\tau}^{n-1}Q_{5,1}\tfrac{\ft}{\fq})\hat{Z}_{\mu_{5,1}\mu_{5,2}}^{\text{SU(2)}},
\end{align}
where
\begin{align}
\hat{Z}_{\mu_{5,1}\mu_{5,2}}^{\text{SU(2)}}\equiv&~\ft^{||\mu_{5,1}^t||^2}\fq^{||\mu_{5,2}||^2}Q_{5}^{|\mu_{5,1}|+|\mu_{5,2}|}\tilde{Z}_{\mu_{5,1}}(\ft,\fq)\tilde{Z}_{\mu_{5,2}}(\ft,\fq)\tilde{Z}_{\mu_{5,1}^t}(\fq,\ft)\tilde{Z}_{\mu_{5,2}^t}(\fq,\ft)\nonumber\\
&\times\!\bigg(\prod_{n=1}^{\infty}\calN_{\mu_{5,2}\mu_{5,1}}\!(Q_{\tau}^nQ_{5,1}^{-1})\calN_{\mu_{5,2}\mu_{5,1}}\!(Q_{\tau}^nQ_{5,1}^{-1}\tfrac{\ft}{\fq})\calN_{\mu_{5,1}\mu_{5,2}}\!(Q_{\tau}^{n-1}Q_{5,1})\nonumber\\
&\times \calN_{\mu_{5,1}\mu_{5,2}}\!(Q_{\tau}^{n-1}Q_{5,1}\tfrac{\ft}{\fq})\calN_{\mu_{5,1}\mu_{5,1}}(Q_{\tau}^n)\calN_{\mu_{5,1}\mu_{5,1}}(Q_{\tau}^n\tfrac{\ft}{\fq})\calN_{\mu_{5,2}\mu_{5,2}}(Q_{\tau}^n)\nn\\
&\times \calN_{\mu_{5,2}\mu_{5,2}}(Q_{\tau}^n\tfrac{\ft}{\fq})\!\bigg)^{-1}\ .
\label{eq:nor_su2}
\end{align}
Next, we compute \eqref{eq:sub_mu1}. 
The numerator of \eqref{eq:sub_mu1} consists of two strips glued by a horizontal edge:
\begin{equation}
\begin{tikzpicture}[scale=.6,thick,baseline = -15]
	\draw (0,0)--++(1,0) coordinate (a)--++(0,1);
	\draw (a)--++(.7,-.7) coordinate (c)--++(0,-.7) coordinate (b)--++(-1.7,0);
	\draw (b)--++(1,-1);
	\draw (c)--++(1,0) coordinate (d)--++(0,1.3);
	\draw (d)--++(.7,-.7);
	\node at (0.4,0.33){$\mu_{5,1}$};
	\node at (0.8,-1.06){$\mu_{5,2}$};
	\node at (2.35,-1.1){$\mu_{1,1}$};
	\draw (0.9,0.8)--(1.1,0.8);
	\draw (2.43,-2.27)--++(0.14,0.14);
	\draw (2.6,0.4-0.2)--++(0.2,0);
	\draw (2.6,0.32-0.2)--++(0.2,0);
	\draw (2.7,-0.7)++(0.57-0.07,-0.57-0.07)--++(0.14,0.14);
	\draw (2.7,-0.7)++(0.51-0.07,-0.51-0.07)--++(0.14,0.14);
	\draw[<->] (-0.2,0)--++(0,-1.4);
	\node at (-0.9,-0.7){$Q_{5,1}$};
	\draw[<->] (1,0.47)--node[midway,anchor=south]{$Q_{1}$}++(1.7,0);
	\draw[<->] (3.55,0.7)--++(0,-2.8);
	\node at (4.15,-0.7){$Q_{\tau}$};
\end{tikzpicture}
=\sum_{\mu_{1,1}}(-Q_{1}Q_{5,1}^{-\frac{1}{2}})^{|\mu_{1,1}|}\times
\begin{tikzpicture}[scale=.6,thick,baseline = -15]
	\draw (0,0)--++(1,0) coordinate (a)--++(0,1);
	\draw (a)--++(.7,-.7) coordinate (c)--++(0,-.7) coordinate (b)--++(-1.7,0);
	\draw (b)--++(1,-1);
	\draw (c)--++(1,0) coordinate (d);		
	\node at (0.4,0.33){$\mu_{5,1}$};
	\node at (0.8,-1.06){$\mu_{5,2}$};
	\node at (2.35,-1.1){$\mu_{1,1}$};
	\draw (0.9,0.8)--(1.1,0.8);
	\draw (2.43,-2.27)--++(0.14,0.14);
	\draw[<->] (-0.2,0)--++(0,-1.4);
	\node at (-0.85,-0.7){$Q_{5,1}$};
	\draw[<->] (3.55-0.6,0.7)--++(0,-2.8);
	\node at (3.55,-0.7){$Q_{\tau}$};
\end{tikzpicture}\times
\begin{tikzpicture}[scale=.6,thick,baseline = -15]
	\draw (1.7,-0.7)--++(1,0) coordinate (d)--++(0,1.3);
	\draw (d)--++(.7,-.7);
	\node at (2.35,-1.1){$\mu_{1,1}$};
	\draw (2.6,0.4)--++(0.2,0);
	\draw (2.6,0.32)--++(0.2,0);
	\draw (2.7,-0.7)++(0.57-0.07,-0.57-0.07)--++(0.14,0.14);
	\draw (2.7,-0.7)++(0.51-0.07,-0.51-0.07)--++(0.14,0.14);
	\draw[<->] (3.55,0.7)--++(0,-2.8);
	\node at (4.15,-0.7){$Q_{\tau}$};
\end{tikzpicture}.
\label{eq:t56}
\end{equation}
We compute the two strips separately:
\begin{align}
&\begin{tikzpicture}[scale=.6,thick,baseline = -15]
	\draw (0,0)--++(1,0) coordinate (a)--++(0,1);
	\draw (a)--++(.7,-.7) coordinate (c)--++(0,-.7) coordinate (b)--++(-1.7,0);
	\draw (b)--++(1,-1);
	\draw (c)--++(1,0) coordinate (d);		
	\node at (0.4,0.33){$\mu_{5,1}$};
	\node at (0.8,-1.06){$\mu_{5,2}$};
	\node at (2.35,-1.1){$\mu_{1,1}$};
	\draw (0.9,0.8)--(1.1,0.8);
	\draw (2.43,-2.27)--++(0.14,0.14);
	\draw[<->] (-0.2,0)--++(0,-1.4);
	\node at (-0.9,-0.7){$Q_{5,1}$};
	\draw[<->] (3.55-0.6,0.7)--++(0,-2.8);
	\node at (3.55,-0.7){$Q_{\tau}$};
\end{tikzpicture}=\sum_{\nu_1\nu_2\nu_3}
\begin{tikzpicture}[scale=.6,thick,baseline = -15]
	\draw (0,0)--++(1,0) coordinate (a)--++(0,1);
	\draw (a)--++(.7,-.7) coordinate (c)--++(0,-.7) coordinate (b)--++(-1.7,0);
	\draw (b)--++(1,-1);
	\draw (c)--++(1,0) coordinate (d);		
	\node at (0.4,0.33){$\mu_{5,1}$};
	\node at (0.6,-1.06){$\mu_{5,2}$};
	\node at (2.35,-1.1){$\mu_{1,1}$};
	\draw (0.9,0.8)--(1.1,0.8);
	\draw (2.43,-2.27)--++(0.14,0.14);
	\draw[<->] (-0.2,0)--++(0,-1.4);
	\node at (-0.85,-0.7){$Q_{5,1}$};
	\draw[<->] (3.55-0.6,0.7)--++(0,-2.8);
	\node at (3.45,-0.7){$Q_{\tau}$};
	\node at (1.4,0.5){$\nu_1$};
	\node at (1.7,-0.2){$\nu_2$};
	\node at (1.4,-1){$\nu_3$};
	\node at (1.9,-2){$\nu_1$};
\end{tikzpicture}\nonumber\\
=&\sum_{\nu_1\nu_2\nu_3}C_{\nu_1\nu_2^t\mu_{5,1}}(\ft,\fq)C_{\nu_3^t\nu_2\mu_{1,1}^t}(\fq,\ft)C_{\nu_3\nu_1^t\mu_{5,2}}(\ft,\fq)(-\frac{Q_{\tau}}{Q_{5,1}})^{|\nu_1|}\tilde{f}_{\nu_1}(\fq,\ft)^{-1}(-Q_{5,1}^{1/2})^{|\nu_2|}(-Q_{5,1}^{1/2})^{|\nu_3|}\nonumber\\
=&\, \fq^{\frac{1}{2}(||\mu_{5,1}||^2+||\mu_{5,2}||^2)}\ft^{\frac{1}{2}||\mu_{1,1}^t||^2}\tilde{Z}_{\mu_{5,1}}(\ft,\fq)\tilde{Z}_{\mu_{5,2}}(\ft,\fq)\tilde{Z}_{\mu_{1,1}^t}(\fq,\ft)\prod_{l=1}^{\infty}\frac{1}{1-Q_{\tau}^l}\nonumber\\
\times &\prod_{n=1}^{\infty}\bigg[\calR_{\mu_{1,1}^t\mu_{5,1}}(Q_{\tau}^nQ_{5,1}^{-1/2})\calR_{\mu_{5,1}^t\mu_{1,1}}(Q_{\tau}^{n-1}Q_{5,1}^{1/2})\calR_{\mu_{5,2}^t\mu_{1,1}}(Q_{\tau}^nQ_{5,1}^{-1/2})\calR_{\mu_{1,1}^t\mu_{5,2}}(Q_{\tau}^{n-1}Q_{5,1}^{1/2})\nn\\
\times & \bigg(\calR_{\mu_{5,2}^t\mu_{5,1}}(Q_{\tau}^nQ_{5,1}^{-1}\sqrt{\frac{\ft}{\fq}})\calR_{\mu_{5,1}^t\mu_{5,2}}(Q_{\tau}^{n-1}Q_{5,1}\sqrt{\frac{\ft}{\fq}})\calR_{\mu_{5,1}^t\mu_{5,1}}(Q_{\tau}^n\sqrt{\frac{\ft}{\fq}})\calR_{\mu_{5,2}^t\mu_{5,2}}(Q_{\tau}^n\sqrt{\frac{\ft}{\fq}})\nn\\
\times & \calR_{\mu_{1,1}^t\mu_{1,1}}(Q_{\tau}^n\sqrt{\frac{\fq}{\ft}})\bigg)^{-1}\bigg]\ ,
\label{eq:t5}\\
&\begin{tikzpicture}[scale=.6,thick,baseline = -15]
	\draw (1.7,-0.7)--++(1,0) coordinate (d)--++(0,1.3);
	\draw (d)--++(.7,-.7);
	\node at (2.35,-1.1){$\mu_{1,1}$};
	\draw (2.6,0.4)--++(0.2,0);
	\draw (2.6,0.32)--++(0.2,0);
	\draw (2.7,-0.7)++(0.57-0.07,-0.57-0.07)--++(0.14,0.14);
	\draw (2.7,-0.7)++(0.51-0.07,-0.51-0.07)--++(0.14,0.14);
	\draw[<->] (3.55,0.7)--++(0,-2.8);
	\node at (4.05,-0.7){$Q_{\tau}$};
\end{tikzpicture}=\sum_{\nu_4}
\begin{tikzpicture}[scale=.6,thick,baseline = -15]
	\draw (1.7,-0.7)--++(1,0) coordinate (d)--++(0,1.3);
	\draw (d)--++(.7,-.7);
	\node at (2.2,-1.1){$\mu_{1,1}$};
	\draw (2.6,0.4)--++(0.2,0);
	\draw (2.6,0.32)--++(0.2,0);
	\draw (2.7,-0.7)++(0.57-0.07,-0.57-0.07)--++(0.14,0.14);
	\draw (2.7,-0.7)++(0.51-0.07,-0.51-0.07)--++(0.14,0.14);
	\draw[<->] (3.55,0.7)--++(0,-2.8);
	\node at (4.05,-0.7){$Q_{\tau}$};
	\node at (3.08,-0.1){$\nu_4$};
	\node at (2.9,-1.6){$\nu_4$};
\end{tikzpicture}\nonumber\\
=&\sum_{\nu_4}C_{\nu_4\nu_4^t\mu_{1,1}}(\ft,\fq)(-Q_{\tau})^{|\nu_4|}\tilde{f}_{\nu_4}(\fq,\ft)^{-1}\nn\\
=&
\fq^{\frac{||\mu_{1,1}||^2}{2}}\tilde{Z}_{\mu_{1,1}}(\ft,\fq)\prod_{l=1}^{\infty}\frac{1}{1-Q_{\tau}^l}\prod_{n=1}^{\infty}\frac{1}{\calR_{\mu_{1,1}^t\mu_{1,1}}(Q_{\tau}^n\sqrt{\frac{\ft}{\fq}})}\ .
\label{eq:t6}
\end{align}
By combining all results, we find
\begin{align}
&Z_{\mu_{5,1}\mu_{5,2}\mu_{1,1}}(Q_{5,1},Q_{\tau},Q_{1})\nn\\
=&\prod_{l=1}^{\infty}\frac{1}{1-Q_{\tau}^l}\prod_{n=1}^{\infty}\frac{\calM(Q_{\tau}^n)\calM(Q_{\tau}^n\frac{\ft}{\fq})}{\calM(Q_{\tau}^nQ_{5,1}^{-1/2}\sqrt{\frac{\ft}{\fq}})^2\calM(Q_{\tau}^{n-1}Q_{5,1}^{1/2}\sqrt{\frac{\ft}{\fq}})^2}\times\hat{Z}_{\mu_{5,1}\mu_{5,2}\mu_{1,1}}(Q_{5,1},Q_{\tau},Q_{1}),
\label{eq:Zofsubmu1}
\end{align}
where
\begin{align}
&\hat{Z}_{\mu_{5,1}\mu_{5,2}\mu_{1,1}}(Q_{5,1},Q_{\tau},Q_{1})\nn\\
\equiv &\fq^{\frac{||\mu_{1,1}||^2}{2}}\ft^{\frac{||\mu_{1,1}^t||^2}{2}}(-Q_{1}Q_{5,1}^{-\frac{1}{2}})^{|\mu_{1,1}|}\tilde{Z}_{\mu_{1,1}}(\ft,\fq)\tilde{Z}_{\mu_{1,1}^t}(\fq,\ft)\nn\\
\times &\prod_{n=1}^{\infty}\bigg[\calN_{\mu_{1,1}\mu_{5,1}}(Q_{\tau}^nQ_{5,1}^{-1/2}\sqrt{\frac{\ft}{\fq}})\calN_{\mu_{5,1}\mu_{1,1}}(Q_{\tau}^{n-1}Q_{5,1}^{1/2}\sqrt{\frac{\ft}{\fq}})\calN_{\mu_{5,2}\mu_{1,1}}(Q_{\tau}^nQ_{5,1}^{-1/2}\sqrt{\frac{\ft}{\fq}})\nn\\
\times &\calN_{\mu_{1,1}\mu_{5,2}}(Q_{\tau}^{n-1}Q_{5,1}^{1/2}\sqrt{\frac{\ft}{\fq}})\bigg(\calN_{\mu_{1,1}\mu_{1,1}}(Q_{\tau}^n)\calN_{\mu_{1,1}\mu_{1,1}}(Q_{\tau}^n\frac{\ft}{\fq})\bigg)^{-1}\bigg]\ .
\label{eq:nor_submu1}
\end{align}
$Z_{\mu_{5,1}\mu_{5,2}\mu_{2,1}}$,$Z_{\mu_{5,1}\mu_{5,2}\mu_{3,1}}$ and $Z_{\mu_{5,1}\mu_{5,2}\mu_{4,1}}$ can be easily obtained from $Z_{\mu_{5,1}\mu_{5,2}\mu_{1,1}}$ according to their definitions.

\paragraph{$\calN$ function to $\varTheta$ function.}
In order to simplify the expression of the partition function above, we replace the $\calN$ functions in \eqref{eq:nor_su2} and \eqref{eq:nor_submu1} by Jacobi theta functions via the following formulas\footnote{The identities $\sum_{(i,j)\in\lambda}\lambda_i=||\lambda||^2$, $\sum_{(i,j)\in\lambda}j=\frac{1}{2}|\lambda|+\frac{1}{2}||\lambda||^2$, $\sum_{(i,j)\in\lambda}\mu_j^t=\sum_{(i,j)\in\mu}\lambda_j^t$ are used to derive these formulas.}$^,$\footnote{Similar formulas appear in \cite{Hohenegger:2013ala} as well as \cite{Haghighat:2013tka}.}
\begin{multline}
	\prod_{n=1}^{\infty}\calN_{\beta\alpha}(Q_{\tau}^nQ^{-1}\tfrac{\ft}{\fq};\ft,\fq)\calN_{\alpha\beta}(Q_{\tau}^{n-1}Q;\ft,\fq)
	=\frac{(-1)^{|\beta|}(Q\sqrt{\frac{\fq}{\ft}})^{\frac{|\beta|+|\alpha|}{2}}\ft^{\frac{||\beta^t||^2-||\alpha^t||^2}{4}}\fq^{\frac{-||\beta||^2+||\alpha||^2}{4}}}{\left(\prod_{n=1}^{\infty}(1-Q_{\tau}^n)\right)^{|\beta|+|\alpha|}}\\
	\times\varTheta_{\beta\alpha}(Q^{-1}\tfrac{\ft}{\fq}|Q_{\tau})\varTheta_{\alpha\beta}(Q|Q_{\tau})\ ,
	\label{eq:Ntotheta1}
\end{multline}
\begin{align}
	\frac{\prod_{n=1}^{\infty}\calN_{\alpha\alpha}(Q_{\tau}^n;\ft,\fq)\calN_{\alpha\alpha}(Q_{\tau}^n\tfrac{\ft}{\fq};\ft,\fq)}{\tilde{Z}_{\alpha}(\ft,\fq)\tilde{Z}_{\alpha^t}(\fq,\ft)}
	=\frac{\ft^{\frac{||\alpha^t||^2}{2}}\fq^{\frac{||\alpha||^2}{2}}}{\left(\prod_{n=1}^{\infty}(1-Q_{\tau}^n)\right)^{2|\alpha|}}\varTheta_{\alpha\alpha}(\tfrac{\ft}{\fq}|Q_{\tau})\varTheta_{\alpha\alpha}(1|Q_{\tau})\ ,
	\label{eq:Ntotheta2}
\end{align}
where we have defined the $\varTheta$ function as follows: 
\begin{align}
	&\varTheta_{\alpha\beta}(Q|Q_{\tau})
	\equiv\prod_{(i,j)\in\alpha}\left((-i Q_{\tau}^{\frac18})^{-1}\theta_1(Q^{-1}\ft^{-\beta_j^t+i}\fq^{-\alpha_i+j-1}|Q_{\tau})\right)\nn\\
	=&\prod_{(i,j)\in\alpha}\left[\sum_{n=0}^{\infty}(-1)^nQ_{\tau}^{\frac{n(n+1)}{2}}\Big((Q^{-1}\ft^{-\beta_j^t+i}\fq^{-\alpha_i+j-1})^{n+\frac12}-(Q^{-1}\ft^{-\beta_j^t+i}\fq^{-\alpha_i+j-1})^{-n-\frac12}\Big)\right] \ ,\label{eq:Theta-definition}
\end{align}
where $\theta_1$ is the Jacobi theta function given as 
\begin{align}\label{eq:Jacobitheta1}
	\theta_1(y|Q_{\tau})=&i\sum_{n=-\infty}^{\infty}(-1)^{n+1}Q_{\tau}^{\frac12(n+\frac12)^2}y^{n+\frac12}=-iQ_{\tau}^{\frac18}\sum_{n=0}^{\infty}(-1)^nQ_{\tau}^{\frac{n(n+1)}{2}}(y^{n+\frac12}-y^{-n-\frac12})\ .
\end{align}
Note that unlike the $\calN_{\alpha\beta}$ function in \eqref{eq:curlyN}, the $\varTheta_{\alpha\beta}$ function has only the product over $\alpha$ in the definition \eqref{eq:Theta-definition}.

In \eqref{eq:nor_su2} and \eqref{eq:nor_submu1}, we encounter the following kind of factor
\begin{equation}
\prod_{n=1}^\infty\calN_{\sigma\rho}\big(Q_{\tau}^{n-1}\frac{Q_{\tau}}{Q_{\Delta}}\frac{\ft/\fq}{X}\big)\, \calN_{\rho\sigma}(Q_{\tau}^{n-1}Q_{\Delta}X),
\label{eq:NtothetaEpl}
\end{equation}
where the vertical position of the brane labeled by Young diagram $\rho$ is higher than that of the brane labeled by Young diagram $\sigma$ in the fundamental region\footnote{In figure \ref{fig:D4_quad}(a), the quadrivalent gluing subdiagrams are vertically periodic, we choose the particular one period shown in the figure as the fundamental region which has parallel NS-charged branes at the ends.} 
of the web diagram. Here, $Q_{\Delta}\equiv e^{-\uupbeta\,d}$ is the vertical distance in K\"ahler parameter between the two branes in the fundamental region where $d$ is the vertical distance, and $X$ represents  $\sqrt{\frac{\ft}{\fq}}$, $\frac{\ft}{\fq}$, or 1. 
When we apply \eqref{eq:Ntotheta1} to \eqref{eq:NtothetaEpl}, we have two choices:
\begin{align}
&Q_{\Delta}X\rightarrow Q\quad\text{and}\quad\rho,\sigma\rightarrow\alpha,\beta,
\label{eq:choice}
\end{align}
or
\begin{align}
&\frac{Q_{\tau}}{Q_{\Delta}}\frac{\ft/\fq}{X}\rightarrow Q\quad\text{and}\quad\sigma,\rho\rightarrow\alpha,\beta\ .
\end{align}
These two choices lead to two different expressions involving $\varTheta$ functions, and we choose 
\eqref{eq:choice}, which turns out to make the expression of the partition function simpler.

\paragraph{Expression of $Z^{D_4}$.}
Before we write down the final expression of the partition function for the $D_4$-type LST, we define some new parameters and symbols so that the result can be expressed in a compact form.

In the fundamental region, let us consider the $k$-th color brane of the $I$-th gauge node whose corresponding Young diagram is denoted by $\mu_{I,k}$. 
If the vertical position of the color brane is $a_{I,k}$, then we name the corresponding K\"ahler parameter as $y_{I,k}$
\begin{equation}\label{eq:defofy}
    y_{I,k}\equiv e^{-\uupbeta a_{I,k}}\ .
\end{equation}
In terms of $y_{I,k}$, we express the periodicity conditions \eqref{eq:D4PDcond1} and \eqref{eq:D4PDcond2} as 
\begin{align}
	y_{1,1}=1,\quad y_{2,1}=1,\quad y_{3,1}=1,\quad y_{4,1}=1,\quad y_{5,1}y_{5,2}=1,
\end{align}
where we have set the vertical position $a_{1,1}=0$. 
Next we define the following symbol
\begin{equation}
    {\textit{\textbf{n}}}_{I,k;J,\ell}\equiv\left\{\begin{array}{ll}
         n-1,&\quad\text{if}\ a_{I,k}>a_{J,\ell},\\
         n,&\quad\text{if}\ a_{I,k}<a_{J,\ell}\ \text{or}\ (I,k)=(J,\ell) \, .
    \end{array}\right. 
\end{equation}

By transforming $\calN$ functions to $\varTheta$ functions, as well as substituting \eqref{eq:D4ZSU21}, \eqref{eq:nor_su2}, \eqref{eq:Zofsubmu1} and \eqref{eq:nor_submu1} into \eqref{eq:Z_D4}, the partition function for the $D_4$-type LST 
$Z^{D_4}$ is expressed as

\begin{equation}
    Z^{D_4}=Z^{D_4}_{\text{pert}}\cdot Z^{D_4}_{\text{inst}}\ ,
\end{equation}
where
\begin{align}\label{eq:D4pert}
Z^{D_4}_{\text{pert}}
&=\prod_{n=1}^{\infty}\prod_{I=1}^4\frac{\calM(\frac{\ft}{\fq}Q_{\tau}^{{\textit{\textbf{n}}}_{I,1;I,1}})\calM(Q_{\tau}^{{\textit{\textbf{n}}}_{I,1;I,1}})}{\prod_{l=1}^2\calM(\frac{y_{I,1}}{y_{5,\ell}}\sqrt{\frac{\ft}{\fq}}Q_{\tau}^{{\textit{\textbf{n}}}_{I,1;5,l}})}\prod_{p=1}^2\frac{\prod_{r=1}^2\calM(\frac{y_{5,p}}{y_{5,r}}\frac{\ft}{\fq}Q_{\tau}^{{\textit{\textbf{n}}}_{5,p;5,r}})\calM(\frac{y_{5,p}}{y_{5,r}}Q_{\tau}^{{\textit{\textbf{n}}}_{5,p;5,r}})}{\prod_{J=1}^4\calM(\frac{y_{5,p}}{y_{J,1}}\sqrt{\frac{\ft}{\fq}}Q_{\tau}^{{\textit{\textbf{n}}}_{5,p;J,1}})}\nn\\
&\quad
\times\prod_{l=1}^{\infty}\frac{1}{(1-Q_{\tau}^l)^6},
\\
\label{eq:D4inst}
Z_{\text{inst}}^{D_4}
=&\sum_{\boldsymbol{\mu}}\prod_{I=1}^4\bigg(Q_{I}^{|\mu_{I,1}|}
\frac{\prod_{l=1}^{2}\varTheta_{\mu_{I,1}\mu_{5,l}}(\frac{y_{I,1}}{y_{5,l}}\sqrt{\tfrac{\ft}{\fq}}|Q_{\tau})}{\varTheta_{\mu_{I,1}\mu_{I,1}}(1|Q_{\tau})\varTheta_{\mu_{I,1}\mu_{I,1}}(\tfrac{\ft}{\fq}|Q_{\tau})}\bigg)\nonumber\\
\qquad
&\times\prod_{p=1}^{2}Q_{5}^{|\mu_{5,p}|}\frac{\prod_{J=1}^4\varTheta_{\mu_{5,p}\mu_{J,1}}(\frac{y_{5,p}}{y_{J,1}}\sqrt{\tfrac{\ft}{\fq}}|Q_{\tau})}{\prod_{r=1}^{2}\varTheta_{\mu_{5,p}\mu_{5,r}}(\frac{y_{5,p}}{y_{5,r}}|Q_{\tau})\varTheta_{\mu_{5,p}\mu_{5,r}}(\frac{y_{5,p}}{y_{5,r}}\tfrac{\ft}{\fq}|Q_{\tau})} \, .
\end{align}
 We will deal with a general higher rank case of LSTs in section \ref{sec:higherRanks},
 where we will write the perturbative part of partition functions in terms of Plethystic exponential (PE).

\subsection{LST on \texorpdfstring{$D_5$}{D5} singularity}
\begin{figure}[htbp]
    \centering
    \includegraphics[scale=0.5]{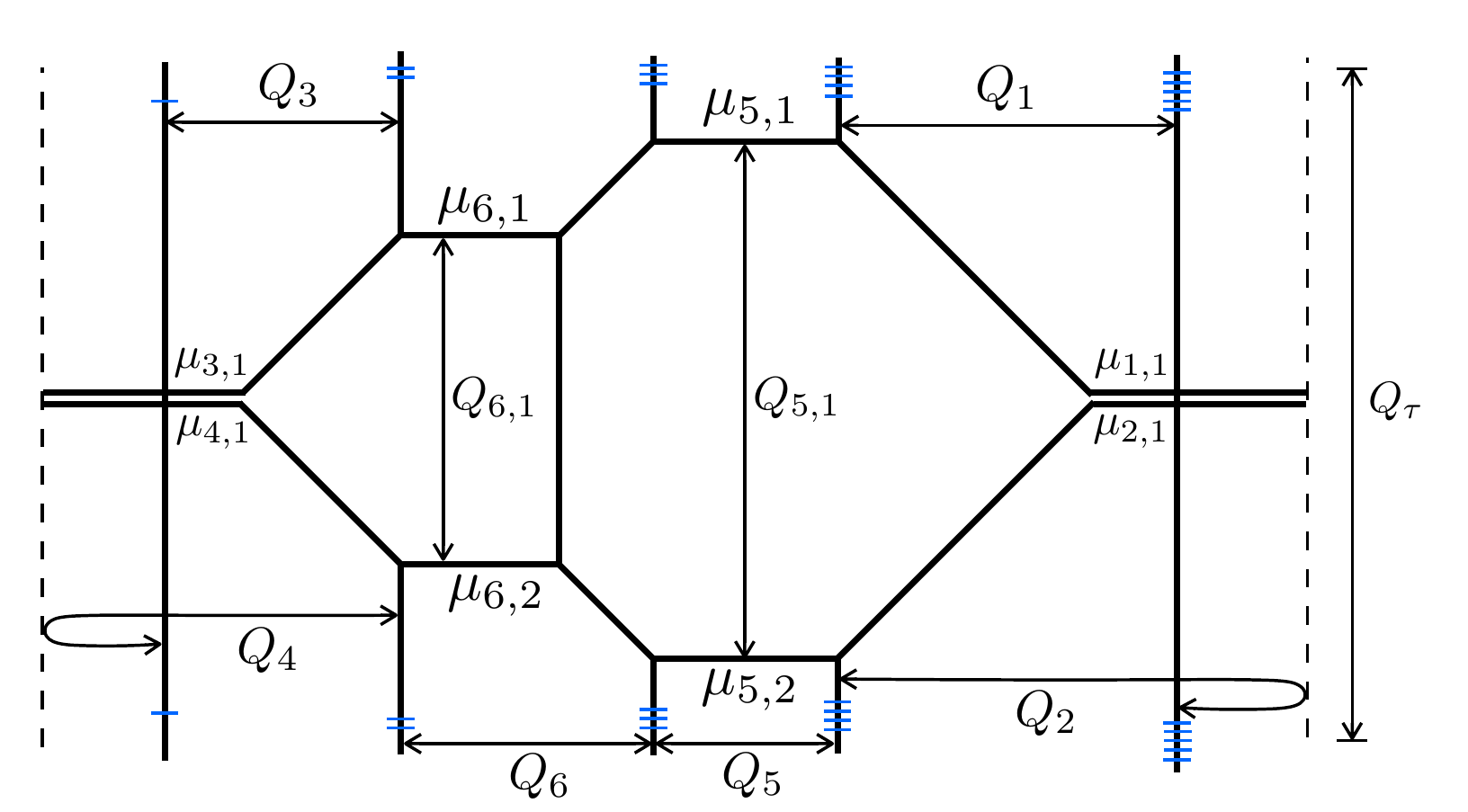}
    \caption{A Type IIB 5-brane web for $D_5$-type LST with two ON-planes. }
    \label{fig:D5wON}
\end{figure}
\begin{figure}[htbp]
    \centering
    \includegraphics[scale=0.5]{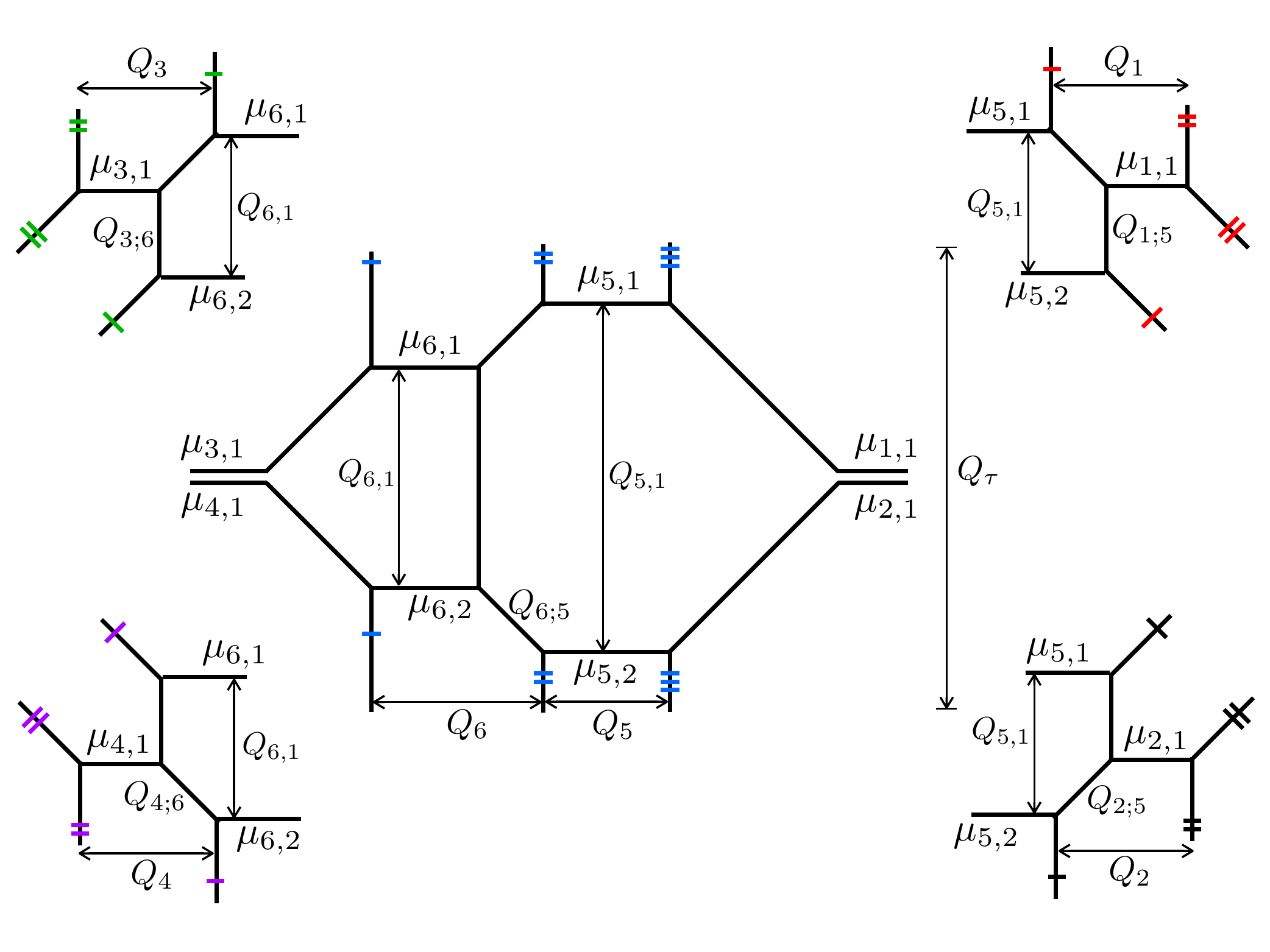}
    \caption{A brane web for $D_5$-type LST as trivalent gluings. The four subdiagrams in the corners correspond to four SU(1) gauge nodes, the middle subdiagram corresponds to SU(2)-SU(2) gauge nodes.}
    \label{fig:D5wQuadra}
\end{figure}

The rank-1 $D_5$-type LST can be realized by a web diagram with two ON-planes as depicted in figure \ref{fig:D5wON}, which describes 6d affine $D_5$ quiver,
\begin{align*}
	\begin{tikzpicture}
		\draw[thick](-1.4,.8)--(-0.6,0.2);
		\draw[thick](-1.4,-.8)--(-0.6,-0.2);
		\draw[thick](0.6+1.8,0.2)--(1.4+1.8,.8);
		\draw[thick](0.6+1.8,-0.2)--(1.4+1.8,-.8);
  		\draw[thick](0.6,0)--(1.2,0);
		\node at (0,0){SU$(2)$};
 		\node at (1.8,0){SU$(2)$};
		\node at (-2,1){SU$(1)$};
		\node at (-2,-1){SU$(1)$};
		\node at (2+1.8,1){SU$(1)$};
		\node at (2+1.8,-1){SU$(1)$};
		\node at (4.8,-1.15){.};
	\end{tikzpicture}
\end{align*}
We propose that the $D_5$-type LST can also be realized by trivalent gluings\footnote{The two SU(1) gauge nodes on the left and the right SU(2) gauge node are trivalently glued at the left SU(2) gauge node, the two SU(1) gauge nodes on the right and the left SU(2) gauge node are trivalently glued at the right SU(2) gauge node. } together with the bivalent self-gluings
as shown in figure \ref{fig:D5wQuadra}, where the 
Young diagram $\mu_{I,j}$ labels the $j$-th color brane of the $I$-th gauge node.

Like the $D_4$-type LST case, in order to do the bivalent self-gluing consistently for each SU(1) subdiagram in figure \ref{fig:D5wQuadra},
we impose the following conditions for vertical periodicity
\begin{align}\label{eq:D5PDcond1}
    Q_{1;5}=Q_{2;5}=Q_{5,1}^{\frac12},\quad Q_{3;6}=Q_{4;6}=Q_{6,1}^{\frac12}\ .
\end{align}
For the middle SU(2)-SU(2) subdiagram in figure \ref{fig:D5wQuadra} to be vertically periodic, the following condition has to be satisfied,
\begin{align}\label{eq:D5PDcond2}
    Q_{6;5}=Q_{5,1}^{\frac12}Q_{6,1}^{-\frac12}\ .
\end{align}
Then like in the $D_4$-type LST case, due to vertical periodicity $Q_1,Q_2,Q_3,Q_4,Q_5,Q_6$ are the instanton factors of the gauge node $1,2,3,4,5,6$ respectively in the 5d theory perspective, and they are also the tensor branch parameters of the 6d $D_5$-type LST. The horizontal period $u$ of the theory due to its affine $D_5$ quiver structure is then given by
\begin{align}\label{eq:uofD5}
	u=Q_1Q_2Q_3Q_4Q_5^2Q_6^2\ .
\end{align}

Using the trivalent gluing diagram in figure \ref{fig:D5wQuadra}, the partition function $Z^{D_5}$ of $D_5$-type LST is given by
\begin{align}
    Z^{D_5}=&\sum_{\mu_{5,1}\, \mu_{5,2}\, \mu_{6,1}\, \mu_{6,2}}  Z^{\text{SU(2)-SU(2)}}_{\mu_{5,1}\mu_{5,2}\mu_{6,1}\mu_{6,2}}\nn\\
    &\times \sum_{\mu_{1,1}}Z_{\mu_{5,1}\mu_{5,2}\mu_{1,1}}(Q_{5,1},Q_{\tau},Q_{1})\sum_{\mu_{2,1}}Z_{\mu_{5,1}\mu_{5,2}\mu_{2,1}}(Q_{5,1},Q_{\tau},Q_{2})\nn\\
    &\times \sum_{\mu_{3,1}}Z_{\mu_{6,1}\mu_{6,2}\mu_{3,1}}(Q_{6,1},Q_{\tau},Q_{3})\sum_{\mu_{4,1}}Z_{\mu_{6,1}\mu_{6,2}\mu_{4,1}}(Q_{6,1},Q_{\tau},Q_{4})\ ,
    \label{eq:ZofD5}
\end{align}
in which the summands are defined as follows: 
\begin{multline}
    Z^{\text{SU(2)-SU(2)}}_{\mu_{5,1}\mu_{5,2}\mu_{6,1}\mu_{6,2}}\equiv\left(-Q_{6}\sqrt{\frac{Q_{6,1}}{Q_{5,1}}}\right)^{|\mu_{6,1}|+|\mu_{6,2}|}(-Q_{5})^{|\mu_{5,1}|+|\mu_{5,2}|}f_{\mu_{5,1}}(\ft,\fq)f_{\mu_{5,2}}(\ft,\fq)^{-1}\\
    \times
    \begin{tikzpicture}[scale=.6,thick,baseline = -15]
	\draw (-0.2,-0.2)--++(-1,0) coordinate (c)--++(0,-1) coordinate (d)--++(1,0);
	\draw (c)--++(-0.9,0.9);
	\draw (d)--++(-0.9,-0.9);
	\node at (-0.6,0.15){$\mu_{6,1}$};
	\node at (-0.6,-1.55){$\mu_{6,2}$};
	\draw[<->] (0,-0.2)--++(0,-1);
	\node at (0.8,-0.7){$Q_{6,1}$};
	\draw[<->] (-2.4,0.7)--++(0,-2.8);
	\node at (-2.85,-0.7){$Q_{\tau}$};
	\draw (-2.43+0.8,-2.27+0.5)++(-0.05,-0.05)--++(-0.14,0.14);
	\draw (-2.43+0.8,0.87-0.5)++(-0.05,0.05)--++(-0.14,-0.14);
\end{tikzpicture}
	\times
	\begin{tikzpicture}[scale=.6,thick,baseline = -11.8]
		\draw (0,0)--++(1,0) coordinate (a)--++(0,-1) coordinate (b)--++(-1,0);
		\draw (a)--++(0.2,0.2) coordinate (c)--++(1.2,0);
		\draw (c)--++(0,0.7);
		\draw (b)--++(0.2,-0.2) coordinate (d)--++(1.2,0);
		\draw (d)--++(0,-0.7);
		\node at (0.5,0.35){$\mu_{6,1}$};
		\node at (0.5,-1.35){$\mu_{6,2}$};
		\node at (1.9,0.55){$\mu_{5,1}$};
		\node at (1.9,-1.55){$\mu_{5,2}$};
		\draw[<->] (-0.2,0)--++(0,-1);
		\node at (-0.85,-0.5){$Q_{6,1}$};
		\draw[<->] (2.6,0.2)--++(0,-1.4);
		\node at (3.3,-0.5){$Q_{5,1}$};
		\draw (1.1,0.7)--++(0.2,0);
		\draw (1.1,-1.7)--++(0.2,0);
		\draw[<->] (-1.55,0.9)--++(0,-2.8);
		\node at (-2,-0.5){$Q_{\tau}$};
	\end{tikzpicture}
	\times
	\begin{tikzpicture}[scale=.6,thick,baseline = -15]
		\draw (0,0)--++(1.4,0) coordinate (a)--++(0,-1.4) coordinate (b)--++(-1.4,0);
		\draw (a)--++(0.7,0.7);
		\draw (b)--++(0.7,-0.7);
		\node at (0.8,0.33){$\mu_{5,1}$};
		\node at (0.8,-1.06){$\mu_{5,2}$};
		\draw[<->] (-0.2,0)--++(0,-1.4);
		\node at (-0.85,-0.7){$Q_{5,1}$};
		\draw (2.43-0.8,-2.27+0.5)++(0.05,-0.05)--++(0.14,0.14);
		\draw (2.43-0.8,0.87-0.5)++(0.05,0.05)--++(0.14,-0.14);
		\draw[<->] (2.4,0.7)--++(0,-2.8);
		\node at (2.9,-0.7){$Q_{\tau}$};
	\end{tikzpicture}\ ,
	\label{eq:D5mid}
\end{multline}
where the edge factors of $\mu_{5,1},\mu_{5,2},\mu_{6,1},\mu_{6,2}$ come from the middle subdiagram of figure \ref{fig:D5wQuadra}. 
The $Z_{\mu_{\boldsymbol{\cdot}}\mu_{\boldsymbol{\cdot}}\mu_{\boldsymbol{\cdot}}}(\boldsymbol{\cdot},\boldsymbol{\cdot},\boldsymbol{\cdot})$ functions in \eqref{eq:ZofD5} which correspond to the four SU(1) subdiagrams of figure \ref{fig:D5wQuadra} are the same function  defined in \eqref{eq:sub_mu1} with indices and arguments accordingly changed.

After computing the three subdiagrams in \eqref{eq:D5mid} and then substituting the results back into \eqref{eq:D5mid}, we find
\begin{multline}
    Z^{\text{SU(2)-SU(2)}}_{\mu_{5,1}\mu_{5,2}\mu_{6,1}\mu_{6,2}}=\prod_{n=1}^{\infty}\calM(Q_{\tau}^n)^4\calM(Q_{\tau}^n\frac{\ft}{\fq})^4\calM(\frac{Q_{\tau}^n}{Q_{5,1}})\calM(\frac{Q_{\tau}^{n}}{Q_{5,1}}\frac{\ft}{\fq})\calM(Q_{\tau}^{n-1}Q_{5,1})\\
    \times\calM(Q_{\tau}^{n-1}Q_{5,1}\frac{\ft}{\fq})\calM(\frac{Q_{\tau}^n}{Q_{6,1}})\calM(\frac{Q_{\tau}^n}{Q_{6,1}}\frac{\ft}{\fq})\calM(Q_{\tau}^{n-1}Q_{6,1})\calM(Q_{\tau}^{n-1}Q_{6,1}\frac{\ft}{\fq})\\
    \times\Big(\calM(Q_{\tau}^nQ_{5,1}^{-\frac{1}{2}}Q_{6,1}^{-\frac{1}{2}}\sqrt{\frac{\ft}{\fq}})^2\calM(Q_{\tau}^{n-1}Q_{5,1}^{\frac{1}{2}}Q_{6,1}^{\frac{1}{2}}\sqrt{\frac{\ft}{\fq}})^2\calM(Q_{\tau}^nQ_{6,1}^{\frac{1}{2}}Q_{5,1}^{-\frac{1}{2}}\sqrt{\frac{\ft}{\fq}})^2\\
    \times\calM(Q_{\tau}^{n-1}Q_{5,1}^{\frac{1}{2}}Q_{6,1}^{-\frac{1}{2}}\sqrt{\frac{\ft}{\fq}})^2\Big)^{-1}\prod_{l=1}^{\infty}\frac{1}{(1-Q_{\tau}^l)^3}\times\hat{Z}^{\text{SU(2)-SU(2)}}_{\mu_{5,1}\mu_{5,2}\mu_{6,1}\mu_{6,2}}\ ,
    \label{eq:ZofD5mid}
\end{multline}
where
\begin{align}\label{eq:ZhatofD5mid}
    &\hat{Z}^{\text{SU(2)-SU(2)}}_{\mu_{5,1}\mu_{5,2}\mu_{6,1}\mu_{6,2}}\nn\\
    =&\fq^{\frac{2||\mu_{5,2}||^2+||\mu_{6,1}||^2+||\mu_{6,2}||^2}{2}}\ft^{\frac{2||\mu_{5,1}^t||^2+||\mu_{6,1}^t||^2+||\mu_{6,2}^t||^2}{2}}Q_{5}^{|\mu_{5,1}|+|\mu_{5,2}|}\left(-\frac{Q_{6}\sqrt{Q_{6,1}}}{\sqrt{Q_{5,1}}}\right)^{|\mu_{6,1}|+|\mu_{6,2}|}\nn\\
   & \times\prod_{I=5}^6\prod_{j=1}^2 Z_{\mu_{I,j}}(\ft,\fq)Z_{\mu_{I,j}^t}(\fq,\ft)\times\prod_{n=1}^{\infty}\calN_{\mu_{6,1}\mu_{5,1}}(Q_{\tau}^nQ_{6,1}^{\frac{1}{2}}Q_{5,1}^{-\frac{1}{2}}\sqrt{\frac{\ft}{\fq}})\calN_{\mu_{5,1}\mu_{6,1}}(Q_{\tau}^{n-1}Q_{5,1}^{\frac{1}{2}}Q_{6,1}^{-\frac{1}{2}}\sqrt{\frac{\ft}{\fq}})\nn\\
&    \times\calN_{\mu_{6,2}\mu_{5,1}}(Q_{\tau}^nQ_{5,1}^{-\frac{1}{2}}Q_{6,1}^{-\frac{1}{2}}\sqrt{\frac{\ft}{\fq}})\calN_{\mu_{5,1}\mu_{6,2}}(Q_{\tau}^{n-1}Q_{5,1}^{\frac{1}{2}}Q_{6,1}^{\frac{1}{2}}\sqrt{\frac{\ft}{\fq}})\calN_{\mu_{5,2}\mu_{6,1}}(Q_{\tau}^nQ_{5,1}^{-\frac{1}{2}}Q_{6,1}^{-\frac{1}{2}}\sqrt{\frac{\ft}{\fq}})\nn\\
  &  \times\calN_{\mu_{6,1}\mu_{5,2}}(Q_{\tau}^{n-1}Q_{5,1}^{\frac{1}{2}}Q_{6,1}^{\frac{1}{2}}\sqrt{\frac{\ft}{\fq}})\calN_{\mu_{5,2}\mu_{6,2}}(Q_{\tau}^nQ_{6,1}^{\frac{1}{2}}Q_{5,1}^{-\frac{1}{2}}\sqrt{\frac{\ft}{\fq}})\calN_{\mu_{6,2}\mu_{5,2}}(Q_{\tau}^{n-1}Q_{5,1}^{\frac{1}{2}}Q_{6,1}^{-\frac{1}{2}}\sqrt{\frac{\ft}{\fq}})\nn\\
   & \times\Big(\calN_{\mu_{5,1}\mu_{5,1}}(Q_{\tau}^n)\calN_{\mu_{5,1}\mu_{5,1}}(Q_{\tau}^n\frac{\ft}{\fq})\calN_{\mu_{5,1}\mu_{5,2}}(Q_{\tau}^{n-1}Q_{5,1})\calN_{\mu_{5,1}\mu_{5,2}}(Q_{\tau}^{n-1}Q_{5,1}\frac{\ft}{\fq})\nn\\
   &\times\calN_{\mu_{5,2}\mu_{5,1}}(\frac{Q_{\tau}^n}{Q_{5,1}})\calN_{\mu_{5,2}\mu_{5,1}}(\frac{Q_{\tau}^n}{Q_{5,1}}\frac{\ft}{\fq})\calN_{\mu_{5,2}\mu_{5,2}}(Q_{\tau}^n)\calN_{\mu_{5,2}\mu_{5,2}}(Q_{\tau}^n\frac{\ft}{\fq})\calN_{\mu_{6,1}\mu_{6,1}}(Q_{\tau}^n)\nn\\
    & \times \calN_{\mu_{6,1}\mu_{6,1}}(Q_{\tau}^n\frac{\ft}{\fq}) \calN_{\mu_{6,1}\mu_{6,2}}(Q_{\tau}^{n-1}Q_{6,1})\calN_{\mu_{6,1}\mu_{6,2}}(Q_{\tau}^{n-1}Q_{6,1}\frac{\ft}{\fq})\calN_{\mu_{6,2}\mu_{6,1}}(\frac{Q_{\tau}^n}{Q_{6,1}})\nn\\
    &\times \calN_{\mu_{6,2}\mu_{6,1}}(\frac{Q_{\tau}^n}{Q_{6,1}}\frac{\ft}{\fq})\calN_{\mu_{6,2}\mu_{6,2}}(Q_{\tau}^n)\calN_{\mu_{6,2}\mu_{6,2}}(Q_{\tau}^n\frac{\ft}{\fq})\Big)^{-1}\ .
\end{align}

\paragraph{Expression of $Z^{D_5}$.}
Using the $y$ parameters defined in \eqref{eq:defofy}, one readily finds that the periodicity conditions in \eqref{eq:D5PDcond1} and \eqref{eq:D5PDcond2} for the $D_5$-type LST become
\begin{align}
    y_{1,1}=1,\quad y_{2,1}=1,\quad y_{3,1}=1,\quad y_{4,1}=1,\quad y_{5,1}y_{5,2}=1,\quad y_{6,1}y_{6,2}=1,
\end{align}
where we have set $a_{1,1}=0$. 
Gathering all the results and rewriting $Z^{D_5}$ suitably with the help of the transformation from $\calN$ to $\varTheta$, we find
\begin{equation}\label{eq:ZofD5insec3}
    Z^{D_5}=Z^{D_5}_{\text{pert}}\cdot Z^{D_5}_{\text{inst}}\ ,
\end{equation}
where
\begin{align}\label{eq:D5pert}
Z^{D_5}_{\text{pert}}
=&\prod_{n=1}^{\infty}\prod_{I=1}^2\frac{\calM(\frac{\ft}{\fq}Q_{\tau}^{{\textit{\textbf{n}}}_{I,1;I,1}})\calM(Q_{\tau}^{{\textit{\textbf{n}}}_{I,1;I,1}})}{\prod_{l=1}^2\calM(\frac{y_{I,1}}{y_{5,l}}\sqrt{\frac{\ft}{\fq}}Q_{\tau}^{{\textit{\textbf{n}}}_{I,1;5,l}})}\prod_{J=3}^4\frac{\calM(\frac{\ft}{\fq}Q_{\tau}^{{\textit{\textbf{n}}}_{J,1;J,1}})\calM(Q_{\tau}^{{\textit{\textbf{n}}}_{J,1;J,1}})}{\prod_{k=1}^2\calM(\frac{y_{J,1}}{y_{6,k}}\sqrt{\frac{\ft}{\fq}}Q_{\tau}^{{\textit{\textbf{n}}}_{J,1;6,k}})}\nn\\
&\times\prod_{p=1}^2\frac{\prod_{c=1}^2\calM(\frac{y_{5,p}}{y_{5,c}}\frac{\ft}{\fq}Q_{\tau}^{{\textit{\textbf{n}}}_{5,p;5,c}})\calM(\frac{y_{5,p}}{y_{5,c}}Q_{\tau}^{{\textit{\textbf{n}}}_{5,p;5,c}})}{\prod_{r=1}^2\calM(\frac{y_{5,p}}{y_{6,r}}\sqrt{\frac{\ft}{\fq}}Q_{\tau}^{{\textit{\textbf{n}}}_{5,p;6,r}})\prod_{K=1}^2\calM(\frac{y_{5,p}}{y_{K,1}}\sqrt{\frac{\ft}{\fq}}Q_{\tau}^{{\textit{\textbf{n}}}_{5,p;K,1}})}\nn\\
&\times\prod_{s=1}^2\frac{\prod_{e=1}^2\calM(\frac{y_{6,s}}{y_{6,e}}\frac{\ft}{\fq}Q_{\tau}^{{\textit{\textbf{n}}}_{6,s;6,e}})\calM(\frac{y_{6,s}}{y_{6,e}}Q_{\tau}^{{\textit{\textbf{n}}}_{6,s;6,e}})}{\prod_{d=1}^2\calM(\frac{y_{6,s}}{y_{5,d}}\sqrt{\frac{\ft}{\fq}}Q_{\tau}^{{\textit{\textbf{n}}}_{6,s;5,d}})\prod_{L=3}^4\calM(\frac{y_{6,s}}{y_{L,1}}\sqrt{\frac{\ft}{\fq}}Q_{\tau}^{{\textit{\textbf{n}}}_{6,s;L,1}})}\nn\\
&\times\prod_{l=1}^{\infty}\frac{1}{(1-Q_{\tau}^l)^7},
\\
\label{eq:D5inst}
Z_{\text{inst}}^{D_5}
=&\sum_{\boldsymbol{\mu}}\prod_{I=1}^2\bigg(Q_{I}^{|\mu_{I,1}|}\frac{\prod_{l=1}^{2}\varTheta_{\mu_{I,1}\mu_{5,l}}(\frac{y_{I,1}}{y_{5,l}}\sqrt{\tfrac{\ft}{\fq}}|Q_{\tau})}{\varTheta_{\mu_{I,1}\mu_{I,1}}(\frac{y_{I,1}}{y_{I,1}}|Q_{\tau})\varTheta_{\mu_{I,1}\mu_{I,1}}(\frac{y_{I,1}}{y_{I,1}}\tfrac{\ft}{\fq}|Q_{\tau})}\bigg)\nn\\
\times&\prod_{J=3}^4\bigg(Q_{J}^{|\mu_{J,1}|}\frac{\prod_{k=1}^{2}\varTheta_{\mu_{J,1}\mu_{6,k}}(\frac{y_{J,1}}{y_{6,k}}\sqrt{\tfrac{\ft}{\fq}}|Q_{\tau})}{\varTheta_{\mu_{J,1}\mu_{J,1}}(\frac{y_{J,1}}{y_{J,1}}|Q_{\tau})\varTheta_{\mu_{J,1}\mu_{J,1}}(\frac{y_{J,1}}{y_{J,1}}\tfrac{\ft}{\fq}|Q_{\tau})}\bigg)\nonumber\\
\times&\prod_{p=1}^{2}\bigg(Q_{5}^{|\mu_{5,p}|}\frac{\prod_{r=1}^{2}\varTheta_{\mu_{5,p}\mu_{6,r}}(\frac{y_{5,p}}{y_{6,r}}\sqrt{\tfrac{\ft}{\fq}}|Q_{\tau})\prod_{K=1}^2\varTheta_{\mu_{5,p}\mu_{K,1}}(\frac{y_{5,p}}{y_{K,1}}\sqrt{\tfrac{\ft}{\fq}}|Q_{\tau})}{\prod_{c=1}^{2}\varTheta_{\mu_{5,p}\mu_{5,c}}(\frac{y_{5,p}}{y_{5,c}}|Q_{\tau})\varTheta_{\mu_{5,p}\mu_{5,c}}(\frac{y_{5,p}}{y_{5,c}}\tfrac{\ft}{\fq}|Q_{\tau})}\bigg)\nn\\
\times&\prod_{s=1}^{2}\bigg(Q_{6}^{|\mu_{6,s}|}\frac{\prod_{d=1}^{2}\varTheta_{\mu_{6,s}\mu_{5,d}}(\frac{y_{6,s}}{y_{5,d}}\sqrt{\tfrac{\ft}{\fq}}|Q_{\tau})\prod_{L=3}^4\varTheta_{\mu_{6,s}\mu_{L,1}}(\frac{y_{6,s}}{y_{L,1}}\sqrt{\tfrac{\ft}{\fq}}|Q_{\tau})}{\prod_{e=1}^{2}\varTheta_{\mu_{6,s}\mu_{6,e}}(\frac{y_{6,s}}{y_{6,e}}|Q_{\tau})\varTheta_{\mu_{6,s}\mu_{6,e}}(\frac{\mu_{6,s}}{\mu_{6,e}}\tfrac{\ft}{\fq}|Q_{\tau})}\bigg).
\end{align}

\subsection{\texorpdfstring{$E$}{E}-type LSTs}\label{sec:E6-type}
Here we apply our trivalent gluing with the bivalent self-gluing to $E$-type LSTs which are engineered from a IIB NS5-brane probing $E_6,E_7,E_8$ singularities. As the simplest example among the $E$-type LSTs, we compute the partition function of $E_6$-type LST and express the partition function in terms of the $\varTheta$ function.

\begin{figure}[H]
    \centering
    \includegraphics[scale=0.7]{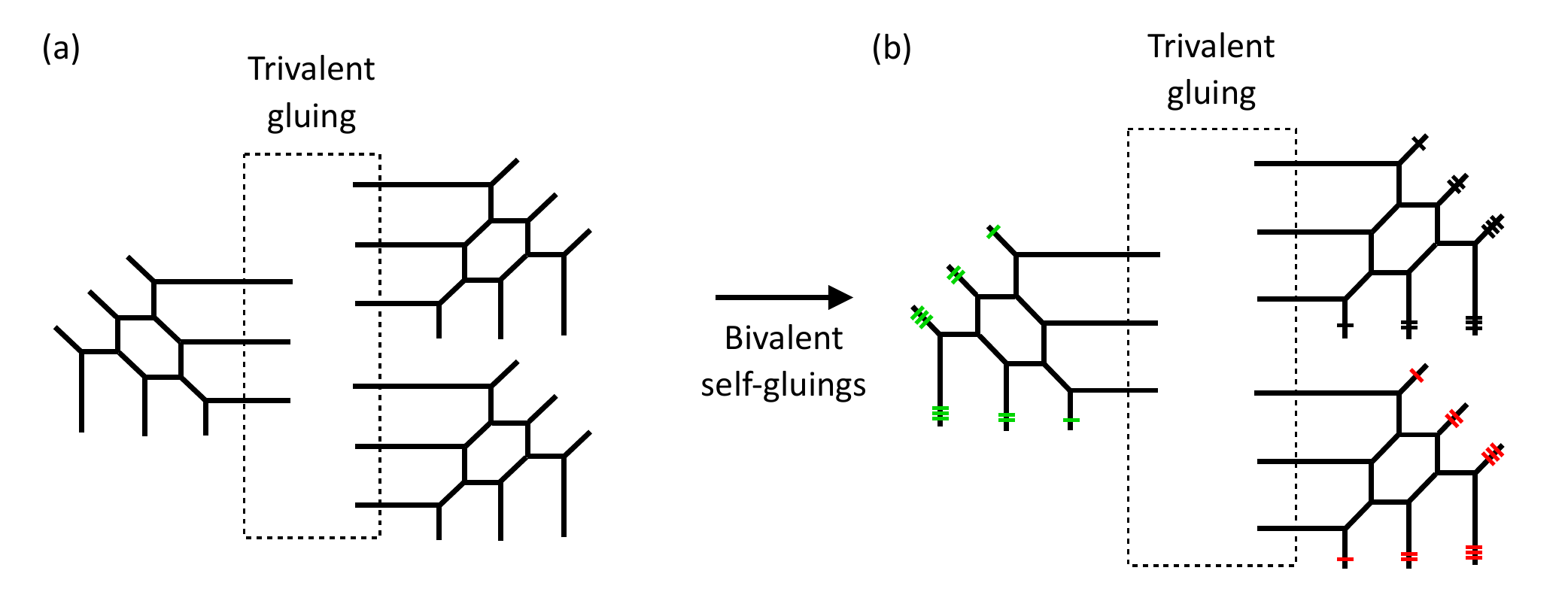}
    \caption{(a) Trivalent gluing of three $T_3$ web diagram realizing the minimal $(E_6,E_6)$ conformal matter on a circle.   
    (b) 5-brane for $E_6$-type LST with the bivalent self-gluing.}
    \label{fig:E6-triv-T3}
\end{figure}

Recall first that the minimal $(E_6, E_6)$ conformal matter theory on a circle is realized as affine $E_6$ quiver~\cite{DelZotto:2014hpa}:
\begin{align}
\label{eq:affineE6web}
\begin{tikzpicture}[scale=1.1]
      \path (0:0cm) node[scale=1.0] (v0) {$SU(3)_0$};
      \path (90:1cm) node (u2) {$SU(2)$};
      \path (90:2cm) node (u1) {$SU(1)$};
      \path (0:2cm) node (r2) {$SU(2)$};
      \path (0:4cm) node (r1) {$SU(1)$};
      \path (180:2cm) node (l2) {$SU(2)$};
      \path (180:4cm) node (l1) {$SU(1)$};
      \draw (v0) -- (u2) -- (u1)
             (v0) -- (r2) -- (r1)
             (v0) -- (l2) -- (l1);
      \node at (4.85,-0.15){.};
\end{tikzpicture}
\end{align}
The corresponding 5-brane web for this theory was proposed via trivalent gluing~\cite{Hayashi:2021pcj} of three $T_3$ web diagrams, as shown in figure~\ref{fig:E6-triv-T3}(a). 
To construct the $E_6$-type LST, we perform the bivalent self-gluings of the external NS-charged branes, which leads to figure \ref{fig:E6-triv-T3}(b).

\begin{figure}[H]
    \centering
    \includegraphics[scale=0.65]{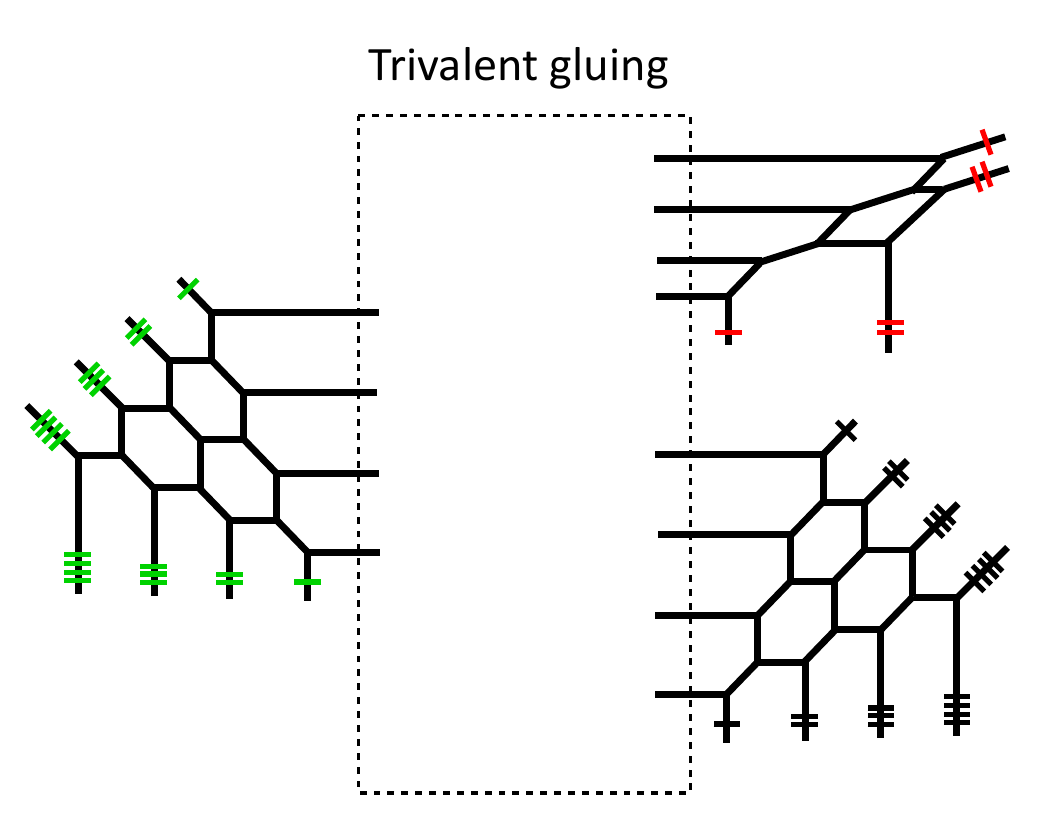}
    \caption{5-brane for $E_7$ LST with the bivalent self-gluing.}
    \label{fig:E7-triv-T4-gluing}
\end{figure}

A construction of brane web for $E_7$-type LST is very similar to the case for $E_6$-type LST.
The minimal $(E_7,E_7)$ conformal matter theory on a circle is realized as affine $E_7$ quiver:
\begin{align}
\label{eq:affineE7web}
\begin{tikzpicture}[scale=1.1]
      \path (0:0cm) node[scale=1.0] (v0) {$SU(4)_0$};
      \path (90:1cm) node (u2) {$SU(2)_0$};
      \path (0:2cm) node (r3) {$SU(3)_0$};
      \path (0:4cm) node (r2) {$SU(2)$};
      \path (0:6cm) node (r1) {$SU(1)$};
      \path (180:2cm) node (l3) {$SU(3)_0$};
      \path (180:4cm) node (l2) {$SU(2)$};
      \path (180:6cm) node (l1) {$SU(1)$};
      \draw (v0) -- (u2) 
             (v0) -- (r3) -- (r2) -- (r1)
             (v0) -- (l3)-- (l2) -- (l1);
      \node at (6.85,-0.15){.};
\end{tikzpicture}
\end{align}
To make a brane web for $E_7$-type LST, we perform the bivalent self-gluings of the external NS-charged branes, as given in figure \ref{fig:E7-triv-T4-gluing}.

\begin{figure}[H]
    \centering
    \includegraphics[width=9cm]{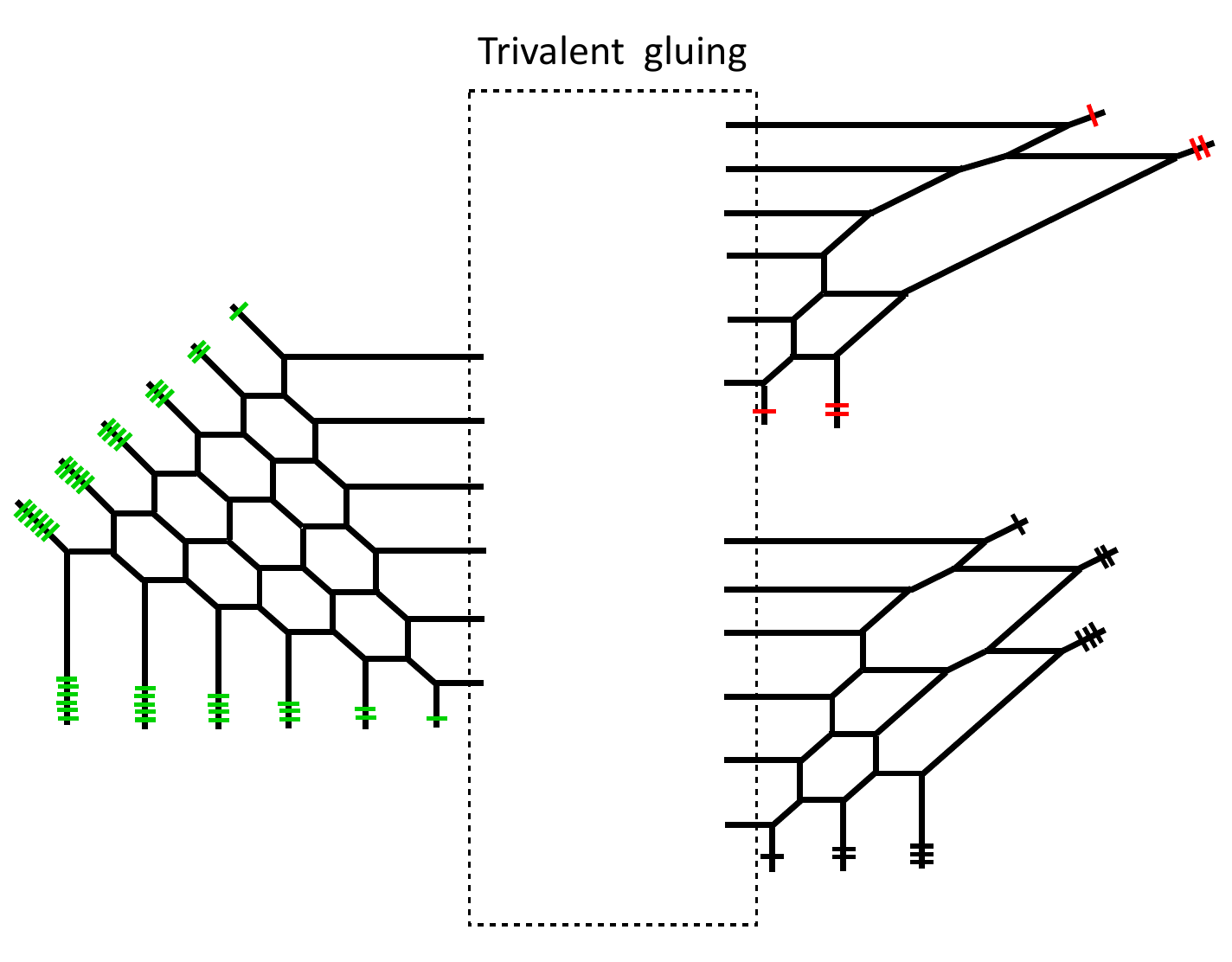}
    \caption{5-brane for $E_8$ LST with the bivalent self-gluings.}
    \label{fig:E8-triv-gluing}
\end{figure}

In the same fashion, to make a brane web for $E_8$-type LST which has the affine $E_8$ quiver:
\begin{align}
\label{eq:affineE8web}
\begin{tikzpicture}[scale=1.1]
      \path (0:0cm) node[scale=1.0] (v0) {$SU(6)_0$};
      \path (90:1cm) node (u2) {$SU(3)_0$};
      \path (0:1.75cm) node (r3) {$SU(4)_0$};
      \path (0:3.5cm) node (r2) {$SU(2)_0$};
      \path (180:1.75cm) node (l6) {$SU(5)_0$};
      \path (180:3.5cm) node (l5) {$SU(4)_0$};
      \path (180:5.25cm) node (l4) {$SU(3)_0$};
      \path (180:7cm) node (l3) {$SU(2)$};
      \path (180:8.65cm) node (l2) {$SU(1)$};
%
      \draw (v0) -- (u2) 
             (v0) -- (r3) -- (r2) 
             (v0) -- (l6) -- (l5) -- (l4)-- (l3)-- (l2) ;
             \node at (4.35,-0.15){,};
\end{tikzpicture}
\end{align}
we perform the bivalent self-gluings of the external NS-charged branes for the web diagram of the minimal $(E_8, E_8)$ conformal matter theory on a circle, as given in figure \ref{fig:E8-triv-gluing}. 


\subsubsection{LST on \texorpdfstring{$E_6$}{E6} singularity}
\begin{figure}[htbp]
    \centering
    \includegraphics[scale=0.5]{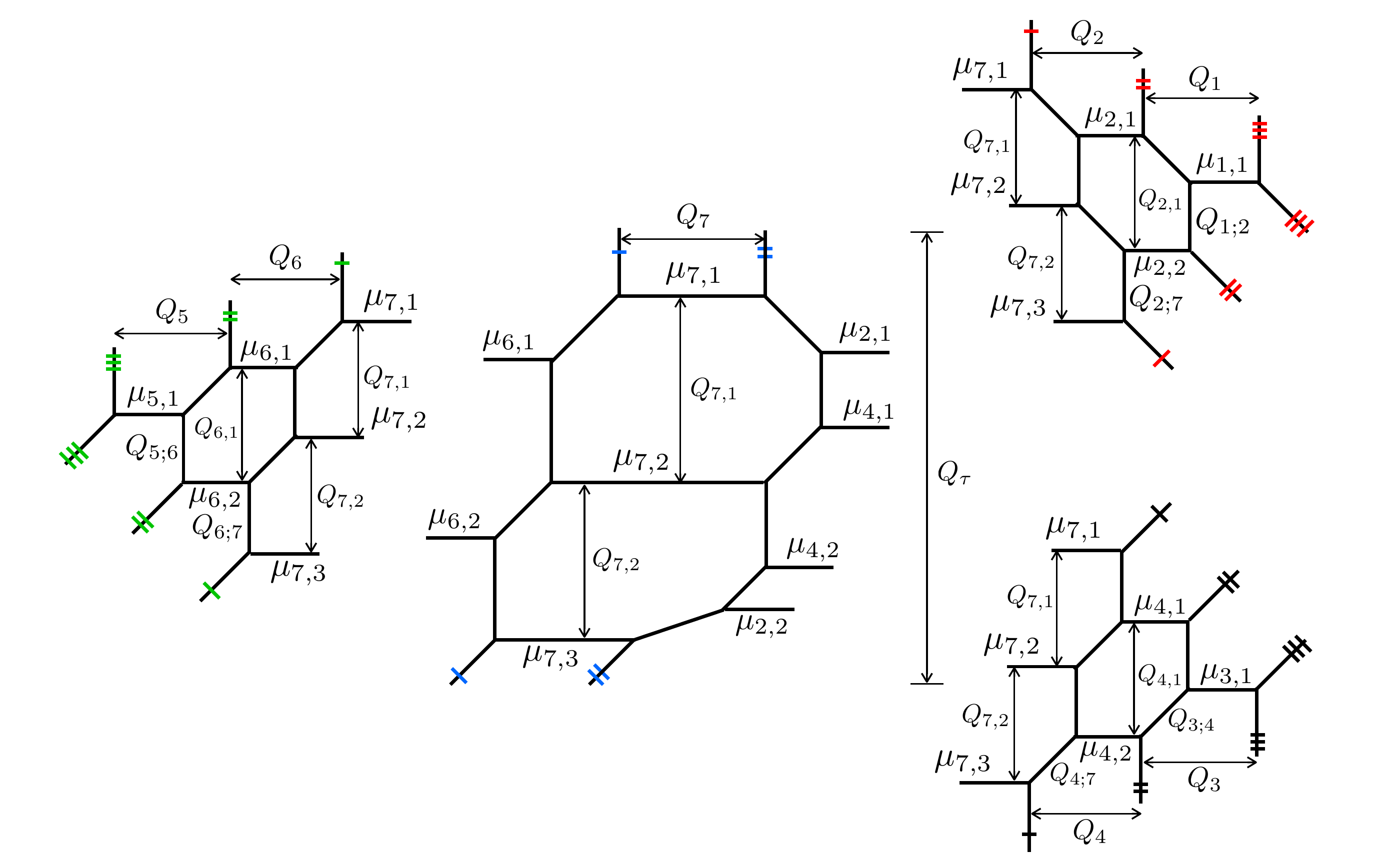}
    \caption{The $E_6$-type LST by trivalent gluing together with the bivalent self-gluing. The three subdiagrams in the corners correspond to SU(2)-SU(1) gauge nodes, the middle subdiagram corresponds to SU(3) gauge node. }
    \label{fig:E6tri}
\end{figure}
\begin{figure}[t]
    \centering
    \includegraphics[scale=0.5]{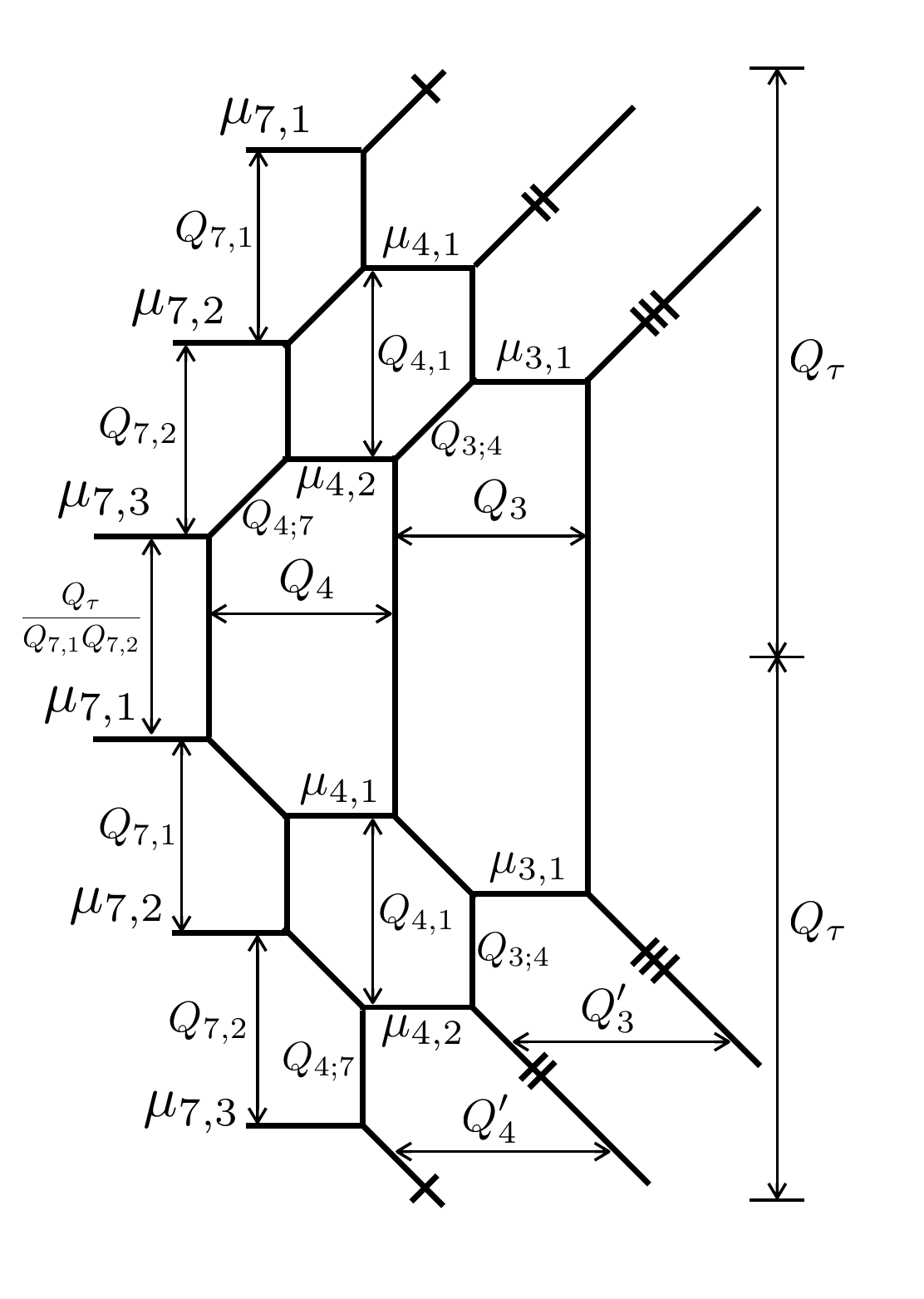}
    \caption{Subdiagram of $E_6$-type LST with two periods.}
    \label{fig:T3sub}
\end{figure}

In this subsection, we compute the partition function of $E_6$-type LST from its trivalent gluing brane webs.
We propose that the $E_6$-type LST can be explicitly realized by the trivalent gluing together with the bivalent self-gluings as given in figure \ref{fig:E6tri}, where the Young diagram $\mu_{I,j}$ label the $j$-th color brane of the $I$-th gauge node. In figure \ref{fig:T3sub}, we also draw two adjacent periods of the bottom right subdiagram of figure \ref{fig:E6tri}. By the periodicity constraints, $Q_{4}=Q'_{4}$ and $Q_{3}=Q'_{3}$, we find that $Q_{4;7}$ and $Q_{3;4}$ have to satisfy the following conditions,
\begin{align}
    Q_{4;7}=Q_{7,1}^{\frac13}Q_{7,2}^{\frac23}Q_{4,1}^{-\frac12},\quad Q_{3;4}=Q_{4,1}^{\frac12}\ .
    \label{eq:t3sub2}
\end{align}
By the same reasoning, for the other two 
SU(2)-SU(1) subdiagrams, we also have
\begin{align}
    &Q_{2;7}=Q_{7,1}^{\frac13}Q_{7,2}^{\frac23}Q_{2,1}^{-\frac12},\qquad Q_{1;2}=Q_{2,1}^{\frac12}\ ,
    \label{eq:t3sub1}\\
    &Q_{6;7}=Q_{7,1}^{\frac13}Q_{7,2}^{\frac23}Q_{6,1}^{-\frac12},\qquad Q_{5;6}=Q_{6,1}^{\frac12}\ .
    \label{eq:t3sub3}
\end{align}
These three periodicity conditions, \eqref{eq:t3sub2}, \eqref{eq:t3sub1}, and \eqref{eq:t3sub3}, also make the middle SU(3) subdiagram of figure \ref{fig:E6tri} vertically periodic.
Then due to the vertical periodicity $Q_1,Q_2,Q_3,Q_4,Q_5,Q_6,Q_7$ are the instanton factors of the gauge node $1,2,3,4,5,6,7$ respectively in the 5d theory perspective, and they are also the tensor branch parameters of the 6d $E_6$-type LST. The horizontal period $u$ of the theory due to its affine $E_6$ quiver structure is then given by
\begin{align}\label{eq:E6horizontal}
	u=Q_1Q_2^2Q_3Q_4^2Q_5Q_6^2Q_7^3\ .
\end{align}

Now we are ready to use the trivalent gluing diagram in figure \ref{fig:E6tri} to obtain the partition function of $E_6$-type LST which takes the following form,
\begin{align}
    Z^{E_6}=\sum_{\boldsymbol{\mu}}Z^{\text{SU(3)}}_{\mu_{7,1}\mu_{7,2}\mu_{7,3}}&\times Z_{\mu_{7,1}\mu_{7,2}\mu_{7,3}\mu_{2,1}\mu_{2,2}\mu_{1,1}}(Q_{7,1},Q_{7,2},Q_{2,1},Q_{1},Q_{2},Q_{\tau})\nn\\
    &\times Z_{\mu_{7,1}\mu_{7,2}\mu_{7,3}\mu_{4,1}\mu_{4,2}\mu_{3,1}}(Q_{7,1},Q_{7,2},Q_{4,1},Q_{3},Q_{4},Q_{\tau})\nn\\
    &\times Z_{\mu_{7,1}\mu_{7,2}\mu_{7,3}\mu_{6,1}\mu_{6,2}\mu_{5,1}}(Q_{7,1},Q_{7,2},Q_{6,1},Q_{5},Q_{6},Q_{\tau}),
    \label{eq:factorsofE6}
\end{align}
with
\begin{align}
    Z^{\text{SU(3)}}_{\mu_{7,1}\mu_{7,2}\mu_{7,3}}\equiv&(-Q_{7})^{|\mu_{7,1}|+|\mu_{7,3}|}f_{\mu_{7,1}}(\ft,\fq)f_{\mu_{7,3}}(\ft,\fq)^{-1}\left(-\frac{Q_{7,1}Q_{7,2}Q_{7}}{Q_{2,1}^{\frac12}Q_{4,1}^{\frac12}Q_{6,1}^{\frac12}}\right)^{|\mu_{7,2}|}\nonumber\\
    &\times
    \begin{tikzpicture}[scale=0.6,thick,baseline=-15]
		\draw (0,0)--++(-1,0) coordinate (a)--++(0,-1) coordinate (b)--++(1,0);
		\draw (a)--++(-0.8,0.8);
		\draw (b)--++(-0.8,-0.8) coordinate (c)--++(1.8,0);
		\draw (c)--++(-0.8,-0.4);
		\node at (-0.4,0.35){$\mu_{7,1}$};
		\node at (-0.4,-0.6){$\mu_{7,2}$};
		\node at (-0.6,-1.45){$\mu_{7,3}$};
		\draw[<->] (0.2,0)--++(0,-1);
		\node at (1,-0.5){$Q_{7,1}$};
		\draw[<->] (0.2,-1)--++(0,-0.8);
		\node at (1,-1.4){$Q_{7,2}$};
		\draw[<->] (-2.8,0.8)--++(0,-3);
		\node at (-3.3,-0.7){$Q_{\tau}$};
		\draw (-1.6+0.07,0.6+0.07)--++(-0.14,-0.14);
		\draw (-2.2-0.0445,-2+0.0895)--++(0.089,-0.179);
	\end{tikzpicture}
    \times
    \begin{tikzpicture}[scale=0.6,thick,baseline=-15]
		\draw (0,0)--++(1,0) coordinate (a)--++(0,-1) coordinate (b)--++(-1,0);
		\draw (a)--++(0.8,0.8);
		\draw (b)--++(0.8,-0.8) coordinate (c)--++(-1.8,0);
		\draw (c)--++(0.8,-0.4);
		\node at (0.5,0.35){$\mu_{7,1}$};
		\node at (0.4,-0.6){$\mu_{7,2}$};
		\node at (0.7,-1.45){$\mu_{7,3}$};
		\draw[<->] (-0.2,0)--++(0,-1);
		\node at (-1,-0.5){$Q_{7,1}$};
		\draw[<->] (-0.2,-1)--++(0,-0.8);
		\node at (-1,-1.4){$Q_{7,2}$};
		\draw[<->] (2.8,0.8)--++(0,-3);
		\node at (3.3,-0.7){$Q_{\tau}$};
		\draw (1.6-0.07,0.6+0.07)--++(0.14,-0.14);
		\draw (2.2+0.0445,-2+0.0895)--++(-0.089,-0.179);
	\end{tikzpicture},
	\label{eq:ZE6mid}
\end{align}
where the edge factors of $\mu_{7,1},\mu_{7,2},\mu_{7,3}$ come from the middle SU(3) diagram in figure \ref{fig:E6tri},
and
\begin{align}   &Z_{\mu_{7,1}\mu_{7,2}\mu_{7,3}\mu_{2,1}\mu_{2,2}\mu_{1,1}}(Q_{7,1},Q_{7,2},Q_{2,1},Q_{1},Q_{2},Q_{\tau})\nonumber\\
    &\qquad\qquad
=\begin{tikzpicture}[scale=0.6,thick,baseline=-18.5]
		\draw (0,0)--++(1,0) coordinate (a)--++(0,-0.5) coordinate (b)--++(-0.5,-0.5) coordinate (c)--++(-0.8,0);
		\draw (c)--++(0,-0.37) coordinate (d)--++(-0.43,-0.43) coordinate (e)--++(-0.6,0);
		\draw (b)--++(0.8,0) coordinate (f)--++(0,-0.44) coordinate (g)--++(-0.43,-0.43) coordinate (h)--(d);
		\draw (a)--++(0.5,0.5);
		\draw (f)--++(0.5,0.5);
		\draw (g)--++(0.6,0) coordinate (i)--++(0.5,0.5);
		\draw (e)--++(0,-0.5);
		\draw (h)--++(0,-0.7);
		\draw (i)--++(0,-1);
		\draw[<->] (-0.75,0)--++(0,-1);
		\draw[<->] (0.07,-2.4)--++(1.3,0);
		\draw[<->] (1.37,-2.2)--++(1.03,0);
		\draw[<->] (1.28,-0.5)--++(0,-0.87);
		\node[scale=0.5] at (0.96,-0.99){$Q_{2,1}$};
		\node at (-1.6,-0.5){$Q_{7,1}$};
		\draw[<->] (-0.75,-1)--++(0,-0.8);
		\node at (-1.6,-1.4){$Q_{7,2}$};
		\draw[<->] (3.2,0.6)--++(0,-3);
		\node at (3.7,-0.7){$Q_{\tau}$};
		\node[scale=0.6] at (0.5,0.23){$\mu_{7,1}$};
		\node[scale=0.6] at (0.1,-0.75){$\mu_{7,2}$};
		\node[scale=0.6] at (-0.15,-1.55){$\mu_{7,3}$};
		\node[scale=0.5] at (0.75,-2.15){$Q_{2}$};
		\node[scale=0.5] at (1.9,-1.95){$Q_{1}$};
		\node[scale=0.6] at (1.45,-0.3){$\mu_{2,1}$};
		\node[scale=0.6] at (0.9,-1.6){$\mu_{2,2}$};
		\node[scale=0.6] at (2.18,-0.75){$\mu_{1,1}$};
		\draw (1.23,0.37)--++(0.14,-0.14);
		\draw (e)++(-0.1,-0.3)--++(0.2,0);
		\draw (f)++(0.23,0.37)--++(0.14,-0.14);
		\draw (f)++(0.33-0.04,0.47-0.04)--++(0.14,-0.14);
		\draw (h)++(-0.1,-0.4)--++(0.2,0);
		\draw (h)++(-0.1,-0.48)--++(0.2,0);
		\draw (i)++(0.23,0.37)--++(0.14,-0.14);
		\draw (i)++(0.23+0.06,0.37+0.06)--++(0.14,-0.14);
		\draw (i)++(0.23-0.06,0.37-0.06)--++(0.14,-0.14);
		\draw (i)++(-0.1,-0.6)--++(0.2,0);
		\draw (i)++(-0.1,-0.68)--++(0.2,0);
		\draw (i)++(-0.1,-0.52)--++(0.2,0);
	\end{tikzpicture}
	\Bigg/
	\begin{tikzpicture}[scale=0.6,thick,baseline=-15]
		\draw (0,0)--++(1,0) coordinate (a)--++(0,-1) coordinate (b)--++(-1,0);
		\draw (a)--++(0.8,0.8);
		\draw (b)--++(0.8,-0.8) coordinate (c)--++(-1.8,0);
		\draw (c)--++(0.8,-0.4);
		\node[scale=0.6] at (0.5,0.23){$\mu_{7,1}$};
		\node[scale=0.6] at (0.5,-0.75){$\mu_{7,2}$};
		\node[scale=0.6] at (0.7,-1.55){$\mu_{7,3}$};
		\draw[<->] (-0.2,0)--++(0,-1);
		\node at (-1.05,-0.5){$Q_{7,1}$};
		\draw[<->] (-0.2,-1)--++(0,-0.8);
		\node at (-1.05,-1.4){$Q_{7,2}$};
		\draw[<->] (2.8,0.8)--++(0,-3);
		\node at (3.3,-0.7){$Q_{\tau}$};
		\draw (1.6-0.07,0.6+0.07)--++(0.14,-0.14);
		\draw (2.2+0.0445,-2+0.0895)--++(-0.089,-0.179);
	\end{tikzpicture}.
	\label{eq:Zt3sub1}
\end{align}
The other two $Z$'s in \eqref{eq:factorsofE6} are also defined in the same way as the one in \eqref{eq:Zt3sub1} with indices and arguments accordingly changed as they correspond to the three SU(2)-SU(1) subdiagrams of figure \ref{fig:E6tri} which are related to each other by $SL(2,\mathbb{Z})$ or a mirror reflection.

We compute the two diagrams in \eqref{eq:ZE6mid}, substitute the results back and obtain
\begin{align}
&Z^{\text{SU(3)}}_{\mu_{7,1}\mu_{7,2}\mu_{7,3}}\nonumber\\
=&\prod_{n=1}^{\infty}\calM(Q_{\tau}^{n-1}Q_{7,1})\calM(Q_{\tau}^{n-1}Q_{7,2})\calM(Q_{\tau}^{n-1}Q_{7,1}Q_{7,2})\calM(Q_{\tau}^{n})^3\calM(\frac{Q_{\tau}^n}{Q_{7,1}})\calM(\frac{Q_{\tau}^n}{Q_{7,2}})\nonumber\\
    \times&\calM(\frac{Q_{\tau}^n}{Q_{7,1}Q_{7,2}})\calM(Q_{\tau}^{n-1}Q_{7,1}\frac{\ft}{\fq})\calM(Q_{\tau}^{n-1}Q_{7,2}\frac{\ft}{\fq})\calM(Q_{\tau}^{n-1}Q_{7,1}Q_{7,2}\frac{\ft}{\fq})\calM(Q_{\tau}^{n}\frac{\ft}{\fq})^3\nonumber\\
    \times&\calM(\frac{Q_{\tau}^n}{Q_{7,1}}\frac{\ft}{\fq})\calM(\frac{Q_{\tau}^n}{Q_{7,2}}\frac{\ft}{\fq})\calM(\frac{Q_{\tau}^n}{Q_{7,1}Q_{7,2}}\frac{\ft}{\fq})\prod_{l=1}^{\infty}\frac{1}{(1-Q_{\tau}^l)^2}\times \hat{Z}^{\text{SU(3)}}_{\mu_{7,1}\mu_{7,2}\mu_{7,3}},
    \label{eq:ZofE6mid}
\end{align}
where
\begin{align}
    &\hat{Z}^{\text{SU(3)}}_{\mu_{7,1}\mu_{7,2}\mu_{7,3}}\nonumber\\
    =&\fq^{\frac{1}{2}||\mu_{7,2}||^2+||\mu_{7,3}||^2}\ft^{||\mu_{7,1}^t||^2+\frac{1}{2}||\mu_{7,2}^t||^2}Q_{7}^{|\mu_{7,1}|+|\mu_{7,3}|}\left(-\frac{Q_{7,1}Q_{7,2}Q_{7}}{Q_{2,1}^{\frac12}Q_{4,1}^{\frac12}Q_{6,1}^{\frac12}}\right)^{|\mu_{7,2}|}\prod_{i=1}^3\tilde{Z}_{\mu_{7,i}}(\ft,\fq)\tilde{Z}_{\mu_{7,i}^t}(\fq,\ft)\nonumber\\
    &\times\prod_{n=1}^{\infty}\calN_{\mu_{7,1}\mu_{7,1}}(Q_{\tau}^n)\calN_{\mu_{7,1}\mu_{7,1}}(Q_{\tau}^n\frac{\ft}{\fq})\calN_{\mu_{7,2}\mu_{7,2}}(Q_{\tau}^n)\calN_{\mu_{7,2}\mu_{7,2}}(Q_{\tau}^n\frac{\ft}{\fq})\calN_{\mu_{7,3}\mu_{7,3}}(Q_{\tau}^n)\nonumber\\
    &\times\calN_{\mu_{7,3}\mu_{7,3}}(Q_{\tau}^n\frac{\ft}{\fq})\calN_{\mu_{7,2}\mu_{7,1}}(\frac{Q_{\tau}^n}{Q_{7,1}}\frac{\ft}{\fq})\calN_{\mu_{7,1}\mu_{7,2}}(Q_{\tau}^{n-1}Q_{7,1})\calN_{\mu_{7,2}\mu_{7,1}}(\frac{Q_{\tau}^n}{Q_{7,1}})\nonumber\\
    &\times\calN_{\mu_{7,1}\mu_{7,2}}(Q_{\tau}^{n-1}Q_{7,1}\frac{\ft}{\fq})\calN_{\mu_{7,3}\mu_{7,1}}(\frac{Q_{\tau}^n}{Q_{7,1}Q_{7,2}}\frac{\ft}{\fq})\calN_{\mu_{7,1}\mu_{7,3}}(Q_{\tau}^{n-1}Q_{7,1}Q_{7,2})\nonumber\\
    &\times\calN_{\mu_{7,3}\mu_{7,1}}(\frac{Q_{\tau}^n}{Q_{7,1}Q_{7,2}})\calN_{\mu_{7,1}\mu_{7,3}}(Q_{\tau}^{n-1}Q_{7,1}Q_{7,2}\frac{\ft}{\fq})\calN_{\mu_{7,3}\mu_{7,2}}(\frac{Q_{\tau}^n}{Q_{7,2}}\frac{\ft}{\fq})\calN_{\mu_{7,2}\mu_{7,3}}(Q_{\tau}^{n-1}Q_{7,2})\nonumber\\
    &\times\calN_{\mu_{7,3}\mu_{7,2}}(\frac{Q_{\tau}^n}{Q_{7,2}})\calN_{\mu_{7,2}\mu_{7,3}}(Q_{\tau}^{n-1}Q_{7,2}\frac{\ft}{\fq}).
    \label{eq:NZofE6mid}
\end{align}
We then compute the two diagrams in \eqref{eq:Zt3sub1} and substitute the results back which yields 
\begin{align}
    &Z_{\mu_{7,1}\mu_{7,2}\mu_{7,3}\mu_{2,1}\mu_{2,2}\mu_{1,1}}(Q_{7,1},Q_{7,2},Q_{2,1},Q_{1},Q_{2},Q_{\tau})\nonumber\\
    =&\prod_{l=1}^{\infty}\frac{1}{(1-Q_{\tau}^l)^2}\prod_{n=1}^{\infty}\calM(Q_{\tau}^{n-1}Q_{2,1})\calM(Q_{\tau}^n)^3\calM(Q_{\tau}^{n}Q_{2,1}^{-1})\calM(Q_{\tau}^{n-1}Q_{2,1}\frac{\ft}{\fq})\nonumber\\
    \times&\calM(Q_{\tau}^n\frac{\ft}{\fq})^3\calM(Q_{\tau}^nQ_{2,1}^{-1}\frac{\ft}{\fq})\times\Bigg(\calM(Q_{\tau}^{n-1}Q_{7,1}^{\frac13}Q_{7,2}^{-\frac13}Q_{2,1}^{\frac12}\sqrt{\frac{\ft}{\fq}})\calM(Q_{\tau}^{n-1}Q_{7,1}^{\frac23}Q_{7,2}^{\frac13}Q_{2,1}^{-\frac12}\sqrt{\frac{\ft}{\fq}})\nonumber\\
    \times&\calM(Q_{\tau}^{n-1}Q_{2,1}^{\frac12}\sqrt{\frac{\ft}{\fq}})^2\calM(Q_{\tau}^{n-1}Q_{7,1}^{\frac{1}{3}}Q_{7,2}^{\frac{2}{3}}Q_{2,1}^{-\frac12}\sqrt{\frac{\ft}{\fq}})\calM(Q_{\tau}^{n-1}Q_{7,1}^{-\frac{1}{3}}Q_{7,2}^{\frac{1}{3}}Q_{2,1}^{\frac12}\sqrt{\frac{\ft}{\fq}})\nonumber\\
    \times&\calM(Q_{\tau}^{n-1}Q_{7,1}^{\frac{2}{3}}Q_{7,2}^{\frac{1}{3}}Q_{2,1}^{\frac12}\sqrt{\frac{\ft}{\fq}})\calM(Q_{\tau}^{n-1}Q_{7,1}^{\frac13}Q_{7,2}^{\frac23}Q_{2,1}^{\frac12}\sqrt{\frac{\ft}{\fq}})\calM(Q_{\tau}^nQ_{7,1}^{-\frac13}Q_{7,2}^{\frac13}Q_{2,1}^{-\frac12}\sqrt{\frac{\ft}{\fq}})\nn\\
    \times&\calM(Q_{\tau}^nQ_{7,1}^{-\frac23}Q_{7,2}^{-\frac13}Q_{2,1}^{\frac12}\sqrt{\frac{\ft}{\fq}})\calM(Q_{\tau}^nQ_{7,1}^{-\frac13}Q_{7,2}^{-\frac23}Q_{2,1}^{-\frac12}\sqrt{\frac{\ft}{\fq}})\calM(Q_{\tau}^nQ_{7,1}^{-\frac{2}{3}}Q_{7,2}^{-\frac{1}{3}}Q_{2,1}^{-\frac12}\sqrt{\frac{\ft}{\fq}})\nn\\
    \times&\calM(Q_{\tau}^nQ_{7,1}^{\frac{1}{3}}Q_{7,2}^{-\frac{1}{3}}Q_{2,1}^{-\frac12}\sqrt{\frac{\ft}{\fq}})\calM(Q_{\tau}^nQ_{7,1}^{-\frac{1}{3}}Q_{7,2}^{-\frac{2}{3}}Q_{2,1}^{\frac12}\sqrt{\frac{\ft}{\fq}})\calM(Q_{\tau}^nQ_{2,1}^{-\frac12}\sqrt{\frac{\ft}{\fq}})^2\Bigg)^{-1}\nonumber\\
    \times&\hat{Z}_{\mu_{7,1}\mu_{7,2}\mu_{7,3}\mu_{2,1}\mu_{2,2}\mu_{1,1}}(Q_{7,1},Q_{7,2},Q_{2,1},Q_{1},Q_{2},Q_{\tau}),
    \label{eq:Zoft3}
\end{align}
where
\begin{align}
    &\hat{Z}_{\mu_{7,1}\mu_{7,2}\mu_{7,3}\mu_{2,1}\mu_{2,2}\mu_{1,1}}(Q_{7,1},Q_{7,2},Q_{2,1},Q_{1},Q_{2},Q_{\tau})\nonumber\\
    =&\fq^{\frac{||\mu_{2,1}||^2+||\mu_{2,2}||^2+||\mu_{1,1}||^2}{2}}\ft^{\frac{||\mu_{2,1}^t||^2+||\mu_{2,2}^t||^2+||\mu_{1,1}^t||^2}{2}}(-Q_{7,1}^{-\frac23}Q_{7,2}^{-\frac13}Q_{2,1}^{\frac12})^{|\mu_{2,1}|}(-Q_{2,1}^{-\frac12}Q_{1})^{|\mu_{1,1}|}\nn\\
    \times&\left(-Q_{7,1}^{-\frac{1}{3}}Q_{7,2}^{-\frac{2}{3}}Q_{2,1}^{\frac12}Q_{2}\right)^{|\mu_{2,2}|}\prod_{i=1}^2\tilde{Z}_{\mu_{2,i}}(\ft,\fq)\tilde{Z}_{\mu_{2,i}^t}(\fq,\ft)\times\tilde{Z}_{\mu_{1,1}}(\ft,\fq)\tilde{Z}_{\mu_{1,1}^t}(\fq,\ft)\nn\\
    \times&\prod_{n=1}^{\infty}\calN_{\mu_{2,1}\mu_{7,1}}(Q_{\tau}^nQ_{7,1}^{-\frac23}Q_{7,2}^{-\frac13}Q_{2,1}^{\frac12}\sqrt{\frac{\ft}{\fq}})\calN_{\mu_{7,1}\mu_{2,1}}(Q_{\tau}^{n-1}Q_{7,1}^{\frac23}Q_{7,2}^{\frac13}Q_{2,1}^{-\frac12}\sqrt{\frac{\ft}{\fq}})\nn\\
    \times&\calN_{\mu_{2,2}\mu_{7,1}}(Q_{\tau}^nQ_{7,1}^{-\frac{2}{3}}Q_{7,2}^{-\frac{1}{3}}Q_{2,1}^{-\frac12}\sqrt{\frac{\ft}{\fq}})\calN_{\mu_{7,1}\mu_{2,2}}(Q_{\tau}^{n-1}Q_{7,1}^{\frac{2}{3}}Q_{7,2}^{\frac{1}{3}}Q_{2,1}^{\frac12}\sqrt{\frac{\ft}{\fq}})\nonumber\\
    \times&\calN_{\mu_{7,2}\mu_{2,1}}(Q_{\tau}^nQ_{7,1}^{-\frac13}Q_{7,2}^{\frac13}Q_{2,1}^{-\frac12}\sqrt{\frac{\ft}{\fq}})\calN_{\mu_{2,1}\mu_{7,2}}(Q_{\tau}^{n-1}Q_{7,1}^{\frac13}Q_{7,2}^{-\frac13}Q_{2,1}^{\frac12}\sqrt{\frac{\ft}{\fq}})\nn\\
    \times&\calN_{\mu_{2,2}\mu_{7,2}}(Q_{\tau}^nQ_{7,1}^{\frac{1}{3}}Q_{7,2}^{-\frac{1}{3}}Q_{2,1}^{-\frac12}\sqrt{\frac{\ft}{\fq}})\calN_{\mu_{7,2}\mu_{2,2}}(Q_{\tau}^{n-1}Q_{7,1}^{-\frac{1}{3}}Q_{7,2}^{\frac{1}{3}}Q_{2,1}^{\frac12}\sqrt{\frac{\ft}{\fq}})\nn\\
    \times&\calN_{\mu_{7,3}\mu_{2,1}}(Q_{\tau}^nQ_{7,1}^{-\frac13}Q_{7,2}^{-\frac23}Q_{2,1}^{-\frac12}\sqrt{\frac{\ft}{\fq}})\calN_{\mu_{2,1}\mu_{7,3}}(Q_{\tau}^{n-1}Q_{7,1}^{\frac13}Q_{7,2}^{\frac23}Q_{2,1}^{\frac12}\sqrt{\frac{\ft}{\fq}})\nonumber\\
    \times&\calN_{\mu_{7,3}\mu_{2,2}}(Q_{\tau}^nQ_{7,1}^{-\frac{1}{3}}Q_{7,2}^{-\frac{2}{3}}Q_{2,1}^{\frac12}\sqrt{\frac{\ft}{\fq}})\calN_{\mu_{2,2}\mu_{7,3}}(Q_{\tau}^{n-1}Q_{7,1}^{\frac{1}{3}}Q_{7,2}^{\frac{2}{3}}Q_{2,1}^{-\frac12}\sqrt{\frac{\ft}{\fq}})\nn\\
    \times&\calN_{\mu_{1,1}\mu_{2,1}}(Q_{\tau}^nQ_{2,1}^{-\frac12}\sqrt{\frac{\ft}{\fq}})\calN_{\mu_{2,1}\mu_{1,1}}(Q_{\tau}^{n-1}Q_{2,1}^{\frac12}\sqrt{\frac{\ft}{\fq}})\calN_{\mu_{2,2}\mu_{1,1}}(Q_{\tau}^nQ_{2,1}^{-\frac12}\sqrt{\frac{\ft}{\fq}})\nn\\
    \times&\calN_{\mu_{1,1}\mu_{2,2}}(Q_{\tau}^{n-1}Q_{2,1}^{\frac12}\sqrt{\frac{\ft}{\fq}})\Bigg(\calN_{\mu_{2,1}\mu_{2,1}}(Q_{\tau}^n)\calN_{\mu_{2,1}\mu_{2,1}}(Q_{\tau}^n\frac{\ft}{\fq})\calN_{\mu_{2,2}\mu_{2,2}}(Q_{\tau}^n)\nn\\
    \times&\calN_{\mu_{2,2}\mu_{2,2}}(Q_{\tau}^n\frac{\ft}{\fq})\calN_{\mu_{1,1}\mu_{1,1}}(Q_{\tau}^n)\calN_{\mu_{1,1}\mu_{1,1}}(Q_{\tau}^n\frac{\ft}{\fq})\calN_{\mu_{2,2}\mu_{2,1}}(Q_{\tau}^nQ_{2,1}^{-1})\nn\\
    \times&\calN_{\mu_{2,1}\mu_{2,2}}(Q_{\tau}^{n-1}Q_{2,1}\frac{\ft}{\fq})\calN_{\mu_{2,2}\mu_{2,1}}(Q_{\tau}^nQ_{2,1}^{-1}\frac{\ft}{\fq})\calN_{\mu_{2,1}\mu_{2,2}}(Q_{\tau}^{n-1}Q_{2,1})\Bigg)^{-1}\ .
    \label{eq:NZoft3}
\end{align}

\paragraph{Expression of $Z^{E_6}$.}
Using the $y$ parameters defined in \eqref{eq:defofy}, one finds that the periodicity condition in \eqref{eq:t3sub2}, \eqref{eq:t3sub1}, and \eqref{eq:t3sub3} for the $E_6$-type LST become
\begin{align}
    &y_{1,1}=1,\qquad y_{2,1}y_{2,2}=1,\qquad y_{3,1}=1,\qquad y_{4,1}y_{4,2}=1,\ \nn\\
    &y_{5,1}=1,\qquad y_{6,1}y_{6,2}=1,\qquad y_{7,1}y_{7,2}y_{7,3}=1,
\end{align}
where we have set $a_{1,1}=0$. 
Substitute \eqref{eq:Zoft3}, \eqref{eq:NZoft3}, \eqref{eq:ZofE6mid}, and \eqref{eq:NZofE6mid} into \eqref{eq:factorsofE6}, and transform the $\calN$ functions into the $\varTheta$ functions, we find
\begin{equation}\label{eq:E6full}
    Z^{E_6}=Z^{E_6}_{\text{pert}}\cdot Z^{E_6}_{\text{inst}}\ ,
\end{equation}
with
\begin{align}\label{eq:E6pert}
	Z^{E_6}_{\text{pert}}=&\prod_{l=1}^{\infty}\frac{1}{(1-Q_{\tau}^l)^8}\prod_{n=1}^{\infty}\prod_{I=1,3,5}\frac{\calM(\frac{\ft}{\fq}Q_{\tau}^{{\textit{\textbf{n}}}_{I,1;I,1}})\calM(Q_{\tau}^{{\textit{\textbf{n}}}_{I,1;I,1}})}{\prod_{l=1}^2\calM(\frac{y_{I,1}}{y_{I+1,l}}\sqrt{\frac{\ft}{\fq}}Q_{\tau}^{{\textit{\textbf{n}}}_{I,1;I+1,l}})}\nn\\
	\times&\prod_{J=2,4,6}\prod_{p=1}^2\frac{\prod_{s=1}^2\calM(\frac{y_{J,p}}{y_{J,s}}\frac{\ft}{\fq}Q_{\tau}^{{\textit{\textbf{n}}}_{J,p;J,s}})\calM(\frac{y_{J,p}}{y_{J,s}}Q_{\tau}^{{\textit{\textbf{n}}}_{J,p;J,s}})}{\calM(\frac{y_{J,p}}{y_{J-1,1}}\sqrt{\frac{\ft}{\fq}}Q_{\tau}^{{\textit{\textbf{n}}}_{J,p;J-1,1}})\prod_{r=1}^3\calM(\frac{y_{J,p}}{y_{7,r}}\sqrt{\frac{\ft}{\fq}}Q_{\tau}^{{\textit{\textbf{n}}}_{J,p;7,r}})}\nn\\
	\times&\prod_{c=1}^3\frac{\prod_{e=1}^3\calM(\frac{y_{7,c}}{y_{7,e}}\frac{\ft}{\fq}Q_{\tau}^{{\textit{\textbf{n}}}_{7,c;7,e}})\calM(\frac{y_{7,c}}{y_{7,e}}Q_{\tau}^{{\textit{\textbf{n}}}_{7,c;7,e}})}{\prod_{L=2,4,6}\prod_{d=1}^2\calM(\frac{y_{7,c}}{y_{L,d}}\sqrt{\frac{\ft}{\fq}}Q_{\tau}^{{\textit{\textbf{n}}}_{7,c;L,d}})}\ ,
\\
\label{eq:E6inst}
Z^{E_6}_{\text{inst}}=&\sum_{\boldsymbol{\mu}}\prod_{I=1,3,5}\bigg(Q_{I}^{|\mu_{I,1}|}\frac{\prod_{l=1}^{2}\varTheta_{\mu_{I,1}\mu_{I+1,l}}(\frac{y_{I,1}}{y_{I+1,l}}\sqrt{\tfrac{\ft}{\fq}}|Q_{\tau})}{\varTheta_{\mu_{I,1}\mu_{I,1}}(\frac{y_{I,1}}{y_{I,1}}|Q_{\tau})\varTheta_{\mu_{I,1}\mu_{I,1}}(\frac{y_{I,1}}{y_{I,1}}\tfrac{\ft}{\fq}|Q_{\tau})}\bigg)\nn\\
    \times&\prod_{J=2,4,6}\prod_{p=1}^2\bigg(Q_{J}^{|\mu_{J,p}|}\frac{\varTheta_{\mu_{J,p}\mu_{J-1,1}}(\frac{y_{J,p}}{y_{J-1,1}}\sqrt{\tfrac{\ft}{\fq}}|Q_{\tau})\prod_{r=1}^3\varTheta_{\mu_{J,p}\mu_{7,r}}(\frac{y_{J,p}}{y_{7,r}}\sqrt{\tfrac{\ft}{\fq}}|Q_{\tau})}{\prod_{s=1}^2\varTheta_{\mu_{J,p}\mu_{J,s}}(\frac{y_{J,p}}{y_{J,s}}|Q_{\tau})\varTheta_{\mu_{J,p}\mu_{J,s}}(\frac{y_{J,p}}{y_{J,s}}\tfrac{\ft}{\fq}|Q_{\tau})}\bigg)\nn\\
    \times&\prod_{c=1}^3\bigg(Q_{7}^{|\mu_{7,c}|}\frac{\prod_{L=2,4,6}\prod_{d=1}^{2}\varTheta_{\mu_{7,c}\mu_{L,d}}(\frac{y_{7,c}}{y_{L,d}}\sqrt{\tfrac{\ft}{\fq}}|Q_{\tau})}{\prod_{e=1}^3\varTheta_{\mu_{7,c}\mu_{7,e}}(\frac{y_{7,c}}{y_{7,e}}|Q_{\tau})\varTheta_{\mu_{7,c}\mu_{7,e}}(\frac{y_{7,c}}{y_{7,e}}\tfrac{\ft}{\fq}|Q_{\tau})}\bigg)\ .
\end{align}

\bigskip
\section{Higher ranks and general expression of partition functions}
\label{sec:higherRanks}

In the previous section, we discussed the partition functions for rank-1 $D_4$, $D_5$, and $E_6$-type LSTs based on tri-/quadri-valent gluings \cite{Hayashi:2017jze,Hayashi:2021pcj}  by further introducing the bivalent self-gluings. In terms of  the $\varTheta$ functions defined in \eqref{eq:Theta-definition}, the partition functions are expressed in a more compact form, which leads to an effective way of organizing or generalizing the partition functions.  In this section, we discuss a generalization of the rank-1 $D_4, D_5$ and $E_6$-type LSTs by increasing the ranks of each gauge node of affine quivers $k$ times, which we refer to as rank-$k$ $D$/$E$-type LSTs. 
We express the partition functions for these higher rank LSTs also in terms of the $\varTheta$ functions. 

To explain how to generalize, let us first recall the procedure of obtaining the partition function of $E_6$-type LST in the previous section. First, we draw the subdiagrams in the trivalent gluing method as shown in figure \ref{fig:E6tri}. Then we  compute the partition function of each subdiagram by the topological vertex formalism. At this step, further corrections to the partition function of each subdiagram are needed:
\begin{enumerate}
\item [(1)] For each of the three SU(2)-SU(1) subdiagrams, we need to divide them by additional factors. For example, in \eqref{eq:Zt3sub1} we remove the vector contributions of SU$(3)$ from the upper right SU(2)-SU(1) subdiagram.
\item [(2)] For the middle subdiagram whose computation is shown in \eqref{eq:ZE6mid}, only the SU$(3)$ vector contributions are counted, while the bifundamental contributions of SU$(3)$-SU$(2)$ and the vector contributions of SU$(2)$ are not counted as they are already taken into account in the partition functions of the three SU(2)-SU(1) subdiagrams. 
\end{enumerate}
There are no bifundamental contributions between two different SU$(2)$'s as they are not connected in the quiver, so there is no string connecting color branes of different SU$(2)$'s in the middle subdiagram of figure \ref{fig:E6tri}. However, such unphysical bifundamental contributions would appear if the middle subdiagram was used directly for computation. So these further corrections to the partition function of each subdiagram ensure that no redundant factors and unphysical factors are counted when we multiply the partition functions of each subdiagram together to get the full partition function. 

We emphasize that, as summarized above, picking up the proper vector and bifundamental contributions for each subdiagram is important for the computation. Roughly speaking, as the $\calR$ factors which are generated in the computation by topological vertex through the formula \eqref{eq:Zofgeneralstrip} represent all the vector and bifundamental contributions, picking up the proper vector and bifundamental contributions for each subdiagram corresponds to picking up the proper $\calR$ factors for each subdiagram. Once we understand this point, we can further simplify the method of obtaining the full partition function of the LST.
\begin{figure}[htbp]
    \centering
    \includegraphics[scale=0.5]{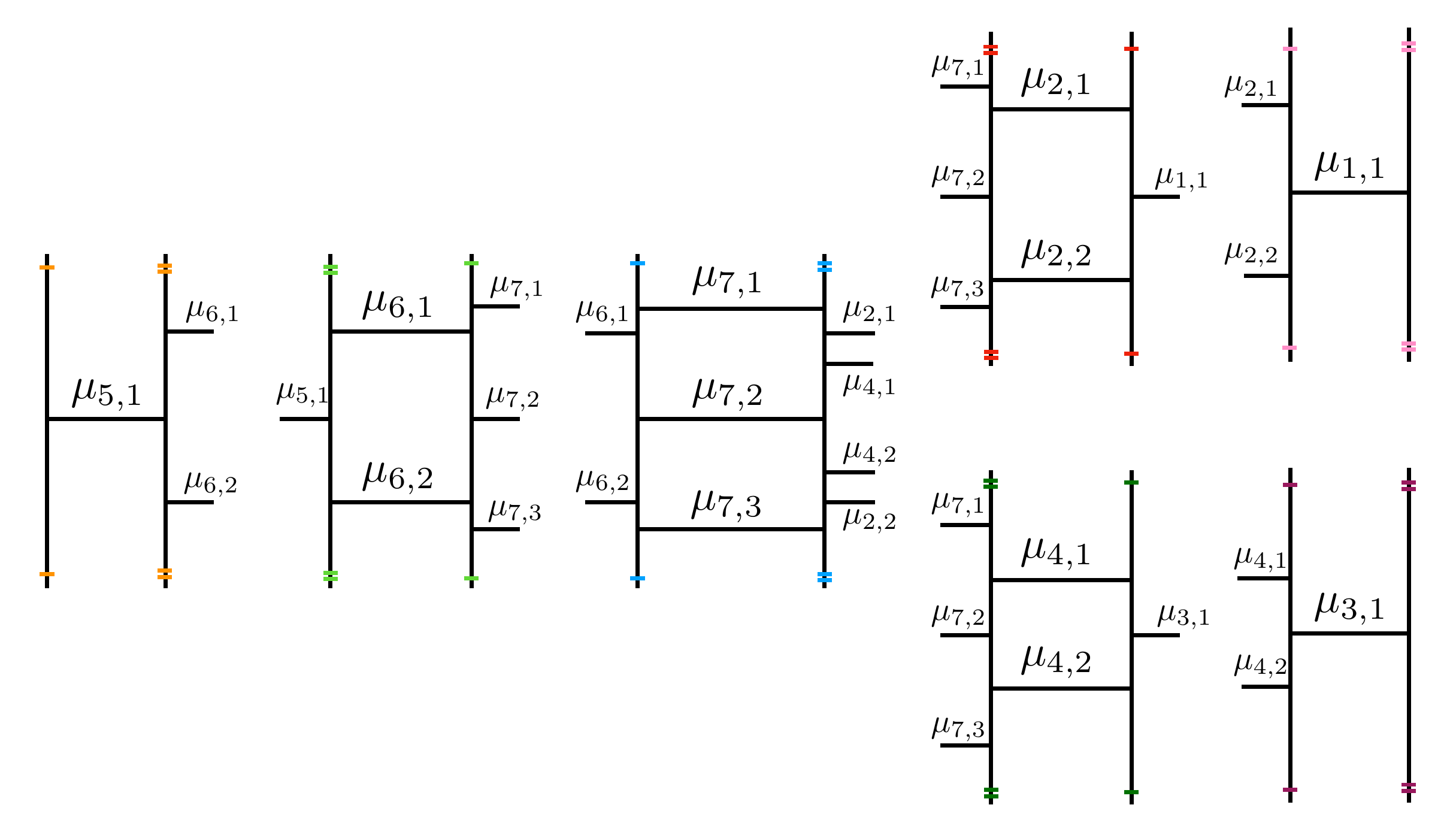}
    \caption{Schematic picture of a decomposition of trivalent gluing subdiagrams of $E_6$-type LST into smaller subdiagrams, where each subdiagram corresponds to a single gauge node.}
    \label{fig:decomposeE6}
\end{figure}

\subsection{Introduction of general node}\label{sec:introGN}
We further decompose the subdiagrams in figure \ref{fig:E6tri} into smaller subdiagrams that correspond to each single node of the affine $E_6$ quiver as is shown in figure \ref{fig:decomposeE6}. Now each subdiagram of figure \ref{fig:decomposeE6} has similar web structure and the subdiagrams differ in the number of color branes in the left, middle and right of the subdiagrams. For a general $D,E$-type LST with arbitrary rank, its tri-/quadri-valent gluing diagrams can also be further decomposed into smaller subdiagrams which correspond to single gauge nodes of the affine quiver. We can treat the subdiagrams of single gauge nodes as the building blocks to construct a general $D,E$-type LST with arbitrary rank. Specifically, we can compute the partition function of each single gauge node and then multiply these partition functions together to obtain the full partition function of the LST.

\begin{figure}[htbp]
    \centering
    \includegraphics[scale=0.6]{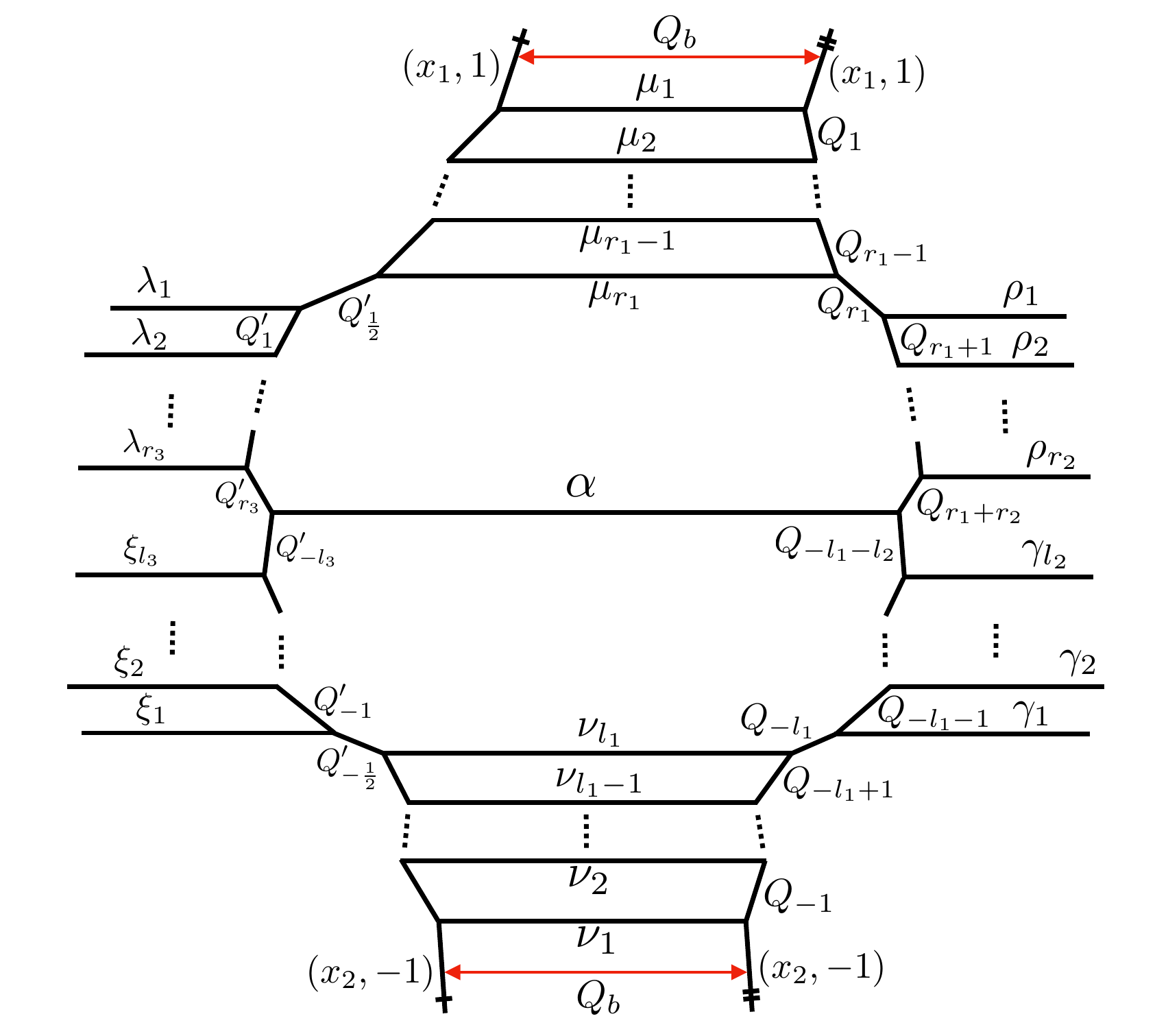}
    \caption{A general subdiagram corresponding to a single gauge node.}
    \label{fig:gaugenode}
\end{figure}

To this end, we first consider a general single gauge node whose subdiagram is illustrated in figure \ref{fig:gaugenode}. Our strategy for obtaining the partition function of this general node is to first compute the contribution to the partition function from one arbitrary color brane inside this node. As we choose this color brane to be arbitrary, we will obtain a formula for the contribution that is applicable for any color brane inside this node.
Then we multiply contributions from all the color branes inside this node to get the partition function of this general node. 

In figure \ref{fig:gaugenode} we pick one arbitrary color brane inside the node and label it by Young diagram $\alpha$, and consider only the bifundamental and vector contributions that involve $\alpha$. For convenience and clarity, we use uppercase Greek letters to denote the name of a color brane whose Young diagram is denoted by the corresponding lowercase Greek letter. For example, we call the color brane labeled by Young diagram $\alpha$ as $A$. In figure \ref{fig:gaugenode}, there are $r_1$ color branes $M_1,\cdots,M_{r_1}$ above $A$ inside the same node which we label by Young diagrams $\mu_1,\cdots,\mu_{r_1}$. There are $l_1$ color branes $N_{l_1},\cdots,N_1$ below $A$ inside the same node which we label by Young diagrams $\nu_{l_1},\cdots,\nu_1$. On the right side of the node, there are $r_2$ color branes $P_1,\cdots,P_{r_2}$ above $A$ which we label by Young diagrams $\rho_1,\cdots,\rho_{r_2}$. There are $l_2$ color branes $\varGamma_{l_2},\cdots,\varGamma_1$ below $A$ which we label by Young diagrams $\gamma_{l_2},\cdots,\gamma_1$. On the left side of the node, there are $r_3$ color branes $\varLambda_1,\cdots,\varLambda_{r_3}$ above $A$ which we label by Young diagrams $\lambda_1,\cdots,\lambda_{r_3}$. There are $l_3$ color branes $\varXi_{l_3},\cdots,\varXi_1$ below $A$ which we label by Young diagrams $\xi_{l_3},\cdots,\xi_1$. If this node is at the junction of the quiver, then there could be more than one adjacent gauge node on the left or right of this node, and correspondingly the color branes on the left or right of the node may belong to different gauge nodes.

In figure \ref{fig:gaugenode}, in order to make this gauge node to be vertically periodic, the top two NS-charged branes have to be parallel to each other and the bottom two NS-charged branes also have to be parallel to each other. We label the corresponding brane charges as $(x_1,1)$ and $(x_2,-1)$.  Also, due to vertical periodicity the horizontal distance between the top two parallel NS-charged branes has to be equal to the horizontal distance between the bottom two parallel NS-charged branes, we denote the distance as $Q_b$ which is also the instanton factor as well as the tensor branch parameter of this gauge node. Following the assignment of the charges of the top and bottom NS-charged branes, we can label all the charges of the remaining NS-charged branes. The charge conservation at the two vertices of color brane $A$ gives the constraints:
\begin{equation}
    \left\{\begin{array}{l}
         x_1+x_2=r_3-r_1+l_3-l_1-1  \\
         x_1+x_2=r_1-r_2+l_1-l_2+1 
    \end{array}\right.\ ,
    \label{eq:x1x2}
\end{equation}
from which we find that
\begin{equation}
    n_2=2n_1\ ,
    \label{eq:paracondi}
\end{equation}
with
\begin{align}
    n_1\equiv r_1+l_1+1,\qquad n_2\equiv r_2+l_2+r_3+l_3.
\end{align} 
Note that $n_1$ is the number of color branes inside the gauge node in figure \ref{fig:gaugenode} and $n_2$ is the number of color branes on the left or right side of the gauge node,  \eqref{eq:paracondi}
 ensures the simultaneous parallelism of the top two and bottom two NS-charged branes. 

From the charge distributions of the branes in figure \ref{fig:gaugenode} with the labelings of K\"ahler parameters, we can find the relations between the K\"ahler parameter $Q_A$ of the brane $A$ and the K\"ahler parameter $Q_{M_1}$ of the brane $M_1$: 
\begin{align}\label{eq:Qup}
    Q_{M_1}=&~Q_{A}Q_1^{x_1-1}\cdots Q_{r_1-1}^{x_1-(r_1-1)}Q_{r_1}^{x_1-r_1}Q_{r_1+1}^{x_1-r_1+1}\cdots Q_{r_1+r_2}^{x_1-r_1+r_2}\nonumber\\
    &\times Q_1^{-x_1-1}\cdots Q_{r_1-1}^{-x_1-(r_1-1)}{Q'_{\frac{1}{2}}}^{-x_1-r_1}{Q'_1}^{-x_1-r_1+1}\cdots {Q'_{r_3}}^{-x_1-r_1+r_3}\nonumber\\
    =&~Q_{A}Q_1^{-1}\cdots Q_{r_1-1}^{-(r_1-1)}Q_{r_1}^{-r_1}Q_{r_1+1}^{-r_1+1}\cdots Q_{r_1+r_2}^{-r_1+r_2}\nonumber\\
    &\times Q_1^{-1}\cdots Q_{r_1-1}^{-(r_1-1)}{Q'_{\frac{1}{2}}}^{-r_1}{Q'_1}^{-r_1+1}\cdots {Q'_{r_3}}^{-r_1+r_3}\nonumber\\
    =&~Q_{A}\frac{Q_{P_{r_2}A}Q_{P_{r_2-1}A}\cdots Q_{P_{1}A}Q_{\varLambda_{r_3}A}Q_{\varLambda_{r_3-1}A}\cdots Q_{\varLambda_{1}A}}{Q_{M_{r_1}A}^2Q_{M_{r_1-1}A}^2\cdots Q_{M_{1}A}^2},
\end{align}
where we have defined, for example, $Q_{\varLambda_1A}\equiv\frac{y_{\varLambda_1}}{y_A}$ in which $y_{\varLambda_1},y_A$ are exponentiated vertical positions of the brane $\varLambda_1$ and $A$ defined in \eqref{eq:defofy}.

By the same reasoning, for the K\"ahler parameter $Q_{N_1}$ of the brane $N_1$, 
\begin{align}\label{eq:Qdown}
    Q_{N_1}=&~Q_{A}Q_{-l_1-l_2}^{x_2-l_1+l_2}\cdots Q_{-l_1-1}^{x_2-l_1+1}Q_{-l_1}^{x_2-l_1}Q_{-l_1+1}^{x_2-l_1+1}\cdots Q_{-1}^{x_2-1}\nonumber\\
    &\times {Q'_{-l_3}}^{l_3-x_2-l_1}\cdots {Q'_{-1}}^{1-x_2-l_1}{Q'_{-\frac{1}{2}}}^{-x_2-l_1}Q_{-l_1+1}^{-x_2-l_1+1}\cdots Q_{-1}^{-x_2-1}\nonumber\\
    =&~Q_{A}\frac{Q_{A\varGamma_{l_2}}Q_{A\varGamma_{l_2-1}}\cdots Q_{A\varGamma_{1}}Q_{A\varXi_{l_3}}Q_{A\varXi_{l_3-1}}\cdots Q_{A\varXi_{1}}}{Q_{AN_{l_1}}^2Q_{AN_{l_1-1}}^2\cdots Q_{AN_{1}}^2}.
\end{align}
In order to have the compactification along the vertical direction, we should have 
\begin{equation}\label{eq:MNb}
	Q_{M_1}=Q_{N_1}=Q_b\ ,
\end{equation}
so
\begin{equation}\label{eq:constraint}
	Q_{\textbf{m}}^{2(r_1+l_1+1)}=Q_{\textbf{r}}^{r_2+l_2}Q_{\textbf{l}}^{r_3+l_3}\ ,
\end{equation}
where
\begin{align}
	Q_{\textbf{m}}\equiv&\left(Q_{M_{1}A}\cdots Q_{M_{r_1}A}Q_{N_{l_1}A}\cdots Q_{N_{1}A}\right)^{\frac{1}{r_1+l_1+1}},\\
	Q_{\textbf{r}}\equiv&\left(Q_{P_{1}A}\cdots Q_{P_{r_2}A}Q_{\varGamma_{l_2}A}\cdots Q_{\varGamma_{1}A}\right)^{\frac{1}{r_2+l_2}},\\
	Q_{\textbf{l}}\equiv&\left(Q_{\varLambda_{1}A}\cdots Q_{\varLambda_{r_3}A}Q_{\varXi_{l_3}A}\cdots Q_{\varXi_{1}A}\right)^{\frac{1}{r_3+l_3}}\ .
\end{align}
$Q_{\textbf{m}}$, $Q_{\textbf{r}}$ and $Q_{\textbf{l}}$ are the relative vertical positions of the centers of color branes in the middle, right, and left of figure \ref{fig:gaugenode} to the color brane $A$.

For a gauge node at one end of an affine $D,E$-type quiver, its adjacent node can only appear on its one side, so the adjacent color branes can only appear either on its right side or its left side. For example, $r_2$ and $l_2$ of such gauge node will be both zero if there are no color branes on the right side of the gauge node. Then, from \eqref{eq:constraint} and \eqref{eq:paracondi}, for this gauge node we have
\begin{equation}
	Q_{\textbf{m}}=Q_{\textbf{l}}\ .
\end{equation}
So the positions of centers of color branes for this node and its adjacent node are the same. Consequently, we find that the vertical positions of the centers of color branes of all nodes in the affine $D,E$-type quiver are the same by using \cref{eq:paracondi,eq:constraint} repeatedly for every node of the quiver, which means the bifundamental masses are all zero and we can choose a common center of Coulomb branch for all the gauge nodes. From now on, we will choose this common center as the zero point of the vertical direction.

\subsection{Partition function of general node}
In figure \ref{fig:gaugenode}, we have picked an arbitrary color brane $A$ inside the gauge node. 
Consider another arbitrary color brane $B$ in figure \ref{fig:gaugenode} which may or may not lie inside the same gauge node. Let us denote the vertical positions of the color branes $A$ and $B$ by $a_A$ and $a_B$, then their exponentiated vertical positions are given by $y_A\equiv e^{-\uupbeta \,a_A}$ and $y_B\equiv e^{-\uupbeta\, a_B}$, respectively. By considering whether the color branes $A$ and $B$ are in the same gauge node or not as well as their relative vertical positions, we discuss and compute vector or bifundamental contribution arising from these color branes.

\paragraph{Bifundamental contributions.}
When the color branes $B$ is not inside but outside the gauge node, the fundamental string connecting brane $A$ and $B$ leads to the bifundamental contributions. Let us first consider the case that their relative vertical position is $a_B>a_A$. Following Appendix \ref{app:formulas}, one readily finds that the corresponding $\calR$ factors are given by 
\begin{equation}\label{eq:bifdR}
	\prod_{n=1}^{\infty}\calR_{\alpha^t\beta}(Q_{\tau}^{n-1}\frac{Q_{\tau}}{Q_{BA}};\ft,\fq)\calR_{\beta^t\alpha}(Q_{\tau}^{n-1}Q_{BA};\ft,\fq).
\end{equation}
These $\calR$ factors are further factorized into $\calM$ factors and $\calN$ factors by the definition in \eqref{eq:RMN}. The $\calM$ factors from \eqref{eq:bifdR} contribute to the perturbative part of the partition function,
\begin{align}\label{eq:bifdM1}
	&\prod_{n=1}^{\infty}\left(\calM(Q_{\tau}^{n-1}Q_{\tau}Q_{AB}\sqrt{\tfrac{\ft}{\fq}};\ft,\fq)\calM(Q_{\tau}^{n-1}Q_{BA}\sqrt{\tfrac{\ft}{\fq}};\ft,\fq)\right)^{-1}\nn\\
	=&\,\prod_{n=1}^{\infty}\text{PE}\left[-\frac{Q_{\tau}^{n-1}Q_{\tau}Q_{AB}\sqrt{\ft\,\fq}}{(1-\ft)(1-\fq)}\right]\text{PE}\left[-\frac{Q_{\tau}^{n-1}Q_{BA}\sqrt{\ft\,\fq}}{(1-\ft)(1-\fq)}\right]\nn\\
	=&\,\text{PE}\left[-\frac{(\sum_{n=1}^{\infty}Q_{\tau}^{n-1})Q_{\tau}Q_{AB}\sqrt{\ft\,\fq}}{(1-\ft)(1-\fq)}\right]\text{PE}\left[-\frac{(\sum_{n=1}^{\infty}Q_{\tau}^{n-1})Q_{BA}\sqrt{\ft\,\fq}}{(1-\ft)(1-\fq)}\right]\nn\\
	=&\, \text{PE}\left[-\frac{\sqrt{\ft\,\fq}}{(1-\ft)(1-\fq)}\frac{1}{1-Q_{\tau}}(Q_{\tau}Q_{AB}+Q_{BA})\right],
\end{align}
where we have assumed $|Q_{\tau}|<1$ and $|\ft|,|\fq|\ll 1$ in order for convergence of the Plethystic exponential and we shall impose this assumption throughout the rest of this paper.

On the other hand, these $\calM$ factors can also be written as
\begin{align}\label{eq:bifdM2}
	&\prod_{n=1}^{\infty}\calM(Q_{\tau}^{n-1}Q_{\tau}Q_{AB}\sqrt{\tfrac{\ft}{\fq}};\ft,\fq)^{-1}\cdot\calM(Q_{BA}\sqrt{\tfrac{\ft}{\fq}};\ft,\fq)^{-1}\prod_{n=1}^{\infty}\calM(Q_{\tau}^{n-1}Q_{\tau}Q_{BA}\sqrt{\tfrac{\ft}{\fq}};\ft,\fq)^{-1}\nn\\
    \rightarrow &\prod_{n=1}^{\infty}\calM(Q_{\tau}^{n-1}Q_{\tau}Q_{AB}\sqrt{\tfrac{\ft}{\fq}};\ft,\fq)^{-1}\cdot\calM(Q_{AB}\sqrt{\tfrac{\ft}{\fq}};\ft,\fq)^{-1}\prod_{n=1}^{\infty}\calM(Q_{\tau}^{n-1}Q_{\tau}Q_{BA}\sqrt{\tfrac{\ft}{\fq}};\ft,\fq)^{-1}\nn\\
	=&\prod_{n=1}^{\infty}\calM(Q_{\tau}^{n-1}Q_{AB}\sqrt{\tfrac{\ft}{\fq}};\ft,\fq)^{-1}\prod_{n=1}^{\infty}\calM(Q_{\tau}^{n-1}Q_{\tau}Q_{BA}\sqrt{\tfrac{\ft}{\fq}};\ft,\fq)^{-1}\nn\\
	=&\, \text{PE}\left[-\frac{\sqrt{\ft\,\fq}}{(1-\ft)(1-\fq)}\frac{1}{1-Q_{\tau}}(Q_{\tau}Q_{BA}+Q_{AB})\right],
\end{align}
where by the $\rightarrow$, we mean an analytic continuation  due to the flop transition of topological string \cite{Iqbal:2004ne,Konishi:2006ev,Taki:2008hb} that we made to the middle $\calM$ by the formula
\begin{align}\label{eq:flop}
	\calM(Q;\ft,\fq)=\text{PE}\left[\frac{Q\fq}{(1-\fq)(1-\ft)}\right]\rightarrow
 \calM(Q^{-1}\tfrac{\ft}{\fq};\ft,\fq)=\text{PE}\left[\frac{Q^{-1}\ft}{(1-\fq)(1-\ft)}\right].
\end{align}
It follows from \eqref{eq:bifdM1} and \eqref{eq:bifdM2} that the product of $\calM$ factors is equal to the following PE up to a flop: 
\begin{align}
    &\prod_{n=1}^{\infty}\left(\calM(Q_{\tau}^{n-1}Q_{\tau}Q_{AB}\sqrt{\tfrac{\ft}{\fq}};\ft,\fq)\calM(Q_{\tau}^{n-1}Q_{BA}\sqrt{\tfrac{\ft}{\fq}};\ft,\fq)\right)^{-1}\nn\\
   & 
   \rightarrow
   \text{PE}\left[-\frac{\sqrt{\ft\,\fq}}{(1-\ft)(1-\fq)}\frac{1+Q_{\tau}}{1-Q_{\tau}}\frac{Q_{AB}}{2}\right]\text{PE}\left[-\frac{\sqrt{\ft\,\fq}}{(1-\ft)(1-\fq)}\frac{1+Q_{\tau}}{1-Q_{\tau}}\frac{Q_{BA}}{2}\right]\ , \label{eq:prodMfactors}
\end{align}
where we have separated the result into two factors, the first PE factor is assigned to the brane $B$ and the second PE factor is assigned to the brane $A$. Since \eqref{eq:prodMfactors} is true also for $a_B<a_A$, we can assign the factors in the same way. So, the factor that is assigned to $A$ is
\begin{align}\label{eq:Msource1}
	\text{PE}\left[-\frac{\sqrt{\ft\,\fq}}{(1-\ft)(1-\fq)}\frac{1+Q_{\tau}}{1-Q_{\tau}}\frac{Q_{BA}}{2}\right]
 \quad\text{for both }a_B>a_A\text{ and }a_B<a_A. 
\end{align}

The $\calN$ factors from \eqref{eq:bifdR} contribute to the instanton part of the partition function which are
\begin{align}\label{eq:bifund}
	&\prod_{n=1}^{\infty}\calN_{\alpha\beta}(Q_{\tau}^{n-1}\frac{Q_{\tau}}{Q_{BA}}\sqrt{\tfrac{\ft}{\fq}};\ft,\fq)\calN_{\beta\alpha}(Q_{\tau}^{n-1}Q_{BA}\sqrt{\tfrac{\ft}{\fq}};\ft,\fq)\\
	&=\frac{(Q_{BA})^{\frac{|\beta|}{2}}{\hat{f}_{\beta}(\ft,\fq)}^{-\frac12}}{W^{|\beta|}}\varTheta_{\beta\alpha}(\tfrac{y_B}{y_A}\sqrt{\tfrac{\ft}{\fq}}|Q_{\tau})\times\frac{(-1)^{|\alpha|}(Q_{BA})^{\frac{|\alpha|}{2}}{\hat{f}_{\alpha}(\ft,\fq)}^{\frac12}}{W^{|\alpha|}}\varTheta_{\alpha\beta}(\tfrac{y_A}{y_B}\sqrt{\tfrac{\ft}{\fq}}|Q_{\tau})\ ,\nonumber
\end{align}
where $\alpha, \beta$ are the Young diagrams associated with the color branes $A, B$, respectively, and $\hat{f}_{\alpha}$ and $W$ are defined as
\begin{align}
    &\hat{f}_{\alpha}(\ft,\fq)\equiv\ft^{\frac{||\alpha^t||^2}{2}}\fq^{-\frac{||\alpha||^2}{2}}\ ,\\
    &W\equiv\prod_{n=1}^{\infty}(1-Q_{\tau}^n)\ .
\end{align}
In \eqref{eq:bifund}, we have separated the result into two parts: the first part is before the multiplication sign which is assigned to $\beta$, and the second part is after the multiplication sign which is assigned to $\alpha$. If $a_B<a_A$, then the factors that should be assigned to $\alpha$ are just the first part with $\alpha,\beta$ exchanged. So, the factors that should be assigned to $\alpha$ are
\begin{equation}
	\left\{\begin{array}{ll}
		\frac{(-1)^{|\alpha|}(Q_{BA})^{\frac{|\alpha|}{2}}{\hat{f}_{\alpha}(\ft,\fq)}^{\frac{1}{2}}}{W^{|\alpha|}}\varTheta_{\alpha\beta}(\frac{y_A}{y_B}\sqrt{\tfrac{\ft}{\fq}}|Q_{\tau}),&\quad\text{if}\ a_{B}>a_{A}\ ,\\
    \frac{(Q_{AB})^{\frac{|\alpha|}{2}}{\hat{f}_{\alpha}(\ft,\fq)}^{-\frac{1}{2}}}{W^{|\alpha|}}\varTheta_{\alpha\beta}(\frac{y_A}{y_B}\sqrt{\tfrac{\ft}{\fq}}|Q_{\tau}),&\quad\text{if}\ a_{B}<a_{A}\ .    
	\end{array}\right.
	\label{eq:sources1}
\end{equation}

\paragraph{Vector contribution I.}
When the color brane $B$(in the case that $B$ is not $A$) is inside the gauge node, the fundamental string connecting brane $A$ and $B$ leads to vector contributions.
If $a_B>a_A$, the corresponding $\calR$ factors based on Appendix \ref{app:formulas} are
\begin{align}\label{eq:vecR}
\prod_{n=1}^{\infty}&\calR_{\alpha^t\beta}(Q_{\tau}^{n-1}\tfrac{Q_{\tau}}{Q_{BA}}\sqrt{\tfrac{\fq}{\ft}};\ft,\fq)\calR_{\beta^t\alpha}(Q_{\tau}^{n-1}Q_{BA}\sqrt{\tfrac{\fq}{\ft}};\ft,\fq)\calR_{\alpha^t\beta}(Q_{\tau}^{n-1}\tfrac{Q_{\tau}}{Q_{BA}}\sqrt{\tfrac{\ft}{\fq}};\ft,\fq)\nn\\	&\times\calR_{\beta^t\alpha}(Q_{\tau}^{n-1}Q_{BA}\sqrt{\tfrac{\ft}{\fq}};\ft,\fq).
\end{align}
The $\calM$ factors from \eqref{eq:vecR} contribute to the perturbative part of the partition function,
\begin{align}\label{eq:vecM1}
	&\prod_{n=1}^{\infty}\calM(Q_{\tau}^{n-1}\tfrac{Q_{\tau}}{Q_{BA}};\ft,\fq)\calM(Q_{\tau}^{n-1}Q_{BA};\ft,\fq)\calM(Q_{\tau}^{n-1}\tfrac{Q_{\tau}}{Q_{BA}}\tfrac{\ft}{\fq};\ft,\fq)\calM(Q_{\tau}^{n-1}Q_{BA}\tfrac{\ft}{\fq};\ft,\fq)\nn\\
	&=\text{PE}\left[\frac{\fq+\ft}{(1-\ft)(1-\fq)}\frac{1}{1-Q_{\tau}}(Q_{\tau}Q_{AB}+Q_{BA})\right]\ .
\end{align}
These $\calM$ factors can also be written as
\begin{align}\label{eq:vecM2}
	&\prod_{n=1}^{\infty}\calM(Q_{\tau}^{n-1}Q_{\tau}Q_{AB};\ft,\fq)\cdot\calM(Q_{BA};\ft,\fq)\prod_{n=1}^{\infty}\calM(Q_{\tau}^nQ_{BA};\ft,\fq)\nn\\
	&\times\prod_{n=1}^{\infty}\calM(Q_{\tau}^{n-1}Q_{\tau}Q_{AB}\tfrac{\ft}{\fq};\ft,\fq)\calM(Q_{BA}\tfrac{\ft}{\fq};\ft,\fq)\prod_{n=1}^{\infty}\calM(Q_{\tau}^nQ_{BA}\tfrac{\ft}{\fq};\ft,\fq)\nn\\
    \rightarrow &\prod_{n=1}^{\infty}\calM(Q_{\tau}^{n-1}Q_{\tau}Q_{AB};\ft,\fq)\cdot\calM(Q_{AB}\tfrac{\ft}{\fq};\ft,\fq)\prod_{n=1}^{\infty}\calM(Q_{\tau}^nQ_{BA};\ft,\fq)\nn\\
	&\times\prod_{n=1}^{\infty}\calM(Q_{\tau}^{n-1}Q_{\tau}Q_{AB}\tfrac{\ft}{\fq};\ft,\fq)\calM(Q_{AB};\ft,\fq)\prod_{n=1}^{\infty}\calM(Q_{\tau}^nQ_{BA}\tfrac{\ft}{\fq};\ft,\fq)\nn\\
	=&\, \prod_{n=1}^{\infty}\calM(Q_{\tau}^{n-1}Q_{AB};\ft,\fq)\prod_{n=1}^{\infty}\calM(Q_{\tau}^nQ_{BA};\ft,\fq)\prod_{n=1}^{\infty}\calM(Q_{\tau}^{n-1}Q_{AB}\tfrac{\ft}{\fq};\ft,\fq)\nn\\
	&\times \prod_{n=1}^{\infty}\calM(Q_{\tau}^nQ_{BA}\tfrac{\ft}{\fq};\ft,\fq)\nn\\
	=&\text{PE}\left[\frac{\fq+\ft}{(1-\ft)(1-\fq)}\frac{1}{1-Q_{\tau}}(Q_{\tau}Q_{BA}+Q_{AB})\right]\ ,
\end{align}
where \eqref{eq:flop} is used when deriving the formula on the right side of the $\rightarrow$.
It follows from \eqref{eq:vecM1} and \eqref{eq:vecM2} that the product of $\calM$ factors is equal to the following PE up to a flop: 
\begin{align}
&\prod_{n=1}^{\infty}\calM(Q_{\tau}^{n-1}\tfrac{Q_{\tau}}{Q_{BA}};\ft,\fq)\calM(Q_{\tau}^{n-1}Q_{BA};\ft,\fq)\calM(Q_{\tau}^{n-1}\tfrac{Q_{\tau}}{Q_{BA}}\tfrac{\ft}{\fq};\ft,\fq)\calM(Q_{\tau}^{n-1}Q_{BA}\tfrac{\ft}{\fq};\ft,\fq)\nn\\
	& \rightarrow 
 \text{PE}\left[\frac{\fq+\ft}{(1-\ft)(1-\fq)}\frac{1+Q_{\tau}}{1-Q_{\tau}}\frac{Q_{AB}}{2}\right]\text{PE}\left[\frac{\fq+\ft}{(1-\ft)(1-\fq)}\frac{1+Q_{\tau}}{1-Q_{\tau}}\frac{Q_{BA}}{2}\right]\ ,
 \label{eq:MMMM2}
\end{align}
where we have separated the result into two factors, the first PE factor is assigned to the brane $B$ and the second PE factor is assigned to the brane $A$. Since \eqref{eq:MMMM2} is true also for $a_B<a_A$, we can assign the factors in the same way. So, the factor that is assigned to $A$ is 
\begin{align}\label{eq:Msource2}
	\text{PE}\left[\frac{\fq+\ft}{(1-\ft)(1-\fq)}\frac{1+Q_{\tau}}{1-Q_{\tau}}\frac{Q_{BA}}{2}\right],\quad\text{for both }a_B>a_A\text{ and }a_B<a_A\ .
\end{align}

The $\calN$ factors from \eqref{eq:vecR} contribute to the instanton part of the partition function,
\begin{align}
	\prod_{n=1}^{\infty}\calN_{\alpha\beta}(Q_{\tau}^{n-1}\tfrac{Q_{\tau}}{Q_{BA}};\ft,\fq)\calN_{\beta\alpha}(Q_{\tau}^{n-1}Q_{BA};\ft,\fq)\calN_{\alpha\beta}(Q_{\tau}^{n-1}\tfrac{Q_{\tau}}{Q_{BA}}\tfrac{\ft}{\fq};\ft,\fq)\calN_{\beta\alpha}(Q_{\tau}^{n-1}Q_{BA}\tfrac{\ft}{\fq};\ft,\fq) \ .
\end{align}
After rewriting the above contributions with the help of \eqref{eq:bifund}, we get
\begin{align}
	\resizebox{0.97\hsize}{!}{$\frac{(Q_{BA})^{|\beta|}{\hat{f}_{\beta}(\ft,\fq)}^{-1}}{W^{2|\beta|}}\varTheta_{\beta\alpha}(\tfrac{y_B}{y_A}|Q_{\tau})\varTheta_{\beta\alpha}(\tfrac{y_B}{y_A}\tfrac{\ft}{\fq}|Q_{\tau})
	\times\frac{(Q_{BA})^{|\alpha|}\hat{f}_{\alpha}(\ft,\fq)}{W^{2|\alpha|}}\varTheta_{\alpha\beta}(\tfrac{y_A}{y_B}|Q_{\tau})\varTheta_{\alpha\beta}(\tfrac{y_A}{y_B}\tfrac{\ft}{\fq}|Q_{\tau})$},
\end{align}
where we have separated the result into two parts, the first part is assigned to the brane $B$ and the second part is assigned to the brane $A$.
If $a_B<a_A$, then we can also get the corresponding result by just exchanging $\alpha,\beta$($A,B$) in the above equation. So, the factors that should be assigned to $\alpha$ are
\begin{equation}
	\left\{
	\begin{array}{ll}
		\left(\frac{(Q_{BA})^{|\alpha|}\hat{f}_{\alpha}(\ft,\fq)}{W^{2|\alpha|}}\varTheta_{\alpha\beta}(\tfrac{y_A}{y_B}|Q_{\tau})\varTheta_{\alpha\beta}(\tfrac{y_A}{y_B}\tfrac{\ft}{\fq}|Q_{\tau})\right)^{-1}&\quad\text{if}\ a_{B}>a_{A}\,,\\
		\left(\frac{(Q_{AB})^{|\alpha|}{\hat{f}_{\alpha}(\ft,\fq)}^{-1}}{W^{2|\alpha|}}\varTheta_{\alpha\beta}(\tfrac{y_A}{y_B}|Q_{\tau})\varTheta_{\alpha\beta}(\tfrac{y_A}{y_B}\tfrac{\ft}{\fq}|Q_{\tau})\right)^{-1}&\quad\text{if}\ a_{B}<a_{A}\,.
	\end{array}
	\right.
	\label{eq:sources2}
\end{equation}

\paragraph{Vector contribution II.}
When brane $B$ is brane $A$,
the fundamental string connecting brane $A$ and 
$B$ also provides vector contributions which are the following $\calR$ factors based on Appendix \ref{app:formulas},
\begin{align}\label{eq:vec2R}
	\prod_{n=1}^{\infty}\calR_{\alpha^t\alpha}(Q_{\tau}^n\sqrt{\tfrac{\fq}{\ft}};\ft,\fq)\calR_{\alpha^t\alpha}(Q_{\tau}^n\sqrt{\tfrac{\ft}{\fq}};\ft,\fq)\ .
\end{align}
 
The $\calM$ factors of \eqref{eq:vec2R} contribute to the perturbative part of the partition function and can be written as the following PE:
\begin{align}\label{eq:vec2M1}
	&\prod_{n=1}^{\infty}\calM(Q_{\tau}^n;\ft,\fq)\calM(Q_{\tau}^n\tfrac{\ft}{\fq};\ft,\fq)\nn\\
	\simeq &\prod_{n=1}^{\infty}\calM(Q_{\tau}^{n-1};\ft,\fq)\calM(Q_{\tau}^n\tfrac{\ft}{\fq};\ft,\fq)
	=\text{PE}\left[\frac{1}{(1-\ft)(1-\fq)}\frac{1}{1-Q_{\tau}}(\fq+\ft\,Q_{\tau})\right]\ ,
\end{align}
where we have included an extra factor $\calM(1;\ft,\fq)$ when getting the right side of the $\simeq$. From here on, we use  the symbol $\simeq$ to denote equality up to extra factors. We can also get another expression by including an extra factor $\calM(\tfrac{\ft}{\fq};\ft,\fq)$ instead: 
\begin{align}\label{eq:vec2M2}
	&\prod_{n=1}^{\infty}\calM(Q_{\tau}^n;\ft,\fq)\calM(Q_{\tau}^n\tfrac{\ft}{\fq};\ft,\fq) \nn\\
	\simeq  &\prod_{n=1}^{\infty}\calM(Q_{\tau}^{n};\ft,\fq)\calM(Q_{\tau}^{n-1}\tfrac{\ft}{\fq};\ft,\fq)=\text{PE}\left[\frac{1}{(1-\ft)(1-\fq)}\frac{1}{1-Q_{\tau}}(\fq\,Q_{\tau}+\ft)\right] \ .
\end{align}
So from \eqref{eq:vec2M1} and \eqref{eq:vec2M2}, the product of $\calM$ factors of \eqref{eq:vec2R} is equal to the following PE up to extra factors:
\begin{align}\label{eq:Msource3}
	&\prod_{n=1}^{\infty}\calM(Q_{\tau}^n;\ft,\fq)\calM(Q_{\tau}^n\tfrac{\ft}{\fq};\ft,\fq)
 \simeq
 \text{PE}\left[\frac{\fq+\ft}{(1-\ft)(1-\fq)}\frac{1+Q_{\tau}}{1-Q_{\tau}}\frac12\right]\ .
\end{align}

Combined with two $\tilde Z$ factors coming from the two vertices at the ends of $A$, the $\calN$ factors from \eqref{eq:vec2R} can be transformed into $\varTheta$ functions by \eqref{eq:Ntotheta2} which contribute to the instanton part of the partition function:
\begin{equation}
    \left(\frac{\ft^{\frac{||\alpha^t||^2}{2}}\fq^{\frac{||\alpha||^2}{2}}}{W^{2|\alpha|}}\varTheta_{\alpha\alpha}(1|Q_{\tau})\varTheta_{\alpha\alpha}(\tfrac{\ft}{\fq}|Q_{\tau})\right)^{-1}\ .
    \label{eq:sources3}
\end{equation}

Apart from the contributions from \cref{eq:sources1,eq:sources2,eq:sources3}, the following factors which come from the leftover factors of the two vertices of $A$ and the edge factor of $A$ also contribute to the instanton partition function of the single node:
\begin{align}
    &\ft^{\frac{||\alpha^t||^2}{2}}\fq^{\frac{||\alpha||^2}{2}}{f_{\alpha}(\ft,\fq)}^{h}(-Q_{A})^{|\alpha|}
    = \ft^{\frac{||\alpha^t||^2}{2}}\fq^{\frac{||\alpha||^2}{2}}(-1)^{(h+1)|\alpha|}{\hat{f}_{\alpha}(\ft,\fq)}^{h}Q_{A}^{|\alpha|}\ ,
    \label{eq:sources4}
\end{align}
with 
\begin{equation}
    h = x_1+x_2+r_1-r_3-l_1+l_2\ .
    \label{eq:eta}
\end{equation}

\paragraph{Computation of the perturbative part.}
We discussed the bifundamental or vector contributions arising from the arbitrary brane $A$ inside the gauge node of figure \ref{fig:gaugenode} and the arbitrary brane $B$, among these contributions there are three types of sources, 
\eqref{eq:Msource1}, \eqref{eq:Msource2} and \eqref{eq:Msource3}, that contribute to the perturbative partition function. Hold $A$ fixed and sum over these three types of sources from all the color branes in figure \ref{fig:gaugenode}, we obtain:
\begin{align}\label{eq:Acontribution}
	\text{PE}\left[\frac{1}{2(1-\ft)(1-\fq)}\frac{1+Q_{\tau}}{1-Q_{\tau}}\bigg(-\sqrt{\ft\,\fq}\sum_{i=n_1+1}^{n_1+n_2}Q_{B_iA}+(\fq+\ft)\sum_{i=1}^{n_1}Q_{B_iA}\bigg)\right] \ ,
\end{align}
where for simplicity we have relabelled the Young diagrams in figure \ref{fig:gaugenode} as
\begin{align}
&\{{\beta_1,\cdots,\beta_{n_1}}\}\equiv\{\mu_1,\cdots,\mu_{r_1},\alpha,\nu_{l_1},\cdots,\nu_1\},
\\
&\{\beta_{n_1+1},\cdots,\beta_{n_1+n_2}\}\equiv\{\rho_1,\cdots,\rho_{r_2},\gamma_{l_2},\cdots,\gamma_1,\lambda_1,\cdots,\lambda_{r_3},\xi_{l_3},\cdots,\xi_1\}\ , 
\end{align}
so the brane $B_i$ is the brane labelled by the Young diagram $\beta_i$.

The contributions to the perturbative partition function coming from the other color branes inside the gauge node can also be obtained from \eqref{eq:Acontribution} by just replacing $A$ with the corresponding color brane. Multiplying all the contributions from all the color branes inside the gauge node together, we get the perturbative partition function of the single gauge node
\begin{align}\label{eq:Zpertofgeneral}
    Z^{\text{node,pert}}=\text{PE}\Bigg[&\frac{1}{2(1-\ft)(1-\fq)}\frac{1+Q_{\tau}}{1-Q_{\tau}}  \cr
    &\times\sum_{j=1}^{n_1}\bigg(-\sqrt{\ft\,\fq}\sum_{i=n_1+1}^{n_1+n_2}Q_{B_iB_j}+(\fq+\ft)\sum_{i=1}^{n_1}Q_{B_iB_j}\bigg)\Bigg]\ .
\end{align}

\paragraph{Computation of the instanton part.}
Among the bifundamental or vector contributions between brane $A$ and $B$, there are four types of sources from \eqref{eq:sources1}, \eqref{eq:sources2}, \eqref{eq:sources3} and \eqref{eq:sources4} that contribute to the instanton partition function. Hold $A$ fixed and sum over these sources from all the color branes in figure \ref{fig:gaugenode}, we can get the contribution of color brane $A$ to the instanton partition function of the single gauge node in which the remaining $W$, $\hat{f}$ and $-1$ factors are 
\begin{align}
W^{-(r_2+r_3+l_2+l_3-2(r_1+l_1+1))}&=1,\label{eq:Wcontri}
\\
\left(\hat{f}_{\alpha}(\ft,\fq)\right)^{h+\frac{1}{2}(r_2+r_3-2r_1+2l_1-l_2-l_3)}&=1,
\\
(-1)^{(h+1+r_2+r_3-2r_1)|\alpha|}=(-1)^{2|\alpha|}&=1,
\end{align}
where we used \eqref{eq:x1x2}, \eqref{eq:paracondi} and \eqref{eq:eta}. On the other hand, by using \cref{eq:Qup,eq:Qdown,eq:MNb}, the net $Q$ factors in the result become
\begin{equation}
    Q_b^{|\alpha|}.
\end{equation}

So the instanton contribution from brane $A$ to the single node partition function can be written as
\begin{align}\label{eq:Ainstcontri}
    Q_b^{|\alpha|}\frac{\prod_{k=n_1+1}^{n_1+n_2}\varTheta_{\alpha\beta_k}(\tfrac{y_A}{y_{B_k}}\sqrt{\tfrac{\ft}{\fq}}|Q_{\tau})}{\prod_{k=1}^{n_1}\varTheta_{\alpha\beta_k}(\tfrac{y_A}{y_{B_k}}|Q_{\tau})\varTheta_{\alpha\beta_k}(\tfrac{y_A}{y_{B_k}}\tfrac{\ft}{\fq}|Q_{\tau})} \ .
\end{align}
The contributions of the other color branes inside the single node can also be obtained from \eqref{eq:Ainstcontri} by replacing $\alpha$ with the corresponding Young diagram of the color brane. Multiply all the contributions from all the color branes inside the single gauge node, we get the instanton partition function of the single node
\begin{align}\label{eq:Zinstofgeneral}
    Z_{\beta_1,\cdots,\beta_{n_1+n_2}}^{\text{node,inst}}=Q_b^{\sum_{\ell=1}^{n_1}|\beta_{\ell}|}\prod_{\ell=1}^{n_1}\frac{\prod_{k=n_1+1}^{n_1+n_2}\varTheta_{\beta_{\ell}\beta_k}(\tfrac{y_{B_{\ell}}}{y_{B_k}}\sqrt{\tfrac{\ft}{\fq}}|Q_{\tau})}{\prod_{k=1}^{n_1}\varTheta_{\beta_{\ell}\beta_k}(\tfrac{y_{B_{\ell}}}{y_{B_k}}|Q_{\tau})\varTheta_{\beta_{\ell}\beta_k}(\tfrac{y_{B_{\ell}}}{y_{B_k}}\tfrac{\ft}{\fq}|Q_{\tau})} \ ,
\end{align}
which is the basic unit that appears in the instanton partition functions of rank-1 $D_4$,$D_5$ and $E_6$-type LSTs in \eqref{eq:D4inst},\eqref{eq:D5inst} and \eqref{eq:E6inst}.

\subsection{Full partition function}\label{sec:generalformula}

We have obtained the partition function of a general single gauge node in the previous subsection.
Now we can easily obtain the full partition function of a general $D,E$-type  LST by multiplying the partition functions of each gauge node together. 
But before that, there is still one more kind of factor that is missing in the perturbative part. 
We further need to include the factor\footnote{The same kind of factor also appears in \eqref{eq:bifund} where we denote it by $W^{-1}$, but the $W$'s there are generated when transforming $\calN$ functions to $\varTheta$ functions by \eqref{eq:Ntotheta1} and \eqref{eq:Ntotheta2}, and we have included them in the instanton part by \eqref{eq:Wcontri}. }
\begin{align}\label{eq:stripfactor}
	\prod_{l=1}^{\infty}\frac{1}{1-Q_{\tau}^l}
\end{align}
which is generated from a single periodic vertical strip by \eqref{eq:Zofgeneralstrip} into the perturbative part. 
This factor can be rewritten\footnote{The two formulas are used: $\log(1-Q)=-\sum_{n=1}^\infty\frac{Q^n}{n}$, $\sum_{l=1}^\infty Q^l=\frac{Q}{1-Q}$. } in terms of PE:
\begin{align}\label{eq:PEstripfactor}
	\prod_{l=1}^{\infty}\frac{1}{1-Q_{\tau}^l}=\exp\left[-\sum_{l=1}^{\infty}\log(1-Q_{\tau}^l)\right]=\exp\left[\sum_{n=1}^\infty\frac{1}{n}\frac{Q_{\tau}^n}{1-Q_{\tau}^n}\right]=\text{PE}\left[\frac{Q_{\tau}}{1-Q_{\tau}}\right]. 
\end{align}
For $D_N$- or $E_N$-type LST which has $N+1$ gauge nodes in the affine quiver, in its trivalent gluing brane webs, a single vertical strip might be shared by more than one gauge node, but we should not double count the strip. 
One can show that the total number of independent vertical strips is $N+2$, so we should include $N+2$ copies of \eqref{eq:stripfactor} into the perturbative part. This is consistent with the corresponding factors in \eqref{eq:D4pert}, \eqref{eq:D5pert} and \eqref{eq:E6pert}. Including this factor, now we can finally obtain the full partition function.

Before providing the explicit form of the partition function, we summarize the notation and convention for the expression of the full partition function:
we use the capital indices $I,J$ to label the nodes in the affine $D_N$- or $E_N$-type Dynkin diagram. The total number of gauge nodes is $N+1$, so $I,J\in\{1,2,\cdots,N+1\}$. The Cartan matrix is denoted as $C_{IJ}$, where $C_{IJ}=+2$ for $I=J$, $C_{IJ}=-1$ if $I$-th node and $J$-th node are connected with each other by a single line, and $C_{IJ}=0$ if $I$-th node and $J$-th node are not connected. Also, we denote the instanton factor of each gauge node as $Q_{I}$. 
Moreover, in order to label the color branes, we use the combination of the capital indices $I,J$ and the small indices $k, \ell$. For example, the combination of the indices $I$ and $k$ indicates the $k$-th color brane inside the $I$-th gauge node. And we use $n_{(I)}$ to denote the total number of color branes in the $I$-th gauge node, so $k\in\{1,2,\cdots,n_{(I)}\}$. We denote the exponentiated  vertical position of the $k$-th color brane inside the $I$-th gauge node as $y_{I,k}$ and the Young diagram assigned to it as $\mu_{I,k}$. 

\paragraph{Partition function of $D_N$- or $E_N$-type LST.}

It then follows from \eqref{eq:Zpertofgeneral} and \eqref{eq:PEstripfactor} that the perturbative part of the full partition function of a $D_N$- or $E_N$-type LST with arbitrary rank is given up a flop by 
\begin{align}\label{eq:unifiedpert}
    Z_{\text{pert}} =\text{PE}\left[\frac{\sqrt{\ft\fq}}{(1-\ft)(1-\fq)}\frac{1+Q_{\tau}}{1-Q_{\tau}}\sum_{I,J}\sum_{k,\ell}\sum_{s=\pm 1}\frac{1}{4}C_{IJ}\frac{y_{J,\ell}}{y_{I,k}}\left(\frac{\ft}{\fq}\right)^{\frac{1}{2}s\,\delta_{IJ}}+\frac{(N+2)Q_{\tau}}{1-Q_{\tau}}\right]\ .
\end{align}
And from 
\eqref{eq:Zinstofgeneral}, one finds that  the instanton part of the full partition function of a $D_N$- or $E_N$-type LST with arbitrary rank is given by
\begin{align}\label{eq:unifiedinst}
    Z_{\text{inst}} = 
    &\sum_{\boldsymbol{\mu}}
    \left( \prod_{I} Q_{I}^{\sum_{k} |\mu_{I,k}|}\right)
    \prod_{I,J} \prod_{k,\ell} \prod_{s=\pm 1} 
    \varTheta_{\mu_{I,k}\mu_{J,\ell}}
    \left( \tfrac{y_{I,k}}{y_{J,\ell}} \left(\tfrac{\ft}{\fq}\right)^{\frac12-\frac12 s \,\delta_{IJ}} \Big| Q_{\tau} \right)^{-\frac{1}{2}C_{IJ}}.
\end{align}
Note that the indices $I$ and $J$ both run all the gauge nodes, and the indices $k$ and $\ell$ run all the color branes in the $I$-th and $J$-th gauge nodes respectively. However, due to the Cartan matrix $C_{IJ}$, there is no contribution unless $I=J$ or $I$-th node and $J$-th node are connected with each other. 
Also, note that we need to assume 
$|Q_{\tau}|<1$ and $|\ft|,|\fq|\ll 1$
in order for convergence of \eqref{eq:unifiedpert}. 

Also remember that the centers of the Coulomb branch of each gauge node have the same vertical position that is set to be zero point which is concluded at the end of section \ref{sec:introGN}, so the $y$ parameters have to satisfy the constraints
\begin{align}\label{eq:yconstraint}
    \prod_{k=1}^{n_{(I)}} y_{I,k}=1,\quad \text{for } I\in\{1,2,\cdots,N+1\}.
\end{align}
If we define the distance between $y_{I,j}$ and $y_{I,j+1}$ as
\begin{align}\label{eq:Qbwys}
    Q_{I,j}\equiv \frac{y_{I,j}}{y_{I,j+1}},
\end{align}
then $Q_{I,j}$'s are independent K\"ahler parameters that solve the constraints \eqref{eq:yconstraint} by
\begin{align}
    y_{I,k}=\frac{\prod_{i=k}^{n_{(I)}-1}Q_{I,i}}{\left(\prod_{i=1}^{n_{(I)}-1}Q_{I,i}^i\right)^{\frac{1}{n_{(I)}}}}\ .
\end{align}

We have confirmed that our formulas \eqref{eq:unifiedpert} and \eqref{eq:unifiedinst} reproduces all the results discussed in section \ref{sec:LSTandselfGluing}. 

\begin{figure}[htbp]
    \centering
    \includegraphics[scale=0.65]{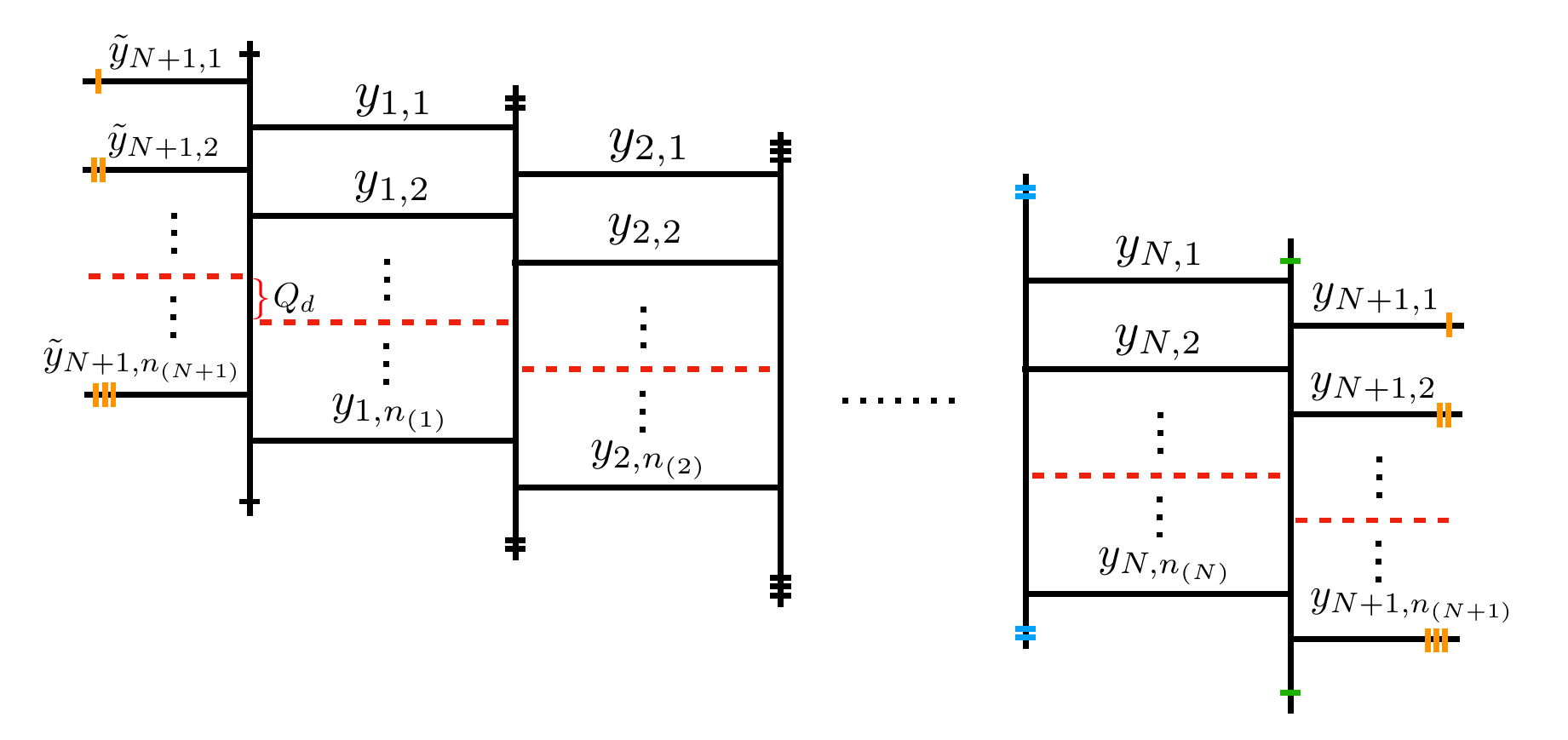}
    \caption{Schematic picture of brane web realization of $A_N$-type LST.}
    \label{fig:sketchANLST}
\end{figure}

\paragraph{Partition function of $A_N$-type LST.}
We can also use the formulas \eqref{eq:unifiedpert} and \eqref{eq:unifiedinst} but with slight modifications to express the partition function of a general $A$-type LST which has affine $A$-type quiver. For $A_{N}$-type LST, we draw the affine $A_N$ Dynkin diagram and number the nodes as the following:
\begin{center}
\begin{tikzpicture}[scale=0.5]
\draw (-1,0) node[anchor=east] {affine $A_N$ ~~~~~};
\draw (0 cm,0) -- (4 cm,0);
\draw (8 cm,0) -- (6 cm,0);
\draw[densely dashed] (4 cm,0) --(6 cm,0);
\draw (0 cm, 0 cm) -- (4 cm,2 cm);
\draw (8 cm, 0 cm) -- (4 cm,2 cm);
\draw[fill=white] (0 cm, 0 cm) circle (.25cm) node[below=4pt]{$1$};
\draw[fill=white] (2 cm, 0 cm) circle (.25cm) node[below=4pt]{$2$};
\draw[fill=white] (4 cm, 0 cm) circle (.25cm) node[below=4pt]{$3$};
\draw[fill=white] (6 cm, 0 cm) circle (.25cm) node[below=6pt,scale=0.8]{$N-1$};
\draw[fill=white] (8 cm, 0 cm) circle (.25cm) node[below=6pt,scale=0.8]{$N$};
\draw[fill=white] (4 cm, 2 cm) circle (.25cm) node[above=4pt,scale=0.8]{$N+1$};
\node at (9.85,-1){.};
\end{tikzpicture}   
\end{center}
The schematic picture of the corresponding brane web is in figure \ref{fig:sketchANLST}, in which we have labelled the vertical positions of all the color branes and the red dashed lines are the centers of Coulomb branch in each gauge node and the leftmost D5-branes are identified with the corresponding rightmost D5-branes to form the $(N+1)$-th gauge node. In order to be vertically periodic, for each gauge node the condition \eqref{eq:paracondi} has to be satisfied, and due to the fact that the number of leftmost color branes is equal to the number of rightmost color branes by the identification, we can derive that the numbers of color branes in each gauge node are the same, that is
\begin{align}
	n_{(1)}=n_{(2)}=\cdots=n_{(N)}=n_{(N+1)}\ .
\end{align}
Then for each gauge node, \eqref{eq:constraint}, which also needs to be satisfied for the vertical periodicity, reduces to
\begin{align}
	Q_{\textbf{m}}^{2}=Q_{\textbf{r}}Q_{\textbf{l}}\ ,
\end{align}
which means that the center of the Coulomb branch of a gauge node is in the middle of the two centers of the Coulomb branch of the two gauge nodes on its left and right. So the distances between the centers of the Coulomb branch of two adjacent gauge nodes are always the same which means the bifundamental masses are all the same, and we denote this distance with $Q_d$ in figure \ref{fig:sketchANLST}. 
Due to the distance $Q_d$ between adjacent gauge nodes, 
the expressions of vertical positions $y_{I,k}$ in terms of independent parameters become
\begin{equation}
	y_{I,k}=Q_d^{-I}\frac{\prod_{i=k}^{n_{(I)}-1}Q_{I,i}}{\left(\prod_{i=1}^{n_{(I)}-1}Q_{I,i}^i\right)^{\frac{1}{n_{(I)}}}},\quad \text{for }I\in\{1,2,\cdots,N+1\}\ ,
\end{equation}
where we have set the leftmost center of Coulomb branch in figure \ref{fig:sketchANLST} to be the zero point.
The vertical positions of the leftmost color branes are shifted by $Q_d^{N+1}$ to the rightmost color branes:
\begin{align}
	\tilde{y}_{N+1,k}=Q_d^{N+1} y_{N+1,k}\ ,\quad \text{for }k\in\{1,2,\cdots,n_{(N+1)}\}, 
\end{align}
so there are ambiguities in the vertical positions of color branes in the $(N+1)$-th gauge node when we apply the formulas \eqref{eq:unifiedpert} and \eqref{eq:unifiedinst}. 
And also for the $A_N$-type LST, the number of independent vertical strips is $N+1$ rather than $N+2$ in the $D_N,E_N$-type case. So the following modifications to formulas \eqref{eq:unifiedpert} and \eqref{eq:unifiedinst} are needed in order to work for $A_N$-type LST:
\begin{enumerate}
	\item [(1)] We need to use $y_{N+1,k}$ for the vertical positions in the $(N+1)$-th gauge node when $I=N,J=N+1$ or $I=N+1,J=N$, and we need to use $\tilde{y}_{N+1,k}$ for the vertical positions in the $(N+1)$-th gauge node when $I=1,J=N+1$ or $I=N+1,J=1$. When $I=J=N+1$, we can use either $y_{N+1,k}$ or $\tilde{y}_{N+1,k}$. 
	\item [(2)] The factor $(N+2)$ in \eqref{eq:unifiedpert} should be replaced by $(N+1)$. 
\end{enumerate} 
Substitute the Cartan matrix $C_{IJ}$ of affine $A$-type quiver into the modified formulas, we can obtain the partition function of $A$-type LST, and we checked that it agrees with the partition function of $A$-type LST computed in \cite{Hohenegger:2013ala}, we have presented their result in Appendix \ref{app:AtypeLittleString}. 

\bigskip

We can interpret our proposal as the Nekrasov partition function for the 6d $ADE$-type quiver gauge theories discussed in \cite{Nekrasov:2012xe, Nekrasov:2013xda}. In our proposal, the contribution from the vector fields and from the bifundamental fields are given in a unified manner.
For example, the contribution from the vector multiplets of the $I_0$-th gauge node,
\begin{align}
\prod_{k,\ell} \left[
    \varTheta_{\mu_{I_0,k}\mu_{I_0,\ell}}
    \left( \tfrac{y_{I_0,k}}{y_{I_0,\ell}} 
    \Big| Q_{\tau} \right)
    \varTheta_{\mu_{I_0,k}\mu_{I_0,\ell}}
    \left( \tfrac{y_{I_0,k}}{y_{I_0,\ell}} \tfrac{\ft}{\fq} \Big| Q_{\tau} \right)
\right]^{-1},
\end{align}
is obtained from $I=J=I_0$ in \eqref{eq:unifiedinst}, where the first factor comes from $s=+1$ while the second factor comes from $s=-1$. Also, the contribution from the bifundamental multiplets on the $I_0$-th gauge node and the $J_0$-th gauge node,
\begin{align}
\prod_{k,\ell} \left[
    \varTheta_{\mu_{I_0,k}\mu_{J_0,\ell}}
    \left( \tfrac{y_{I_0,k}}{y_{J_0,\ell}} 
    \sqrt{\tfrac{\ft}{\fq}}
    \Big| Q_{\tau} \right)
    \varTheta_{\mu_{J_0,\ell}\mu_{I_0,k}}
    \left( \tfrac{y_{J_0,\ell}}{y_{I_0,k}} \sqrt{\tfrac{\ft}{\fq}} \Big| Q_{\tau} \right)
\right],
\end{align}
is obtained from $I=I_0, J=J_0$, which gives the first factor, and from $I=J_0, J=I_0$, which gives the second factor.
The product for $s$ is performed trivially and the dummy indices $k$ and $\ell$ are exchanged for the second factor.

Finally, we comment on some 6d, 5d and 4d SCFTs limits. By taking the limit $Q_{\tau} \to 0$, we obtain the partition functions for 5d affine quiver gauge theories which are also dual to 6d conformal matters on a circle. The perturbative part of the partition function is given up to flop by
\begin{align}
    Z_{\text{pert}}^{\text{5d}} =\text{PE}\left[\frac{\sqrt{\ft\fq}}{(1-\ft)(1-\fq)}\sum_{I,J}\sum_{k,\ell}\sum_{s=\pm 1}\frac{1}{4}C_{IJ}\frac{y_{J,\ell}}{y_{I,k}}\left(\frac{\ft}{\fq}\right)^{\frac{1}{2}s\,\delta_{IJ}}\right]\ .
\end{align}
 And the instanton part is given by
\begin{align}
    Z_{\text{inst}}^{\text{5d}} = 
    &\sum_{\boldsymbol{\mu}}
    \left( \prod_{I} Q_{I}^{\sum_{k} |\mu_{I,k}|}\right)
    \prod_{I,J} \prod_{k,\ell} \prod_{s=\pm 1} \prod_{(i,j) \in \mu_{I,k}}
    \cr
    & \qquad 
    \left[
    \left( \frac{y_{J,\ell}}{y_{I,k}} \ft^{-\left( \mu^t_{J,\ell} \right)_j + i - \frac{1}{2} + \frac{1}{2} s \delta_{IJ}} \fq^{-\left( \mu_{I,k} \right)_i+j-\frac{1}{2}-\frac{1}{2} s \delta_{IJ}} \right)^{\frac{1}{2}}
    \right.
    \cr
    & \qquad \left.
    - \left( \frac{y_{J,\ell}}{y_{I,k}} \ft^{-\left( \mu^t_{J,\ell} \right)_j + i - \frac{1}{2} + \frac{1}{2} s \delta_{IJ}} \fq^{-\left( \mu_{I,k} \right)_i+j-\frac{1}{2}-\frac{1}{2} s \delta_{IJ}} \right)^{-\frac{1}{2}}
    \right]^{-\frac{1}{2}C_{IJ}}
\cr
    = &\sum_{\boldsymbol{\mu}}
    \left( \prod_{I} Q_{I}^{\sum_{k} |\mu_{I,k}|}\right)
    \prod_{I,J} \prod_{k,\ell} \prod_{s=\pm 1} \prod_{(i,j) \in \mu_{I,k}}
    \cr
    & \qquad \left[ 2\sinh \left(\frac{\boldsymbol{\beta}}{2} \Big( a_{I,k}-a_{J,\ell} + \big( -( \mu^t_{J,\ell} )_j + i - \frac{1}{2} + \frac{1}{2} s \delta_{IJ} \big) \epsilon_2 
    \right. \right.
    \cr
    & \left. \left. \qquad \qquad \qquad \qquad \qquad
    + \big( (\mu_{I,k})_i-j+\frac{1}{2}+\frac{1}{2} s \delta_{IJ} \big) \epsilon_1 \Big) \right) \right]^{-\frac{1}{2}C_{IJ}}.
\end{align}

By further taking the limit ${\boldsymbol{\beta}} \to 0$ while keeping $Q_{I}$ as they are, 
we obtain the partition function for 4d affine quiver gauge theories. The perturbative part of the partition function is given by
\begin{align}
    Z_{\text{pert}}^{\text{4d}} 
    = \prod_{I,J} \prod_{k,\ell} \prod_{s=\pm 1}  \Gamma_{\epsilon_1, \epsilon_2}\left( a_{I,k} - a_{J,\ell} + \frac{1}{2} ( 1 + s \delta_{IJ} ) (\epsilon_1 + \epsilon_2)
    \right)^{-\frac{1}{2}C_{IJ}}\ ,
\end{align}
up to regularization, where $\Gamma_{\epsilon_1, \epsilon_2}$ is the Barnes double gamma function \cite{Barnes},
 which can be regarded as a regularized infinite product
\begin{align}
\Gamma_{\epsilon_1, \epsilon_2} (x)
\propto 
\prod_{m,n \ge 1} \left( x - m \epsilon_1 + (n-1) \epsilon_2 \right)\ ,
\end{align}
in the region $\epsilon_1<0$, $\epsilon_2>0$.
The instanton part is given by
\begin{align}
    Z_{\text{inst}}^{\text{4d}} = 
    &\sum_{\boldsymbol{\mu}}
    \left( \prod_{I} Q_{I}^{\sum_{k} |\mu_{I,k}|}\right)
    \prod_{I,J} \prod_{k,\ell} \prod_{s=\pm 1} \prod_{(i,j) \in \mu_{I,k}}
    \cr
    & \quad \times\left[ a_{I,k}-a_{J,\ell} + \Big( -( \mu^t_{J,\ell} )_j + i - \frac{1}{2} + \frac{1}{2} s \delta_{IJ} \Big) \epsilon_2 
    \right. 
    \cr
    & \left. \qquad \qquad \qquad \qquad \qquad
    + \Big( (\mu_{I,k})_i-j+\frac{1}{2}+\frac{1}{2} s \delta_{IJ} \Big) \epsilon_1  \right]^{-\frac{1}{2}C_{IJ}},
\end{align} 
in which we have used $\sinh ({\boldsymbol{\beta}}x)\rightarrow {\boldsymbol{\beta}}x$ in the limit ${\boldsymbol{\beta}} \to 0$ and the ${\boldsymbol{\beta}}$ factors are extracted out and all cancelled in a similar way that leads to the cancellation of $W$ factors in \eqref{eq:Wcontri}.

From the $(1,0)$ $ADE$-type LSTs, we can also obtain 6d $(1,0)$ $ADE$-type SCFTs \cite{Gadde:2015tra,Haghighat:2017vch} by taking the limit $Q_{I_0}\rightarrow 0$ where $I_0$ denotes the affine node of the affine $ADE$ quiver, which means taking the length of the instanton factor of the affine gauge node to be infinite. 
Then the Young diagrams $\mu_{I_0,k}$ of the affine node are restricted to be empty sets in order to have nonzero contributions to the partition function. We compared the instanton partition function in \eqref{eq:unifiedinst} at this limit with the computation of elliptic genus of the 6d $(1,0)$ $ADE$-type SCFTs in \cite{Haghighat:2017vch} where the unrefined limit is taken in the result and the tensor branch parameters of all the gauge nodes are turned off, our result agrees with theirs when we also take unrefined limit and turn off tensor branch parameters. 


\section{Partition function and symmetry}\label{sec:Symmetry}
In this section, we will see the symmetries of $D$-type and $E$-type LSTs in the form of characters by expanding their partition functions  which are obtained in the previous section.  Although we have constructed brane webs for general $D$-type and $E$-type LSTs, we still need to determine the appropriate parameters to expand the partition function correctly and to represent the fugacities of the symmetry group. Surprisingly, we find that the brane web has all the information of the Weyl group symmetry that the LSTs should have.  Then, from the Weyl group symmetry in the brane web, we can determine the appropriate parameters for the fugacities of the symmetry group as well as the appropriate expansion parameters. Moreover, the Weyl group symmetry is also helpful in determining the coefficients of the expansion of the partition function. 
In the following, we first give a brief introduction about Weyl group symmetry hidden in brane webs, and then we derive the expansion forms of the partition functions of $D_4$, $D_5$ and $E_6$-type LSTs by utilizing the Weyl group symmetry. 

\subsection{Weyl group symmetry in brane web}
One of the ways to see the (global) symmetry of a gauge theory explicitly is to expand its partition function with the expansion parameters being the Coulomb branch parameters, and then the coefficients of the Coulomb branch parameters will turn out to be expressed by the characters of the symmetry group. 
In this process, we have to make sure that the Coulomb branch parameters are not affected by the Weyl reflections of the symmetry group, so we have to properly choose the so called invariant Coulomb branch parameters \cite{Mitev:2014jza, Hayashi:2019jvx} as the expansion parameters which are invariant under the Weyl reflection. Thus the partition function has Weyl group symmetry as we expect. We will see such Weyl group symmetry after we obtain the expansion of partition function. 
However, the Weyl group symmetry can also be seen just from the brane web without computing the partition functions.

For example, a part of the Weyl reflections can be identified as the fiber-base duality \cite{Katz:1997eq, Aharony:1997bh, Bao:2011rc} 
in the case of 5d rank-1 gauge theories \cite{Mitev:2014jza}. Another simpler example is exchanging parallel flavor branes on the same side, which corresponds to the Weyl group symmetry of exchanging mass parameters. These examples illustrate that the Weyl reflections of the symmetry group can be seen from the brane web. In this paper, we are dealing with the affine $D$-type and $E$-type quiver gauge theories, and we will find that all the independent Weyl reflections of the symmetry groups of these quiver theories can be identified with certain symmetries of the corresponding brane webs. In the following subsections, we will show that the partition functions have the same symmetries as the web diagrams. 

\subsection{Expansion of \texorpdfstring{$D_4$}{D4}-type LST}
\begin{figure}[htbp]
	\centering
	\includegraphics[scale=0.5]{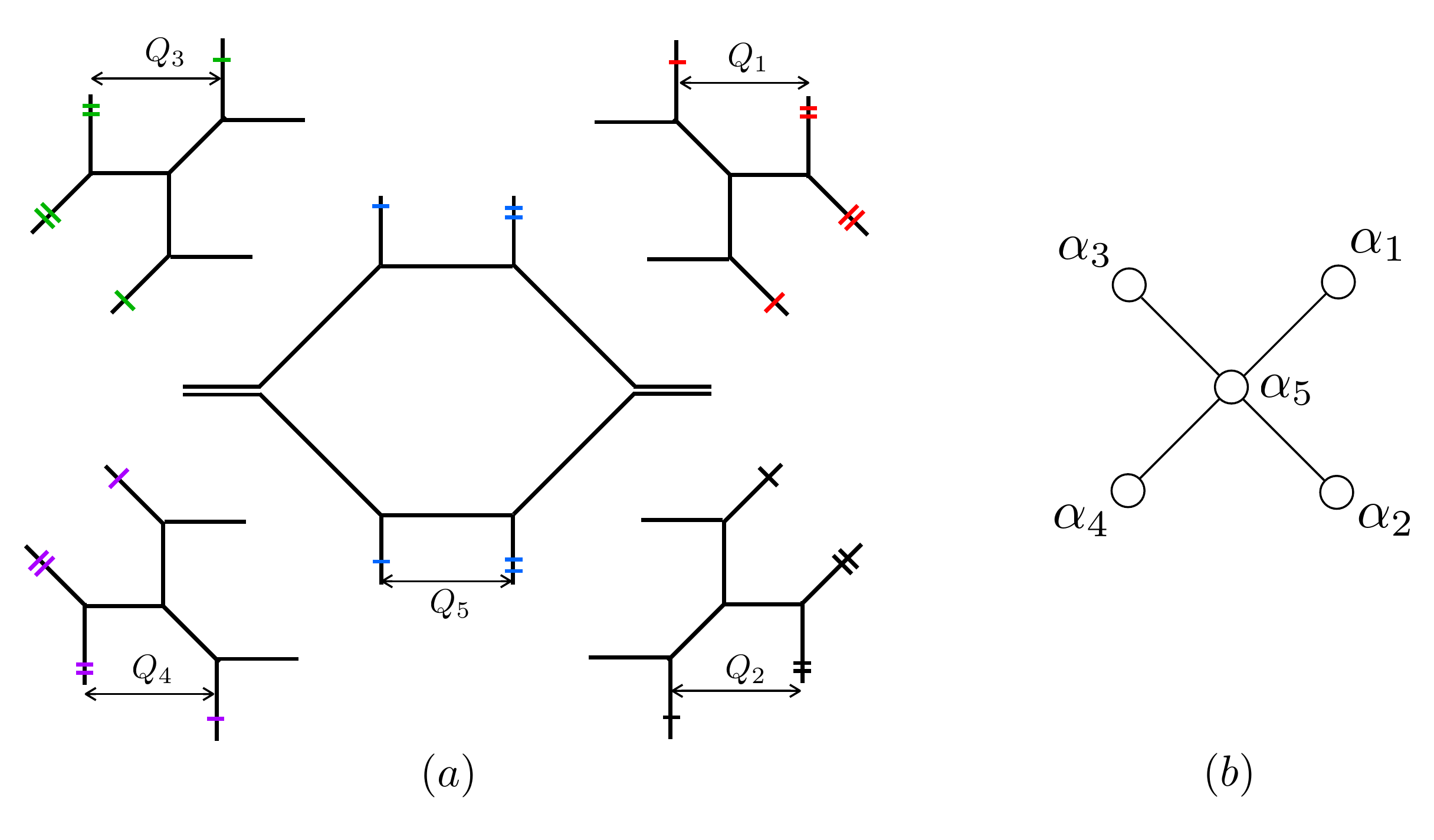}
	\caption{(a) A quadrivalent gluing brane web of $D_4$-type LST. (b) An affine Dynkin diagram of affine $D_4$. }
	\label{fig:D4Dynkin}
\end{figure}
For the brane web of $D_4$-type LST, as shown in figure \ref{fig:D4Dynkin}(a), each gauge node has two parallel external NS-charged branes\footnote{Note that we have identified the top and bottom NS-charged branes, so they are just counted as a single brane. }, so the configuration of the brane web is invariant if we swap the two NS-charged branes. In this swapping, the K\"ahler parameters will accordingly be transformed. For the $D_4$-type LST, we have five gauge nodes, and we will have five K\"ahler parameter transformations that correspond to the swapping of external NS-charged branes for each of the five nodes. 
The five transformations are as follows:
\begin{equation}\label{eq:D4WeylinQaff}
	\left\{\begin{array}{l}
	Q_1\rightarrow Q_1^{-1},\ Q_5\rightarrow Q_5Q_1;\\
	Q_2\rightarrow Q_2^{-1},\ Q_5\rightarrow Q_5Q_2;\\
	Q_3\rightarrow Q_3^{-1},\ Q_5\rightarrow Q_5Q_3;\\
	Q_4\rightarrow Q_4^{-1},\ Q_5\rightarrow Q_5Q_4;\\
	Q_5\rightarrow Q_5^{-1},\ Q_1\rightarrow Q_1Q_5,\ Q_2\rightarrow Q_2Q_5,\ Q_3\rightarrow Q_3Q_5,\ Q_4\rightarrow Q_4Q_5;	
	\end{array}\right.
\end{equation}
where we have omitted the trivial transformations such as $Q_2\rightarrow Q_2$.
Due to the structure of the brane web in figure \ref{fig:D4Dynkin}(a), $Q_1$ is only connected to $Q_5$. Then, when $Q_1$ is transformed to $Q_1^{-1}$, only $Q_5$ is affected. However, $Q_5$ is connected to $Q_1,Q_2,Q_3,Q_4$, so when $Q_5$ is transformed to $Q_5^{-1}$, $Q_1,Q_2,Q_3,Q_4$ are all affected. 

On the other hand, we consider the Dynkin diagram of affine $D_4$ given in figure \ref{fig:D4Dynkin}(b), where we have labeled the five affine simple roots whose indices are in accord with the $Q$'s in figure \ref{fig:D4Dynkin}(a). Via the following formula for (affine) Weyl reflection
\begin{align}\label{eq:affineWeylref}
	r_{\alpha_i}\cdot \alpha_j=\alpha_j-\frac{2(\alpha_j,\alpha_i)}{(\alpha_i,\alpha_i)}\alpha_i,
\end{align}
where $r_{\alpha_i}$ is the (affine) Weyl reflection corresponding to the simple root $\alpha_i$ and it acts on the simple root $\alpha_j$, 
we can easily obtain the five affine Weyl reflections that correspond to the five simple roots:
\begin{equation}\label{eq:D4Weylinalpha}
	\left\{\begin{array}{l}
	r_{\alpha_1}:\ \ \alpha_1\rightarrow -\alpha_1,\ \alpha_5\rightarrow \alpha_5+\alpha_1;\\
	r_{\alpha_2}:\ \ \alpha_2\rightarrow -\alpha_2,\ \alpha_5\rightarrow \alpha_5+\alpha_2;\\
	r_{\alpha_3}:\ \ \alpha_3\rightarrow -\alpha_3,\ \alpha_5\rightarrow \alpha_5+\alpha_3;\\
	r_{\alpha_4}:\ \ \alpha_4\rightarrow -\alpha_4,\ \alpha_5\rightarrow \alpha_5+\alpha_4;\\
	r_{\alpha_5}:\ \ \alpha_5\rightarrow -\alpha_5,\ \alpha_1\rightarrow \alpha_1+\alpha_5,\ \alpha_2\rightarrow \alpha_2+\alpha_5,\ \alpha_3\rightarrow \alpha_3+\alpha_5,\ \alpha_4\rightarrow \alpha_4+\alpha_5;	
	\end{array}\right.
\end{equation}
where we have omitted the trivial transformations such as $\alpha_2\rightarrow \alpha_2$. 
By comparing \eqref{eq:D4WeylinQaff} with \eqref{eq:D4Weylinalpha}, we realize that the $Q$'s are just the fugacities of affine $D_4$ in $\alpha$-basis and the five transformations in \eqref{eq:D4WeylinQaff} are actually the five Weyl reflections in \eqref{eq:D4Weylinalpha}.

For usual brane web which is just a single web diagram, we know that its partition function computed by topological vertex formalism is flop invariant up to possible analytic continuation, where the ``flop" includes flop transitions which change the configuration of the brane web as well as only swappings of parallel branes which do not change the brane web configuration. 
Although we use trivalent and quadrivalent gluing of brane webs in this paper which are not usual brane webs, we will still assume that the partition function of trivalently and quadrivalently glued brane webs is flop invariant up to possible analytic continuation.  Now if we start with the brane webs in figure \ref{fig:D4Dynkin}(a) and swap the external NS-charged branes, the partition function $Z^{D_4}_{\text{swapped}}$ of the swapped brane webs will be the same as the partition function $Z^{D_4}_{\text{original}}$ of the original brane webs up to possible analytic continuation: 
\begin{equation}\label{eq:flopinvariance}
	Z^{D_4}_{\text{original}}(Q_1,Q_2,Q_3,Q_4,Q_5)= Z^{D_4}_{\text{swapped}}\ .
\end{equation}
On the other hand, the partition function $Z^{D_4}_{\text{swapped}}$ of the swapped brane webs can also be obtained by just replacing the K\"ahler parameters in the partition function $Z^{D_4}_{\text{original}}$ of the original brane web by the transformed K\"ahler parameters via \eqref{eq:D4WeylinQaff} due to the invariance of the configuration of the brane webs under swapping external NS-charged branes:
\begin{equation}\label{eq:configinvariance}
	Z^{D_4}_{\text{swapped}}=Z^{D_4}_{\text{original}}(Q_1',Q_2',Q_3',Q_4',Q_5')\ ,
\end{equation}
where $Q_1',Q_2',Q_3',Q_4',Q_5'$ are the transformed K\"ahler parameters, for example, if we swap the two NS-charged branes of node 1 in figure \ref{fig:D4Dynkin}, then the transformed K\"ahler parameters are $Q_1'=Q_1^{-1},Q_2'=Q_2,Q_3'=Q_3,Q_4'=Q_4,Q_5'=Q_5Q_1$ according to the first line in \eqref{eq:D4WeylinQaff} which is the Weyl reflection of $Q_1$. Then from \eqref{eq:flopinvariance} and \eqref{eq:configinvariance}, up to analytic continuation we get
\begin{equation}\label{eq:D4argument}
	Z^{D_4}_{\text{original}}(Q_1,Q_2,Q_3,Q_4,Q_5)= Z^{D_4}_{\text{original}}(Q_1',Q_2',Q_3',Q_4',Q_5')\ .
\end{equation}
Thus we conclude that: 
	The partition function of $D_4$-type LST is invariant up to possible analytic continuation under the transformations in \eqref{eq:D4WeylinQaff} which are all the independent affine Weyl reflections of affine $D_4$, so \textit{the partition function of $D_4$-type LST has affine $D_4$ symmetry.}

Note that in the above analysis we focus on $Q_1,\cdots,Q_5$ which are the fugacities of affine $D_4$ in $\alpha$-basis, but aside from these parameters there are also  other parameters such as Coulomb branch parameters, usually such parameters are also transformed under Weyl reflections. In order to see the symmetry manifestly, we need to choose the proper expansion parameters that are invariant under Weyl reflections and expand the partition function with respect to these invariant expansion parameters, then we will see the characters of the symmetry group in the coefficients of the expansion \cite{Hayashi:2019jvx}. In principle, if we choose the expansion parameters to be invariant under the affine $D_4$ Weyl reflections, we will see affine $D_4$ characters in the expansion of partition function. But the affine characters are infinite polynomials that contain infinitely high orders of fugacities, in order to see these infinitely high orders we need to sum over all the horizontal Young diagrams in the brane webs up to infinity which is demanding. So, instead we try to just see the $D_4$ symmetry whose characters are finite polynomials. Then we should choose the expansion parameters to be invariant under $D_4$ Weyl reflections. There are two ways to reduce affine $D_4$ to $D_4$ in the $D_4$-type LST which correspond to two different choices of selecting four Weyl reflections out of the five affine $D_4$ Weyl reflections.

\paragraph{Choice I.} One choice is to select the Weyl reflections corresponding to $Q_1,Q_2,Q_3,Q_4$.
The remaining parameter $Q_5$ can not be used as a proper expansion parameter since it transforms under these four Weyl reflections non-trivially, so we replace $Q_5$ by the horizontal period $u$ in \eqref{eq:horiofD4}
which is Weyl reflection invariant. Now the four chosen affine Weyl reflections reduce to the following transformations:
\begin{equation}\label{eq:D4Weylun}
	\left\{\begin{array}{l}
	Q_1\rightarrow Q_1^{-1};\\
	Q_2\rightarrow Q_2^{-1};\\
	Q_3\rightarrow Q_3^{-1};\\
	Q_4\rightarrow Q_4^{-1}.
	\end{array}\right.
\end{equation}
They are the Weyl reflections of SU(2) in each SU(1) gauge node, but they do not form the Weyl reflections of $D_4$. By looking at the brane web in figure \ref{fig:D4Dynkin}(a), we can see that there is a permutation symmetry among $Q_1,Q_2,Q_3,Q_4$,
\begin{equation}\label{eq:permute}
	\left\{\begin{array}{l}
	Q_1\rightarrow Q_2,\ Q_2\rightarrow Q_1;\\
	Q_2\rightarrow Q_3,\ Q_3\rightarrow Q_2;\\
	Q_3\rightarrow Q_4,\ Q_4\rightarrow Q_3;
	\end{array}\right.
\end{equation}
 because the four SU(1) subdiagrams share the same Coulomb branch parameter with the middle SU(2) subdiagram, and they only differ by their instanton factors $Q_1,Q_2,Q_3,Q_4$. If we pick only one of the transformations in \eqref{eq:D4Weylun}, we can use the permutation symmetry to obtain the other three transformations in \eqref{eq:D4Weylun} by using \eqref{eq:permute}. 
 So the SU(2) Weyl reflection and the permutation symmetry form the following symmetry of the theory
 \begin{equation}\label{eq:fundsymofD4rk1}
 	\left\{\begin{array}{l}
	Q_1\rightarrow Q_1^{-1};\\
	Q_1\rightarrow Q_2,\ Q_2\rightarrow Q_1;\\
	Q_2\rightarrow Q_3,\ Q_3\rightarrow Q_2;\\
	Q_3\rightarrow Q_4,\ Q_4\rightarrow Q_3.
	\end{array}\right.
 \end{equation}
From this symmetry, we can obtain the following transformations
\begin{equation}\label{eq:D4Weylrefl}
	\left\{\begin{array}{l}
	Q_1\rightarrow Q_2,\ Q_2\rightarrow Q_1;\\
	Q_2\rightarrow Q_3,\ Q_3\rightarrow Q_2;\\
	Q_3\rightarrow Q_4,\ Q_4\rightarrow Q_3;\\
	Q_3\rightarrow Q_4^{-1},\ Q_4\rightarrow Q_3^{-1}.
	\end{array}\right.
\end{equation}
They are the Weyl reflections of $D_4$ in orthonormal basis fugacities $Q_1,Q_2,Q_3,Q_4$. 

Note that this method of obtaining the $D_4$ symmetry is applicable only for the rank-1 $D_4$-type LST. If we consider higher rank $D_4$-type LST, there is no permutation symmetry that only involves $Q_1,Q_2,Q_3,Q_4$ because each one of the four gauge nodes in the corners will have different Coulomb branch parameters in the higher rank case, and the permutation symmetry will involve both instanton factors and Coulomb branch parameters of the four gauge nodes in the corners.

For convenience in computations, we use the following new Coulomb branch parameter
\begin{align}
	w_5\equiv Q_{5,1}^{1/2}\ .
\end{align}
$w_5$ is invariant under the Weyl reflections in \eqref{eq:D4Weylrefl}, so  it is an invariant Coulomb branch parameter and can be used as an expansion parameter. Note that $w_5$ is not affected by the four SU(2) Weyl reflections in \eqref{eq:D4Weylun} either, so the coefficients of $w_5$ in the expansion of the partition function will also have SU(2) Weyl group symmetries. $Q_{\tau}$ is Weyl reflection invariant, so it is also an expansion parameter.

\paragraph{Choice II.} The other choice is to select the Weyl reflections\footnote{Of course we can also choose $Q_2,Q_3,Q_4,Q_5$ and so on. They are equivalent choices. } correspond to $Q_1,Q_2,Q_3,Q_5$, and we replace $Q_4$ by the horizontal period $u$ in \eqref{eq:horiofD4} as the expansion parameter. 
The four chosen affine Weyl reflections reduce to the following:
\begin{equation}\label{eq:unaffD4WeylinQ}
	\left\{\begin{array}{l}
	Q_3\rightarrow Q_3^{-1},\ Q_5\rightarrow Q_5Q_3;\\
	Q_5\rightarrow Q_5^{-1},\ Q_1\rightarrow Q_1Q_5,\ Q_2\rightarrow Q_2Q_5,\ Q_3\rightarrow Q_3Q_5;\\
	Q_1\rightarrow Q_1^{-1},\ Q_5\rightarrow Q_5Q_1;\\
	Q_2\rightarrow Q_2^{-1},\ Q_5\rightarrow Q_5Q_2.
	\end{array}\right.
\end{equation}
They are the $D_4$ Weyl reflections in the basis $\{\alpha_3,\alpha_5,\alpha_1,\alpha_2\}$ where the $\alpha$'s as well as the corresponding $Q$'s are ordered in a conventional way for $D_4$ simple roots.
\begin{figure}[htbp]
	\centering
	\includegraphics[scale=0.6]{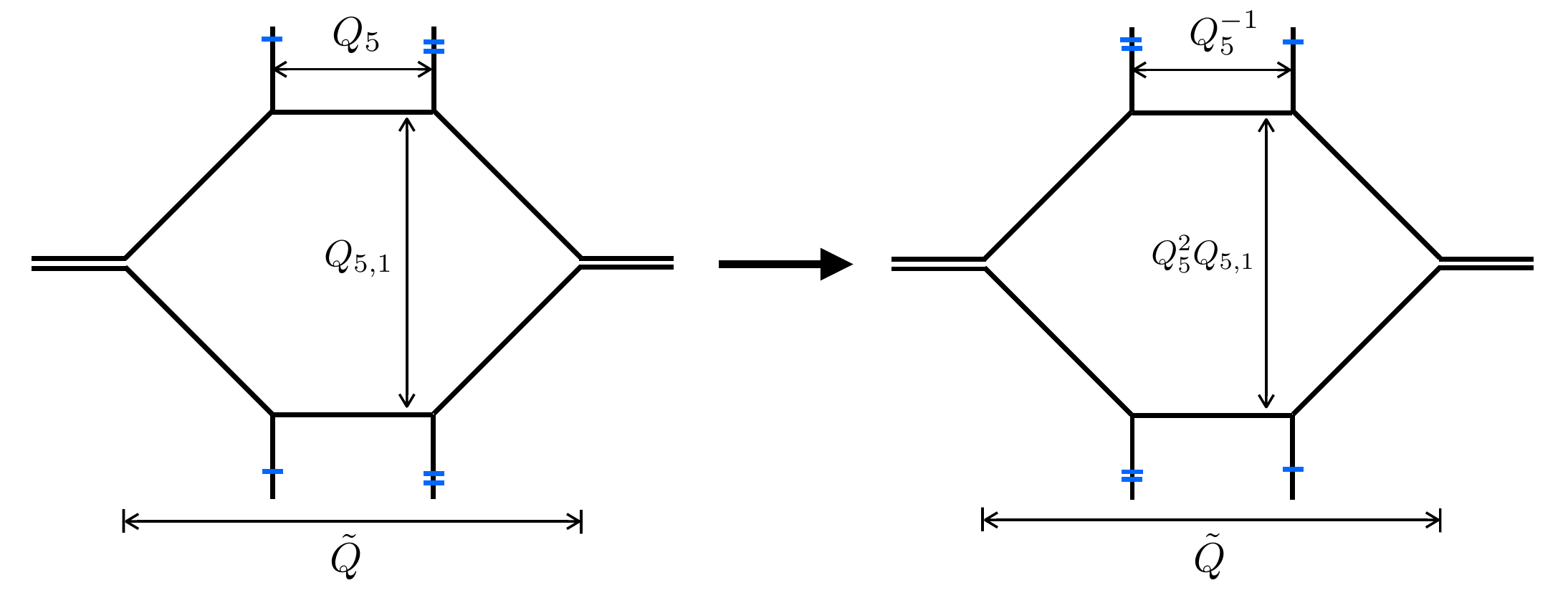}
	\caption{Transformation of Coulomb branch parameter in gauge node 5.}
	\label{fig:D4Coulomb}
\end{figure}

The Coulomb branch parameter $w_5$ does not change under the Weyl reflections of $Q_3,Q_1,Q_2$ , but it changes under the Weyl reflection of $Q_5$ due to the invariance of $\tilde{Q}$, as shown in figure \ref{fig:D4Coulomb}.  Including Coulomb branch parameter $w_5$, the Weyl reflections become as follows:
\begin{equation}\label{eq:unaffD4WeylinQwithw}
	\left\{\begin{array}{l}
	Q_3\rightarrow Q_3^{-1},\ Q_5\rightarrow Q_5Q_3;\\
	Q_5\rightarrow Q_5^{-1},\ Q_1\rightarrow Q_1Q_5,\ Q_2\rightarrow Q_2Q_5,\ Q_3\rightarrow Q_3Q_5,\ w_5\rightarrow Q_5 w_5;\\
	Q_1\rightarrow Q_1^{-1},\ Q_5\rightarrow Q_5Q_1;\\
	Q_2\rightarrow Q_2^{-1},\ Q_5\rightarrow Q_5Q_2.
	\end{array}\right.
\end{equation}
The invariant Coulomb branch parameter will take the general form of 
\begin{equation}
	Q_1^a Q_2^b Q_3^c Q_5^d w_5\ .
\end{equation}
Then, by substituting the four Weyl reflections in \eqref{eq:unaffD4WeylinQwithw} into the general form respectively and requiring the invariance under these transformations, we can get four linear equations of $a,b,c,d$, and obtain one solution. According to this solution, we get the invariant Coulomb branch parameter $A$
\begin{align}
	A=Q_1Q_2Q_3Q_5^2w_5\ .
\end{align}
If we switch from the $\alpha$-basis fugacities $Q$'s to the orthonormal basis fugacities $B$'s, we have
\begin{align}\label{eq:D4orthonormal}
Q_3=\frac{B_1}{B_2},\quad
Q_5=\frac{B_2}{B_3},\quad
Q_1=\frac{B_3}{B_4},\quad
Q_2=B_3 B_4\ .
\end{align}
Then in the orthonormal basis, the $D_4$ Weyl reflections in \eqref{eq:unaffD4WeylinQ} become as follows:
\begin{equation}\label{eq:D4WeylinB}
	\left\{\begin{array}{l}
	B_1\rightarrow B_2,\ B_2\rightarrow B_1;\\
	B_2\rightarrow B_3,\ B_3\rightarrow B_2;\\
	B_3\rightarrow B_4,\ B_4\rightarrow B_3;\\
	B_3\rightarrow B_4^{-1},\ B_4\rightarrow B_3^{-1}.
	\end{array}\right.
\end{equation}

\bigskip

In the following part of this subsection, we will compute the expansion of  the partition function of the $D_4$-type LST. There are two different ways of doing the expansion due to the two different choices of Weyl reflections. 

From \eqref{eq:D4inst}, we have
\begin{align}\label{eq:oldD4inst}
	Z^{D_4}_{\text{inst}}=\sum_{\mu_{5,1},\mu_{5,2}}\bar{Z}_{\mu_{5,1}\mu_{5,2}}^{\text{SU(2)}}(w_5,Q_5)&\sum_{\mu_{1,1}}\bar{Z}_{\mu_{5,1}\mu_{5,2}\mu_{1,1}}(w_5,Q_1)\sum_{\mu_{2,1}}\bar{Z}_{\mu_{5,1}\mu_{5,2}\mu_{2,1}}(w_5,Q_2)\nn\\
	\times &\sum_{\mu_{3,1}}\bar{Z}_{\mu_{5,1}\mu_{5,2}\mu_{3,1}}(w_6,Q_3)\sum_{\mu_{4,1}}\bar{Z}_{\mu_{5,1}\mu_{5,2}\mu_{4,1}}(w_6,Q_4),
\end{align}
where
\begin{align}
	&\bar{Z}_{\mu_{5,1}\mu_{5,2}}^{\text{SU(2)}}(w_5,Q_5)\equiv\frac{Q_5^{|\mu_{5,1}|+|\mu_{5,2}|}}{\prod_{p=1}^{2}\prod_{r=1}^{2}\varTheta_{\mu_{5,p}\mu_{5,r}}(\frac{y_{5,p}}{y_{5,r}}|Q_{\tau})\varTheta_{\mu_{5,p}\mu_{5,r}}(\frac{y_{5,p}}{y_{5,r}}\tfrac{\ft}{\fq}|Q_{\tau})},\label{eq:zsu2ofD4}\\
	&\bar{Z}_{\mu_{5,1}\mu_{5,2}\mu_{1,1}}(w_5,Q_1)\equiv Q_{1}^{|\mu_{1,1}|}
    \frac{\prod_{l=1}^{2}\varTheta_{\mu_{1,1}\mu_{5,l}}(\frac{y_{1,1}}{y_{5,l}}\sqrt{\tfrac{\ft}{\fq}}|Q_{\tau})\prod_{p=1}^2\varTheta_{\mu_{5,p}\mu_{1,1}}(\frac{y_{5,p}}{y_{1,1}}\sqrt{\tfrac{\ft}{\fq}}|Q_{\tau})}{\varTheta_{\mu_{1,1}\mu_{1,1}}(1|Q_{\tau})\varTheta_{\mu_{1,1}\mu_{1,1}}(\tfrac{\ft}{\fq}|Q_{\tau})}.\label{eq:zsu1ofD4}
\end{align}
We can observe from \eqref{eq:zsu2ofD4} and \eqref{eq:zsu1ofD4} that
\begin{align}
	\bar{Z}_{\mu_{5,1}\mu_{5,2}}^{\text{SU(2)}}(w_5,Q_5)=&\bar{Z}_{\mu_{5,2}\mu_{5,1}}^{\text{SU(2)}}(w_5^{-1},Q_5)\ , \\
	\bar{Z}_{\mu_{5,1}\mu_{5,2}\mu_{1,1}}(w_5,Q_1)=&\bar{Z}_{\mu_{5,2}\mu_{5,1}\mu_{1,1}}(w_5^{-1},Q_1)\ ,
\end{align}
which reveals the Weyl reflection $w_5\rightarrow w_5^{-1}$ of the SU(2) gauge group and the invariance of the brane web configuration under this Weyl reflection.

We extract the zeroth order term of $Q_5$ from \eqref{eq:oldD4inst} and combine it with $Z^{D_4}_{\text{pert}}$ to form the zeroth order of $Q_5$ which we denote as $Z^{D_4}_{Q_5^{\bf 0}}$ for the full partition function. Then the full partition function of the $D_4$-type LST can be rewritten as
\begin{equation}\label{eq:ZD4decompose}
	Z^{D_4}=Z^{D_4}_{Q_5^{\bf 0}}\cdot Z^{D_4}_{Q_5} \ , 
\end{equation}
where 
\begin{align}\label{eq:nZ_D4pert}
	Z^{D_4}_{Q_5^{\bf 0}}\equiv Z^{D_4}_{\text{pert}}\cdot &\sum_{\mu_{1,1}}\bar{Z}_{\oo\oo\mu_{1,1}}(w_5,Q_1) \sum_{\mu_{2,1}}\bar{Z}_{\oo\oo\mu_{2,1}}(w_5,Q_2)\nonumber\\
	\times &\sum_{\mu_{3,1}}\bar{Z}_{\oo\oo\mu_{3,1}}(w_5,Q_3)\sum_{\mu_{4,1}}\bar{Z}_{\oo\oo\mu_{4,1}}(w_5,Q_4)\ ,
\end{align}
\begin{align}
Z^{D_4}_{Q_5}
\equiv &\sum_{\mu_{5,1}\mu_{5,2}}\bar{Z}_{\mu_{5,1}\mu_{5,2}}^{\text{SU(2)}}(w_5,Q_5)\frac{\sum_{\mu_{1,1}}\bar{Z}_{\mu_{5,1}\mu_{5,2}\mu_{1,1}}(w_5,Q_{1})}{\sum_{\mu_{1,1}}\bar{Z}_{\oo\oo\mu_{1,1}}(w_5,Q_1)}\frac{\sum_{\mu_{2,1}}\bar{Z}_{\mu_{5,1}\mu_{5,2}\mu_{2,1}}(w_5,Q_{2})}{\sum_{\mu_{2,1}}\bar{Z}_{\oo\oo\mu_{2,1}}(w_5,Q_2)}\nn\\
&\times\frac{\sum_{\mu_{3,1}}\bar{Z}_{\mu_{5,1}\mu_{5,2}\mu_{3,1}}(w_5,Q_{3})}{\sum_{\mu_{3,1}}\bar{Z}_{\oo\oo\mu_{3,1}}(w_5,Q_3)}\frac{\sum_{\mu_{4,1}}\bar{Z}_{\mu_{5,1}\mu_{5,2}\mu_{4,1}}(w_5,Q_{4})}{\sum_{\mu_{4,1}}\bar{Z}_{\oo\oo\mu_{4,1}}(w_5,Q_4)}\quad. \label{eq:nZ_D4inst}
\end{align}
$Z^{D_4}_{Q_5}$ is the part of partition function that depends on $Q_5$. 

\subsubsection{\texorpdfstring{$D_4$}{D4} symmetry from expansion I}\label{sec:choice1}
In choice I, we use $Q_1,Q_2,Q_3,Q_4$ as the fugacities and $w_5,u,Q_{\tau}$ as the expansion parameters. Replacing $Q_5$ by $u$, $Z^{D_4}_{Q_5^{\bf 0}}$ becomes the zeroth order of $u$ which we denote as $Z^{D_4}_{u^{\bf 0}}$, and $Z^{D_4}_{Q_5}$ becomes the $u$-dependent part $Z^{D_4}_{u}$, 
\begin{align}
	&Z^{D_4}_{u^{\bf 0}}=Z^{D_4}_{Q_5^{\bf 0}},\qquad  Z^{D_4}_{u}=Z^{D_4}_{Q_5}\ ,\label{eq:ZuD4}\\
	&Z^{D_4}=Z^{D_4}_{u^{\bf 0}}\cdot Z^{D_4}_u\ .
\end{align}

\paragraph{$Z^{D_4}_{u^{\bf 0}}$ part.}
From the general formula in \eqref{eq:unifiedpert}, we can obtain the perturbative part of the partition function of $D_4$-type LST,
\begin{align}\label{eq:D4pertPE}
	Z^{D_4}_{\text{pert}}\nn\\
	=\text{PE}&\bigg[\frac{1}{(1-\ft)(1-\fq)}\frac{1+Q_{\tau}}{1-Q_{\tau}}\Big(3(\ft+\fq)-4\sqrt{\ft\fq}(Q_{5,1}^{\frac12}+Q_{5,1}^{-\frac12})+\frac{1}{2}(\ft+\fq)(Q_{5,1}+Q_{5,1}^{-1})\Big)\nn\\
	&+\frac{6Q_{\tau}}{1-Q_{\tau}}\bigg].
\end{align}
We expand \eqref{eq:D4pertPE} up to first order of $Q_{\tau}$ and get
\begin{align}\label{eq:zoldpertd4}
	Z^{D_4}_{\text{pert}} 
	=~\text{PE}\bigg[&\frac{1}{(1-\ft)(1-\fq)}\bigg(-8\sqrt{\ft\fq}w_5+(\ft+\fq)(3+w_5^2)+Q_{\tau}\Big(6(1+\ft\fq)\nn\\
	&-8\sqrt{\ft\fq}(w_5+w_5^{-1})
	+(\ft+\fq)(w_5^2+w_5^{-2})\Big)+\calO(Q_{\tau}^2)\bigg)\bigg],
\end{align}
where we have flipped the negative power of $w_5$ by \eqref{eq:flop} in the zeroth order of $Q_{\tau}$. 

The function $\bar{Z}_{\oo\oo\mu_{1,1}}$ is proportional to $Q_1^{|\mu_{1,1}|}$, so $\sum_{\mu_{1,1}}\bar{Z}_{\oo\oo\mu_{1,1}}$ is an infinite polynomial of $Q_1$. We expand it up to $Q_{\tau}Q_1^4$ inside PE,
\begin{align}
&\sum_{\mu_{1,1}}\bar{Z}_{\oo\oo\mu_{1,1}}(w_5,Q_1)\\
=&~\text{PE}\bigg[\frac{Q_1\big(\fq+\ft-\sqrt{\fq\ft}(w_5+w_5^{-1})\big)}{(1-\fq)(1-\ft)}+\frac{Q_{\tau}}{(1-\fq)(1-\ft)}\bigg(\frac{Q_1}{\fq\ft}(\fq+\ft)P_0(w_5)
 \cr
& +\frac{Q_1^2}{\fq^2\ft^2}(\fq^3+\fq^2\ft+\fq\ft^2+\ft^3)P_0(w_5)+\frac{Q_1^3}{\fq^3\ft^3}(\fq^5+\fq^4\ft+\fq^3\ft^2+\fq^2\ft^3+\fq\ft^4+\ft^5)P_0(w_5)
\cr 
&+\frac{Q_1^4}{\fq^4\ft^4}(\fq^7+\fq^6\ft+\fq^5\ft^2+\fq^4\ft^3+\fq^3\ft^4+\fq^2\ft^5+\fq\ft^6+\ft^7)P_0(w_5)+\calO(Q_1^5)\bigg)+\calO(Q_{\tau}^2)\bigg],\nn
\end{align}
where
\begin{equation}
    P_0(w_5)=w_5^{-2}(\sqrt{\fq}w_5-\sqrt{\ft})(\sqrt{\ft}w_5-\sqrt{\fq})(w_5-\sqrt{\fq\ft})(\sqrt{\fq\ft}w_5-1).
\end{equation}
We find that the coefficient of $Q_{\tau}$ is an infinite polynomial that has a pattern of a geometric series and can be analytically summed:
\begin{align}
	&\frac{Q_1}{\fq\ft}(\fq+\ft)P_0(w_5)+\frac{Q_1^2}{\fq^2\ft^2}(\fq^3+\fq^2\ft+\fq\ft^2+\ft^3)P_0(w_5)\nn\\
    &+\frac{Q_1^3}{\fq^3\ft^3}(\fq^5+\fq^4\ft+\fq^3\ft^2+\fq^2\ft^3+\fq\ft^4+\ft^5)P_0(w_5)\nn\\
	&+\frac{Q_1^4}{\fq^4\ft^4}(\fq^7+\fq^6\ft+\fq^5\ft^2+\fq^4\ft^3+\fq^3\ft^4+\fq^2\ft^5+\fq\ft^6+\ft^7)P_0(w_5)+\calO(Q_1^5)\nn\\
	&=\sum_{n=1}^{\infty}\left(\frac{Q_1^n}{\fq^n\ft^n}\sum_{i=0}^{2n-1}\fq^{2n-1-i}\ft^i\right)P_0(w_5)=\frac{Q_1(\fq+\ft)}{(Q_1\fq-\ft)(Q_1\ft-\fq)}P_0(w_5),
\end{align}
so
\begin{align}\label{eq:zsubu1}
	\sum_{\mu_{1,1}}\bar{Z}_{\oo\oo\mu_{1,1}}(w_5,Q_1)=\text{PE}&\Bigg[\frac{\frac12(\fq+\ft)(Q_1+Q_1^{-1})-\sqrt{\fq\ft}(Q_1+Q_1^{-1})w_5}{(1-\fq)(1-\ft)}\nn\\
    & +\frac{Q_{\tau}P_0(w_5)}{(1-\fq)(1-\ft)}\frac{Q_1(\fq+\ft)}{(Q_1\fq-\ft)(Q_1\ft-\fq)}+\calO(Q_{\tau}^2)\Bigg]\ ,
\end{align}
where we have used \eqref{eq:flop} in the zeroth order term of $Q_{\tau}$. 
From \eqref{eq:zsubu1}, we can see that $\sum_{\mu_{1,1}}\bar{Z}_{\oo\oo\mu_{1,1}}$ is invariant under $Q_1\rightarrow Q_1^{-1}$ at the zeroth order and first order of $Q_{\tau}$ since it has the Weyl group symmetry of SU(2) in \eqref{eq:D4Weylun}.

Substituting \eqref{eq:zoldpertd4} and \eqref{eq:zsubu1} into \eqref{eq:nZ_D4pert}, we obtain the zeroth order of $u$ in terms of $\chi_{\bf 1}, \chi_{\bf 2}, \chi_{\bf s}$ and $\chi_{\bf c}$ which are characters of fundamental weights of $D_4$ whose expressions are in \eqref{eq:D4fundamentals} but with the orthonormal basis being $\{Q_1,Q_2,Q_3,Q_4\}$ : 
\begin{align}\label{eq:PEZD4withu0}
	Z^{D_4}_{u^{\bf 0}}=&~\text{PE}\bigg[\frac12(\chi_{\bf 1}+6)\left[0,\frac12\right]+(\chi_{\bf 1}+8)[0,0]w_5+\left[0,\frac12\right]w_5^2\nn\\
	&+Q_{\tau}\Big(c_0+c_1\big(w_5^{-1}+w_5\big)+c_2\big(w_5^{-2}+w_5^2\big)\Big)+\calO(Q_{\tau}^2)\bigg],
\end{align}
with 
\begin{align}
	c_0=&6\left[\frac12,0\right]+\bigg((-2\chi_{\bf 1}+2\chi_{\bf c}\chi_{\bf s})\bigg[0,\frac12\bigg]-4\chi_{\bf 2}\bigg[0,\frac32\bigg]+6\chi_{\bf 1}\bigg[0,\frac52\bigg]-8\bigg[0,\frac72\bigg]\nn\\
	&+(-\chi_{\bf 1}+\chi_{\bf c}\chi_{\bf s})\bigg[\frac12,0\bigg]+(-\chi_{\bf 1}-2\chi_{\bf 2}+\chi_{\bf c}\chi_{\bf s})\bigg[\frac12,1\bigg]+(3\chi_{\bf 1}-2\chi_{\bf 2})\bigg[\frac12,2\bigg]\nn\\
	&+(-4+3\chi_{\bf 1})\bigg[\frac12,3\bigg]-4\bigg[\frac12,4\bigg]\bigg)\bigg/D,\label{eq:c0}\\
	c_1=&8[0,0]+\bigg((-\chi_{\bf 1}+\chi_{\bf c}\chi_{\bf s})[0,0]+(-\chi_{\bf 1}-2\chi_{\bf 2}+\chi_{\bf c}\chi_{\bf s})[0,1]+(3\chi_{\bf 1}-2\chi_{\bf 2})[0,2]\nn\\
	&+(-4+3\chi_{\bf 1})[0,3]-4[0,4]+(-\chi_{\bf 1}+\chi_{\bf c}\chi_{\bf s})\left[\frac12,\frac12\right]-2\chi_{\bf 2}\bigg[\frac12,\frac32\bigg]\nn\\
	&+3\chi_{\bf 1}\bigg[\frac12,\frac52\bigg]-4\bigg[\frac12,\frac72\bigg]\bigg)\bigg/D,\label{eq:c1}\\
	c_2=&\left[0,\frac12\right]+\bigg((-\chi_{\bf 1}+\chi_{\bf c}\chi_{\bf s})\bigg[0,\frac12\bigg]-2\chi_{\bf 2}\bigg[0,\frac32\bigg]+3\chi_{\bf 1}\bigg[0,\frac52\bigg]-4\bigg[0,\frac72\bigg]\bigg)\bigg/D, \label{eq:c2}\\
	D=&\prod_{i=1}^4\f{(Q_i\fq-\ft)(Q_i\ft-\fq)}{Q_i\fq\ft}\nn\\
	=&-\frac{1}{[0,0]}\bigg((2+\chi_{\bf1}+2\chi_{\bf2}-\chi_{\bf c}^2-\chi_{\bf c}\chi_{\bf s}-\chi_{\bf s}^2)\left[0,0\right]+(-\chi_{\bf 1}+\chi_{\bf 2}+\chi_{\bf c}\chi_{\bf s})\left[0,1\right]\nn\\
	&+(-\chi_{\bf 1}-\chi_{\bf 2})\left[0,2\right]+(1+\chi_{\bf 1})\left[0,3\right]-\left[0,4\right]\bigg),\label{eq:DofD4}
\end{align}
where 
\begin{equation}\label{eq:spincontent}
	\left[j_l,j_r\right]\equiv\f{(-1)^{2j_l+2j_r+1}\left[(\ft\fq)^{-j_l}+\cdots+(\ft\fq)^{j_l}\right]\left[\left(\f{\ft}{\fq}\right)^{-j_r}+\cdots+\left(\f{\ft}{\fq}\right)^{j_r}\right]}{(\ft^{1/2}-\ft^{-1/2})(\fq^{1/2}-\fq^{-1/2})}\ . 
\end{equation}
We note that $\chi_{\bf 1},\chi_{\bf 2}$ respect the SU(2) Weyl group symmetries in \eqref{eq:D4Weylun}, but $\chi_{\bf s},\chi_{\bf c}$ do not. However the combination such as $\chi_{\bf s}\chi_{\bf c},\chi_{\bf s}^2+\chi_{\bf c}^2,\chi_{\bf s}+\chi_{\bf c}$ respect the SU(2) Weyl group symmetries. That is why $\chi_{\bf s},\chi_{\bf c}$ always appear as such combined form in \eqref{eq:c0}, \eqref{eq:c1}, \eqref{eq:c2}, and \eqref{eq:DofD4}. Notice also that $c_0$, $c_1$ and $c_2$ in \eqref{eq:c0}, \eqref{eq:c1}, and \eqref{eq:c2} which are coefficients at $Q_{\tau}^1$ order all contain the same factor $D$ in the denominators, we will soon see that the common denominator $D$ also appears in $Z^{D_4}_{u}$ at $Q_{\tau}^1$ order.

\paragraph{$Z^{D_4}_u$ part.}
Firstly we define 
\begin{equation}\label{eq:defZQ1}
	\calZ_{\mu_{5,1}\mu_{5,2}}(w_5,Q_1)\equiv\frac{\sum_{\mu_{1,1}}\bar{Z}_{\mu_{5,1}\mu_{5,2}\mu_{1,1}}(w_5,Q_{1})}{\sum_{\mu_{1,1}}\bar{Z}_{\oo\oo\mu_{1,1}}(w_5,Q_1)}\quad.
\end{equation}
Combine \eqref{eq:nZ_D4inst} with \eqref{eq:ZuD4} and switch the parameter $Q_5$ to $u$, we get
\begin{align}\label{eq:zD4instp}
&Z^{D_4}_{u}\nn\\
=&\resizebox{0.95\hsize}{!}{$\sum_{\mu_{5,1}\mu_{5,2}}\bar{Z}_{\mu_{5,1}\mu_{5,2}}^{\text{SU(2)}}(w_5,u^{\frac12})\frac{\calZ_{\mu_{5,1}\mu_{5,2}}(w_5,Q_1)}{Q_1^{\frac12(|\mu_{5,1}|+|\mu_{5,2}|)}}\frac{\calZ_{\mu_{5,1}\mu_{5,2}}(w_5,Q_2)}{Q_2^{\frac12(|\mu_{5,1}|+|\mu_{5,2}|)}}\frac{\calZ_{\mu_{5,1}\mu_{5,2}}(w_5,Q_3)}{Q_3^{\frac12(|\mu_{5,1}|+|\mu_{5,2}|)}}\frac{\calZ_{\mu_{5,1}\mu_{5,2}}(w_5,Q_4)}{Q_4^{\frac12(|\mu_{5,1}|+|\mu_{5,2}|)}}$} .
\end{align}

By the definition \eqref{eq:defZQ1}, we need to sum over the Young diagrams $\mu_{1,1}$ to infinity.
Because both the functions $\bar{Z}$ in the numerator and denominator are proportional to $Q_1^{|\mu_{1,1}|}$, the numerator and denominator will both be infinite polynomials of $Q_1$ with the leading order being zeroth order. As an example, we compute $\calZ_{\oo,\{1\}}$ expanded up to $Q_{\tau}Q_1^4$: 
\begin{align}\label{eq:Z01}
	\calZ_{\oo,\{1\}}(w_5,Q_1)=&\frac{(\sqrt{\ft}w_5-\sqrt{\fq})(1+Q_1)}{(\fq\ft)^{1/4}\sqrt{w_5}}+Q_{\tau}\bigg[\nn\\
	&-\frac{\ft^{3/2}w_5^3-\fq^{3/2}}{(\fq\ft)^{3/4}w_5^{3/2}}(1+Q_1)+\frac{Q_1}{(\fq\ft)^{5/4}}P_1(w_5)+\frac{Q_1^2}{(\fq\ft)^{9/4}}(\fq^2+\fq\ft+\ft^2)P_1(w_5)\nn\\
	&+\frac{Q_1^3}{(\fq\ft)^{13/4}}(\fq^4+\fq^3\ft+\fq^2\ft^2+\fq\ft^3+\ft^4)P_1(w_5)+\frac{Q_1^4}{(\fq\ft)^{17/4}}(\fq^6+\fq^5\ft+\fq^4\ft^2\nn\\
	&+\fq^3\ft^3+\fq^2\ft^4+\fq\ft^5+\ft^6)P_1(w_5)+\calO(Q_1^5)\bigg]+\calO(Q_{\tau}^2)\, ,
\end{align}
where
\begin{align}
	P_1(w_5)=w_5^{-5/2}(\sqrt{\ft}w_5-\sqrt{\fq})^2(\sqrt{\fq}w_5-\sqrt{\ft})(w_5-\sqrt{\fq\ft})(\sqrt{\fq\ft}w_5-1)\, .
\end{align}
From \eqref{eq:Z01}, we see that the zeroth order coefficient of $Q_{\tau}$ is a finite polynomial of $Q_1$ while the first order coefficient of $Q_{\tau}$ is not finite. By looking at the polynomial of $Q_1$ in the coefficient of $Q_{\tau}$, we find that it also has a pattern of a geometric series starting from the second term in the square bracket which can be analytically summed:
\begin{align}
	&\frac{Q_1}{(\fq\ft)^{5/4}}P_1(w_5)+\frac{Q_1^2}{(\fq\ft)^{9/4}}(\fq^2+\fq\ft+\ft^2)P_1(w_5)
	+\frac{Q_1^3}{(\fq\ft)^{13/4}}(\fq^4+\fq^3\ft+\fq^2\ft^2+\fq\ft^3+\ft^4)\nn\\
	&\times P_1(w_5)+\frac{Q_1^4}{(\fq\ft)^{17/4}}(\fq^6+\fq^5\ft+\fq^4\ft^2
	+\fq^3\ft^3+\fq^2\ft^4+\fq\ft^5+\ft^6)P_1(w_5)+\calO(Q_1^5)\nn\\
	=&\sum_{n=1}^{\infty}\left(\f{Q_1^{n}}{(\fq\ft)^{n+\frac14}}\sum_{i=0}^{2n-2}\fq^{2n-2-i}\ft^i\right)P_1(w_5)=\frac{Q_1(1+Q_1)}{(\fq\ft)^{1/4}(Q_1\fq-\ft)(Q_1\ft-\fq)}P_1(w_5),
\end{align}
 so
\begin{align}\label{eq:calZ01}
	\calZ_{\oo,\{1\}}(w_5,Q_1)=&\frac{(\sqrt{\ft}w_5-\sqrt{\fq})(1+Q_1)}{(\fq\ft)^{1/4}\sqrt{w_5}}+Q_{\tau}\bigg[\nn\\
	&-\frac{\ft^{3/2}w_5^3-\fq^{3/2}}{(\fq\ft)^{3/4}w_5^{3/2}}(1+Q_1)+\frac{Q_1(1+Q_1)}{(\fq\ft)^{1/4}(Q_1\fq-\ft)(Q_1\ft-\fq)}P_1(w_5)\bigg]\nn\\
	&+\calO(Q_{\tau}^2).
\end{align}
We can easily check that $Q_1^{-1/2}\calZ_{\oo,\{1\}}(w_5,Q_1)$ is invariant under $Q_1\rightarrow Q_1^{-1}$ at the zeroth order and the first order of $Q_{\tau}$, which is also due to the SU(2) Weyl group symmetry in \eqref{eq:D4Weylun}. 

 In this way, we can also compute other $\calZ$ functions.
 As an example, we list $\calZ_{\{1\},\oo}$ below,
\begin{align}\label{eq:calZ10}
	\calZ_{\{1\},\oo}(w_5,Q_1)=&-\frac{(\sqrt{\fq}w_5-\sqrt{\ft})(1+Q_1)}{(\fq\ft)^{1/4}\sqrt{w_5}}+Q_{\tau}\bigg[\nn\\
	&\frac{(\fq^{3/2}w_5^3-\ft^{3/2})(1+Q_1)}{(\fq\ft)^{3/4}w_5^{3/2}}-\frac{Q_1(1+Q_1)}{(\fq\ft)^{1/4}(Q_1\fq-\ft)(Q_1\ft-\fq)}P_2(w_5)\bigg]+\calO(Q_{\tau}^2),
\end{align} 
where
\begin{align}
	P_2(w_5)=&w_5^{-5/2}(\sqrt{\fq}w_5-\sqrt{\ft})^2(\sqrt{\ft}w_5-\sqrt{\fq})(w_5-\sqrt{\fq\ft})(\sqrt{\fq\ft}w_5-1).
\end{align}
Then we can plug these $\calZ$ functions into \eqref{eq:zD4instp} to obtain the $u$-dependent part of the partition function. We express the $u$-dependent part in terms of PE 
\begin{equation}\label{eq:PEZD4withu}
	Z^{D_4}_{u}=\text{PE}\left[\sum_{\ell,n,r}d_{\ell,n,r}w_5^{\ell}u^{\frac{n}{2}}Q_{\tau}^{r}\right],
\end{equation}
and compute the expansion up to $w_5^1 u^{\frac12} Q_{\tau}^1$ order. The corresponding nonzero coefficients are listed below 
\begin{align}
	d_{0,1,0}=&(\chi_{\bf c}+\chi_{\bf s})\left[0,\frac{1}{2}\right],\\
	d_{1,1,0}=&(8\chi_{\bf c}+8\chi_{\bf s})\left[0,0\right],\\
	d_{-2,1,1}=&\bigg((\chi_{\bf1}\chi_{\bf c}+\chi_{\bf1}\chi_{\bf s})\left[0,\f{1}{2}\right]+(-3\chi_{\bf c}-4\chi_{\bf 2}\chi_{\bf c}+\chi_{\bf c}^3-3\chi_{\bf s}-4\chi_{\bf 2}\chi_{\bf s}+\chi_{\bf c}^2\chi_{\bf s}\nn\\
	&+\chi_{\bf c}\chi_{\bf s}^2+\chi_{\bf s}^3)\left[0,\f{3}{2}\right]+(4\chi_{\bf 1}\chi_{\bf c}+4\chi_{\bf 1}\chi_{\bf s}-\chi_{\bf c}^2\chi_{\bf s}-\chi_{\bf c}\chi_{\bf s}^2)\left[0,\f{5}{2}\right]\nn\\
	&+(-4\chi_{\bf c}+\chi_{\bf 2}\chi_{\bf c}-4\chi_{\bf s}+\chi_{\bf 2}\chi_{\bf s})\left[0,\f{7}{2}\right]+(-\chi_{\bf 1}\chi_{\bf c}-\chi_{\bf 1}\chi_{\bf s})\left[0,\f{9}{2}\right]\nn\\
	&+(\chi_{\bf c}+\chi_{\bf s})\left[0,\f{11}{2}\right]\bigg)\bigg/D,\\
	d_{-1,1,1}=&\bigg(-(32\chi_{\bf c}+23\chi_{\bf 1}\chi_{\bf c}+8\chi_{\bf 2}\chi_{\bf c}+32\chi_{\bf s}+23\chi_{\bf 1}\chi_{\bf s}+8\chi_{\bf 2}\chi_{\bf s}+\chi_{\bf c}^2\chi_{\bf s}\nn\\
	&+\chi_{\bf c}\chi_{\bf s}^2)\left[0,0\right]+(48\chi_{\bf c}+17\chi_{\bf 1}\chi_{\bf c}+34\chi_{\bf 2}\chi_{\bf c}-8\chi_{\bf c}^3+48\chi_{\bf s}+17\chi_{\bf 1}\chi_{\bf s}\nn\\
	&+34\chi_{\bf 2}\chi_{\bf s}-9\chi_{\bf c}^2\chi_{\bf s}-9\chi_{\bf c}\chi_{\bf s}^2-8\chi_{\bf s}^3)[0,1]+(-24\chi_{\bf c}-35\chi_{\bf 1}\chi_{\bf c}+2\chi_{\bf 2}\chi_{\bf c}\nn\\
	&-24\chi_{\bf s}-35\chi_{\bf 1}\chi_{\bf s}+2\chi_{\bf 2}\chi_{\bf s}+8\chi_{\bf c}^2\chi_{\bf s}+8\chi_{\bf c}\chi_{\bf s}^2)[0,2]+(36\chi_{\bf c}-3\chi_{\bf 1}\chi_{\bf c}\nn\\
	&-8\chi_{\bf 2}\chi_{\bf c}+36\chi_{\bf s}-3\chi_{\bf 1}\chi_{\bf s}-8\chi_{\bf 2}\chi_{\bf s})[0,3]+(4\chi_{\bf c}+8\chi_{\bf 1}\chi_{\bf c}+4\chi_{\bf s}\nn\\
	&+8\chi_{\bf 1}\chi_{\bf s})[0,4]-(8\chi_{\bf c}+8\chi_{\bf s})[0,5]+(\chi_{\bf 1}\chi_{\bf c}+\chi_{\bf 1}\chi_{\bf s}-\chi_{\bf c}^2\chi_{\bf s}-\chi_{\bf c}\chi_{\bf s}^2)\left[\f{1}{2},\f{1}{2}\right]\nn\\
	&+(2\chi_{\bf 2}\chi_{\bf c}+2\chi_{\bf 2}\chi_{\bf s})\left[\f{1}{2},\f{3}{2}\right]-(3\chi_{\bf 1}\chi_{\bf c}+3\chi_{\bf 1}\chi_{\bf s})\left[\f{1}{2},\f{5}{2}\right]\nn\\
	&+(4\chi_{\bf c}+4\chi_{\bf s})\left[\f{1}{2},\f{7}{2}\right]\bigg)\bigg/D,\\
	d_{0,1,1}=&\bigg((-60\chi_{\bf c}-16\chi_{\bf 1}\chi_{\bf c}-90\chi_{\bf 2}\chi_{\bf c}+30\chi_{\bf c}^3-60\chi_{\bf s}-16\chi_{\bf 1}\chi_{\bf s}-90\chi_{\bf 2}\chi_{\bf s}+47\chi_{\bf c}^2\chi_{\bf s}\nn\\
	&+47\chi_{\bf c}\chi_{\bf s}^2+30\chi_{\bf s}^3)\left[0,\f{1}{2}\right]+(-3\chi_{\bf c}+60\chi_{\bf 1}\chi_{\bf c}-38\chi_{\bf 2}\chi_{\bf c}+\chi_{\bf c}^3-3\chi_{\bf s}\nn\\
	&+60\chi_{\bf 1}\chi_{\bf s}-38\chi_{\bf 2}\chi_{\bf s}-29\chi_{\bf c}^2\chi_{\bf s}-29\chi_{\bf c}\chi_{\bf s}^2+\chi_{\bf s}^3)\left[0,\f{3}{2}\right]+(-30\chi_{\bf c}+55\chi_{\bf 1}\chi_{\bf c}\nn\\
	&+30\chi_{\bf 2}\chi_{\bf c}-30\chi_{\bf s}+55\chi_{\bf 1}\chi_{\bf s}+30\chi_{\bf 2}\chi_{\bf s}-\chi_{\bf c}^2\chi_{\bf s}-\chi_{\bf c}\chi_{\bf s}^2)\left[0,\f{5}{2}\right]+(-72\chi_{\bf c}\nn\\
	&-30\chi_{\bf 1}\chi_{\bf c}+\chi_{\bf 2}\chi_{\bf c}-72\chi_{\bf s}-30\chi_{\bf 1}\chi_{\bf s}+\chi_{\bf 2}\chi_{\bf s})\left[0,\f{7}{2}\right]+(30\chi_{\bf c}-\chi_{\bf 1}\chi_{\bf c}+30\chi_{\bf s}\nn\\
	&-\chi_{\bf 1}\chi_{\bf s})\left[0,\f{9}{2}\right]+(\chi_{\bf c}+\chi_{\bf s})\left[0,\f{11}{2}\right]+(30\chi_{\bf c}+15\chi_{\bf 1}\chi_{\bf c}+13\chi_{\bf 2}\chi_{\bf c}+\chi_{\bf c}^3\nn\\
	&+30\chi_{\bf s}+15\chi_{\bf 1}\chi_{\bf s}+13\chi_{\bf 2}\chi_{\bf s}+10\chi_{\bf c}^2\chi_{\bf s}+10\chi_{\bf c}\chi_{\bf s}^2+\chi_{\bf s}^3)\left[\f{1}{2},0\right]+(-34\chi_{\bf c}\nn\\
	&-23\chi_{\bf 1}\chi_{\bf c}-21\chi_{\bf 2}\chi_{\bf c}+\chi_{\bf c}^3-34\chi_{\bf s}-23\chi_{\bf 1}\chi_{\bf s}-21\chi_{\bf 2}\chi_{\bf s}+\chi_{\bf c}^2\chi_{\bf s}+\chi_{\bf c}\chi_{\bf s}^2\nn\\
	&+\chi_{\bf s}^3)\left[\f{1}{2},1\right]+(31\chi_{\bf c}+29\chi_{\bf 1}\chi_{\bf c}-\chi_{\bf 2}\chi_{\bf c}+31\chi_{\bf s}+29\chi_{\bf 1}\chi_{\bf s}-\chi_{\bf 2}\chi_{\bf s}-\chi_{\bf c}^2\chi_{\bf s}\nn\\
	&-\chi_{\bf c}\chi_{\bf s}^2)\left[\f{1}{2},2\right]+(-37\chi_{\bf c}+2\chi_{\bf 1}\chi_{\bf c}+\chi_{\bf 2}\chi_{\bf c}-37\chi_{\bf s}+2\chi_{\bf 1}\chi_{\bf s}+\chi_{\bf 2}\chi_{\bf s})\left[\f{1}{2},3\right]\nn\\
	&-(3\chi_{\bf c}+3\chi_{\bf s}+\chi_{\bf 1}\chi_{\bf c}+\chi_{\bf 1}\chi_{\bf s})\left[\f{1}{2},4\right]+(\chi_{\bf c}+\chi_{\bf s})\left[\f{1}{2},5\right]\bigg)\bigg/D,\\
	d_{1,1,1}=&\bigg((153\chi _{\bf c}^2\chi _{\bf s} + 153\chi _{\bf c}\chi _{\bf s}^2 + 72\chi _{\bf c}^3 - 
   57\chi _ {\bf 1}\chi _{\bf c} - 144\chi _ {\bf 2}\chi _{\bf c} - 112\chi _{\bf c} + 
   72\chi _{\bf s}^3 - 57\chi _ {\bf 1}\chi _{\bf s} \nn\\
   &- 144\chi _ {\bf 2}\chi _{\bf s} - 
   112\chi _{\bf s})[0, 0]+ (-39\chi _{\bf c}^2\chi _{\bf s} - 39\chi _{\bf c}\chi _{\bf s}^2 + 16\chi _{\bf c}^3 + 
   47\chi _ {\bf 1}\chi _{\bf c} - 154\chi _ {\bf 2}\chi _{\bf c} \nn\\
   &- 64\chi _{\bf c} + 16\chi _{\bf s}^3 + 
   47\chi _ {\bf 1}\chi _{\bf s} - 154\chi _ {\bf 2}\chi _{\bf s} - 64\chi _{\bf s} )[0,1]+ (-16\chi _{\bf c}^2\chi _{\bf s} - 16\chi _{\bf c}\chi _{\bf s}^2\nn\\
   & + 
   163\chi _ {\bf 1}\chi _{\bf c} + 38\chi _ {\bf 2}\chi _{\bf c} + 16\chi _{\bf c} + 
   163\chi _ {\bf 1}\chi _{\bf s} + 38\chi _ {\bf 2}\chi _{\bf s} + 16\chi _{\bf s} )[0,2]+ (-21\chi _{\bf 1}\chi _{\bf c} \nn\\
   &+ 16\chi _ {\bf 2}\chi _{\bf c} - 172\chi _{\bf c} - 
   21\chi _ {\bf 1}\chi _{\bf s} + 16\chi _ {\bf 2}\chi _{\bf s} - 172\chi _{\bf s} )[0,3]+ (-16\chi _ {\bf 1}\chi _{\bf c} + 4\chi _{\bf c} \nn\\
   &- 16\chi _ {\bf 1}\chi _{\bf s} + 
   4\chi _{\bf s} )[0,4]+ (16\chi _{\bf c} + 16\chi _{\bf s} )[0,5]+ (25\chi _{\bf c}^2\chi _{\bf s} + 25\chi _{\bf c}\chi _{\bf s}^2 + 8\chi _{\bf c}^3 \nn\\
   &- 
   17\chi _ {\bf 1}\chi _{\bf c} - 24\chi _ {\bf 2}\chi _{\bf c} - 16\chi _{\bf c} + 8\chi _{\bf s}^3 - 
   17\chi _ {\bf 1}\chi _{\bf s} - 24\chi _ {\bf 2}\chi _{\bf s} - 16\chi _{\bf s} )\left[\frac {1} {2}, \frac {1} {2}\right]\nn\\
   &+(-8\chi _{\bf c}^2\chi _{\bf s} - 8\chi _{\bf c}\chi _{\bf s}^2 + 16\chi _ {\bf 1}\chi _{\bf c} - 
   34\chi _ {\bf 2}\chi _{\bf c} + 16\chi _ {\bf 1}\chi _{\bf s} - 34\chi _ {\bf 2}\chi _{\bf s} )\left[\frac {1} {2}, \frac {3} {2}\right]\nn\\
   &+ (51\chi _ {\bf 1}\chi _{\bf c} + 8\chi _ {\bf 2}\chi _{\bf c} - 8\chi _{\bf c} + 
   51\chi _ {\bf 1}\chi _{\bf s} + 8\chi _ {\bf 2}\chi _{\bf s} - 8\chi _{\bf s} )\left[\frac {1} {2}, \frac {5} {2}\right]+ (-8\chi _ {\bf 1}\chi _{\bf c} \nn\\
   &- 68\chi _{\bf c} - 8\chi _ {\bf 1}\chi _{\bf s} - 
   68\chi _{\bf s} )\left[\frac {1} {2}, \frac {7} {2}\right]+ (8\chi _{\bf c} + 8\chi _{\bf s} )\left[\frac {1} {2}, \frac {9} {2}\right]\bigg)\bigg/D,
\end{align}
where $\chi_{\bf 1}, \chi_{\bf 2}, \chi_{\bf s}$ and $\chi_{\bf c}$ are the characters of fundamental weights of $D_4$ whose expressions are in \eqref{eq:D4fundamentals} but with the orthonormal basis being $\{Q_1,Q_2,Q_3,Q_4\}$, and $D$ is the one defined in \eqref{eq:DofD4}. 

From the above listed expressions for $d_{\ell,n,r}$, we see that the form of $d_{\ell,n,1}$ is  quite different from $d_{\ell,n,0}$, there is a common denominator $D$ that appears in $d_{\ell,n,1}$ which are the coefficients of $Q_{\tau}^1$ order terms. We have also encountered similar situation in \eqref{eq:PEZD4withu0} for $Z^{D_4}_{u^{\bf 0}}$. The coefficients of the $Q_{\tau}^0$ order terms in $Z^{D_4}_{u^{\bf 0}}$ and $Z^{D_4}_u$ can be understood as Gopakumar-Vafa invariants \cite{Gopakumar:1998ii, Gopakumar:1998jq}, but the $Q_{\tau}^1$ order terms do not look like Gopakumar-Vafa invariants at first sight because of the denominator $D$. This is because we are dealing with $D_4$-type LST whose tensor branch parameters $Q_1,Q_2,Q_3,Q_4$ are not parameters for global symmetry but parameters for local symmetry. In the limit $Q_{\tau}\rightarrow 0$, the vertical direction of the $D_4$-type LST is decompactified, and the theory reduces to E-string theory, so the $Q_{\tau}^0$ terms also coincide with the corresponding terms in the partition function of E-string theory \cite{Kim:2022dbr}. But the $Q_{\tau}^1$ order terms contain the information of LST at nonzero $Q_{\tau}$, so $Q_1,Q_2,Q_3,Q_4$ appear as parameters for local symmetry in the $Q_{\tau}^1$ terms. In order to see the Gopakumar-Vafa invariants at the $Q_{\tau}^1$ order, we have to further expand the $Q_{\tau}^1$ terms with respect to $Q_1,Q_2,Q_3,Q_4$. We expand $c_0,c_1,c_2,d_{-2,1,1},d_{-1,1,1},d_{0,1,1},d_{1,1,1}$ which are all the $Q_{\tau}^1$ order coefficients in $Z^{D_4}_{u^{\bf 0}}$ and $Z^{D_4}_u$ we have obtained with respect to $Q_1,Q_2,Q_3,Q_4$ and arrange $\ft,\fq$ in the form of spin content in \eqref{eq:spincontent}, and find that the coefficients of the expansion terms are all positive integers, 
so it is consistent with the fact that the Gopakumar-Vafa invariants should be positive integers.

The partition function of rank-1 $D_4$-type LST was computed in \cite{Kim:2017xan},  where
  the authors expanded the partition function in terms of $Q_1,$ $Q_2,$ $Q_3,$ $Q_4,$ $Q_5,$ $w_5,$$Q_{\tau}$ up to $Q_1^2Q_2^2Q_3^2Q_4^2Q_5^2w_5^2Q_{\tau}^1$ order. 
    In order to compare our result with \cite{Kim:2017xan}, we re-expanded $Z^{D_4}$     and we checked our result is consistent with the result of \cite{Kim:2017xan}, up to the order that they computed.


\subsubsection{\texorpdfstring{$D_4$}{D4} symmetry from expansion II}\label{sec:choice2}
In choice II, we use $Q_1,Q_2,Q_3,Q_5$ as the fugacities and $A,u,Q_{\tau}$ as the expansion parameters. We call these sets of parameters as new parameters compared with the old parameters $Q_1,Q_2,Q_3,Q_4,Q_5,w_5,Q_{\tau}$.
 However, in actual computation of the partition function, it is much more convenient to use the old parameters to do expansions. So we will keep using these old parameters in the intermediate process of computation, but in the final step we will switch to the new parameters in order to see the $D_4$ symmetry manifestly. 
 
 By \eqref{eq:ZD4decompose}, we compute $Z^{D_4}_{Q_5^{\bf 0}}$ and $Z^{D_4}_{Q_5}$ separately. 

\paragraph{$Z^{D_4}_{Q_5^{\bf 0}}$ part.}
We rewrite \eqref{eq:zsubu1} by \eqref{eq:flop} as
\begin{align}\label{eq:zsubu1rev}
	&\sum_{\mu_{1,1}}\bar{Z}_{\oo\oo\mu_{1,1}}(w_5,Q_1)\nn\\
	=&\text{PE}\left[\frac{Q_1(\fq+\ft)-\sqrt{\fq\ft}(Q_1+Q_1^{-1})w_5}{(1-\fq)(1-\ft)}+Q_{\tau}P_0(w_5)\frac{Q_1(\fq+\ft)}{(Q_1\fq-\ft)(Q_1\ft-\fq)}+\calO(Q_{\tau}^2)\right].
\end{align}
By substituting \eqref{eq:zsubu1rev} and \eqref{eq:zoldpertd4} back to \eqref{eq:nZ_D4pert}, and expanding up to $Q_4 Q_{\tau}$ order,
we get
\begin{align}
	&Z^{D_4}_{Q_5^{\bf 0}}\nn\\
	=&\text{PE}\bigg[\frac{1}{(1-\fq)(1-\ft)}\bigg((\fq+\ft)(3+Q_1+Q_2+Q_3)-\sqrt{\fq\ft}w_5\big(8+Q_1+\frac{1}{Q_1}+Q_2\nn\\
	&+\frac{1}{Q_2}+Q_3+\frac{1}{Q_3}\big)+(\fq+\ft)w_5^2+Q_4\big(\fq+\ft-\sqrt{\fq\ft}(w_5+\frac{1}{w_5})\big)+\calO(Q_4^2)\bigg)\nn\\
	&+\frac{Q_{\tau}}{(1-\fq)(1-\ft)}\bigg(6(1+\fq\ft)+\big(\ft+\fq^2\ft+\fq(1+\ft)^2\big)(\fq+\ft)\Big(\frac{Q_1}{(Q_1\fq-\ft)(Q_1\ft-\fq)}\nn\\
	&+\frac{Q_2}{(Q_2\fq-\ft)(Q_2\ft-\fq)}+\frac{Q_3}{(Q_3\fq-\ft)(Q_3\ft-\fq)}\Big)-\sqrt{\fq\ft}\big(w_5+\frac{1}{w_5}\big)\Big(8\nn\\
	&+\frac{Q_1(\fq+\ft)(1+\fq)(1+\ft)}{(Q_1\fq-\ft)(Q_1\ft-\fq)}+\frac{Q_2(\fq+\ft)(1+\fq)(1+\ft)}{(Q_2\fq-\ft)(Q_2\ft-\fq)}+\frac{Q_3(\fq+\ft)(1+\fq)(1+\ft)}{(Q_3\fq-\ft)(Q_3\ft-\fq)}\Big)\nn\\
	&+(\fq+\ft)\big(w_5^2+\frac{1}{w_5^2}\big)\Big(1+\frac{Q_1\fq\ft}{(Q_1\fq-\ft)(Q_1\ft-\fq)}+\frac{Q_2\fq\ft}{(Q_2\fq-\ft)(Q_2\ft-\fq)}\nn\\
	&+\frac{Q_3\fq\ft}{(Q_3\fq-\ft)(Q_3\ft-\fq)}\Big)+Q_4\Big(\frac{(\fq+\ft)\big(\ft+\fq^2\ft+\fq(1+\ft)^2\big)}{\fq\ft}-\frac{(\fq+\ft)(1+\fq)(1+\ft)}{\sqrt{\fq\ft}}\nn\\
	&\times\big(w_5+\frac{1}{w_5}\big)+(\fq+\ft)\big(w_5^2+\frac{1}{w_5^2}\big)\Big)+\calO(Q_4^2)\bigg)+\calO(Q_{\tau}^2)\bigg]\ .
\end{align}

\paragraph{$Z^{D_4}_{Q_5}$ part and rescaling.}
From \eqref{eq:nZ_D4inst} and \eqref{eq:defZQ1}, we have
\begin{align}\label{eq:zD4Q5}
Z^{D_4}_{Q_5}
=\sum_{\mu_{5,1},\mu_{5,2}} &\bar{Z}_{\mu_{5,1}\mu_{5,2}}^{\text{SU(2)}}(w_5,Q_5)\, 
\calZ_{\mu_{5,1}\mu_{5,2}}(w_5,Q_1)\, 
\calZ_{\mu_{5,1}\mu_{5,2}}(w_5,Q_2)\cr
& \times
\calZ_{\mu_{5,1}\mu_{5,2}}(w_5,Q_3)\,
\calZ_{\mu_{5,1}\mu_{5,2}}(w_5,Q_4)\  .
\end{align}
The summation over $\mu_{5,1},\mu_{5,2}$ up to infinity is difficult as the computation will be more and more time consuming when the upper bound of the total Young diagram box number increases, so we expand $Z^{D_4}_{Q_5}$ with respect to $Q_5$ whose order is restricted by $|\mu_{5,1}|+|\mu_{5,2}|$. 
As $Q_4$ is proportional to $u$, we also expand $Z^{D_4}_{Q_5}$ with respect to $Q_4$ to obtain the expansion of the partition function with respect to $u$. 
Even with the two expansion parameters $Q_5,Q_4$, the computation is still time-consuming, and we need to further expand with respect to $w_5$ in order to reduce the heavy computation. We first need to expand each of the five factors of the summand in \eqref{eq:zD4Q5} with respect to $w_5$ and then multiply the expansions together to obtain the expansion of $Z^{D_4}_{Q_5}$ with respect to $w_5$. 
However, unlike $Q_4,Q_5$ and $Q_{\tau}$ with respect to which the expansions of each of the five factors only have non-negative power terms, there are negative powers appearing in the expansions with respect to $w_5$ for each of these five factors which will make the total expansion with respect to $w_5$ for $Z^{D_4}_{Q_5}$ difficult to truncate. 
We need to rescale parameters and the five factors in order to have non-negative power expansions with $w_5$. We find that the following rescalings,
\begin{align}\label{eq:D4rescalingrules}
    \calZ_{\mu_{5,1}\mu_{5,2}}&~\rightarrow~~ w_5^{\frac12(|\mu_{5,1}|+|\mu_{5,2}|)}\calZ_{\mu_{5,1}\mu_{5,2}},\nn\\ \bar{Z}^{\text{SU(2)}}_{\mu_{5,1}\mu_{5,2}}&~\rightarrow~~  w_5^{-2(|\mu_{5,1}|+|\mu_{5,2}|)}\bar{Z}^{\text{SU(2)}}_{\mu_{5,1}\mu_{5,2}},\nn\\
    Q_{\tau}&~\rightarrow ~~ w_5^2 Q_{\tau},
\end{align}
make $\bar{Z}^{\text{SU(2)}}$ and $\calZ$ functions up to $Q_{\tau}^1$ order have non-negative power expansions with respect to $w_5$. Then through multiplication and truncation, we can obtain non-negative power expansion of $Z^{D_4}_{Q_5}$ with respect to $w_5$ and transform it into the corresponding expansion inside PE. Finally, we switch back to the original parameter $Q_{\tau}$ inside the PE. 

\paragraph{Evaluation of the partition function by Weyl group  symmetry.}
Before we dive into detailed computation, we would like to discuss the general problems that we would encounter when doing the computation. 

If we finally obtain the expansion of $Z^{D_4}$ inside PE with the expansion parameters being $A,u$ and $Q_{\tau}$, the term $A^\ell u^n Q_{\tau}^r$ will be of the following form,
\begin{align}
	b_{\ell,n,r}(Q_1,Q_2,Q_3,Q_5)A^\ell u^n Q_{\tau}^r,
\end{align}
where $b_{\ell,n,r}$ is function of $Q_1,Q_2,Q_3,Q_5$.
From the results in \eqref{eq:PEZD4withu0} and \eqref{eq:PEZD4withu} in the last subsection, we know that the $Q_{\tau}^0$ order coefficients of $Z^{D_4}_{u^{\bf 0}}$ and $Z^{D_4}_{u}$ are composed of finite polynomials of $D_4$ characters, whereas the $Q_{\tau}^1$ order coefficients of $Z^{D_4}_{u^{\bf 0}}$ and $Z^{D_4}_{u}$ are in fractional forms with both numerators and denominators being composed of finite polynomials of $D_4$ characters. We explained that this is due to the fact that the $D_4$ fugacities appear as parameters for local symmetry in the $Q_{\tau}^1$ order terms. We expect that similar things also happen to $b_{\ell,n,r}$.

\paragraph{(1)}
For $r=0$ which is at $Q_{\tau}^0$ order,  we expect $b_{\ell,n,0}$ to be also finite polynomials of $D_4$ characters with $Q_1,Q_2,Q_3,Q_5$ as the fugacities. If we focus on $Q_5$ which is the intermediate expansion parameter, $b_{\ell,n,0}$ can be regarded as Laurent polynomials of $Q_5$ with both negative and non-negative powers.  Since 
\begin{align}
	\quad A=Q_1Q_2Q_3Q_5^2 w_5,\quad u=Q_1Q_2Q_3Q_4Q_5^2,
\end{align}
the $A^{\ell}u^n$ term will be of the following form,
\begin{align}\label{eq:Auterm}
	b_{\ell,n,0}(Q_1,Q_2,Q_3,Q_5)A^{\ell}u^n=b_{\ell,n,0}(Q_1,Q_2,Q_3,Q_5)(Q_1Q_2Q_3)^{\ell+n}Q_5^{2\ell+2n}w_5^{\ell}Q_4^n.
\end{align}
We know that the contribution to $w_5^{\ell}Q_4^n$ term inside PE from the $Z^{D_4}_{Q_5}$ part with $Q_5$ expanded up to $Q_5^k$ order takes the following form,
\begin{align}\label{eq:Q5expansion}
	\Big(``Q_5+Q_5^2+\cdots+Q_5^k+\cdots"\Big) \, w_5^\ell Q_4^n\ .
\end{align}
Here we introduced a notation ``\hspace{5mm}" which means that we ignore the coefficients of the polynomial of $Q_5$ which should be functions of $Q_1,Q_2,Q_3$, and  if some of the coefficients are accidentally zero, the corresponding terms should disappear from \eqref{eq:Q5expansion}.
In the remaining part of this section, we will use this notation to ignore such coefficients in order to avoid complicated expressions.

In \eqref{eq:Q5expansion}, $n$ is always a non-negative integer whereas $\ell$ could be any integer. For $n=0$, we need to flop $w_5^\ell$ terms with $\ell\leq 0$ by \eqref{eq:flop} and combine them with $\ell\geq 0$ terms, in particular, $\ell=0$ term will have $\frac12$ factor due to the flop and combination. These will bring negative power terms of $Q_5$ to the $w_5^\ell$ terms with $\ell\geq 0$. There are also possible $Q_5^0$ order contributions to the $w_5^{\ell}Q_4^n$ term from the $Z^{D_4}_{Q_5^{\bf 0}}$ part. In total, the $w_5^{\ell}Q_4^n$ term inside the PE of $Z^{D_4}$ will be of the following form,
\begin{align}\label{eq:w5Q4term}
	\Big(``\cdots+1+Q_5+Q_5^2+\cdots+Q_5^k+\cdots"\Big)\, w_5^\ell Q_4^n\ ,
\end{align}
where the first $\cdots$ means possible negative orders of $Q_5$ because of flop. 

Because the right hand side of \eqref{eq:Auterm} is the general form of $w_5^\ell Q_4^n$ term that we expect, it should equal to \eqref{eq:w5Q4term}, thus we have
\begin{align}\label{eq:compare}
	b_{\ell,n,0}(Q_1,Q_2,Q_3,Q_5)(Q_1Q_2Q_3)^{\ell+n}Q_5^{2\ell+2n}=``\cdots+1+Q_5+Q_5^2+\cdots+Q_5^k+\cdots"\ .
\end{align}
As $b_{\ell,n,0}$ is Laurent polynomial of $Q_5$ which is finite, the right hand side of \eqref{eq:compare} should also be finite. 
So the coefficient of $w_5^\ell Q_4^n$ does not go up to infinite power of $Q_5$ but stops at some finite order of $Q_5$. 
If the highest order of $Q_5$ in $b_{\ell,n,0}$ is $Q_5^j$, then we just need to expand $Q_5$ up to $Q_5^{j+2\ell+2n}$ on the right hand side and the result will be exact, and from the right hand side result, we can then determine $b_{\ell,n,0}$ on the left hand side. 
Then the issue is how can we know the highest order $j$ of $Q_5$ in $b_{\ell,n,0}$. Instead of considering the highest order $j$, we consider how to determine $b_{\ell,n,0}$ without expanding $Q_5$ up to the highest order on the right hand side. 

As we said, $b_{\ell,n,0}$ is a finite polynomial of $D_4$ characters, so we can divide the finite polynomial $b_{\ell,n,0}$ into a summation of Weyl group orbits\footnote{A Weyl group orbit of $D_4$ can be constructed as the following: First we start with a monomial $Q_1^{i_0} Q_2^{j_0} Q_3^{k_0} Q_5^{l_0}$, then we act all possible Weyl reflections on the monomial and it will generate new monomials. All the different monomials(including the original one) form a Weyl group obit of $D_4$, we also call the corresponding polynomial a Weyl group orbit.} of $D_4$ where each Weyl group orbit can be regarded as a Laurent polynomial of $Q_5$. If we know one monomial from a Weyl group orbit, we can generate the full Weyl group orbit by Weyl group symmetry. So if we know at least one monomial for each Weyl group orbit of $b_{\ell,n,0}$, we can then determine the full polynomial $b_{\ell,n,0}$. As each Weyl group orbit of $b_{\ell,n,0}$ is a Laurent polynomial of $Q_5$ that contains $Q_5^0$ order term, if we can obtain the complete $Q_5^0$ order term of $b_{\ell,n,0}$, we can completely determine the full polynomial $b_{\ell,n,0}$ by Weyl group symmetry.  Because of the factor $Q_5^{2\ell+2n}$ on the left hand side of \eqref{eq:compare}, the $Q_5^0$ order term of $b_{\ell,n,0}$ is shifted to $Q_5^{2\ell+2n}$ order on the right hand side. 
So we can just compute up to $Q_5^{2\ell+2n}$ order on the right hand side and use the $Q_5^{2\ell+2n}$ order term to determine the $Q_5^0$ order term of $b_{\ell,n,0}$. Then we can use Weyl group symmetry to generate all the terms of $b_{\ell,n,0}$. 
After that, we will know what is the highest order $j$ of $Q_5$ in $b_{\ell,n,0}$, and we can then compute $Q_5$ up to $Q_5^{j+2\ell+2n}$ order on the right hand side to confirm the correctness of $b_{\ell,n,0}$ that we have obtained by Weyl group symmetry. 

\paragraph{(2)}
For $r=1$ which is at $Q_{\tau}^1$ order, we expect that $b_{\ell,n,1}$ to be of the following fractional form
\begin{align}
	b_{\ell,n,1}(Q_1,Q_2,Q_3,Q_5)=\frac{\tilde{b}_{\ell,n,1}(Q_1,Q_2,Q_3,Q_5)}{\tilde{D}},
\end{align}
where both $\tilde{b}_{\ell,n,1}$ and $\tilde{D}$ are finite polynomials of $D_4$ characters which is similar to the form of $d_{\ell,n,1}$ in \eqref{eq:PEZD4withu}. The form of $D$ in $d_{\ell,n,1}$ is determined and given in \eqref{eq:DofD4}, and it actually comes from the exact summations of the geometric series of $Q_1,Q_2,Q_3,Q_4$ that we encountered in the last subsection \ref{sec:choice1}. Here, we also need to compute the geometric series that involve $Q_1,Q_2,Q_3,Q_5$ in order to know the explicit form of $\tilde{D}$, but this is difficult for the current case because we have to compute the expansion of $Q_5$ up to very high order in order to see the pattern of geometric series. There is a more convenient method to determine the explicit form of $\tilde{D}$, but before that let us illustrate how this method works for determining the form of $D$.

Let us first write down the denominator $D$ that we found in the choice I case,
\begin{align}\label{eq:denominatorC1}
	D=&\prod_{i=1}^4\f{(Q_i\fq-\ft)(Q_i\ft-\fq)}{Q_i\fq\ft}\ .
\end{align}
It is not difficult to explain why it should be in the above form. 
First, it has to respect the symmetry in \eqref{eq:fundsymofD4rk1} which includes both SU(2) Weyl group symmetry and $D_4$ Weyl group symmetry. 
Second, the $\ft,\fq$ in the expression after the expansion should have appropriate powers so that the expression can be written as the numerators of the spin contents in \eqref{eq:spincontent}. 
As we said, $D$ comes from exact summations of the geometric series of $Q_1,Q_2,Q_3,Q_4$, in particular, the geometric series of $Q_1$ contributes the following factor
\begin{align}\label{eq:denoQ1}
	\frac{(Q_1\fq-\ft)(Q_1\ft-\fq)}{Q_1}\ ,
\end{align}
which respects the SU(2) Weyl group symmetry $Q_1\rightarrow Q_1^{-1}$. Now, by requiring the highest orders and lowest orders of $\ft,\fq$ to have the same absolute values, we need to divide \eqref{eq:denoQ1} by $\fq\ft$, so we get
\begin{align}\label{eq:denofactor}
	\frac{(Q_1\fq-\ft)(Q_1\ft-\fq)}{Q_1\fq\ft}.
\end{align}
By further requiring the permutation symmetry among $Q_1,\cdots,Q_4$, we can finally reproduce \eqref{eq:denominatorC1}. The lesson that we learnt is that the SU(2) Weyl group symmetry of $Q_1$ together with the permutation symmetry among $Q_1,\cdots,Q_4$ generate the group orbit of $Q_1$,
\begin{align}
	\text{orbit}(Q_1)=\big\{ Q_1,Q_1^{-1},Q_2,Q_2^{-1},Q_3,Q_3^{-1},Q_4,Q_4^{-1 }\big\}.
\end{align}
For each pair of elements $\{x,x^{-1}\}\subseteq \text{orbit}(Q_1)$, there is a corresponding factor whose form is like \eqref{eq:denofactor} with $Q_1$  replaced by $x$, and the denominator $D$ is just the product of all such factors. 

Using this method, now we can also determine the form of $\tilde{D}$. First, we have the same factor \eqref{eq:denoQ1} that comes from the exact summation of the geometric series of $Q_1$. Then, by the Weyl group symmetry of the current case from \eqref{eq:unaffD4WeylinQ}, the Weyl group orbit of $Q_1$ is
\begin{align}
	\text{orbit}(Q_1)=&\{Q_1,Q_1^{-1},Q_2,Q_2^{-1},Q_3,Q_3^{-1},Q_5,Q_5^{-1},Q_1Q_5,(Q_1Q_5)^{-1},Q_2Q_5,(Q_2Q_5)^{-1},\nn\\
	&Q_3Q_5,(Q_3Q_5)^{-1},Q_1Q_2Q_5,(Q_1Q_2Q_5)^{-1},Q_1Q_3Q_5,(Q_1Q_3Q_5)^{-1},Q_2Q_3Q_5,\nn\\
	&(Q_2Q_3Q_5)^{-1},Q_1Q_2Q_3Q_5,(Q_1Q_2Q_3Q_5)^{-1},Q_1Q_2Q_3Q_5^2,(Q_1Q_2Q_3Q_5^2)^{-1}\},
\end{align}
so we have
\begin{align}
	\tilde{D}=&\frac{1}{\fq^{12}\ft^{12}Q_1^6Q_2^6Q_3^6Q_5^{10}}(Q_1\fq-\ft)(Q_1\ft-\fq)(Q_2\fq-\ft)(Q_2\ft-\fq)(Q_3\fq-\ft)(Q_3\ft-\fq)\nn\\
	&\times(Q_5\fq-\ft)(Q_5\ft-\fq)(Q_1Q_5\fq-\ft)(Q_1Q_5\ft-\fq)(Q_2Q_5\fq-\ft)(Q_2Q_5\ft-\fq)\nn\\
	&\times(Q_3Q_5\fq-\ft)(Q_3Q_5\ft-\fq)(Q_1Q_2Q_5\fq-\ft)(Q_1Q_2Q_5\ft-\fq)(Q_1Q_3Q_5\fq-\ft)\nn\\
	&\times(Q_1Q_3Q_5\ft-\fq)(Q_2Q_3Q_5\fq-\ft)(Q_2Q_3Q_5\ft-\fq)(Q_1Q_2Q_3Q_5\fq-\ft)\nn\\
	&\times(Q_1Q_2Q_3Q_5\ft-\fq)(Q_1Q_2Q_3Q_5^2\fq-\ft)(Q_1Q_2Q_3Q_5^2\ft-\fq).
\end{align}
Transforming to orthonormal basis by \eqref{eq:D4orthonormal}, $\tilde{D}$ takes the following form
\begin{align}
	\tilde{D}=\prod_{i=1}^4\prod_{j>i}^4\frac{(B_iB_j\fq-\ft)(B_iB_j\ft-\fq)(B_i\fq-\ft)(B_i\ft-\fq)}{(B_iB_j\fq\ft)^2}.
\end{align}

Now for $A^{\ell}u^nQ_{\tau}^1$ term, we have
\begin{align}\label{eq:D4AuQtauterm2}
	b_{\ell,n,1}(Q_1,Q_2,Q_3,Q_5)A^{\ell}u^nQ_{\tau}^1=\frac{\tilde{b}_{\ell,n,1}(Q_1,Q_2,Q_3,Q_5)}{\tilde{D}}(Q_1Q_2Q_3)^{\ell+n}Q_5^{2\ell+2n}w_5^{\ell}Q_4^nQ_{\tau}^1,
\end{align}
and also like \eqref{eq:w5Q4term}, the $w_5^{\ell}Q_4^n Q_{\tau}^1$ term should be
\begin{align}\label{eq:D4w5Q4Qtauterm2}
	\big(``1+Q_5+Q_5^2+\cdots+Q_5^k+\cdots"\big)\, w_5^{\ell}Q_4^nQ_{\tau}^1.
\end{align}
Note that there is no negative order of $Q_5$ in the above polynomial of $Q_5$ because the power of $Q_{\tau}$ is now positive and so we do not do any flop. 

By equating the right hand side of \eqref{eq:D4AuQtauterm2} with \eqref{eq:D4w5Q4Qtauterm2}, we have
\begin{align}\label{eq:D4Qtauorder}
	&\tilde{b}_{\ell,n,1}(Q_1,Q_2,Q_3,Q_5)\fq^{12}\ft^{12}(Q_1Q_2Q_3)^{6+\ell+n}Q_5^{10+2\ell+2n}\nn\\
	=&\big(``1+Q_5+Q_5^2+\cdots+Q_5^k+\cdots"\big)\,(Q_1\fq-\ft)(Q_1\ft-\fq)(Q_2\fq-\ft)(Q_2\ft-\fq)(Q_3\fq-\ft)\nn\\
	&\times(Q_3\ft-\fq)(Q_5\fq-\ft)(Q_5\ft-\fq)(Q_1Q_5\fq-\ft)(Q_1Q_5\ft-\fq)(Q_2Q_5\fq-\ft)(Q_2Q_5\ft-\fq)\nn\\
	&\times(Q_3Q_5\fq-\ft)(Q_3Q_5\ft-\fq)(Q_1Q_2Q_5\fq-\ft)(Q_1Q_2Q_5\ft-\fq)(Q_1Q_3Q_5\fq-\ft)\nn\\
	&\times(Q_1Q_3Q_5\ft-\fq)(Q_2Q_3Q_5\fq-\ft)(Q_2Q_3Q_5\ft-\fq)(Q_1Q_2Q_3Q_5\fq-\ft)\nn\\
	&\times(Q_1Q_2Q_3Q_5\ft-\fq)(Q_1Q_2Q_3Q_5^2\fq-\ft)(Q_1Q_2Q_3Q_5^2\ft-\fq).
\end{align}
From \eqref{eq:D4Qtauorder}, in order to obtain the $Q_5^0$ order terms of $\tilde{b}_{\ell,n,1}$, the expansion of $Q_5$ has to be taken up to $k=10+2\ell+2n$. 
After that, we can determine the complete form of $\tilde{b}_{\ell,n,1}$ by computing the Weyl group orbits of the $Q_5^0$ order terms of $\tilde{b}_{\ell,n,1}$. But due to the difficulty of computation up to higher order of $Q_5$, obtaining exact $\tilde{b}_{\ell,n,1}$ is difficult, and we will only compute $Q_{\tau}^0$ order for the partition function.

\paragraph{Final result of $Z^{D_4}$.}
Combining $Z^{D_4}_{Q_5^{\bf 0}}$ with $Z^{D_4}_{Q_5}$, 
we express $Z^{D_4}$ as PE form with the expansion parameters being $A,u,Q_{\tau}$,
\begin{align}
	Z^{D_4}=Z^{D_4}_{Q_5^{\bf 0}}Z^{D_4}_{Q_5}=\text{PE}\bigg[\sum_{\ell,n,r}b_{\ell,n,r}A^\ell u^n Q_{\tau}^r\bigg]\ ,
\end{align}
and compute up to $A^1u^1Q_{\tau}^0$ order with the expansion of $Q_5$ up to $Q_5^4$ order. The corresponding coefficients are listed in the following,
\begin{align}
    &b_{0,0,0}=\frac12(2+\chi_{\bf 2})\left[0,\frac12\right],\\
    &b_{1,0,0}=(\chi_{\bf 1}^2+2\chi_{\bf 2}+\chi_{\bf s}^2+\chi_{\bf c}^2)[0,0]+\bigg[\frac12,\frac12\bigg],\\
    &b_{-1,1,0}=[0,0],\\
    &b_{0,1,0}=\chi_{\bf 2}\bigg[0,\frac12\bigg]+2\bigg[\frac12,0\bigg],\\
    &b_{1,1,0}=(-3+\textcolor{red}{2\chi_{\bf 1}^2-4\chi_{\bf 2}+\chi_{\bf 2}^2+6\chi_{\bf 1}\chi_{\bf s}\chi_{\bf c}+2\chi_{\bf s}^2+2\chi_{\bf c}^2})[0,0]\nn\\
	&\qquad\quad+(1+\textcolor{red}{\chi_{\bf 1}^2+2\chi_{\bf 2}+\chi_{\bf s}^2+\chi_{\bf c}^2})\bigg[\frac12,\frac12\bigg]+[1,1],\label{eq:coefofAu}
\end{align}
where $\chi_{\bf 1},\chi_{\bf 2},\chi_{\bf s},\chi_{\bf c}$ are the characters of fundamental weights of $D_4$ which will be in the form of \eqref{eq:D4fundamentals} if we switch to the orthonormal basis by \eqref{eq:D4orthonormal}. 
We remark here that in \eqref{eq:coefofAu}, we wrote some terms in red to denote that those are terms obtained by the Weyl group symmetry rather than the output of a direct computation. 
Due to demanding computations, we restrict ourselves to the expansion of $Q_5$ up to order $Q_5^4$ which is just enough to determine the $Q_5^0$ order terms in $b_{1,1,0}$. The highest order among these characters in red is $Q_5^4$, so by \eqref{eq:compare} we need to compute the $Q_5$ expansion up to $Q_5^8$ to confirm these characters in red, which is computationally demanding. 

\subsection{Expansion of \texorpdfstring{$D_5$}{D5}-type LST}
\begin{figure}[htbp]
    \centering
    \includegraphics[scale=0.5]{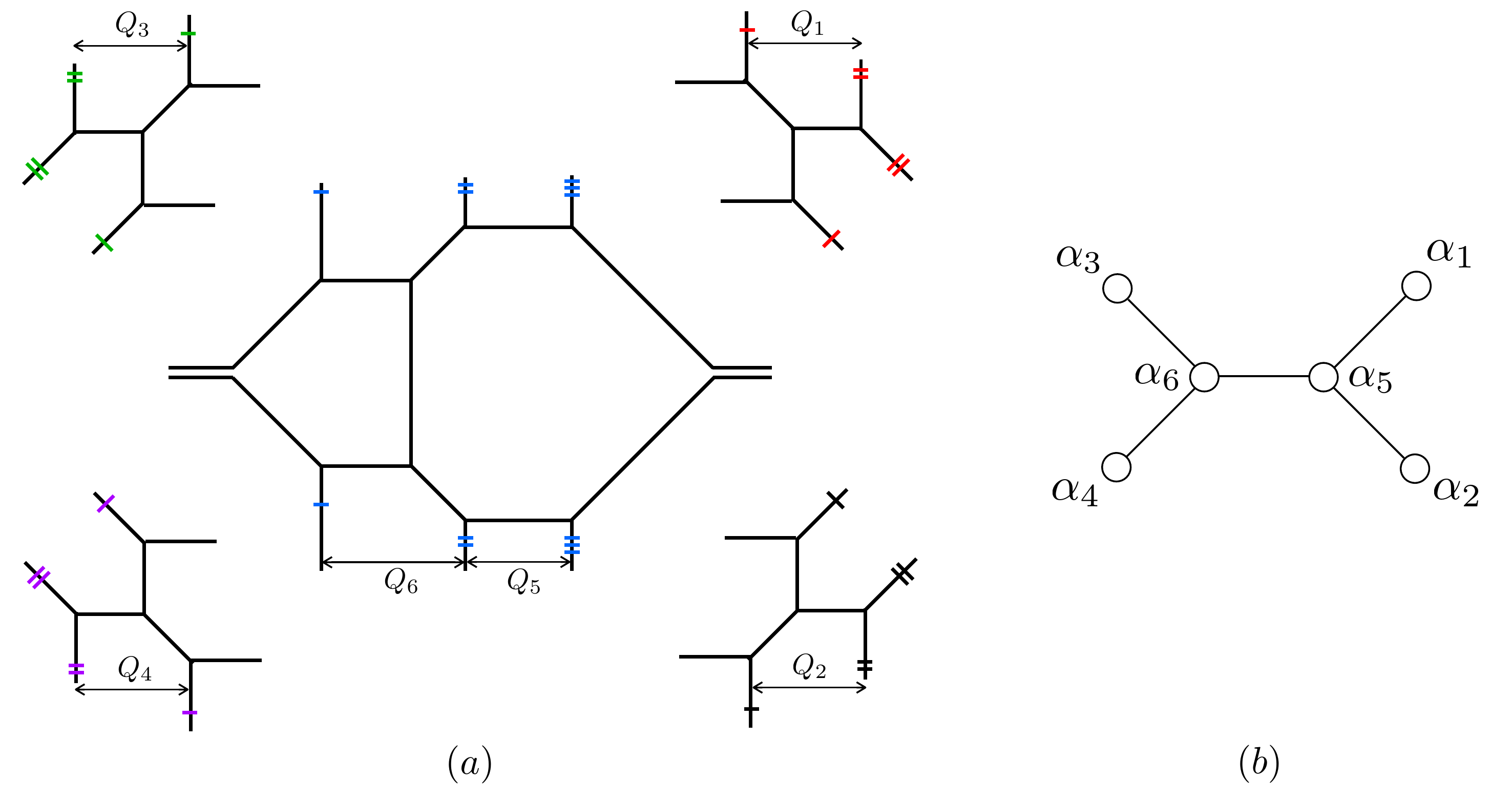}
    \caption{(a) Trivalent gluing brane web of $D_5$-type LST. (b) Affine Dynkin diagram of affine $D_5$. }
    \label{fig:D5Dynkin}
\end{figure}

Just like $D_4$-type LST, swapping the external NS-charged branes of each gauge node of the brane web of $D_5$-type LST corresponds to the Weyl reflection of the corresponding simple root in the affine $D_5$ Dynkin diagram. The Weyl reflections from the brane web of $D_5$-type LST as shown in figure \ref{fig:D5Dynkin}(a) are as follows:
\begin{equation}\label{eq:D5WeylinQaff}
	\left\{\begin{array}{l}
	Q_1\rightarrow Q_1^{-1},\ Q_5\rightarrow Q_5Q_1;\\
	Q_2\rightarrow Q_2^{-1},\ Q_5\rightarrow Q_5Q_2;\\
	Q_3\rightarrow Q_3^{-1},\ Q_6\rightarrow Q_6Q_3;\\
	Q_4\rightarrow Q_4^{-1},\ Q_6\rightarrow Q_6Q_4;\\
	Q_5\rightarrow Q_5^{-1},\ Q_1\rightarrow Q_1Q_5,\ Q_2\rightarrow Q_2Q_5,\ Q_6\rightarrow Q_6Q_5;\\
	Q_6\rightarrow Q_6^{-1},\ Q_3\rightarrow Q_3Q_6,\ Q_4\rightarrow Q_4Q_6,\ Q_5\rightarrow Q_5Q_6.	
	\end{array}\right.
\end{equation}
By similar argument like the one that leads to \eqref{eq:D4argument} in the $D_4$-type LST case, we conclude that \textit{the partition function of $D_5$-type LST has affine $D_5$ symmetry}. Again, in order to see the affine $D_5$ characters manifestly in the expansion of the partition function, we have to sum over all the Young diagrams of all the color branes up to infinity, which is very demanding. So instead, we try to see the $D_5$ characters by using five of the six independent affine $D_5$ Weyl reflections. But unlike the rank-1 $D_4$-type LST case, here there is only one choice to reduce affine $D_5$ Weyl reflections to $D_5$ Weyl reflections. We pick the affine Weyl reflections\footnote{Of course we can also pick $Q_1,Q_2,Q_4,Q_5,Q_6$ and so on which are equivalent choices. } of $Q_1,Q_2,Q_3,Q_5,Q_6$, and then $Q_4$ will be replaced by the horizontal period $u$ in \eqref{eq:uofD5}
as Weyl reflection invariant parameter. The five chosen affine $D_5$ Weyl reflections reduce to the following forms:
\begin{equation}\label{eq:D5WeylinQunaff}
	\left\{\begin{array}{l}
	Q_3\rightarrow Q_3^{-1},\ Q_6\rightarrow Q_6Q_3;\\
	Q_6\rightarrow Q_6^{-1},\ Q_3\rightarrow Q_3Q_6,\ Q_5\rightarrow Q_5Q_6;\\
	Q_5\rightarrow Q_5^{-1},\ Q_1\rightarrow Q_1Q_5,\ Q_2\rightarrow Q_2Q_5,\ Q_6\rightarrow Q_6Q_5;\\
	Q_1\rightarrow Q_1^{-1},\ Q_5\rightarrow Q_5Q_1;\\
	Q_2\rightarrow Q_2^{-1},\ Q_5\rightarrow Q_5Q_2 .
	\end{array}\right.
\end{equation}
They are the $D_5$ Weyl reflections in the basis $\{\alpha_3,\alpha_6,\alpha_5,\alpha_1,\alpha_2\}$ where the $\alpha$'s as well as the corresponding $Q$'s are ordered in a conventional way for $D_5$ simple roots.

\begin{figure}[htbp]
    \centering
    \includegraphics[scale=0.5]{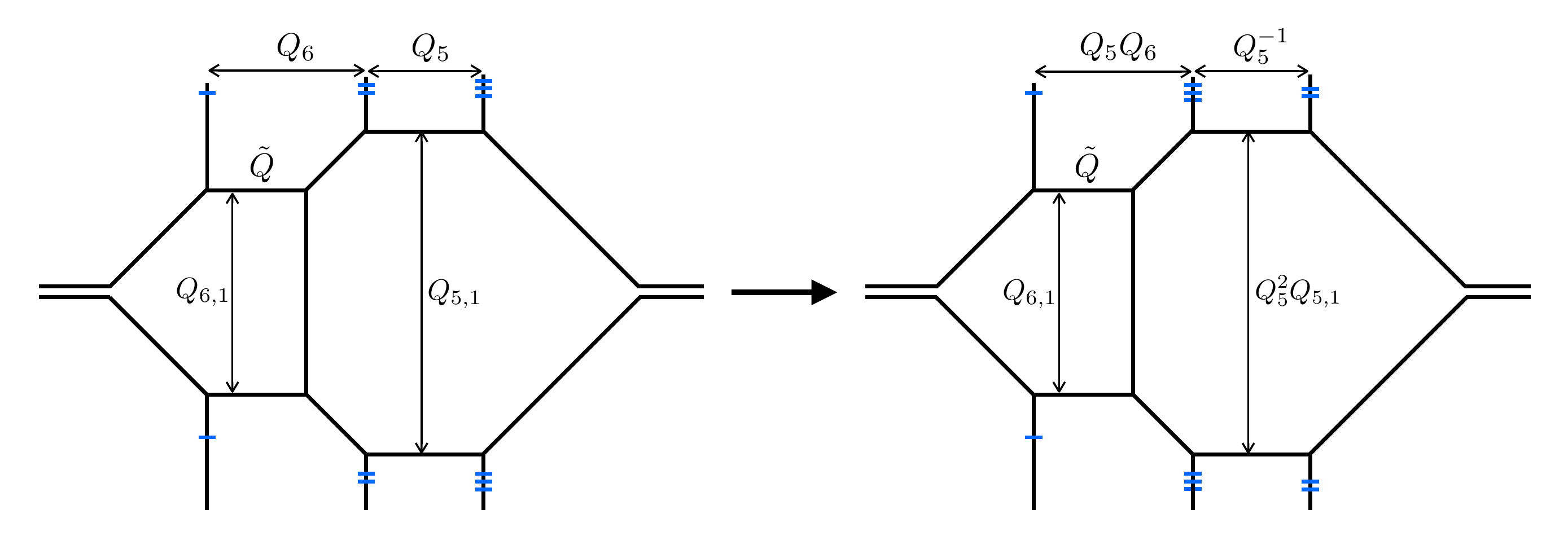}
    \caption{Transformation of gauge node 5.}
    \label{fig:Q5transf}
\end{figure}

\begin{figure}[htbp]
    \centering
    \includegraphics[scale=0.5]{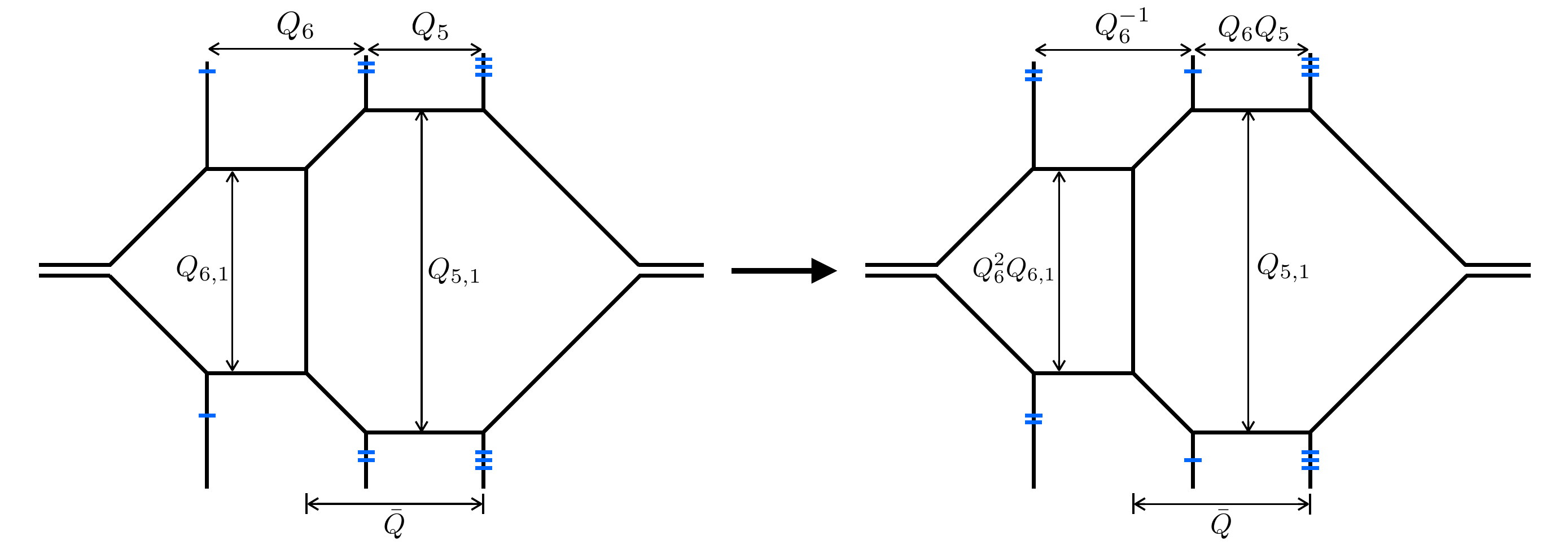}
    \caption{Transformation of gauge node 6.}
    \label{fig:Q6transf}
\end{figure}

For $D_5$-type LST, we have two Coulomb branch parameters $Q_{5,1},Q_{6,1}$ as shown in figure \ref{fig:D5wQuadra}. For convenience in computation, we use the following new Coulomb branch parameters $w_5$ and $w_6$, 
\begin{align}
	w_5\equiv Q_{5,1}^{1/2},\qquad w_6\equiv Q_{6,1}^{1/2}. 
\end{align}
$w_5$ does not change under the Weyl reflections of $Q_1,Q_2,Q_3,Q_6$ but it changes under the Weyl reflection of $Q_5$ due to the invariance of $\tilde{Q}$ and $Q_{6,1}$ as shown in figure \ref{fig:Q5transf}, whereas $w_6$ does not change under the Weyl reflections of $Q_1,Q_2,Q_3,Q_5$ but it changes under the Weyl reflection of $Q_6$ due to the invariance of $\bar{Q}$ and $Q_{5,1}$ as shown in figure \ref{fig:Q6transf}. 
Including the Coulomb branch parameters $w_5$ and $w_6$, the Weyl reflections become as follows:
\begin{equation}
	\left\{\begin{array}{l}
	Q_3\rightarrow Q_3^{-1},\ Q_6\rightarrow Q_6Q_3;\\
	Q_6\rightarrow Q_6^{-1},\ Q_3\rightarrow Q_3Q_6,\ Q_5\rightarrow Q_5Q_6,\ w_6\rightarrow Q_6w_6;\\
	Q_5\rightarrow Q_5^{-1},\ Q_1\rightarrow Q_1Q_5,\ Q_2\rightarrow Q_2Q_5,\ Q_6\rightarrow Q_6Q_5,\ w_5\rightarrow Q_5w_5;\\
	Q_1\rightarrow Q_1^{-1},\ Q_5\rightarrow Q_5Q_1;\\
	Q_2\rightarrow Q_2^{-1},\ Q_5\rightarrow Q_5Q_2.
	\end{array}\right.
\end{equation}
The above transformations involve $Q_1,Q_2,Q_3,Q_5,Q_6,w_5,w_6$, so the invariant Coulomb branch parameters should be a combination of these parameters which take the general form of
\begin{equation}
 Q_1^aQ_2^bQ_3^cQ_5^dQ_6^ew_5^fw_6^g. 
 \end{equation}
 By requiring this general form to be invariant under the above five transformations, we obtain five linear equations about $a,b,c,d,e,f,g$. There are two independent invariant Coulomb branch parameters that are proportional to $w_5$ (i.e., $f=1,g=0$) and $w_6$ (i.e., $f=0,g=1$) respectively. For each case we can solve the five linear equations and obtain one solution, so we get the following solutions which are the invariant Coulomb branch parameters
\begin{align}\label{eq:invariantCBofD5}
	A_1=Q_1^{\frac32}Q_2^{\frac32}Q_3Q_5^3Q_6^2w_5,\quad A_2=Q_1Q_2Q_3Q_5^2Q_6^2w_6\ . 
\end{align}
If we switch from the $\alpha$-basis fugacities $Q$'s to the orthonormal basis fugacities $B$'s, we have
\begin{align}\label{eq:D5orthonormaltran}
	Q_3=\frac{B_1}{B_2},\ Q_6=\frac{B_2}{B_3},\ Q_5=\frac{B_3}{B_4},\ Q_1=\frac{B_4}{B_5},\ Q_2=B_4B_5.
\end{align}
Then in the orthonormal basis, the $D_5$ Weyl reflections in \eqref{eq:D5WeylinQunaff} become as follows:
\begin{equation}\label{eq:D5WeylinB}
	\left\{\begin{array}{l}
	B_1\rightarrow B_2,\ B_2\rightarrow B_1;\\
	B_2\rightarrow B_3,\ B_3\rightarrow B_2;\\
	B_3\rightarrow B_4,\ B_4\rightarrow B_3;\\
 	B_4\rightarrow B_5,\ B_5\rightarrow B_4;\\
	B_4\rightarrow B_5^{-1},\ B_5\rightarrow B_4^{-1}.
	\end{array}\right.
\end{equation} 

Like in section \ref{sec:choice2}, it is most convenient to keep using the old parameters $Q_1,Q_2,Q_3,Q_4,Q_5,Q_6,w_5,w_6,Q_{\tau}$ when computing the partition function of $D_5$-type LST in the intermediate process and in particular pick $Q_5,Q_6$ as the main expansion parameters, and switch to the new set of parameters $Q_1,Q_2,Q_3,Q_5,Q_6,A_1,A_2,u,Q_{\tau}$ in the final step in order to see the $D_5$ characters manifestly. 

We separate the partition function into $Z^{D_5}_{Q_5^{\bf 0}Q_6^{\bf 0}}$ part which is the $Q_5^0 Q_6^0$ order term in the full partition function and $Z^{D_5}_{Q_5Q_6}$ part which is the remaining part that depends on $Q_5,Q_6$ after extracting out $Z^{D_5}_{Q_5^{\bf 0}Q_6^{\bf 0}}$,
\begin{align}
	Z^{D_5}=Z^{D_5}_{Q_5^{\bf 0}Q_6^{\bf 0}}Z^{D_5}_{Q_5Q_6}\ .
\end{align}
From \eqref{eq:ZofD5insec3} and \eqref{eq:D5inst}, the $Z^{D_5}_{Q_5^{\bf 0}Q_6^{\bf 0}}$ part that does not depend on $Q_5,Q_6$ is
\begin{align}\label{eq:nZ_D5pert}
	Z^{D_5}_{Q_5^{\bf 0}Q_6^{\bf 0}}=Z^{D_5}_{\text{pert}}\cdot &\sum_{\mu_{1,1}}\bar{Z}_{\oo\oo\mu_{1,1}}(w_5,Q_1) \sum_{\mu_{2,1}}\bar{Z}_{\oo\oo\mu_{2,1}}(w_5,Q_2)\nonumber\\
	\times &\sum_{\mu_{3,1}}\bar{Z}_{\oo\oo\mu_{3,1}}(w_6,Q_3)\sum_{\mu_{4,1}}\bar{Z}_{\oo\oo\mu_{4,1}}(w_6,Q_4)\, ,
\end{align}
the $Z^{D_5}_{Q_5Q_6}$ part that depend on $Q_5,Q_6$ is
\begin{align}\label{eq:ZD5Q5Q6}
	Z^{D_5}_{Q_5Q_6}=\sum_{\mu_{5,1},\mu_{5,2},\mu_{6,1},\mu_{6,2}}&\bar{Z}^{\text{SU(2)-SU(2)}}_{\mu_{5,1}\mu_{5,2}\mu_{6,1}\mu_{6,2}}(w_5,w_6,Q_5,Q_6)\calZ_{\mu_{5,1}\mu_{5,2}}(w_5,Q_1)\calZ_{\mu_{5,1}\mu_{5,2}}(w_5,Q_2)\nn\\
	&\times\calZ_{\mu_{6,1}\mu_{6,2}}(w_6,Q_3)\calZ_{\mu_{6,1}\mu_{6,2}}(w_6,Q_4),
\end{align}
where 
\begin{align}
	&\bar{Z}^{\text{SU(2)-SU(2)}}_{\mu_{5,1}\mu_{5,2}\mu_{6,1}\mu_{6,2}}(w_5,w_6,Q_5,Q_6)\nn\\
	&\equiv Q_{5}^{|\mu_{5,1}|+|\mu_{5,2}|}\prod_{p=1}^{2}\bigg(\frac{\prod_{r=1}^{2}\varTheta_{\mu_{5,p}\mu_{6,r}}(\frac{y_{5,p}}{y_{6,r}}\sqrt{\tfrac{\ft}{\fq}}|Q_{\tau})}{\prod_{c=1}^{2}\varTheta_{\mu_{5,p}\mu_{5,c}}(\frac{y_{5,p}}{y_{5,c}}|Q_{\tau})\varTheta_{\mu_{5,p}\mu_{5,c}}(\frac{y_{5,p}}{y_{5,c}}\tfrac{\ft}{\fq}|Q_{\tau})}\bigg)\nn\\
    &~\times Q_{6}^{|\mu_{6,1}|+|\mu_{6,2}|}\prod_{s=1}^{2}\bigg(\frac{\prod_{d=1}^{2}\varTheta_{\mu_{6,s}\mu_{5,d}}(\frac{y_{6,s}}{y_{5,d}}\sqrt{\tfrac{\ft}{\fq}}|Q_{\tau})}{\prod_{e=1}^{2}\varTheta_{\mu_{6,s}\mu_{6,e}}(\frac{y_{6,s}}{y_{6,e}}|Q_{\tau})\varTheta_{\mu_{6,s}\mu_{6,e}}(\frac{\mu_{6,s}}{\mu_{6,e}}\tfrac{\ft}{\fq}|Q_{\tau})}\bigg)\ ,
\end{align}
and the $\calZ$'s are the same functions that is defined in \eqref{eq:defZQ1}. 

\paragraph{$Z^{D_5}_{Q_5^{\bf 0}Q_6^{\bf 0}}$ part.}
From the general formula in \eqref{eq:unifiedpert}, we get the perturbative part of the partition function of $D_5$-type LST,
\begin{align}\label{eq:D5pertPE}
	Z^{D_5}_{\text{pert}}=&~\text{PE}\bigg[\frac{1}{(1-\ft)(1-\fq)}\frac{1+Q_{\tau}}{1-Q_{\tau}}\Big((\fq+\ft)\big(4+\frac12(w_5^2+\frac{1}{w_5^2})+\frac12(w_6^2+\frac{1}{w_6^2})\big)
	\nn\\
	&-\sqrt{\ft\fq}\big(2(w_5+\frac{1}{w_5})+2(w_6+\frac{1}{w_6})+w_5w_6+\frac{1}{w_5w_6}+\frac{w_5}{w_6}+\frac{w_6}{w_5}\big)\Big)\nn\\
	&+7\frac{Q_{\tau}}{1-Q_{\tau}}\bigg]\ .
\end{align}
Substitute \eqref{eq:D5pertPE} and \eqref{eq:zsubu1rev} into \eqref{eq:nZ_D5pert} and expand up to $Q_{\tau}Q_4$ order, we get
\begin{align}
	Z^{D_5}_{Q_5^{\bf 0}Q_6^{\bf 0}}=&~\text{PE}\bigg[\frac{1}{(1-\ft)(1-\fq)}\bigg((\fq+\ft)(Q_1+Q_2+Q_3+4)-\sqrt{\ft \fq}\Big((Q_1+\frac{1}{Q_1}+Q_2\nn\\
	&+\frac{1}{Q_2}+4)w_5+(Q_3+\frac{1}{Q_3}+4)w_6+2 w_5w_6+2\frac{w_5}{w_6}\Big)+Q_4\Big((\fq+\ft)\nn\\
	&-\sqrt{\fq\ft}(w_6+\frac{1}{w_6})\Big)+\calO(Q_4^2)\bigg)+\frac{Q_{\tau}}{(1-\ft)(1-\fq)}\bigg(7(\fq\ft+1)+\fq+\ft\nn\\
	&+(\fq+\ft)\big(\ft+\fq^2\ft+\fq(1+\ft)^2\big)\Big(\frac{Q_1}{(Q_1\fq-\ft)(Q_1\ft-\fq)}+\frac{Q_2}{(Q_2\fq-\ft)(Q_2\ft-\fq)}\nn\\
	&+\frac{Q_3}{(Q_3\fq-\ft)(Q_3\ft-\fq)}\Big)-\sqrt{\fq\ft}\Big((w_5+\frac{1}{w_5})\big(4+\frac{Q_1(1+\fq)(1+\ft)(\fq+\ft)}{(Q_1\fq-\ft)(Q_1\ft-\fq)}\nn\\
	&+\frac{Q_2(1+\fq)(1+\ft)(\fq+\ft)}{(Q_2\fq-\ft)(Q_2\ft-\fq)}\big)+(w_6+\frac{1}{w_6})\big(4+\frac{Q_3(1+\fq)(1+\ft)(\fq+\ft)}{(Q_3\fq-\ft)(Q_3\ft-\fq)}\big)\nn\\
	&+2(w_5w_6+\frac{1}{w_5w_6}+\frac{w_5}{w_6}+\frac{w_6}{w_5})\Big)+(\fq+\ft)\Big((w_5^2+\frac{1}{w_5^2})\big(\frac{Q_1\fq\ft}{(Q_1\fq-\ft)(Q_1\ft-\fq)}\nn\\
	&+\frac{Q_2\fq\ft}{(Q_2\fq-\ft)(Q_2\ft-\fq)}+1\big)+(w_6^2+\frac{1}{w_6^2})\big(\frac{Q_3\fq\ft}{(Q_3\fq-\ft)(Q_3\ft-\fq)}+1\big)\Big)\nn\\
	&+Q_4\Big(\frac{(\fq+\ft)\big(\ft+\fq^2\ft+\fq(1+\ft)^2\big)}{\fq\ft}-\frac{(1+\fq)(1+\ft)(\fq+\ft)}{\sqrt{\fq\ft}}(w_6+\frac{1}{w_6})\nn\\
	&+(\fq+\ft)(w_6^2+\frac{1}{w_6^2})\Big)+\calO(Q_4^2)\bigg)+\calO(Q_{\tau}^2)\bigg].
\end{align}

\paragraph{$Z^{D_5}_{Q_5Q_6}$ part and rescaling.}
To compute $Z^{D_5}_{Q_5Q_6}$ in \eqref{eq:ZD5Q5Q6}, we expand with respect to $Q_5,Q_6$ whose orders are restricted by $|\mu_{5,1}|+|\mu_{5,2}|,|\mu_{6,1}|+|\mu_{6,2}|$ and with respect to $Q_4$ which is proportional to $u$. 
Like in the $D_4$-type LST case, in order to reduce the heavy computation 
we need to further expand $Z^{D_5}_{Q_5Q_6}$ with respect to $w_5,w_6$ up to finite order. To obtain such expansion of $Z^{D_5}_{Q_5Q_6}$, it is easier to expand the five factors in the summand of \eqref{eq:ZD5Q5Q6} first and then multiply and truncate the expansions. 
In order to make the truncation easy, we need to rescale $\bar{Z}^{\text{SU(2)-SU(2)}}$, $\calZ$ functions, and parameters in \eqref{eq:ZD5Q5Q6}.  
We find that the following rescalings,
\begin{align}\label{eq:calZD5rescaling}
	&\calZ_{\mu_{5,1}\mu_{5,2}}\rightarrow w_5^{\frac12(|\mu_{5,1}|+|\mu_{5,2}|)}\calZ_{\mu_{5,1}\mu_{5,2}},\cr
    &\calZ_{\mu_{6,1}\mu_{6,2}}\rightarrow w_6^{\frac12(|\mu_{6,1}|+|\mu_{6,2}|)}\calZ_{\mu_{6,1}\mu_{6,2}},\nn\\
	&\bar{Z}^{\text{SU(2)-SU(2)}}_{\mu_{5,1}\mu_{5,2}\mu_{6,1}\mu_{6,2}}\rightarrow w_5^{-(|\mu_{5,1}|+|\mu_{5,2}|)}w_6^{-(|\mu_{6,1}|+|\mu_{6,2}|)}\bar{Z}^{\text{SU(2)-SU(2)}}_{\mu_{5,1}\mu_{5,2}\mu_{6,1}\mu_{6,2}},\nn\\
	&Q_5\rightarrow w_6Q_5,\qquad Q_6\rightarrow w_5Q_6,\qquad Q_{\tau}\rightarrow w_5^2 w_6^2 Q_{\tau},
\end{align}
make $\bar{Z}^{\text{SU(2)-SU(2)}}$ and $\calZ$ functions have non-negative power expansions with respect to $w_5,w_6$. After obtaining the PE form of $Z^{D_5}_{Q_5Q_6}$ expanded with respect to $w_5,w_6$ under the rescalings \eqref{eq:calZD5rescaling}, we switch back to the original parameters $Q_5,Q_6,Q_{\tau}$ in the PE form. 

\paragraph{Evaluation of the partition function by Weyl group symmetry.}
For the $A_1^\ell A_2^m u^n Q_{\tau}^r$ term in the PE of $Z^{D_5}$, it will be of the following form,
\begin{align}\label{eq:D5generalterm}
	b_{\ell,m,n,r}(Q_1,Q_2,Q_3,Q_5,Q_6)A_1^\ell A_2^m u^n Q_{\tau}^r.
\end{align}
\paragraph{(1)}
For $r=0$ which is at $Q_{\tau}^0$ order, we expect $b_{\ell,m,n,0}$ to be finite polynomials of $D_5$ characters with $Q_1,Q_2,Q_3,Q_5,Q_6$ as the fugacities. If we focus on $Q_5, Q_6$ which are the intermediate expansion parameters, $b_{\ell,m,n,0}$ can be regarded as Laurent polynomials of $Q_5, Q_6$ with both negative and non-negative powers. In particular, all the Weyl group orbits of the polynomial $b_{\ell,m,n,0}$ are also Laurent polynomials of $Q_5, Q_6$ and they all contain $Q_5^0 Q_6^0$ order terms. So if we can get the complete $Q_5^0 Q_6^0$ order terms of $b_{\ell,m,n,0}$, we can completely determine the full polynomial $b_{\ell,m,n,0}$ by Weyl group symmetry.

In \eqref{eq:D5generalterm}, we switch to the parameters $w_5,w_6,Q_4$ by \eqref{eq:invariantCBofD5} and \eqref{eq:uofD5}, then we get
\begin{align}\label{eq:blmnofD5}
	&b_{\ell,m,n,0}(Q_1,Q_2,Q_3,Q_5,Q_6)A_1^\ell A_2^m u^n\nn\\
	=&b_{\ell,m,n,0}(Q_1,Q_2,Q_3,Q_5,Q_6)(Q_1Q_2)^{\frac32\ell+m+n}Q_3^{\ell+m+n}Q_5^{3\ell+2m+2n}Q_6^{2\ell+2m+2n}w_5^{\ell}w_6^m Q_4^n\ .
\end{align}
On the other hand, from the actual computation, we would get 
\begin{align}\label{eq:D5expansioncf}
    \big(``\cdots+1+Q_5+Q_6+\cdots+Q_5^{k_1} Q_6^{k_2}+\cdots"\big)\, w_5^{\ell}w_6^m Q_4^n,
\end{align}
if we sum over the Young diagrams up to $|\mu_{5,1}|+|\mu_{5,2}|\leq k_1,|\mu_{6,1}|+|\mu_{6,2}|\leq k_2$. 
By equating the right hand side of \eqref{eq:blmnofD5} with \eqref{eq:D5expansioncf}, we have 
\begin{align}\label{eq:D5judge}
	&b_{\ell,m,n,0}(Q_1,Q_2,Q_3,Q_5,Q_6)(Q_1Q_2)^{\frac32\ell+m+n}Q_3^{\ell+m+n}Q_5^{3\ell+2m+2n}Q_6^{2\ell+2m+2n}\nn\\
	=&``\cdots+1+Q_5+Q_6+\cdots+Q_5^{k_1} Q_6^{k_2}+\cdots" \, .
\end{align}
In order to obtain the $Q_5^0 Q_6^0$ order terms of $b_{\ell,m,n,0}$, we need to expand $Q_5,Q_6$ on the right hand side of \eqref{eq:D5judge} up to $k_1=3\ell+2m+2n,k_2=2\ell+2m+2n$. Then we can use Weyl group symmetry to obtain the full terms of $b_{\ell,m,n,0}$ from its $Q_5^0 Q_6^0$ order terms. After that we will know the highest order of $Q_5,Q_6$ in $b_{\ell,m,n,0}$, and we can increase the expansion orders of $Q_5, Q_6$ to necessary orders on the right hand side of \eqref{eq:D5judge}  to confirm the correctness of the result of $b_{\ell,m,n,0}$ that is obtained by Weyl group symmetry. 

\paragraph{(2)}
For $r=1$ which is at $Q_{\tau}^1$ order, like in the $D_4$-type LST case, we expect that there will be a common denominator $D$ appearing in the $Q_{\tau}^1$ order, and $b_{\ell,m,n,1}$ will be in the form of a finite polynomial $\tilde{b}_{\ell,m,n,1}$ divided by $D$,
\begin{align}
	b_{\ell,m,n,1}=\frac{\tilde{b}_{\ell,m,n,1}}{D}.
\end{align}
The denominator $D$ of the current $D_5$-type LST case has the same factor \eqref{eq:denoQ1} which comes from the exact summation of the geometric series of $Q_1$, so we can start from it and use the $D_5$ Weyl group symmetry \eqref{eq:D5WeylinQunaff} to generate all the factors of $D$. The Weyl group orbit of $Q_1$ is 
\begin{align}
	&\text{orbit}(Q_1)\nn\\
	=&\{Q_1,Q_1^{-1},Q_2,Q_2^{-1},Q_3,Q_3^{-1},Q_5,Q_5^{-1},Q_6,Q_6^{-1},Q_1Q_5,(Q_1Q_5)^{-1},Q_2Q_5,(Q_2Q_5)^{-1},\nn\\
	&Q_3Q_6,(Q_3Q_6)^{-1},Q_5Q_6,(Q_5Q_6)^{-1},Q_1Q_2Q_5,(Q_1Q_2Q_5)^{-1},Q_1Q_5Q_6,(Q_1Q_5Q_6)^{-1},\nn\\
	&Q_2Q_5Q_6,(Q_2Q_5Q_6)^{-1},Q_3Q_5Q_6,(Q_3Q_5Q_6)^{-1},Q_1Q_2Q_5Q_6,(Q_1Q_2Q_5Q_6)^{-1},\nn\\
	&Q_1Q_3Q_5Q_6,(Q_1Q_3Q_5Q_6)^{-1},Q_2Q_3Q_5Q_6,(Q_2Q_3Q_5Q_6)^{-1},Q_1Q_2Q_5^2Q_6,(Q_1Q_2Q_5^2Q_6)^{-1},\nn\\
	&Q_1Q_2Q_3Q_5Q_6,(Q_1Q_2Q_3Q_5Q_6)^{-1},Q_1Q_2Q_3Q_5^2Q_6,(Q_1Q_2Q_3Q_5^2Q_6)^{-1},Q_1Q_2Q_3Q_5^2Q_6^2,\nn\\
	&(Q_1Q_2Q_3Q_5^2Q_6^2)^{-1}\},
\end{align}
so we find
\begin{align}
	D=&\frac{1}{\fq^{20}\ft^{20}Q_1^{10}Q_2^{10}Q_3^8Q_5^{18}Q_6^{14}}(Q_1\fq-\ft)(Q_1\ft-\fq)(Q_2\fq-\ft)(Q_2\ft-\fq)(Q_3\fq-\ft)(Q_3\ft-\fq)\nn\\
	&\times(Q_5\fq-\ft)(Q_5\ft-\fq)(Q_1Q_5\fq-\ft)(Q_1Q_5\ft-\fq)(Q_2Q_5\fq-\ft)(Q_2Q_5\ft-\fq)\nn\\
	&\times(Q_1Q_2Q_5\fq-\ft)(Q_1Q_2Q_5\ft-\fq)(Q_6\fq-\ft)(Q_6\ft-\fq)(Q_3Q_6\fq-\ft)(Q_3Q_6\ft-\fq)\nn\\
	&\times(Q_5Q_6\fq-\ft)(Q_5Q_6\ft-\fq)(Q_1Q_5Q_6\fq-\ft)(Q_1Q_5Q_6\ft-\fq)(Q_2Q_5Q_6\fq-\ft)\nn\\
	&\times(Q_2Q_5Q_6\ft-\fq)(Q_1Q_2Q_5Q_6\fq-\ft)(Q_1Q_2Q_5Q_6\ft-\fq)(Q_3Q_5Q_6\fq-\ft)\nn\\
	&\times(Q_3Q_5Q_6\ft-\fq)(Q_1Q_3Q_5Q_6\fq-\ft)(Q_1Q_3Q_5Q_6\ft-\fq)(Q_2Q_3Q_5Q_6\fq-\ft)\nn\\
	&\times(Q_2Q_3Q_5Q_6\ft-\fq)(Q_1Q_2Q_3Q_5Q_6\fq-\ft)(Q_1Q_2Q_3Q_5Q_6\ft-\fq)(Q_1Q_2Q_5^2Q_6\fq-\ft)\nn\\
	&\times(Q_1Q_2Q_5^2Q_6\ft-\fq)(Q_1Q_2Q_3Q_5^2Q_6\fq-\ft)(Q_1Q_2Q_3Q_5^2Q_6\ft-\fq)\nn\\
	&\times(Q_1Q_2Q_3Q_5^2Q_6^2\fq-\ft)(Q_1Q_2Q_3Q_5^2Q_6^2\ft-\fq).
\end{align}
Transformed into the orthonormal basis by \eqref{eq:D5orthonormaltran}, $D$ takes the following form,
\begin{align}
	D=\prod_{i=1}^5\prod_{j>i}^5\frac{(B_iB_j\fq-\ft)(B_iB_j\ft-\fq)(B_i\fq-B_j\ft)(B_i\ft-B_j\fq)}{(B_iB_j\fq\ft)^2}.
\end{align}
Like in the $D_4$-type LST case, to obtain the coefficients of $Q_{\tau}^1$ order, the expansion of $Q_5,Q_6$ has to be taken up to very high order which is difficult, so we will only compute the $Q_{\tau}^0$ order for $Z^{D_5}$. 

\paragraph{Final result of $Z^{D_5}$.}
Combining $Z^{D_5}_{Q_5^{\bf 0}Q_6^{\bf 0}}$ with $Z^{D_5}_{Q_5Q_6}$, we express $Z^{D_5}$ as PE form with the expansion parameters being $A_1,A_2,u,Q_{\tau}$,
\begin{align}
	Z^{D_5}=Z^{D_5}_{Q_5^{\bf 0}Q_6^{\bf 0}}\cdot Z^{D_5}_{Q_5Q_6}=\text{PE}\bigg[\sum_{\ell,m,n,r}b_{\ell,m,n,r}A_1^\ell A_2^m u^n Q_{\tau}^r\bigg]\ ,
\end{align}
and list a few $b_{\ell,m,n,r}$ that we could determine with the expansion of $Q_5,Q_6$ up to $Q_5^3 Q_6^2$ order,
\begin{align}
    &b_{0,0,0,0}=\frac12(3+\chi_{\bf 2})\left[0,\frac12\right],\nn\\
    &b_{1,-1,0,0}=2\chi_{\bf 1}[0,0],\nn\\
    &b_{1,0,0,0}=(\textcolor{red}{\chi_{\bf s}^2+\chi_{\bf c}^2})[0,0]+\textcolor{red}{2\chi_{\bf 1}}\bigg[0,\frac12\bigg],\nn\\
    &b_{0,1,0,0}=(1+\textcolor{red}{\chi_{\bf 1}^2+2\chi_{\bf 2}})[0,0]+[0,1]+\bigg[\frac12,\frac12\bigg],\nn\\
    &b_{0,0,1,0}=(1+\textcolor{red}{\chi_{\bf 2}})\bigg[0,\frac12\bigg]+2\bigg[\frac12,0\bigg],\nn\\
    &b_{1,-1,1,0}=\textcolor{red}{2\chi_{\bf 1}}[0,0]\ ,\nn\\
    &b_{-1,1,1,0}=2\chi_{\bf 1}[0,0]\ ,
\end{align}
where 
$\chi_{\bf 1}, \chi_{\bf 2}, \chi_{\bf 3}, \chi_{\bf s}$ and $\chi_{\bf c}$ are characters of fundamental weights of $D_5$ whose expressions are in \eqref{eq:D5fundamentals} when transformed to orthonormal basis by \eqref{eq:D5orthonormaltran}. For $b_{1,0,0,0}, b_{0,1,0,0}, b_{0,0,1,0}$, and $b_{1,-1,1,0}$, we denote the characters in red which means that these characters are obtained by Weyl group symmetry but not yet confirmed by direct computation, because we only compute the expansion of $Q_5, Q_6$ up to $Q_5^3Q_6^2$ order which is just enough to determine the $Q_5^0Q_6^0$ order of these characters.  As the highest order among these characters is $Q_5^3Q_6^2$, by \eqref{eq:D5judge} one needs to compute the $Q_5, Q_6$ expansion up to $Q_5^6Q_6^4$ to confirm these characters in red, which is computationally demanding.

The partition function of rank-1 $D_5$-type LST computed in \cite{Kim:2017xan} is expanded in terms of $Q_1,$ $Q_2,$ $Q_3,$ $Q_4,$$Q_5,$$Q_6,$$w_6,$$Q_{\tau}$ up to $Q_1^1Q_2^1Q_3^1Q_4^1Q_5^2Q_6^2w_6^2Q_{\tau}^1$ order. Analogous to the $D_4$-case, we also re-expanded $Z^{D_5}$ and our result is consistent with the result of \cite{Kim:2017xan} up to the order that they computed.

\subsection{Expansion of \texorpdfstring{$E_6$}{E6}-type LST}
\begin{figure}[htbp]
	\centering
	\includegraphics[scale=0.45]{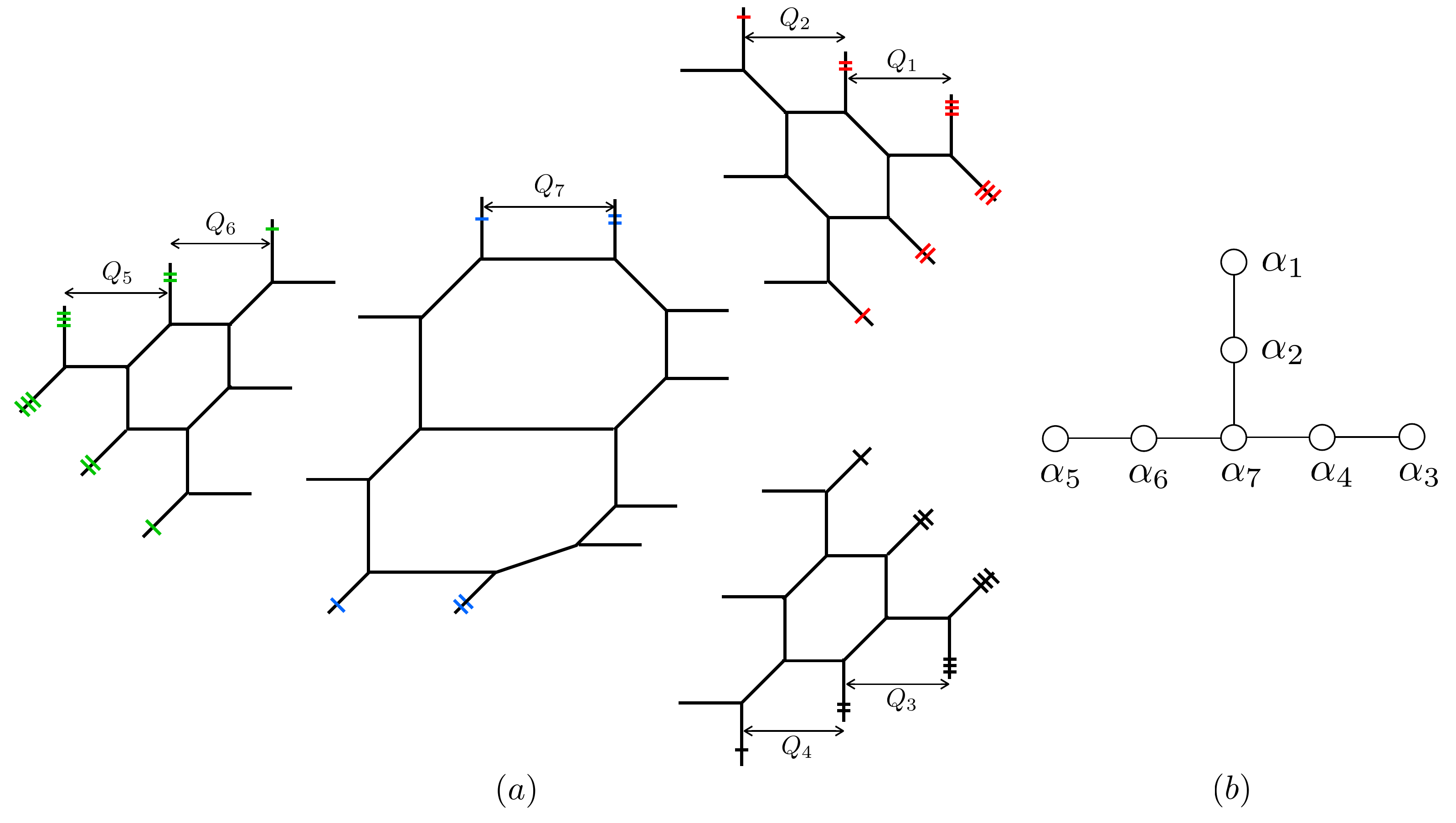}
	\caption{(a) Trivalent gluing brane web of $E_6$-type LST. (b) Affine Dynkin diagram of affine $E_6$. }
	\label{fig:E6Dynkin}
\end{figure}
For $E_6$-type LST, it is also easy to show that the swapping of external NS-charged branes of each gauge node of $Q_i$ in the trivalent gluing brane web in figure \ref{fig:E6Dynkin}(a) corresponds to the affine Weyl reflection of each affine simple root $\alpha_i$ in the affine $E_6$ Dynkin diagram in figure \ref{fig:E6Dynkin}(b). So by similar argument like in the $D_4$-type LST case, we conclude that \textit{the partition function of $E_6$-type LST has affine $E_6$ symmetry}.

The affine $E_6$ Weyl reflections from the brane web in figure \ref{fig:E6Dynkin}(a) are as following forms:
\begin{equation}
	\left\{\begin{array}{l}
	Q_1\rightarrow Q_1^{-1},\ Q_2\rightarrow Q_2Q_1;\\
	Q_2\rightarrow Q_2^{-1},\ Q_1\rightarrow Q_1Q_2,\ Q_7\rightarrow Q_7Q_2;\\
	Q_3\rightarrow Q_3^{-1},\ Q_4\rightarrow Q_4Q_3;\\
	Q_4\rightarrow Q_4^{-1},\ Q_3\rightarrow Q_3Q_4,\ Q_7\rightarrow Q_7Q_4;\\
	Q_5\rightarrow Q_5^{-1},\ Q_6\rightarrow Q_6Q_5;\\
	Q_6\rightarrow Q_6^{-1},\ Q_5\rightarrow Q_5Q_6,\ Q_7\rightarrow Q_7Q_6;\\
	Q_7\rightarrow Q_7^{-1},\ Q_2\rightarrow	 Q_2Q_7,\ Q_4\rightarrow Q_4Q_7,\ Q_6\rightarrow Q_6Q_7 .
	\end{array}\right.
\end{equation}
Again like in the $D_4,D_5$-type LST cases, seeing the affine $E_6$ characters in the expansion of partition function is difficult, so we try to reduce them to the $E_6$ characters by using only six of the seven affine Weyl reflections. We pick the affine Weyl reflections of $Q_2,Q_3,Q_4,Q_5,Q_6,Q_7$, and then $Q_1$ will be replaced by the horizontal period $u$ in \eqref{eq:E6horizontal}
as Weyl reflection invariant parameter. The six chosen affine $E_6$ Weyl reflections reduce to the following forms:
\begin{equation}\label{eq:uaffE6WeylinQ}
	\left\{\begin{array}{l}
	Q_5\rightarrow Q_5^{-1},\ Q_6\rightarrow Q_6Q_5;\\
 	Q_6\rightarrow Q_6^{-1},\ Q_5\rightarrow Q_5Q_6,\ Q_7\rightarrow Q_7Q_6;\\
  	Q_7\rightarrow Q_7^{-1},\ Q_2\rightarrow	 Q_2Q_7,\ Q_4\rightarrow Q_4Q_7,\ Q_6\rightarrow Q_6Q_7;\\
   	Q_4\rightarrow Q_4^{-1},\ Q_3\rightarrow Q_3Q_4,\ Q_7\rightarrow Q_7Q_4;\\
   	Q_3\rightarrow Q_3^{-1},\ Q_4\rightarrow Q_4Q_3;\\
	Q_2\rightarrow Q_2^{-1},\ Q_7\rightarrow Q_7Q_2 .
	\end{array}\right.
\end{equation}
They are the $E_6$ Weyl reflections in the basis $\{\alpha_5,\alpha_6,\alpha_7,\alpha_4,\alpha_3,\alpha_2\}$ where the $\alpha$'s as well as the corresponding $Q$'s are ordered in a conventional way for $E_6$ simple roots.
\begin{figure}[htbp]
	\centering
	\includegraphics[scale=0.5]{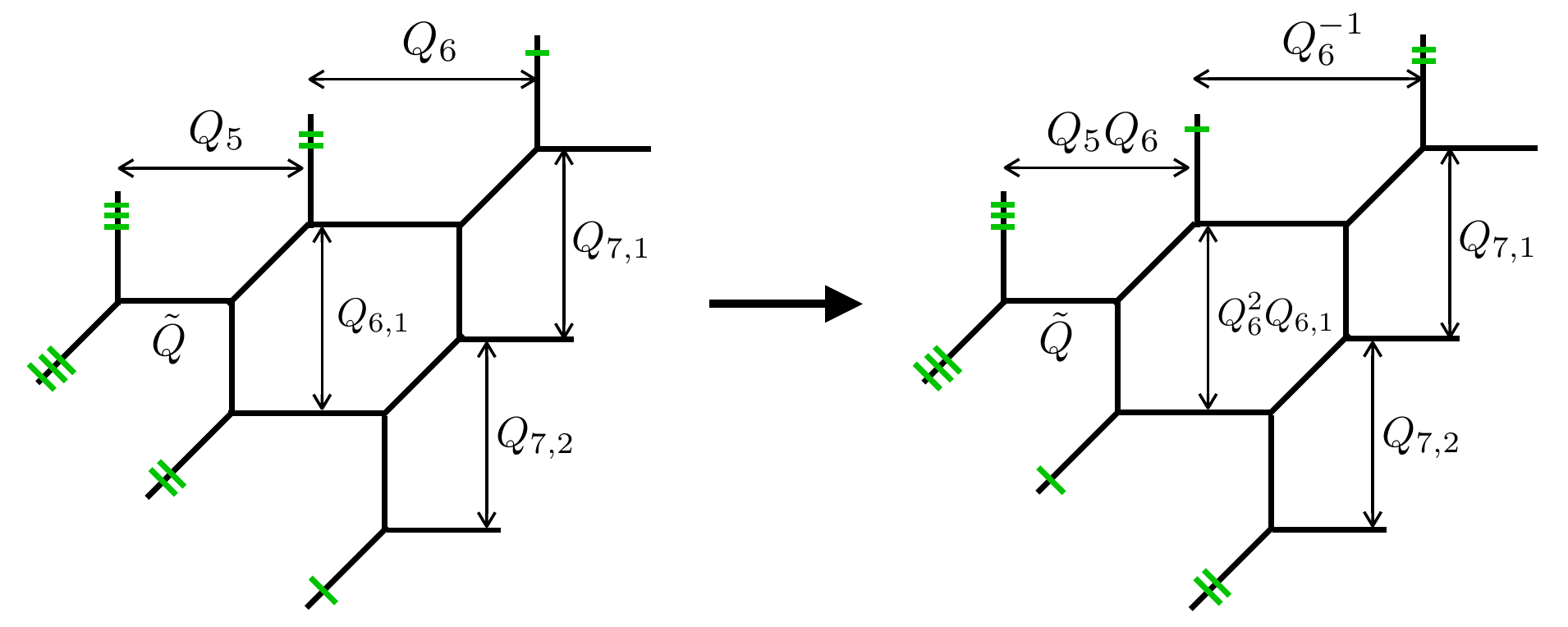}
	\caption{
 Transformation of node 6. The nodes 2 and 4 also transform similarly. }
	\label{fig:E6Q6trans}
\end{figure}
\begin{figure}[htbp]
	\centering
	\includegraphics[scale=0.5]{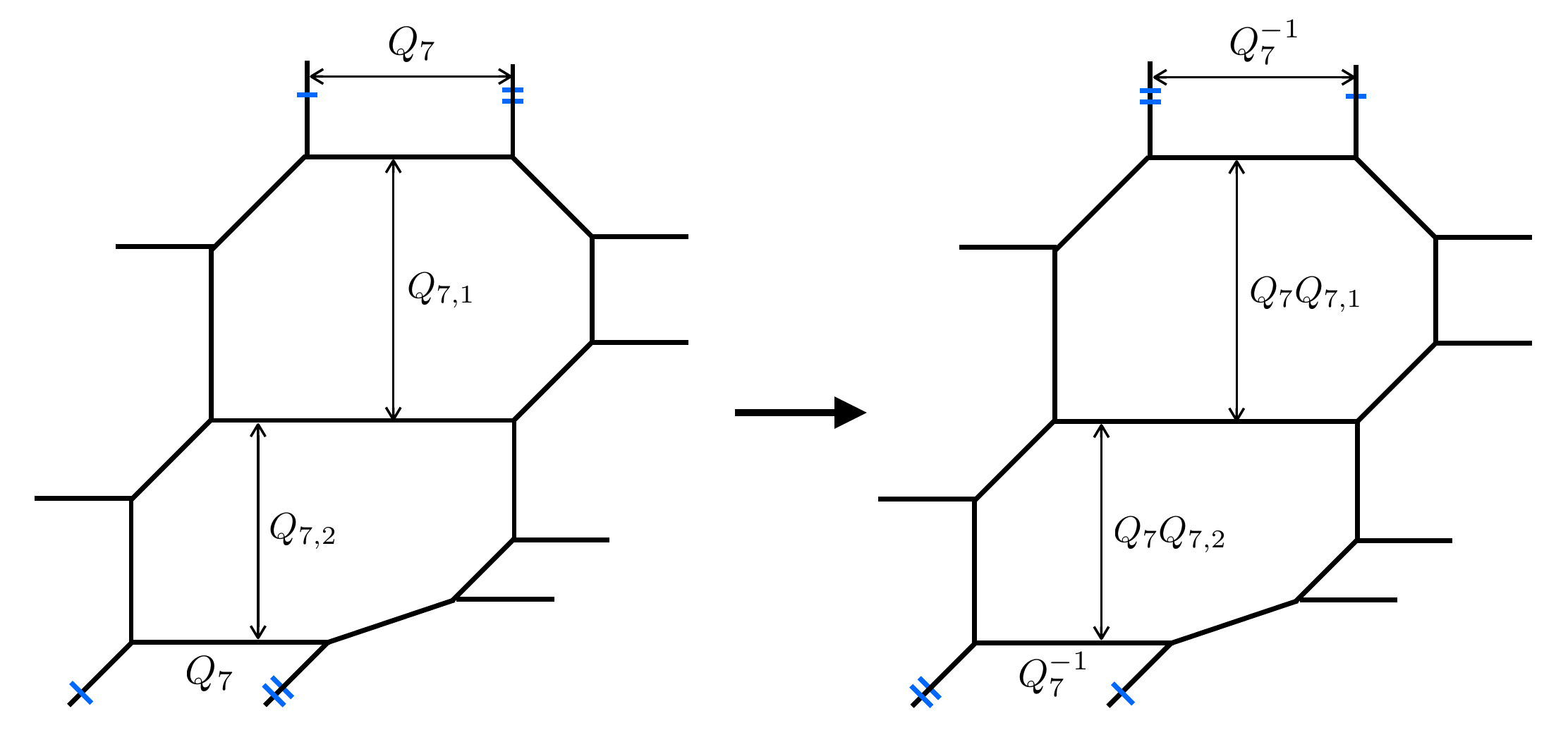}
	\caption{Transformation of node 7.}
	\label{fig:E6Q7trans}
\end{figure}

For $E_6$-type LST, we have five Coulomb branch parameters $Q_{7,1},Q_{7,2},Q_{2,1},Q_{4,1}$ and $Q_{6,1}$ as shown in figure \ref{fig:E6tri}, and these Coulomb branch parameters also change under the $E_6$ Weyl reflections as shown in figure \ref{fig:E6Q6trans} and figure \ref{fig:E6Q7trans}. For convenience in the later computation, we transform them into the following new Coulomb branch parameters $v_1,v_2,w_2,w_4,w_6,$
\begin{align}
	Q_{7,1}=v_1^3,\quad Q_{7,2}=v_2^3,\quad Q_{2,1}=w_2^2,\quad Q_{4,1}=w_4^2,\quad Q_{6,1}=w_6^2.
\end{align}
The $E_6$ Weyl reflections including the Coulomb branch parameters now take the following forms:
\begin{equation}\label{eq:E6WeylwithCB}
	\left\{\begin{array}{l}
	Q_5\rightarrow Q_5^{-1},\ Q_6\rightarrow Q_6Q_5;\\
 	Q_6\rightarrow Q_6^{-1},\ Q_5\rightarrow Q_5Q_6,\ Q_7\rightarrow Q_7Q_6,\ w_6\rightarrow Q_6w_6;\\
  	Q_7\rightarrow Q_7^{-1},\ Q_2\rightarrow	 Q_2Q_7,\ Q_4\rightarrow Q_4Q_7,\ Q_6\rightarrow Q_6Q_7,\ v_1\rightarrow Q_7^{\frac13}v_1,\ v_2\rightarrow Q_7^{\frac13}v_2;\\
   	Q_4\rightarrow Q_4^{-1},\ Q_3\rightarrow Q_3Q_4,\ Q_7\rightarrow Q_7Q_4,\ w_4\rightarrow Q_4 w_4;\\
	Q_3\rightarrow Q_3^{-1},\ Q_4\rightarrow Q_4Q_3;\\
	Q_2\rightarrow Q_2^{-1},\ Q_7\rightarrow Q_7Q_2,\ w_2\rightarrow Q_2 w_2 .
	\end{array}\right.
\end{equation}
The invariant Coulomb branch parameters should take the general form of
\begin{align}
	Q_2^aQ_3^bQ_4^cQ_5^dQ_6^eQ_7^fv_1^gv_2^hw_2^iw_4^jw_6^k\ .
\end{align}
By requiring this general form to be invariant under all the Weyl reflections in \eqref{eq:E6WeylwithCB}, we obtain six linear equations about $a,b,c,d,e,f,g,h,i,j,k$. There are five independent invariant Coulomb branch parameters which are proportional to $v_1$(i.e. $g=1,h=i=j=k=0$), $v_2$(i.e. $h=1,g=i=j=k=0$), $w_2$(i.e. $i=1,g=h=j=k=0$), $w_4$(i.e. $j=1,g=h=i=k=0$) and $w_6$(i.e. $k=1,g=h=i=j=0$) respectively.
For each case, we can solve the six linear equations and obtain one solution. 
Totally, we get the following five invariant Coulomb branch parameters,
\begin{align}\label{eq:E6invCB}
	&A_1=Q_2Q_3^{\frac23}Q_4^{\frac43}Q_5^{\frac23}Q_6^{\frac43}Q_7^{2}v_1,\ A_2=Q_2Q_3^{\frac23}Q_4^{\frac43}Q_5^{\frac23}Q_6^{\frac43}Q_7^2v_2,\ A_3=Q_2^2Q_3Q_4^2Q_5Q_6^2Q_7^3w_2,\nn\\
	&A_4=Q_2^2Q_3^{\frac53}Q_4^{\frac{10}{3}}Q_5^{\frac43}Q_6^{\frac83}Q_7^4w_4,\ A_5=Q_2^2Q_3^{\frac43}Q_4^{\frac83}Q_5^{\frac53}Q_6^{\frac{10}{3}}Q_7^4w_6\ .
\end{align}

If we switch from $E_6$ $\alpha$-basis fugacities $Q$'s to the parameters $B$'s which corresponds to the embedding $E_6\supset \text{ SO}(10)\times \text{U}(1)$ 
\begin{align}\label{eq:E6embed}
Q_5=\frac{B_1}{B_2},~ Q_6=\frac{B_2}{B_3}, ~
    Q_7=\frac{B_3}{B_4}, ~
    Q_4=B_4 B_5, ~
    Q_3=\frac{\sqrt{B_6}}{\sqrt{B_1 B_2 B_3 B_4 B_5}}, ~ 
    Q_2=\frac{B_4}{B_5}, 
 \end{align} 
the $E_6$ Weyl reflections in \eqref{eq:uaffE6WeylinQ} become as follows:
\begin{equation}\label{eq:E6WeylinB}
	\left\{\begin{array}{l}
	B_1\rightarrow B_2,\ B_2\rightarrow B_1;\\
	B_2\rightarrow B_3,\ B_3\rightarrow B_2;\\
	B_3\rightarrow B_4,\ B_4\rightarrow B_3;\\
 	B_4\rightarrow B_5^{-1},\ B_5\rightarrow B_4^{-1};\\
  	B_i\rightarrow \frac{B_iB_6^{\frac14}}{(B_1B_2B_3B_4B_5)^{\frac14}}\ \text{for}\ i=1,2,3,4,5,\ B_6\rightarrow (B_1B_2B_3B_4B_5)^{\frac34}B_6^{\frac14};\\
	B_4\rightarrow B_5,\ B_5\rightarrow B_4.
	\end{array}\right.
\end{equation}

Like the previous subsections, it is most convenient to keep using the old parameters $Q_1,Q_2,Q_3,Q_4,Q_5,Q_6,Q_7,v_1,v_2,w_2,w_4,w_6,Q_{\tau}$ when computing the partition function of $E_6$-type LST in the intermediate process and in particular pick $Q_7$ as a main expansion parameter, and switch to the new set of parameters $Q_2,Q_3,Q_4,Q_5,Q_6,$ $Q_7,A_1,A_2,A_3,A_4,A_5,u,Q_{\tau}$ in the final step in order to see the $E_6$ characters manifestly.

We separate the full partition function $Z^{E_6}$ into $Z^{E_6}_{Q_7^{\bf 0}}$ part which is the $Q_7^0$ order term in the full partition function and $Z^{E_6}_{Q_7}$ part which is the remaining part that depend on $Q_7$ after extracting out $Z^{E_6}_{Q_7^{\bf 0}}$,
\begin{align}
	Z^{E_6}=Z^{E_6}_{Q_7^{\bf 0}}\cdot Z^{E_6}_{Q_7},
\end{align}
where by \eqref{eq:E6full} and \eqref{eq:E6inst}
\begin{align}\label{eq:ZofE6Q70}
	Z^{E_6}_{Q_7^{\bf 0}}=Z^{E_6}_{\text{pert}}&\sum_{\mu_{1,1}}\bar{Z}_{\oo\oo\mu_{1,1}}(w_2,Q_1)\sum_{\mu_{2,1},\mu_{2,2}}\bar{Z}^{\text{SU(2)}}_{\oo\oo\oo\mu_{2,1}\mu_{2,2}}(v_1,v_2,w_2,Q_2)\calZ_{\mu_{2,1}\mu_{2,2}}(w_2,Q_1)\nn\\
	\times & \sum_{\mu_{3,1}}\bar{Z}_{\oo\oo\mu_{3,1}}(w_4,Q_3)\sum_{\mu_{4,1},\mu_{4,2}}\bar{Z}^{\text{SU(2)}}_{\oo\oo\oo\mu_{4,1}\mu_{4,2}}(v_1,v_2,w_4,Q_4)\calZ_{\mu_{4,1}\mu_{4,2}}(w_4,Q_3)\nn\\
	\times & \sum_{\mu_{5,1}}\bar{Z}_{\oo\oo\mu_{5,1}}(w_6,Q_5)\sum_{\mu_{6,1},\mu_{6,2}}\bar{Z}^{\text{SU(2)}}_{\oo\oo\oo\mu_{6,1}\mu_{6,2}}(v_1,v_2,w_6,Q_6)\calZ_{\mu_{6,1}\mu_{6,2}}(w_6,Q_5),
\end{align}
\begin{align}\label{eq:ZE6Q7}
	Z^{E_6}_{Q_7}=&\sum_{\mu_{7,1},\mu_{7,2},\mu_{7,3}}\bar{Z}^{\text{SU(3)}}_{\mu_{7,1}\mu_{7,2}\mu_{7,3}}(v_1,v_2,Q_7)\nn\\
	&\times\frac{\sum_{\mu_{2,1},\mu_{2,2}}\bar{Z}^{\text{SU(2)}}_{\mu_{7,1}\mu_{7,2}\mu_{7,3}\mu_{2,1}\mu_{2,2}}(v_1,v_2,w_2,Q_2)\calZ_{\mu_{2,1}\mu_{2,2}}(w_2,Q_1)}{\sum_{\mu_{2,1},\mu_{2,2}}\bar{Z}^{\text{SU(2)}}_{\oo\oo\oo\mu_{2,1}\mu_{2,2}}(v_1,v_2,w_2,Q_2)\calZ_{\mu_{2,1}\mu_{2,2}}(w_2,Q_1)}\nn\\
	&\times\frac{\sum_{\mu_{4,1},\mu_{4,2}}\bar{Z}^{\text{SU(2)}}_{\mu_{7,1}\mu_{7,2}\mu_{7,3}\mu_{4,1}\mu_{4,2}}(v_1,v_2,w_4,Q_4)\calZ_{\mu_{4,1}\mu_{4,2}}(w_4,Q_3)}{\sum_{\mu_{4,1},\mu_{4,2}}\bar{Z}^{\text{SU(2)}}_{\oo\oo\oo\mu_{4,1}\mu_{4,2}}(v_1,v_2,w_4,Q_4)\calZ_{\mu_{4,1}\mu_{4,2}}(w_4,Q_3)}\nn\\
	&\times\frac{\sum_{\mu_{6,1},\mu_{6,2}}\bar{Z}^{\text{SU(2)}}_{\mu_{7,1}\mu_{7,2}\mu_{7,3}\mu_{6,1}\mu_{6,2}}(v_1,v_2,w_6,Q_6)\calZ_{\mu_{6,1}\mu_{6,2}}(w_6,Q_5)}{\sum_{\mu_{6,1},\mu_{6,2}}\bar{Z}^{\text{SU(2)}}_{\oo\oo\oo\mu_{6,1}\mu_{6,2}}(v_1,v_2,w_6,Q_6)\calZ_{\mu_{6,1}\mu_{6,2}}(w_6,Q_5)}.
\end{align}
with
\begin{align}
	&\bar{Z}^{\text{SU(3)}}_{\mu_{7,1}\mu_{7,2}\mu_{7,3}}(v_1,v_2,Q_7)\equiv\frac{Q_{7}^{|\mu_{7,1}|+|\mu_{7,2}|+|\mu_{7,3}|}}{\prod_{c=1}^3\prod_{e=1}^3\varTheta_{\mu_{7,c}\mu_{7,e}}(\frac{y_{7,c}}{y_{7,e}}|Q_{\tau})\varTheta_{\mu_{7,c}\mu_{7,e}}(\frac{y_{7,c}}{y_{7,e}}\tfrac{\ft}{\fq}|Q_{\tau})},\\
	&\bar{Z}^{\text{SU(2)}}_{\mu_{7,1}\mu_{7,2}\mu_{7,3}\mu_{2,1}\mu_{2,2}}(v_1,v_2,w_2,Q_2)\nn\\
	 &~~\equiv Q_{2}^{|\mu_{2,1}|+|\mu_{2,2}|}\prod_{p=1}^2\bigg(\frac{\prod_{r=1}^3\varTheta_{\mu_{2,p}\mu_{7,r}}(\frac{y_{2,p}}{y_{7,r}}\sqrt{\tfrac{\ft}{\fq}}|Q_{\tau})\varTheta_{\mu_{7,r}\mu_{2,p}}(\frac{y_{7,r}}{y_{2,p}}\sqrt{\tfrac{\ft}{\fq}}|Q_{\tau})}{\prod_{s=1}^2\varTheta_{\mu_{2,p}\mu_{2,s}}(\frac{y_{2,p}}{y_{2,s}}|Q_{\tau})\varTheta_{\mu_{2,p}\mu_{2,s}}(\frac{y_{2,p}}{y_{2,s}}\tfrac{\ft}{\fq}|Q_{\tau})}\bigg).\label{eq:E6ZSU2def}
\end{align}
By defining
\begin{align}
	&\calZ_{\mu_{7,1}\mu_{7,2}\mu_{7,3}}(v_1,v_2,w_2,Q_2,Q_1)\nn\\
	&\,\equiv \frac{\sum_{\mu_{2,1},\mu_{2,2}}\bar{Z}^{\text{SU(2)}}_{\mu_{7,1}\mu_{7,2}\mu_{7,3}\mu_{2,1}\mu_{2,2}}(v_1,v_2,w_2,Q_2)\calZ_{\mu_{2,1}\mu_{2,2}}(w_2,Q_1)}{\sum_{\mu_{2,1},\mu_{2,2}}\bar{Z}^{\text{SU(2)}}_{\oo\oo\oo\mu_{2,1}\mu_{2,2}}(v_1,v_2,w_2,Q_2)\calZ_{\mu_{2,1}\mu_{2,2}}(w_2,Q_1)},
\end{align}
we can rewrite \eqref{eq:ZE6Q7} as
\begin{align}\label{eq:ZofE6Q7new}
	Z^{E_6}_{Q_7}=&\sum_{\mu_{7,1},\mu_{7,2},\mu_{7,3}}\bar{Z}^{\text{SU(3)}}_{\mu_{7,1}\mu_{7,2}\mu_{7,3}}(v_1,v_2,Q_7)\calZ_{\mu_{7,1}\mu_{7,2}\mu_{7,3}}(v_1,v_2,w_2,Q_2,Q_1)\nn\\
	&\times \calZ_{\mu_{7,1}\mu_{7,2}\mu_{7,3}}(v_1,v_2,w_4,Q_4,Q_3) \calZ_{\mu_{7,1}\mu_{7,2}\mu_{7,3}}(v_1,v_2,w_6,Q_6,Q_5).
\end{align}
Like the $\calZ_{\mu_{5,1}\mu_{5,2}}$ defined in \eqref{eq:defZQ1}, we can also compute $\calZ_{\mu_{7,1}\mu_{7,2}\mu_{7,3}}$ by expansion with respect to $Q_2$. Then we can substitute the expansions of $\calZ$'s into \eqref{eq:ZofE6Q7new} to obtain $Z^{E_6}_{Q_7}$. 

\paragraph{Evaluation of the partition function by Weyl group symmetry.}
Among all the terms of a character of $E_6$ in an arbitrary representation in the $\alpha$-basis, the terms $Q_7^0Q_2^0Q_3^i Q_4^j Q_5^k Q_6^\ell$ with $i,j,k,\ell\leq 0$ include at least one element of each Weyl orbit in the character, so these terms are enough to determine all the terms of the character by Weyl group symmetry. We can utilize this to determine the coefficients of the expansion of the partition function.  
In the computation of $Z^{E_6}_{Q_7^{\bf 0}}$ and $Z^{E_6}_{Q_7}$, we need to sum over enough Young diagrams in order to go up to higher orders of $Q_2,Q_4,Q_6,Q_7$. As we go up to higher orders, it becomes more time consuming. By combining a lower order computation with the Weyl group symmetry analysis, we find that up to $Q_2^2Q_4^2Q_6^2Q_7^3$ order, the coefficients of the following terms inside PE of $Z^{E_6}$ can be determined by Weyl group symmetry
\begin{align}\label{eq:E6terms}
    1,\ \frac{A_3^2}{A_1A_2^2},\ \frac{A_3^2}{A_1^2A_2},\ \frac{A_4^3}{A_1^3A_2^3},\ \frac{A_5^3}{A_1^3A_2^3},\ A_3,\ \frac{A_1A_3}{A_2},\ \frac{A_2A_3}{A_1},\ u, 
\end{align}
where $1$ means the zeroth order of all the expansion parameters. Note that there might be more terms whose coefficients can also be determined by Weyl group symmetry when the expansion is taken up to $Q_2^2Q_4^2Q_6^2Q_7^3$ order, we did not try to exhaust all such possible terms.  In the following computation of $Z^{E_6}_{Q_7^{\bf 0}}$ and $Z^{E_6}_{Q_7}$, we will only consider these terms in \eqref{eq:E6terms} which are all at $Q_{\tau}^0$ order, and all the other terms in the PE will be ignored and just put into ``$\cdots$". Switching to $v_1,v_2,w_2,w_4,w_6,Q_1$, these  terms are
\begin{align}\label{eq:E6termsold}
    1,\ \frac{w_2^2}{v_1v_2^2},\ \frac{w_2^2}{v_1^2v_2},\ \frac{w_4^3}{v_1^3v_2^3},\ \frac{w_6^3}{v_1^3v_2^3},\ w_2,\ \frac{v_1w_2}{v_2},\ \frac{v_2w_2}{v_1},\ Q_1.
\end{align}

\paragraph{$Z^{E_6}_{Q_7^{\bf 0}}$ part.}
We first compute the summation over $\mu_{2,1},\mu_{2,2}$ in \eqref{eq:ZofE6Q70} by the definition in \eqref{eq:E6ZSU2def} and \eqref{eq:defZQ1}. In order to obtain the exact form, we have to sum over $\mu_{2,1},\mu_{2,2}$ up to infinity, but if we further expand with respect to $v_1,v_2,w_2$ up to low  orders, its expansion coefficients turn out to be finite polynomials of $Q_2$ which only depend on a few Young diagrams, and the result is given as follows:
\begin{align}\label{eq:ZbarSU21}
	&\sum_{\mu_{2,1},\mu_{2,2}}\bar{Z}^{\text{SU(2)}}_{\oo\oo\oo\mu_{2,1}\mu_{2,2}}(v_1,v_2,w_2,Q_2)\calZ_{\mu_{2,1}\mu_{2,2}}(w_2,Q_1)\nn\\
	&=~\text{PE}\bigg[\frac{(\fq+\ft)Q_2}{(1-\fq)(1-\ft)}+\frac{(\fq+\ft)Q_2}{(1-\fq)(1-\ft)}\frac{w_2^2}{v_1v_2^2}+\frac{(\fq+\ft)Q_2}{(1-\fq)(1-\ft)}\frac{w_2^2}{v_1^2v_2}\nn\\
 &~~-\frac{2\sqrt{\fq\ft}Q_2}{(1-\fq)(1-\ft)} w_2-\frac{\sqrt{\fq\ft}Q_2}{(1-\fq)(1-\ft)}\frac{v_1w_2}{v_2}-\frac{\sqrt{\fq\ft}Q_2}{(1-\fq)(1-\ft)}\frac{v_2w_2}{v_1}\nn\\
 &~~+\frac{(\fq+\ft)Q_2}{(1-\fq)(1-\ft)}Q_1+\cdots\bigg].
\end{align}
In this way, we also compute the summation over $\mu_{4,1},\mu_{4,2}$ and $\mu_{6,1},\mu_{6,2}$ in \eqref{eq:ZofE6Q70} and obtain,
\begin{align}\label{eq:ZbarSU22}
	&\sum_{\mu_{4,1},\mu_{4,2}}\bar{Z}^{\text{SU(2)}}_{\oo\oo\oo\mu_{4,1}\mu_{4,2}}(v_1,v_2,w_4,Q_4)\calZ_{\mu_{4,1}\mu_{4,2}}(w_4,Q_3)\nn\\
	&=~\text{PE}\bigg[\frac{(\fq+\ft)(1+Q_3)Q_4}{(1-\fq)(1-\ft)}-\frac{(\fq^2+\fq\ft+\ft^2)Q_3Q_4^2}{(1-\fq)(1-\ft)\sqrt{\fq\ft}}\frac{w_4^3}{v_1^3v_2^3}+\cdots\bigg],
\end{align}
\begin{align}\label{eq:ZbarSU23}
	&\sum_{\mu_{6,1},\mu_{6,2}}\bar{Z}^{\text{SU(2)}}_{\oo\oo\oo\mu_{6,1}\mu_{6,2}}(v_1,v_2,w_6,Q_6)\calZ_{\mu_{6,1}\mu_{6,2}}(w_6,Q_5)\nn\\
	&=~\text{PE}\bigg[\frac{(\fq+\ft)(1+Q_5)Q_6}{(1-\fq)(1-\ft)}-\frac{(\fq^2+\fq\ft+\ft^2)Q_5Q_6^2}{(1-\fq)(1-\ft)\sqrt{\fq\ft}}\frac{w_6^3}{v_1^3v_2^3}+\cdots\bigg].
\end{align}
Substituting \eqref{eq:ZbarSU21}, \eqref{eq:ZbarSU22}, \eqref{eq:ZbarSU23}, \eqref{eq:zsubu1rev}, and \eqref{eq:unifiedpert} into \eqref{eq:ZofE6Q70}, we get
\begin{align}\label{eq:ZE6Q7zero}
	Z^{E_6}_{Q_7^{\bf 0}}=&~\text{PE}\bigg[\frac{\fq+\ft}{(1-\fq)(1-\ft)}\Big(6+Q_2+Q_3+Q_4+Q_5+Q_6+Q_3Q_4+Q_5Q_6\Big)\nn\\&+\frac{\fq+\ft}{(1-\fq)(1-\ft)}Q_2\frac{w_2^2}{v_1v_2^2}+\frac{\fq+\ft}{(1-\fq)(1-\ft)}Q_2\frac{w_2^2}{v_1^2v_2}-\frac{(\fq^2+\fq\ft+\ft^2)Q_3Q_4^2}{(1-\fq)(1-\ft)\sqrt{\fq\ft}}\frac{w_4^3}{v_1^3v_2^3}\nn\\&-\frac{(\fq^2+\fq\ft+\ft^2)Q_5Q_6^2}{(1-\fq)(1-\ft)\sqrt{\fq\ft}}\frac{w_6^3}{v_1^3v_2^3}-\frac{2\sqrt{\fq\ft}(1+Q_2)}{(1-\fq)(1-\ft)} w_2-\frac{\sqrt{\fq\ft}(1+Q_2)}{(1-\fq)(1-\ft)}\frac{v_1w_2}{v_2}\nn\\
 &-\frac{\sqrt{\fq\ft}(1+Q_2)}{(1-\fq)(1-\ft)}\frac{v_2w_2}{v_1}+\frac{(\fq+\ft)(1+Q_2)}{(1-\fq)(1-\ft)}Q_1+\cdots\bigg],
\end{align}
where terms containing $w_2^{-1}$ are flopped.

\paragraph{$Z^{E_6}_{Q_7}$ part and rescaling.}
To compute $Z^{E_6}_{Q_7}$ in \eqref{eq:ZofE6Q7new}, we expand with respect to $Q_7,Q_2,Q_4,Q_6$ whose orders are restricted by Young diagram box numbers and with respect to $Q_1$ which is proportional to $u$.  
Like in the previous cases, in order to reduce the heavy computation we need to further expand $Z^{E_6}_{Q_7}$ with respect to 
 $v_1,v_2,w_2,w_4,w_6$ up to finite order. And we need to rescale $\bar{Z}^{\text{SU(3)}}$, $\calZ$ functions and parameters in \eqref{eq:ZofE6Q7new}  to make the expansions of $\bar{Z}^{\text{SU(3)}}$, $\calZ$ with respect to $v_1,v_2,w_2,w_4,w_6$ have all non-negative orders which can be fulfilled by the following rescalings up to $Q_{\tau}^0$ order:
\begin{align}
	&\calZ_{\mu_{7,1}\mu_{7,2}\mu_{7,3}}(w_2)\rightarrow (v_1^2v_2^2w_2)^{|\mu_{7,1}|+|\mu_{7,2}|+|\mu_{7,3}|}\calZ_{\mu_{7,1}\mu_{7,2}\mu_{7,3}}(w_2),\nn\\
	&\calZ_{\mu_{7,1}\mu_{7,2}\mu_{7,3}}(w_4)\rightarrow (v_1^2v_2^2w_4)^{|\mu_{7,1}|+|\mu_{7,2}|+|\mu_{7,3}|}\calZ_{\mu_{7,1}\mu_{7,2}\mu_{7,3}}(w_4),\nn\\
	&\calZ_{\mu_{7,1}\mu_{7,2}\mu_{7,3}}(w_6)\rightarrow (v_1^2v_2^2w_6)^{|\mu_{7,1}|+|\mu_{7,2}|+|\mu_{7,3}|}\calZ_{\mu_{7,1}\mu_{7,2}\mu_{7,3}}(w_6),\nn\\
	&\bar{Z}^{\text{SU(3)}}_{\mu_{7,1}\mu_{7,2}\mu_{7,3}}\rightarrow (v_1^6 v_2^6 w_2w_4w_6)^{-(|\mu_{7,1}|+|\mu_{7,2}|+|\mu_{7,3}|)}\bar{Z}^{\text{SU(3)}}_{\mu_{7,1}\mu_{7,2}\mu_{7,3}}, \nn\\
	&Q_2\rightarrow v_1v_2Q_2,\quad Q_4\rightarrow v_1v_2Q_4,\quad Q_6\rightarrow v_1v_2Q_6,\quad Q_7\rightarrow v_1^6 v_2^6 w_2w_4w_6 Q_7\ .
\end{align}
After obtaining the PE form of $Z^{E_6}_{Q_7}$ expanded with $v_1,v_2,w_2,w_4,w_6$ under the rescalings, we switch back to the original parameters in the PE form.

\paragraph{Final result of $Z^{E_6}$.}
We compute the terms \eqref{eq:E6termsold} in the PE of $Z^{E_6}_{Q_7}$ and then combine it with \eqref{eq:ZE6Q7zero} to obtain the full partition function $Z^{E_6}$. 
We express $Z^{E_6}$ as PE form with the expansion parameters being $A_1,A_2,A_3,A_4,A_5,u,Q_{\tau},$
\begin{align}
    Z^{E_6}=Z^{E_6}_{Q_7^{\bf 0}}\cdot Z^{E_6}_{Q_7}=\text{PE}\left[\sum_{i,j,k,\ell,m,n,r}b_{i,j,k,\ell,m,n,r}A_1^i A_2^j A_3^k A_4^\ell A_5^m u^n Q_{\tau}^r\right],
\end{align}
and list the coefficients of the terms \eqref{eq:E6terms} which we could determine from the expansion of $Q_2,Q_4,Q_6,Q_7$ up to $Q_2^2Q_4^2Q_6^2Q_7^3$ order,
\begin{align}
    &b_{0,0,0,0,0,0,0}=\frac12(6+\chi_{\bf 6})\bigg[0,\frac12\bigg],\nn\\
    &b_{-1,-2,2,0,0,0,0}=b_{-2,-1,2,0,0,0,0}=\bigg[0,\frac12\bigg],\nn\\
    &b_{-3,-3,0,3,0,0,0}=b_{-3,-3,0,0,3,0,0}=[0,1],\nn\\
    &b_{0,0,1,0,0,0,0}=(6+\textcolor{red}{2\chi_{\bf 6}})[0,0]+2[0,1]+3\bigg[\frac12,\frac12\bigg],\nn\\
    &b_{1,-1,1,0,0,0,0}=b_{-1,1,1,0,0,0,0}=(2+\textcolor{red}{\chi_{\bf 6}})[0,0]+\bigg[\frac12,\frac12\bigg],\nn\\
    &b_{0,0,0,0,0,1,0}=(2+\textcolor{red}{\chi_{\bf 6}})\bigg[0,\frac12\bigg]+4\bigg[\frac12,0\bigg],
\end{align}
where $\chi_{\bf 6}$ is for short the character $\chi_{\bf 6}^{E_6}$ of the adjoint representation of $E_6$ in \eqref{eq:E6fundamentals}. Note that for $b_{0,0,1,0,0,0,0}$, $b_{1,-1,1,0,0,0,0}$, $b_{-1,1,1,0,0,0,0}$ and $b_{0,0,0,0,0,1,0}$ we denote the characters in red which means the characters are obtained by Weyl group symmetry but not yet confirmed by direct computation of higher orders.

\subsection{\texorpdfstring{$E_7$}{E7} and \texorpdfstring{$E_8$}{E8}-type LSTs}
Due to the quiver structures of the brane webs of $E_7$ and $E_8$ LSTs, it is also easy to figure out the affine Weyl group symmetry in the brane webs which can confirm that the partition functions of these theories also have affine $E_7$ and affine $E_8$ symmetries. 
We can also follow the similar procedure of the $E_6$-type LST to reduce affine $E_7$ and affine $E_8$ symmetries to $E_7$ and $E_8$ symmetries, and determine the corresponding invariant Coulomb branch parameters,
then use the general formulas \eqref{eq:unifiedpert} and \eqref{eq:unifiedinst} to compute the partition functions and do expansions to see the $E_7$ and $E_8$ characters. 
But as we have experienced time-consuming in the computation of $E_6$-type LST, we expect that it will be more time-consuming in the computation of $E_7$ and $E_8$ LSTs, so we do not compute $E_7$ and $E_8$ cases, but the computation is essentially straightforward. 

\bigskip


\section{Conclusion}\label{sec:conclusion}
In this paper, we proposed a novel way of computing the supersymmetric partition functions for 6d $\mathcal{N}=(1,0)$ little string theories (LSTs) engineered from type IIB NS5-branes probing $D_N$ or $E_N$ singularity. The effective gauge theory description of these LSTs on tensor branch is affine $D_N$ or $E_N$ quiver gauge theories and we propose 5-brane configurations realizing these quiver gauge theories based on tri-/quadri-valent gluings of 5-brane webs \cite{Hayashi:2017jze,Hayashi:2021pcj}, by identifying  external 5-branes which gives rise to a compact direction. This can be summarized as follows. We first utilize 5-brane configurations of the tri-/quadri-valent gluings so that the resulting configuration describes 6d theories on a circle, and then we further introduce the bivalent self-gluing of the external 5-branes in such a way that all external 5-branes are glued which makes another compact direction.

5-brane configurations are known for little string theories of $A$-type or those engineered from type IIA NS5-brane probing $D$-type singularities. These 5-brane webs consist of a circular quiver. Our brane construction is, however, different from such conventional configurations, as we enclose the external 5-branes of Type IIB configuration that arise from trivalent (or quadrivalent) gluings for 6d $(D_N,D_N)$ and $(E_N,E_N)$ conformal matters. Our construction is therefore a natural generalization making those little string theories whose conformal theory limit gives rise to a $D$- and $E$-type conformal matter. With these glued 5-brane webs, we computed the supersymmetric partition functions of these LSTs, using topological vertex, and we obtained compact formulas for the partition function of a general $D$- or $E$-type LST with arbitrary rank. And the general formulas also work for $A$-type LST upon slight modifications.

For $D$-type LSTs, it is well-known that one can realize the LSTs using type IIA brane configurations with two ON$^0$-planes as they give rise to affine $D_N$-type quiver theories \cite{Hanany:1999sj}. Upon T-duality of this IIA configuration, one can find a type IIB description for this $D_N$-type LSTs with 5-brane webs with two ON-planes where NS5-branes are compactified along the T-dual circle. We showed that such 5-brane description with two ON-planes can be consistently mapped to a trivalent  or quadrivalent gluing of 5-brane webs and hence they are equivalent. Unlike $D$-type LSTs, there are no known brane descriptions for $E$-type LSTs. Our proposal  gives  5-brane configurations that can describe $E$-type LSTs.

We then tried to expand the partition functions of the LSTs to see their symmetries. We discovered that the symmetries of the affine quiver structures of these LSTs are hidden in various flops of the brane webs. These flops are just exchanges of external NS-charged branes which keep the brane configurations invariant, and each such flop correspond to a Weyl reflection of the symmetry. Thus we conclude that the $D$- and $E$-type LSTs have corresponding affine $D$ and $E$ symmetries. These affine symmetries are demanding to see in the expansion because their characters are infinite polynomials that contain infinitely high orders which means the Young diagram sums need to be taken up to infinity, so we reduce the affine $D$ and $E$ symmetries to $D$ and $E$ symmetries and choose appropriate parameters under these reduced symmetries. We successfully obtained the expansions of the partition functions of $D_4,D_5,E_6$-type LSTs up to low orders and saw the $D_4,D_5,E_6$ characters in the coefficients.

Our computation for the expansions of partition functions is for rank-1 LSTs engineered from one NS5-brane probing $D,E$-type singularities.  The partition functions for rank-$k$ LSTs engineered from $k$ NS5-branes probing $D,E$-type singularities can be generalized in a straightforward manner in our setup, though the computation itself would be demanding.

There are various directions to pursue. One of them would be the T-duality of $E$-type LSTs. The T-duality of LSTs can be realized as an S-duality in the corresponding type IIB 5-brane configurations. As discussed, $D$-type LSTs, engineered from IIB NS5-brane probing $D$-type singularity, have a 5-brane description with two ON-planes and then taking the S-duality leads another 5-brane configuration with two O5-planes which give rise to a circular quiver. This corresponds to IIA NS5-brane probing the same $D$-type singularity. For $E$-type LSTs, in particular, no 5-brane webs are known. Checking T-duality for $E$-type LSTs would be very challenging. As we proposed a new 5-brane configuration for $D, E$-type LSTs as a tri-/quadri-valent gluing, it would be interesting to study the T-duality within the framework of glued branes. For instance, the corresponding brane configurations for this S-duality would be interesting. Also, once we have them, we can take the decompactification limit which yields a 6d SCFT, which would give a new 5-brane realization of the corresponding SCFT.

\bigskip

\acknowledgments
We thank Joonho Kim, Hee-Cheol Kim, and Kimyeong Lee for useful discussions.  We acknowledge the hospitality at APCTP, Fudan university, USTC, YMSC Tsinghua university where part of this work was done.  SK is supported by the NSFC grant No. 12250610188 and also partially supported by the Fundamental Research Funds for the Central Universities 2682021ZTPY043. FY is supported by the NSFC grant No. 11950410490, by Start-up research grant YH1199911312101, and in part by Fundamental Research Funds for the Central Universities 2682021ZTPY043. The research of YS is supported by Samsung Science and Technology Foundation under Project Number SSTF-BA2002-05 and by the National Research Foundation of Korea (NRF) grant funded by the Korea government (MSIT) (No. 2018R1D1A1B07042934). XW is partially supported by NSFC grant No. 11950410490.

\bigskip
\appendix
\section{Gluings over nonparallel external 5-branes}\label{app:non-parallel}
\begin{figure}[htbp]
\centering
 \begin{tikzpicture}[thick,baseline=(current bounding box.center)]
	\footnotesize
	\node at (-2.2,1.2) {(a)};
	\node at (-1.4,.8) {$Q_f$}; 	\node at (-1.0,-.1) {$Q$}; \node at (-.0,.1) {$Q_b$};
	\draw (-1,1) -- ++(0, -.7) -- ++(.5, -.5) -- ++(0, -.7); 
	\draw (-1, .3) -- ++(-.7,0); \draw (-.5,-.2) -- ++(1.0,0);
	\draw (1,-.7) -- ++(.7,0);
	\draw (-.9,.8) -- ++(-.2,0); \draw (-.9,.7) -- ++(-.2,0); 
	\draw (-.4,-.7) -- ++(-.2,0); \draw (-.4,-.6) -- ++(-.2,0); 
	\begin{scope}[shift={(1.5,-.5)}]
	\draw (-1,1) -- ++(0, -.7) -- ++(.5, -.5) -- ++(0, -.7);
	\draw (-.9,.8) -- ++(-.2,0); 
	\draw (-.4,-.7) -- ++(-.2,0);
	\end{scope}
	\begin{scope}[shift={(7,0)}]
	\node at (-2.8,1.2) {(b)};
	\node at (-1.8,.8) {$Q_f Q^{-1}$}; 	\node at (-1.0,-.1) {$Q$}; \node at (-.0,.0) {$Q_b$}; \node at (-.7,-.5) {$Q$}; \node at (1.1,.3) {$Q_f Q$};
	\draw (-1,1) -- ++(0, -.7) -- ++(.5, -.5) -- ++(0, -.7) -- ++ (.5,-.5); 
	\draw (-1, .3) -- ++(-.7,0); \draw (-.5,-.2) -- ++(1.0,0); \draw (-.5,-.9) -- ++(-.7,0);
	\draw[shift={(1.5,-.5)}] (-1,1) -- ++(0, -.7) -- ++(.5,-.5);
	\draw (-.9,.8) -- ++(-.2,0); \draw (-.9,.7) -- ++(-.2,0); 
	\draw (-.3,-1.3) -- ++ (.2,.2); \draw (-.23,-1.37) -- ++ (.2,.2);  
	\draw[shift={(1.5,-.5)}]  (-.9,.8) -- ++(-.2,0); 
	\draw (.75,-.65) -- ++ (.2,.2); 
	\end{scope}
	\nonumber
\normalsize
\end{tikzpicture}
\caption{(a) A 5-brane web for 6d $[1]-SU(1)-[1]$. (b) A 5-brane web for 6d $[2]-SU(1)$.}
\label{fig:6dU1-2F-web}
\end{figure}
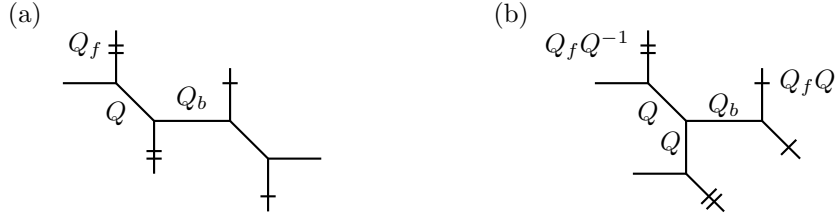

In this appendix, we discuss the validity of non-parallel identification of 5-branes which  appeared in the main text. Type IIB description of 6d SU(1) gauge theory with 2 fundamental hypermultiplets and with a tensor multiplet is depicted in figure \ref{fig:6dU1-2F-web}. In the first figure (a), one hyper is located on the left of the first NS5-brane and the other hyper is located on the right of the second NS5-brane, which we denote by ($[1]-SU(1)-[1]$). These two NS5 branes are identified (or glued) to yield a 5-brane configuration for a 6d theory on a circle, which is, in fact, a well-known (S-dual configuration of) 5-brane setup for rank-1 M-string theory on a circle \cite{Haghighat:2013gba}.

In figure \ref{fig:6dU1-2F-web}(a), the 5-brane configuration is equivalent to  a 5-brane where two external 5-branes are attached to two different D7-branes whose 7-brane monodromy cuts are pointing the opposite direction such that the 7-brane on the left points $(-1,0)$-direction (to the left) where the 7-brane on the right points $(1,0)$-direction (to the right). One then can take a Hanany-Witten move with one of these D7-branes, say we take the 7-brane on the right to the left which gives rise to the 5-brane web given in figure \ref{fig:6dU1-2F-web}(b), where the monodromy cuts for both D7-branes are pointing the $(-1, 0)$-direction. As a result, the vertically identified NS5-branes in figure  \ref{fig:6dU1-2F-web}(a) are now identified in a non-parallel way. This resultant configuration represents 6d $[2]-SU(1)$ on a circle. One can think of this 5-brane configuration as a 5-brane web for 5d KK theory where non-parallel 5-branes of the same NS5-brane charges are identified. In other words, one can start with a 5-brane web for $[2]-SU(1)$ where two 5-branes of non-zero NS5-brane charges are bivalently self-glued. As the Hanany-Witten moves do not affect gauge theories, we claim that the partition function based on the 5-brane web in figure \ref{fig:6dU1-2F-web}(a) is equivalent to that based on figure \ref{fig:6dU1-2F-web}(b).

To check our claim, we compute the partition function for each web diagram and compare two partition functions. The partition function for the left diagram, call it $Z^{(a)}$, is already derived in \cite{Haghighat:2013gba} and given by 
\be\ba
Z^{(a)} =
&\prod_{k=1}^\infty \frac{1}{(1-Q_\tau^k)^2}
\prod_{k=1}^\infty \prod_{i,j=1}^\infty 
\frac{
(1-Q Q_\tau^{k-1} \ft^{i-\f{1}{2}}\fq^{j-\f{1}{2}})^2 (1-Q^{-1} Q_\tau^{k} \ft^{i-\f{1}{2}}\fq^{j-\f{1}{2}})^2
}{
(1-Q_\tau^k \ft^{i-1} \fq^{j})^2 (1-Q_\tau^k \ft^{i} \fq^{j-1})^2}
\\
&\times \sum_{\mu_b} (-Q Q_b)^{|\mu_b|} \prod_{(i,j)\in\mu_b} \frac{
\theta_1(Q^{-1} \ft^{-\mu_{b,i}+j-\f{1}{2}} \fq^{i-\f{1}{2}})\theta_1(Q^{-1} \ft^{\mu_{b,i}-j+\f{1}{2}} \fq^{-i+\f{1}{2}})
}{
\theta_1(\ft^{-\mu_{b,i}+j-1} \fq^{-\mu_{b,j}^t+i}) \theta_1(\ft^{-\mu_{b,i}+j} \fq^{-\mu_{b,j}^t+i-1})
},
\ea\ee
where $Q_\tau = Q_f Q$.
As the partition function based on the right diagram is not known, we compute the partition function of the right diagram and call it $Z^{(b)}$.
To this end, we first divide the diagram (b) into two parts, left and right building blocks:
\be\ba
\centering
\footnotesize
 \begin{tikzpicture}[thick,baseline=(current bounding box.center)]
	\draw (-1,1) -- ++(0, -.7) -- ++(.5, -.5) -- ++(0, -.7) -- ++ (.5,-.5); 
	\draw (-1, .3) -- ++(-.7,0); \draw (-.5,-.2) -- ++(1.0,0); \draw (-.5,-.9) -- ++(-.7,0);
	\draw[shift={(1.5,-.5)}] (-1,1) -- ++(0, -.7) -- ++(.5,-.5);
	\draw (-.9,.8) -- ++(-.2,0); \draw (-.9,.7) -- ++(-.2,0); 
	\draw (-.3,-1.3) -- ++ (.2,.2); \draw (-.23,-1.37) -- ++ (.2,.2);  
	\draw[shift={(1.5,-.5)}]  (-.9,.8) -- ++(-.2,0); 
	\draw (.75,-.65) -- ++ (.2,.2); 
	\node at (2.8,-.2) {${\displaystyle=\quad \sum_{\mu_b} (-Q_b)^{|\mu_b|}}$};
	\begin{scope}[shift={(6,0)}] 
	\node at (-.1,.0) {$\mu_b$}; \node at (.6,.1) {$\mu_b$};
	\draw (-1,1) -- ++(0, -.7) -- ++(.5, -.5) -- ++(0, -.7) -- ++ (.5,-.5); 
	\draw (-1, .3) -- ++(-.7,0); \draw (-.5,-.2) -- ++(.5,0); \draw (-.5,-.9) -- ++(-.7,0);
	\draw[shift={(2.0,-.5)}] (-1,1) -- ++(0, -.7) -- ++(.5,-.5);
	\draw (.5,-.2) -- +(.5,0);
	\draw (-.9,.8) -- ++(-.2,0); \draw (-.9,.7) -- ++(-.2,0); 
	\draw (-.3,-1.3) -- ++ (.2,.2); \draw (-.23,-1.37) -- ++ (.2,.2);  
	\draw[shift={(1.5,-.5)}]  (-.4,.8) -- ++(-.2,0); 
	\draw (1.25,-.65) -- ++ (.2,.2); 
	\end{scope}
\end{tikzpicture}\ .
\normalsize
\label{buildMstring}
\ea\ee
We label the topological vertex computation of the left and the right building blocks respectively as $Z^{{\rm L}}(\mu_b)$ and $Z^{{\rm R}}(\mu_b)$, where the middle edges are associated with Young diagram $\mu_b$, and the partition function is given as a sum of this $\mu_b$,  
\begin{align}
    Z^{(b)} & = \sum_{\mu_b}\, (-Q_b)^{|\mu_b|}\, Z^{\rm L}(\mu_b) Z^{\rm R}(\mu_b) \ , 
\end{align}
where each piece is given by 
\begin{subequations}
\begin{align}
Z^{\rm L}(\mu_b) &= \sum_{\nu_{1,2,3}} (-Q)^{|\nu_1|+|\nu_2|} (-Q_f Q^{-1})^{|\nu_3|} 
\nonumber \\
&\hspace{13mm}\times
\tilde{f}^{-1}_{\nu_3}(\fq,\ft) C_{\nu_3 \nu_1^t \Emptyset}(\ft,\fq) C_{\nu_2^t \nu_1 \mu_b^t}(\fq,\ft) C_{\nu_2 \nu_3^t \Emptyset}(\ft,\fq)\ ,
\\
Z^{\rm R}(\mu_b) &= \sum_{\nu} (-Q_{\tau})^{|\nu|} \tilde{f}_\nu^{-1}(\fq,\ft) C_{\nu \nu^t \mu_b}(\ft,\fq)\ .
\end{align}
\end{subequations}
A little calculation\footnote{See Appendix \ref{app:formulas} for how to compute the partition function of a vertical periodic strip. } leads that two partition functions are equivalent up to an (unphysical) extra factor which does not depend on the Coulomb branch parameters, 
\begin{align}
\frac{Z^{(b)}}{Z^{{\rm extra}}} = Z^{(a)}\ ,    \label{eq:uptoextrafactor}
\end{align}
where 
\be\ba
Z^{{\rm extra}} =  \prod_{k=1}^\infty \prod_{i,j=1}^\infty  \frac{1}{(1-Q^2 Q_\tau^{k-1} \ft^{i} \fq^{j-1})(1-Q^{-2} Q_\tau^{k} \ft^{i} \fq^{j-1})}\ .
\ea\ee
The extra factor $Z^{{\rm extra}}$ typically arises when there are external parallel 5-branes in a given 5-brane configuration. It follows from \eqref{eq:uptoextrafactor} that the physical part of two partition functions $Z^{(a)}$ and $Z^{(b)}$ are the same. Hence, non-parallel identification leads to a physically equivalent result compared with the partition function with parallel identification. We note that such non-parallel identification of external edges in 5-brane webs of two O5-planes are also discussed in \cite{Kim:2021cua} where the partition functions for KK theories of 6d $SO(8)$ and $SU(3)$ theories on a circle with a $\mathbb{Z}_2$ outer automorphism twist were computed.  
\bigskip


\section{Sum over Young diagrams along a periodic strip}\label{app:formulas}
In computing the partition functions for LSTs, we need to sum over Young diagrams along periodic strips. This involves a successive use of the Cauchy identities \eqref{eq:cauchy2}--\eqref{eq:Cauchy-new2} and leads to Jacobi theta functions, as discussed in \cite{Haghighat:2013gba,Kim:2021cua}. 

We encounter products of the Schur functions whose Young diagram indices form a closed loop. As an example, we consider the following which comes from \eqref{eq:rt_half},
\begin{align}
\sum_{\boldsymbol{\nu,\eta}}&\left(Q_{\tau}Q_f^{-1}\sqrt{\frac{\ft}{\fq}}\right)^{|\nu_1|}\left(Q_f\sqrt{\frac{\ft}{\fq}}\right)^{|\nu_2|}\left(\frac{\fq}{\ft}\right)^{\frac{|\eta_1|}{2}}\left(\frac{\fq}{\ft}\right)^{\frac{|\eta_2|}{2}}\nonumber\\
    &\times s_{\nu_1^t/\eta_1}(\ft^{-\rho}\fq^{-\lambda_1})s_{\nu_2^t/\eta_1}(\fq^{-\rho}\ft^{-\lambda_1^t})s_{\nu_2^t/\eta_2}(\ft^{-\rho}\fq^{-\lambda_2})s_{\nu_1^t/\eta_2}(\fq^{-\rho}\ft^{-\lambda_2^t})\ ,\label{eq:SchurApp}
\end{align}
which can be decomposed into three parts: the first part is the product of the factors with the power $|\nu_i|$, the second part is the product of the factors with the power $|\eta_i|$, the third part is the product of Schur functions. Because the $\nu$'s are before the $\eta$'s in the indices of Schur functions, we first sum over $\nu$'s by the extended Cauchy identities \eqref{eq:Cauchy-new1} and \eqref{eq:Cauchy-new2}.  We can see that the factor $Q^{|\lambda|}$ does not change before and after the equal sign, so the first part in the above formula also remains the same after summation over the $\nu$'s. Then we get some $\calR$ factors generated from the extended Cauchy identities and the indices of Schur functions now change to the form in which $\eta$'s are before $\nu$'s. Then we sum over the $\eta$'s by the extended Cauchy identities. By the same reasoning as above, 
the second part of the above formula does not change and the indices of Schur functions become the form in which $\nu$'s are before $\eta$'s again. After repeating another cycle of this process, one finds that \eqref{eq:SchurApp} is expressed as
\begin{align}
    &\prod_{n=1,2}\frac{1}{\calR_{\lambda_2^t\lambda_1}(Q_{\tau}^nQ_f^{-1}\sqrt{\frac{\ft}{\fq}})\calR_{\lambda_1^t\lambda_2}(Q_{\tau}^{n-1}Q_f\sqrt{\frac{\ft}{\fq}})\calR_{\lambda_1^t\lambda_1}(Q_{\tau}^n\sqrt{\frac{\ft}{\fq}})\calR_{\lambda_2^t\lambda_2}(Q_{\tau}^n\sqrt{\frac{\ft}{\fq}})}\nn\\
    &\times\sum_{\boldsymbol{\nu,\eta}}\left(Q_{\tau}Q_f^{-1}\sqrt{\frac{\ft}{\fq}}\right)^{|\nu_1|}\left(Q_f\sqrt{\frac{\ft}{\fq}}\right)^{|\nu_2|}\left(\frac{\fq}{\ft}\right)^{\frac{|\eta_1|}{2}}\left(\frac{\fq}{\ft}\right)^{\frac{|\eta_2|}{2}}\nonumber\\
    &\times s_{\nu_1^t/\eta_1}(Q_{\tau}\ft^{-\rho}\fq^{-\lambda_1})s_{\nu_2^t/\eta_1}(Q_{\tau}\fq^{-\rho}\ft^{-\lambda_1^t})s_{\nu_2^t/\eta_2}(Q_{\tau}\ft^{-\rho}\fq^{-\lambda_2})s_{\nu_1^t/\eta_2}(Q_{\tau}\fq^{-\rho}\ft^{-\lambda_2^t})\ .
\end{align}
Note that the arguments of Schur functions are increased by $Q_{\tau}$ with additional $\calR$ functions in the front. If we call this process one full cycle, then after $m$ full cycles, we have
\begin{align}
    &\prod_{n=1}^{2m}\frac{1}{\calR_{\lambda_2^t\lambda_1}(Q_{\tau}^nQ_f^{-1}\sqrt{\frac{\ft}{\fq}})\calR_{\lambda_1^t\lambda_2}(Q_{\tau}^{n-1}Q_f\sqrt{\frac{\ft}{\fq}})\calR_{\lambda_1^t\lambda_1}(Q_{\tau}^n\sqrt{\frac{\ft}{\fq}})\calR_{\lambda_2^t\lambda_2}(Q_{\tau}^n\sqrt{\frac{\ft}{\fq}})}\nn\\
    &\times\sum_{\boldsymbol{\nu,\eta}}\left(Q_{\tau}Q_f^{-1}\sqrt{\frac{\ft}{\fq}}\right)^{|\nu_1|}\left(Q_f\sqrt{\frac{\ft}{\fq}}\right)^{|\nu_2|}\left(\frac{\fq}{\ft}\right)^{\frac{|\eta_1|}{2}}\left(\frac{\fq}{\ft}\right)^{\frac{|\eta_2|}{2}}\nonumber\\
    &\times s_{\nu_1^t/\eta_1}(Q_{\tau}^m\ft^{-\rho}\fq^{-\lambda_1})s_{\nu_2^t/\eta_1}(Q_{\tau}^m\fq^{-\rho}\ft^{-\lambda_1^t})s_{\nu_2^t/\eta_2}(Q_{\tau}^m\ft^{-\rho}\fq^{-\lambda_2})s_{\nu_1^t/\eta_2}(Q_{\tau}^m\fq^{-\rho}\ft^{-\lambda_2^t})\ .
\end{align}
Because $0<Q_{\tau}<1$, in the limit $m\rightarrow\infty$, only when $\nu_1^t=\nu_2^t=\eta_1=\eta_2$ the contribution is nonzero.
So the above expression becomes
\begin{align}
    &\bigg(\prod_{n=1}^{\infty}\frac{1}{\calR_{\lambda_2^t\lambda_1}(Q_{\tau}^nQ_f^{-1}\sqrt{\frac{\ft}{\fq}})\calR_{\lambda_1^t\lambda_2}(Q_{\tau}^{n-1}Q_f\sqrt{\frac{\ft}{\fq}})\calR_{\lambda_1^t\lambda_1}(Q_{\tau}^n\sqrt{\frac{\ft}{\fq}})\calR_{\lambda_2^t\lambda_2}(Q_{\tau}^n\sqrt{\frac{\ft}{\fq}})}\bigg)\sum_{\nu_1}Q_{\tau}^{|\nu_1|}\nn\\
    =&\bigg(\prod_{n=1}^{\infty}\frac{1}{\calR_{\lambda_2^t\lambda_1}(Q_{\tau}^nQ_f^{-1}\sqrt{\frac{\ft}{\fq}})\calR_{\lambda_1^t\lambda_2}(Q_{\tau}^{n-1}Q_f\sqrt{\frac{\ft}{\fq}})\calR_{\lambda_1^t\lambda_1}(Q_{\tau}^n\sqrt{\frac{\ft}{\fq}})\calR_{\lambda_2^t\lambda_2}(Q_{\tau}^n\sqrt{\frac{\ft}{\fq}})}\bigg)\prod_{l=1}^{\infty}\frac{1}{1-Q_{\tau}^l}\ .
\end{align}
\begin{figure}
    \centering
    \includegraphics[scale=0.7]{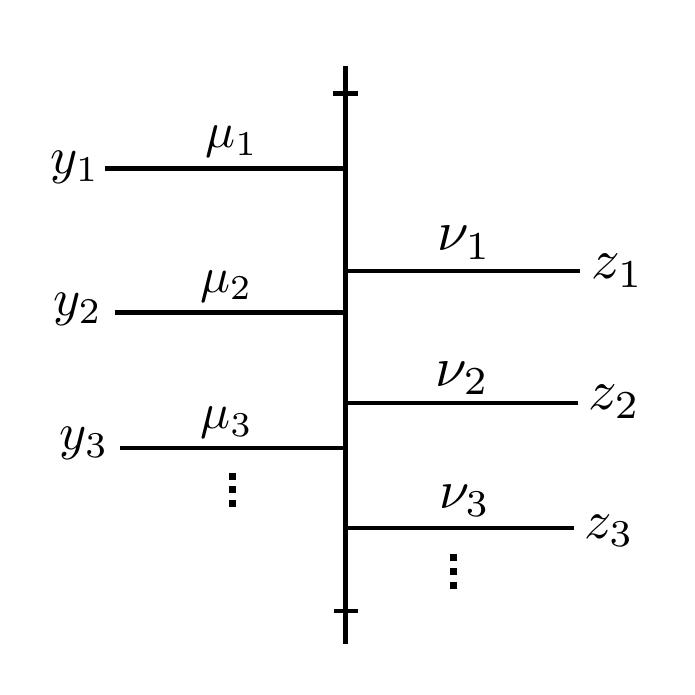}
    \caption{A general periodic strip, $y_i$ and $z_i$ are the exponentiated vertical positions of the corresponding branes.}
    \label{fig:generalstrip}
\end{figure}

More general cases of products of Schur functions whose Young diagram indices form a closed loop can also be computed via this method. Now we write only the result of the topological string partition function that corresponds to a general periodic vertical strip shown in figure \ref{fig:generalstrip}, 
\begin{align}\label{eq:Zofgeneralstrip}
    &Z^{\text{strip}}\nn\\
    =&\prod_{i}\fq^{\frac{||\mu_i||^2}{2}}\tilde{Z}_{\mu_i}(\ft,\fq)\prod_{j}\ft^{\frac{||\nu_j^t||^2}{2}}\tilde{Z}_{\nu_j^t}(\fq,\ft)\bigg[\prod_{n=1}^{\infty}\prod_{i<j}\left(\calR_{\mu_i^t\mu_j}(Q_{\tau}^{n-1}\frac{y_i}{y_j}\sqrt{\tfrac{\ft}{\fq}})\calR_{\mu_j^t\mu_i}(Q_{\tau}^n\frac{y_j}{y_i}\sqrt{\tfrac{\ft}{\fq}})\right)^{-1}\nn\\
    &\times\prod_i\left(\calR_{\mu_i^t\mu_i}(Q_{\tau}^n\sqrt{\tfrac{\ft}{\fq}}\right)^{-1}\prod_{i<j}\left(\calR_{\nu_i^t\nu_j}(Q_{\tau}^{n-1}\frac{z_i}{z_j}\sqrt{\tfrac{\fq}{\ft}})\calR_{\nu_j^t\nu_i}(Q_{\tau}^n\frac{z_j}{z_i}\sqrt{\tfrac{\fq}{\ft}})\right)^{-1}\nn\\
    &\times\prod_{i}\left(\calR_{\nu_i^t\nu_i}(Q_{\tau}^n\sqrt{\tfrac{\fq}{\ft}})\right)^{-1}\prod_{i,j}X_{\mu_i^t\nu_j}(y_i,z_j)X_{\nu_j^t\mu_i}(z_j,y_i)\bigg]\prod_{l=1}^{\infty}\frac{1}{1-Q_{\tau}^l}\ ,
\end{align}
with
\begin{equation}
    X_{\mu^t\nu}(y,z)\equiv\left\{
    \begin{array}{ll}
    \calR_{\mu^t\nu}(Q_{\tau}^{n-1}\frac{y}{z}),&\quad\text{if}\ y<z\\
    \calR_{\mu^t\nu}(Q_{\tau}^n\frac{y}{z}),&\quad\text{if}\ y>z 
    \end{array}
    \right.\ .
\end{equation}

\bigskip

\section{\texorpdfstring{$A$}{}-type little string theory}\label{app:AtypeLittleString}
\begin{figure}[htbp]
    \centering
    \includegraphics[scale=0.65]{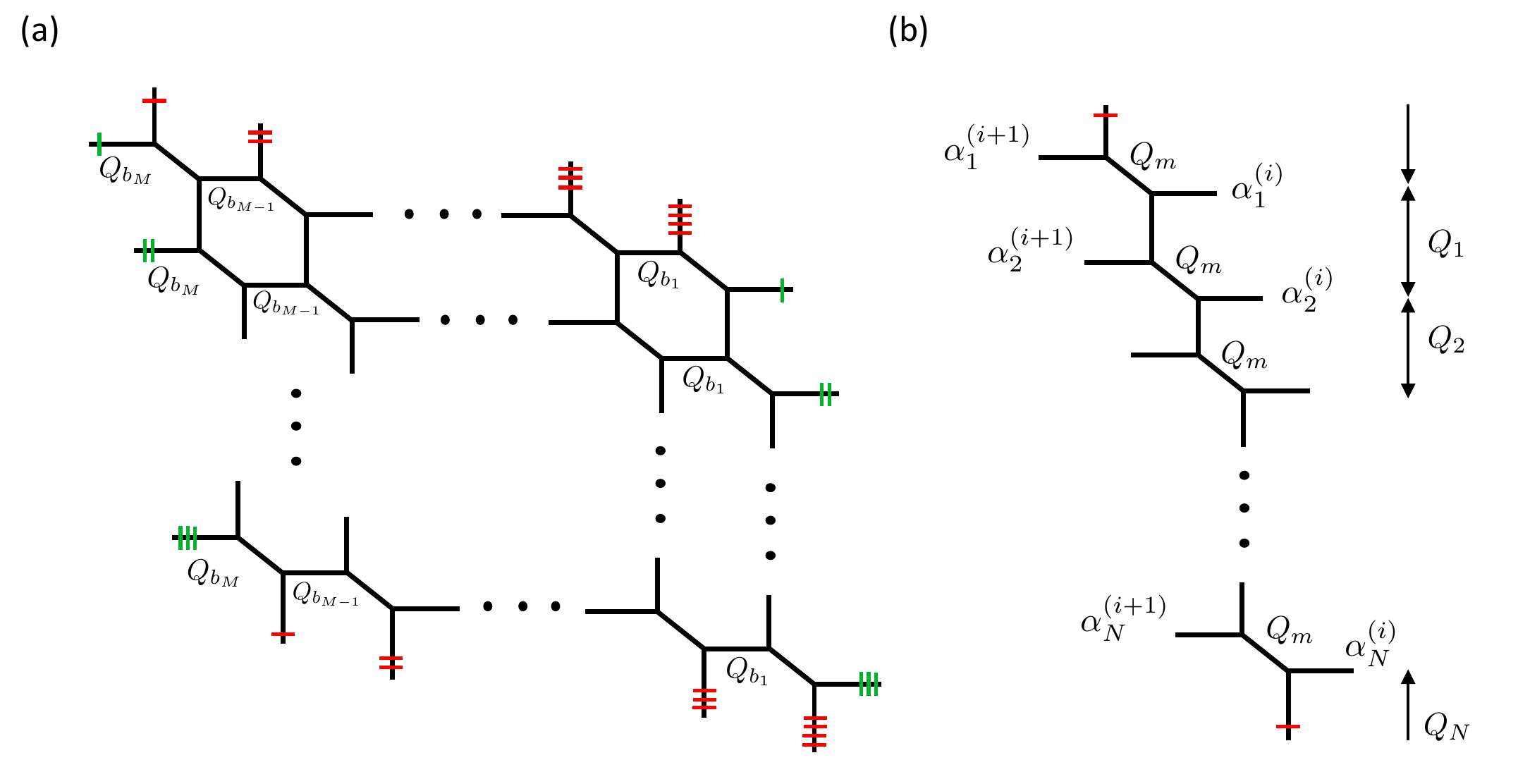}
    \caption{(a) The web diagram describing $A_{M-1}$-type little string theory. (b) Its building block.}
    \label{fig-ANLST}
\end{figure}

The $A$-type little string theory has been considered in the literature, e.g., \cite{Hohenegger:2013ala}.
For rank-$N$ $A_{M-1}$-type LST with the same set of Coulomb branch parameters in each gauge node, the web diagram description is given in figure \ref{fig-ANLST}, and the partition function can be obtained by the topological vertex formalism \cite{Hohenegger:2013ala}. Here we write their result as it is,
\begin{align}\label{eq:Atypepert}
	Z^{(M,N)}_{\text{pert}}=&\bigg(\frac{1}{\widehat{\eta}(Q_U)}\prod_{i,j,k=1}^{\infty}\prod_{r,\ell=1}^N\Big[\frac{1-Q_U^{k-1}Q_{r,r+\ell}Q_m^{-1}\ft^{i-\frac12}\fq^{j-\frac12}}{1-Q_U^{k-1}Q_{r,r+\ell}\ft^{i-1}\fq^j}\nn\\
	&\times\frac{1-Q_U^{k-1}Q_{r,r+\ell-1}Q_m\ft^{i-\frac12}\fq^{j-\frac12}}{1-Q_U^{k-1}Q_{r,r+\ell}\ft^i\fq^{j-1}}\Big]\bigg)^M\ ,
\end{align}
\begin{align}\label{eq:Atypeinst}
	Z^{(M,N)}_{\text{inst}}=&\sum_{\alpha^{(i)}_a}\Big(\prod_{i=1}^M Q_{B_i}^{|\alpha^{(i)}|}\Big)\prod_{i=1}^M\prod_{a=1}^N \frac{\vartheta_{\alpha^{(i+1)}_a \alpha^{(i)}_a}(Q_m)}{\vartheta_{\alpha^{(i)}_a\alpha^{(i)}_a}(\sqrt{\ft/\fq})}\nn\\
	&\times\prod_{1\leq a<b\leq N}\prod_{i=1}^M\frac{\vartheta_{\alpha^{(i)}_a\alpha^{(i+1)}_b}(Q_{ab}Q_m^{-1})\vartheta_{\alpha^{(i+1)}_a\alpha^{(i)}_b}(Q_{ab}Q_m)}{\vartheta_{\alpha^{(i)}_a\alpha^{(i)}_b}(Q_{ab}\sqrt{\ft/\fq})\vartheta_{\alpha^{(i)}_a\alpha^{(i)}_b}(Q_{ab}\sqrt{\fq/\ft})}\ ,
\end{align}
where $\widehat{\eta}(Q_U)=\prod_{k=1}^{\infty}(1-Q_U^k)$ and $Q_U=Q_1Q_2\cdots Q_{N-1}Q_N$, $Q_U$ is the vertical period,
$Q_{B_i}=Q_{b_i}Q_m$ for $i=1,2,\cdots,M$, 
$\alpha^{(a)}$ is a set of $N$ Young diagrams $\{\alpha^{(a)}_1,\cdots,\alpha^{(a)}_N\}$ and $|\alpha^{(a)}|=\sum_{b=1}^N|\alpha^{(a)}_b|$, and $M+1\equiv 1, N+1\equiv 1$ due to the horizontal and vertical periodicities. 
Here,  $Q_{ab}$ is defined as
\be\ba
Q_{ab} = 
\left\{
\begin{array}{ll}
    Q_a Q_{a+1} \cdots Q_{b-1},&\quad\text{for }a<b\ ,  \\
    Q_UQ_{ba}^{-1},&\quad\text{for }a>b\ , \\
    1,&\quad\text{for }a=b\ .
\end{array}
\right.
\ea\ee
The $\vartheta$ function is defined as
\begin{align}
	\resizebox{0.9\hsize}{!}{$\vartheta_{\mu\nu}(x|Q_U)=\prod_{(i,j)\in \mu}\vartheta(x^{-1}\ft^{-\nu^t_j+i-\frac12}\fq^{-\mu_i+j-\frac12}|Q_U)\prod_{(i,j)\in \nu}\vartheta(x^{-1}\ft^{\mu^t_j-i+\frac12}\fq^{\nu_i-j+\frac12}|Q_U)$},
\end{align}
with 
\begin{equation}
	\vartheta(x|Q_U)=\frac{\theta_1(x|Q_U)}{(-i Q_U^{\frac18})\prod_{k=1}^{\infty}(1-Q_U^k)}\ ,
\end{equation}
where $\theta_1$ is the Jacobi theta function defined in \eqref{eq:Jacobitheta1}. 

The $\vartheta$ function can be expressed by the $\varTheta$ function defined in \eqref{eq:Theta-definition},
\begin{equation}
	\vartheta_{\mu\nu}(x|Q_U)=\frac{(-1)^{|\nu|}}{\big(\prod_{k=1}^{\infty}(1-Q_U^k)\big)^{|\mu|+|\nu|}}\varTheta_{\mu\nu}\Big(x\sqrt{\tfrac{\ft}{\fq}}\ \big|Q_U\Big)\varTheta_{\nu\mu}\Big(x^{-1}\sqrt{\tfrac{\ft}{\fq}}\ \big|Q_U\Big),
\end{equation}
which makes it easy to compare \eqref{eq:Atypeinst} with the formula for instanton part of the partition function of $A$-type LST in section \ref{sec:generalformula}. 


\bigskip

\section{Characters}\label{app:characters}
Here, we list the characters of symmetry groups that we used in the main text. 
 The characters of the fundamental weights of $D_4$ are given as follows in an orthonormal basis $\{B_1,B_2,B_3,B_4\}$:
\begin{align}\label{eq:D4fundamentals}
	\chi_{\bf1}
	&=\sum_{i=1}^8X_i^2,\nonumber\\
	\chi_{\bf2}
	&=\sum_{i=1}^8\sum_{j>i}^8X_i^2X_j^2,\nonumber\\
	%
	\chi_{\text{\bf s}}
	&=\frac{1}{2}\left(\prod_{i=1}^4\left(X_i+X_{i+4}\right)+\prod_{i=1}^4\left(X_i-X_{i+4}\right)\right),\nonumber\\
	\chi_{\text{\bf c}}
	&=\frac{1}{2}\left(\prod_{i=1}^4\left(X_i+X_{i+4}\right)-\prod_{i=1}^4\left(X_i-X_{i+4}\right)\right),
\end{align}
in which $X_i\in\big\{B_1^{1/2},\dots,B_4^{1/2},B_1^{-1/2},\dots,B_4^{-1/2}\big\}$.

The characters of the fundamental weights of $D_5$ in orthonormal basis $\{B_1,B_2,B_3$, $B_4,B_5\}$ are given as follows:
\begin{align}\label{eq:D5fundamentals}
	\chi_{\bf1}
	&=\sum_{i=1}^{10}X_i^2,\nonumber\\
	\chi_{\bf2}
	&=\sum_{i=1}^{10}\sum_{j>i}^{10}X_i^2X_j^2,\nonumber\\
	%
	\chi_{\bf 3}&=\sum_{i=1}^{10}\sum_{j>i}^{10}\sum_{k>j}^{10}X_i^2X_j^2X_k^2,\nn\\
	\chi_{\text{\bf s}}
	&=\frac{1}{2}\left(\prod_{i=1}^5\left(X_i+X_{i+5}\right)+\prod_{i=1}^5\left(X_i-X_{i+5}\right)\right),\nonumber\\
	\chi_{\text{\bf c}}
	&=\frac{1}{2}\left(\prod_{i=1}^5\left(X_i+X_{i+5}\right)-\prod_{i=1}^5\left(X_i-X_{i+5}\right)\right),
\end{align}
in which $X_i\in\big\{B_1^{1/2},\dots,B_5^{1/2},B_1^{-1/2},\dots,B_5^{-1/2}\big\}$.

The characters of the fundamental weights of $E_6$ 
can be obtained from the embedding $E_6\supset\text{SO}(10)\times \text{U}(1)$ in \eqref{eq:E6embed}. The Dynkin diagram and the assignment of the fundamental weights are given as
\begin{center}
\begin{tikzpicture}[scale=0.5]
\draw (-1,0) node[anchor=east] {$E_6$ ~~~~~};
\draw (0 cm,0) -- (8 cm,0);
\draw (4 cm, 0 cm) -- +(0,2 cm);
\draw[fill=white] (0 cm, 0 cm) circle (.25cm) node[below=4pt]{$1$};
\draw[fill=white] (2 cm, 0 cm) circle (.25cm) node[below=4pt]{$2$};
\draw[fill=white] (4 cm, 0 cm) circle (.25cm) node[below=4pt]{$3$};
\draw[fill=white] (6 cm, 0 cm) circle (.25cm) node[below=4pt]{$4$};
\draw[fill=white] (8 cm, 0 cm) circle (.25cm) node[below=4pt]{$5$};
\draw[fill=white] (4 cm, 2 cm) circle (.25cm) node[right=3pt]{$6$};
\node at (9.85,-1){.};
\end{tikzpicture}    
\end{center}
The characters  $\chi_i^{E_6}$ of the fundamental weights associated with the node $i$  of the Dynkin diagram are given as follows, 
\begin{align}\label{eq:E6fundamentals}
    &\chi_{\bf 1}^{E_6}=B_6^{-\frac23}+\chi_{\bf s}^{D_5}B_6^{-\frac16}+\chi_{\bf 1}^{D_5}B_6^{\frac13},\nn\\
    &\chi_{\bf 2}^{E_6}=\chi_{\bf s}^{D_5}B_6^{-\frac56}+(\chi_{\bf 1}^{D_5}+\chi_{\bf 3}^{D_5})B_6^{-\frac13}+\chi_{\bf 1}^{D_5}\chi_{\bf s}^{D_5}B_6^{\frac16}+\chi_{\bf 2}^{D_5}B_6^{\frac23},\nn\\
    &\chi_{\bf 3}^{E_6}=\chi_{\bf 3}^{D_5}B_6^{-1}+\chi_{\bf 2}^{D_5}\chi_{\bf c}^{D_5}B_6^{-\frac12}+\chi_{\bf 2}^{D_5}+\chi_{\bf 1}^{D_5}\chi_{\bf 3}^{D_5}+\chi_{\bf 2}^{D_5}\chi_{\bf s}^{D_5}B_6^{\frac12}+\chi_{\bf 3}^{D_5}B_6,\nn\\
    &\chi_{\bf 4}^{E_6}=\chi_{\bf 2}^{D_5}B_6^{-\frac23}+\chi_{\bf 1}^{D_5}\chi_{\bf c}^{D_5}B_6^{-\frac16}+(\chi_{\bf 1}^{D_5}+\chi_{\bf 3}^{D_5})B_6^{\frac13}+\chi_{\bf c}^{D_5}B_6^{\frac56},\nn\\
    &\chi_{\bf 5}^{E_6}=\chi_{\bf 1}^{D_5}B_6^{-\frac13}+\chi_{\bf c}^{D_5}B_6^{\frac16}+B_6^{\frac23},\nn\\
    &\chi_{\bf{6}}^{E_6}=\chi_{\bf c}^{D_5}B_6^{-\frac12}+1+\chi_{\bf 2}^{D_5}+\chi_{\bf s}^{D_5}B_6^{\frac12},
\end{align}
where the $\chi^{D_5}$'s are the characters of the fundamental weights of $D_5$ given in  \eqref{eq:D5fundamentals}.
In a similar fashion, one can obtain the characters for the fundamental weights of $E_7$ and $E_8$ groups.

\bibliographystyle{JHEP}
\bibliography{ref}
\end{document}